\documentclass[%
reprint,
 amsmath,amssymb,
 aps,
rmp,
floatfix, longbibliography
]{revtex4-2}

\usepackage{graphicx}% Include figure files
\usepackage{dcolumn}% Align table columns on decimal point
\usepackage{bm}% bold math
\usepackage{upgreek}
\usepackage[dvipsnames]{xcolor}      % use if color is used in text
\usepackage{dcolumn}% Align table columns on decimal point
\usepackage{bm}% bold math
\usepackage{amsmath}
\usepackage{multirow}
\usepackage{graphicx}
\usepackage{float}
\usepackage[colorlinks=true,citecolor=blue,bookmarks=false]{hyperref}
\usepackage{longtable}
\usepackage{soul}

\newcommand{\mean}[1]{\mbox{$\langle{#1}\rangle$}}

\newcommand{\astra}{{\sc Astra }}

\usepackage{url}
\usepackage{CJKutf8}
\usepackage[english]{babel}
\usepackage{siunitx}

\begin{document}

\preprint{APS/123-QED}

\title{Bunch Shaping in Electron Linear Accelerators}% Force line breaks with \\

\begin{CJK*}{UTF8}{gbsn}
\author{G. Ha}%
%\email{gha@anl.gov}
\affiliation{%
Argonne National Laboratory, Argonne, Illinois 60439, USA
}%
\author{K.-J. Kim}%
%\email{kwangje@anl.gov}
\affiliation{%
Argonne National Laboratory, Argonne, Illinois 60439, USA
}%
\author{P. Piot}%
%\email{ppiot@niu.edu}
\affiliation{%
Northern Illinois University, Department of Physics and  
Northern Illinois Center for Accelerator \& Detector Development, DeKalb Illinois 60115, USA
}%
\affiliation{%
Argonne National Laboratory, Argonne, Illinois 60439, USA
}%
\author{J.G. Power}%
\email{jp@anl.gov}
\affiliation{%
Argonne National Laboratory, Argonne, Illinois 60439, USA
}%
\author{Y. Sun (孙银娥)}%
%\email{yinesun@anl.gov}
\affiliation{%
Argonne National Laboratory, Argonne, Illinois 60439, USA
}%

\date{\today}% It is always \today, today,

\begin{abstract}
Modern electron linear accelerators are often designed to produce smooth bunch distributions characterized by their macroscopic ensemble-average moments.  However, an increasing number of accelerator applications call for finer control over the beam distribution, e.g., by requiring specific shapes for its projection along one coordinate. Ultimately, the control of the beam distribution at the single-particle level could enable new opportunities in accelerator science.   This review discusses the recent progress toward controlling electron beam distributions on the ``mesoscopic" scale with an emphasis on shaping the beam or introducing complex correlations required for some applications.  This review emphasizes experimental and theoretical developments of electron-bunch shaping methods based on bounded external electromagnetic fields or via interactions with the  self-generated velocity and radiation fields.

\end{abstract}
\maketitle
\end{CJK*}
 
{ 
\hypersetup{linkcolor=blue}
\tableofcontents
}
\setcounter{secnumdepth}{5}
\setcounter{tocdepth}{5}

%%%%%%%%%%%%%%%%%%%%%%%%%%%%%%%%%%%%%%%%%%%%%%%%%%%%%%%%%%%%%%%%%%%%%%%%%
%
%\onecolumngrid
\section{Introduction~\label{sec1}}

%The ability to control the electron beam distribution produced by an linear accelerator (linac) evolved during the 20th century. 
The first generation of electron linacs (e-linacs) that begun with Wideroe's invention~\cite{wideroe-1928-a} produced continuous streams of electrons by placing a cathode in an electrostatic gap and therefore had no control over the longitudinal distribution and only modest control over the transverse distribution, based on the size of the hole in the anode plate. Control over the longitudinal distribution began when RF power generators, developed for radar applications, became available after World War II. At that time, Luis Alvarez ~\cite{alvarez-1946-a} proposed an accelerator based on a linear array of drift tubes enclosed in resonant cavities, and the second generation of e-linacs (operating with DC electron guns) was born~\cite{ginzton-1948-a}. This generation of e-linacs culminated in the construction of the 100-GeV electron-positron SLAC linear collider~\cite{neal-1968-a}. Electron bunches in these linacs can be approximated by a Gaussian distribution in phase space which is characterized by its second order moments. The next significant progression in the control over the bunch distribution in e-linacs took place in the early 1990s with the widespread adaptation of the RF photocathode gun~\cite{fraser-1985-a} and development of the magnetic chicane compressor~\cite{carlsten-1996-a}. This progress facilitated, for example, the development of X-ray free-electron lasers (FELs); for a review see~\cite{kim-2017-a}.  

\begin{figure} 
\includegraphics[width=0.48\textwidth]{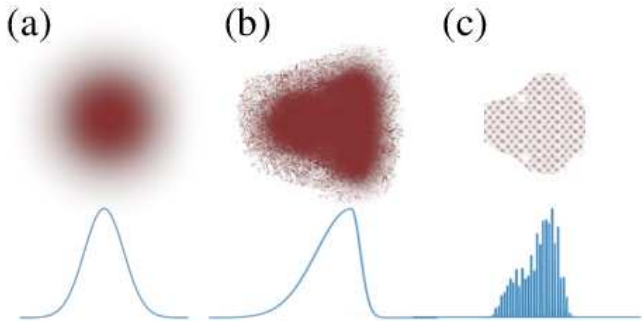}
\caption{Definition of the level of control on a beam distribution: macroscopic  (a), mesoscopic (b), and microscopic (c) scales. The red-shaded distributions depict the projected beam distribution in a plane defined by any pair of coordinates associated with the beam.  The blue traces describe the projection of the distribution along one arbitrary axis. \label{fig:sec1:mesoscopic}}
\end{figure}

Despite the progress that was made during the 20th century, ever more demanding accelerator applications continued to appear. The e-linac community responded to these challenges and is now on the verge of taking the next step in the evolution of control over the bunch distribution. There are two aspects in this effort: Towards a control at a level finer than the macroscopic scale but coarser than the microscopic scale, which will be referred to as {\textit{mesoscopic} level, and towards multi-dimensional beam shaping for distributions that can no longer be characterized by the second order moments; see Fig.~\ref{fig:sec1:mesoscopic}(b). The next step in beam shaping will in general involve both of these aspects. 

The ultimate challenge for beam control~\cite{nagaitsev-2021-a} is to produce interesting and useful distribution at the finest level -- the {\em microscopic scale}  where the distribution is described by a ``granular" Klimontovich-distribution function~\cite{klimontovich-1995-a}. Such an ultimate degree of control would open the path toward producing structured beams, for example,  Wigner-crystal beams~\cite{wigner-1934-a} with arbitrary shapes; see Fig.~\ref{fig:sec1:mesoscopic}(c).

The development of multi-dimensional shaping began in the late 1990s and early 2000s. The flat-beam generation~\cite{derbenev-1998-a, brinkmann-2001-a} and emittance exchange between transverse and longitudinal phase spaces~\cite{cornacchia-2002-a, kim-2006-a} were introduced in this period. Experimental demonstration followed; the flat beam generation by~\cite{edwards-2000-a, piot-2006-a} and emittance exchange by~\cite{ruan-2011-a}. 
The emittance exchange opened up the possibility of shaping a beam in transverse dimension and  transferring it to the temporal dimension, and vice versa. The technique was used to produce a train of sub pico-second bunchlets  \cite{sun-2010-b} and bunches with linearly-ramped current} profile~\cite{ha-2017-b}

%In this review, we focus on beam manipulations capable of tailoring the distribution at a level finer than its macroscopic scale but coarser than the microscopic scale. Throughout this review, we refer to this intermediary scale as the {\em mesoscopic scale}. Specifically, we focus on methods capable of imprinting controlled variation over the beam distribution at a scale smaller than the second-order moment associated with the considered coordinate. We discussed techniques to shape the projected distribution over a specific coordinate but also capable of imprinting correlations within a given degree of freedom and beyond. The later class of control also enables a precise macroscopic control (e.g. partitioning of emittances, introducing complex correlation, or providing a finer control to reach specific ensemble-averaged parameters). 
%%

%Mesoscopic control over the 
Advanced beam phase-space shaping is needed to enable many accelerator applications, e.g., improving the efficiency of beam-driven advanced acceleration techniques~\cite{bane-1985-a}, improved X-ray free-electron laser interaction~\cite{emma-2006-a}, or the development of compact accelerator-based radiation sources~\cite{gover-2019-a}. Tailored electron beams are also used as a tool to manipulate hadron beams by exerting nonlinear focusing~\cite{shiltsev-2016-a} and cooling~\cite{blaskiewicz-2014-a}. Finally beam distributions assuming known continuous function can also be used to mitigate beam degradation arising from collective effects, such as space-charge force~\cite{kapchinskij-1959-a,kellogg-1967-a} and self-interaction via radiative effects ~\cite{seeman-1992-a,derbenev-1995-a}

The present review is mainly devoted to phase-space shaping techniques employing bounded external electromagnetic fields or via interactions with self-generated velocity and radiation fields.  Techniques that couple lasers with electron beams  are not included here since  they been reviewed in ~\cite{hemsing-2014-a}. 

We categorize these techniques into three broad classes. The first category includes techniques that control the electron distribution from the electron gun. This includes shaping the distribution of the emission-triggering laser in photoemission electron sources or engineering the cathode properties or surface to control the emitted electron distribution. The second class of manipulation consists of shaping systems that operate within one degree of freedom, e.g., those based on the use of external and internal fields to control the distribution in one of the three phase-space planes. Finally, the third category involves shaping techniques that also use the external and internal fields to introduce correlations between two degrees of freedom. An example of such a manipulation is the transverse-to-longitudinal phase-space exchange beamline. It maps a tailored horizontal profile (e.g., obtained with a shaped collimator) to the longitudinal plane, thereby enabling the formation of temporally shaped electron bunches.\\

This article is organized as follows. Section~\ref{sec2} introduces the fundamental concepts necessary to the understanding of phase-space manipulations, and the subsequent Sections~\ref{sec3}, \ref{sec4}, \ref{sec5}, and \ref{sec6} discuss the various classes of manipulation mentioned above. Finally, Section~\ref{sec7} offers some perspective on likely research directions motivated by recent developments.

\section{General Principles}\label{sec2}

Particle motion is described by Hamiltonian mechanics. As a corollary, the beam distribution in phase space is constrained by Liouville's theorem, which states that the phase-space density is invariant along a physical trajectory. This is true when the force is due to an external electromagnetic field.  This is also true when the force is due to the beam-generated electromagnetic field, if the  discreetness of the particles can be neglected so that the phase-space distribution can be approximated as a continuous function.

Section \ref{sec2a} is devoted to basic constraints of the Hamiltonian system; Section \ref{sec2b} contains examples of  external EM fields and  particle motion under their influence; Section \ref{sec2c} discusses how beam distributions change when constrained by Liouville's theorem; and Section \ref{sec2d} contains a discussion on how the collective forces of beams are computed and how these forces affect the beam distribution.

\subsection{Hamiltonian formalism}\label{sec2a}
\subsubsection{Equation of motion under EM fields}\label{Ham form}

We consider the motion of a charged particle under electromagnetic field $\mathbf E$ and $\mathbf B$ satisfying Maxwell equations:
\begin{eqnarray}\label{Maxwell}
\nabla \cdot \mathbf E &=&\frac{1}{\varepsilon _{0} } \rho_S , \nonumber\\ \nabla \cdot \mathbf B &=& 0 , \nonumber\\ \nabla \times \mathbf E &=& -\frac{\partial }{\partial t} \mathbf B , \nonumber\\ \nabla \times \mathbf B&=& \mu _{0} \mathbf {J}_S+\frac{1}{c^{2} } \frac{\partial }{\partial t} \mathbf E
\end{eqnarray}
Here $\rho_S$ and $\mathbf J_S$ are the charge density and current density of the sources, respectively, consisting of the external sources and the beam itself. External sources are not present within the beam pipes. Thus $\rho_S=0$ and $\mathbf J_S=0$ until we consider the beam generated fields in \ref{sec2d}.  We will use the MKS units throughout this paper. Introducing the vector and the scalar potential, $\mathbf A$ and $\phi$ , respectively, the electromagnetic fields can be written as follows:
\begin{equation}\label{EM to pot}
\mathbf B=\mathbf{\nabla} \times \mathbf A \mbox{~and~}\mathbf E=-\frac{\partial}{\partial t}\mathbf A -\mathbf \nabla \phi.
\end{equation}
The Hamiltonian $H$ for a particle of mass $m$ and charge $e$  is 
 \begin{equation}\label{Hamiltonian}
 H(\mathbf x,\mathbf p,t)=\sqrt{m^2 c^4+c^2(\mathbf p-e\mathbf A(\mathbf x,t))^2}+e\phi(\mathbf x,t).
 \end{equation}
 Here $c$ is the velocity of light, $\mathbf x $ the coordinate vector, and $\mathbf p$ is the canonical momentum conjugate to $\mathbf x $. 

The Hamiltonian equations of motion are
\begin{eqnarray}\label{H eq m}
\frac{d \mathbf x}{dt} &=&\frac{\partial H}{\partial \mathbf p}, \nonumber \\ \frac{d \mathbf p}{dt} &=&  -\frac{\partial H}{\partial \mathbf x}. 
\end{eqnarray}
 The Hamilton's equations reproduce the Lorentz force equation:
\begin{equation}\label{Lorentz}
\frac{d {\mathbf p}_{kin}}{dt}=e\left(\mathbf E +\mathbf{v}\times \mathbf B \right).
\end{equation}
Here, $\mathbf {p}_{kin}$ is the kinetic momentum related to the canonical momentum $\mathbf{p}$ as follows:
\begin{equation}\label{ kin and can}
\mathbf{p}_{kin}\equiv m\gamma\frac{d \mathbf x}{dt} = \mathbf{p}-e\mathbf{A},
\end{equation}
where $\gamma$ = relativistic kinetic energy$/m c^2 =1/\sqrt{1-\beta^2}$, $\beta=v/c$, $v=|d\mathbf{x} /dt|$.
%If the magnetic field is transverse to the longitudinal direction $z$, that is, $\mathbf B = \left(B_x, %B_y,0 \right)$, then if follows from Eq. (\ref{EM to pot}) that $\mathbf A =\left(0,0, A_z)$. The 6D phase %space spanned by the canonically conjugate variables $\mathbf x , \mathbf p)$ is endowed with special %properties originating from the structure of Hamiltonian form of Equation \ref{H eq m}.

%From the action principle, it can be seen readily that the transformation $H \rightarrow -p_z, %t\rightarrow -z, (x,y,z)\rightarrow (x,y,-z), (p_x,p_y,p_z)\rightarrow (p_x,p_y,p_z) $ is a canonical %transformation.
\subsubsection{Curvilinear coordinates}\label{sec2a2 }
\begin{figure}
 \includegraphics[width=.85\linewidth]{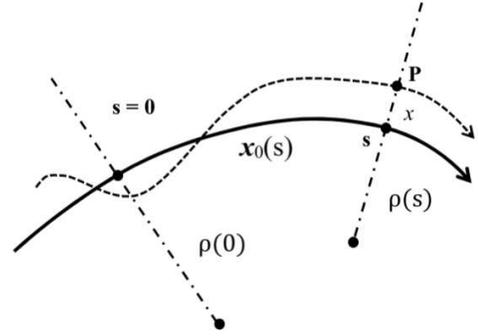}
\caption{\label{fig:sec2:coordinateSystem} Curvilinear coordinate system. The reference trajectory $\mathbf{x}_0(s)$ is indicated by the solid curve, assumed to be on the horizontal plane, which is the plane of the page. The position on the reference trajectory is specified by the arc distance $s$ from an initial point. The transverse position $\mathbf{P}$ of an arbitrary particle is specified by a 2D Cartesian coordinate system by the curvature line (dot-dash) as the x-axis and the vertical line as the $y$-axis. Since the trajectory of an arbitrary particle will in general not be on the horizontal plane, the dotted curve represents its projection onto the horizontal plane.}
\end{figure}

In a beam, particles are bunched in a small region of space and stay together while moving. Therefore, it makes sense to introduce a ``reference" particle as the one that is at the beam ``center". Its trajectory is referred to as the  reference orbit, which may be curved but will be assumed to lie on a plane referred to as the horizontal plane. The reference orbit is parametrized as $\mathbf {x}_0 (s)$ where $s$ is the arc length along the reference orbit. Then the position of any particle in the beam can be represented as:
\begin{equation}\label{curv coord}
\mathbf x =\mathbf{x}_0 (s) + x \mathbf{e}_{\rho}+y \mathbf{e}_{y},
\end{equation}
where $\mathbf{e}_{\rho}$ is the unit vector normal to the reference orbit at $s$ on the horizontal plane and $\mathbf{e}_{y}$ is the unit vector normal in the vertical direction. The set $(x,y,s)$ constitutes the curvilinear coordinate system shown in Fig.~\ref{fig:sec2:coordinateSystem}.  The figure is drawn on  the horizontal plane on which the reference trajectory $\mathbf{x}_0 (s) $ (solid line) lies. The unit vector $\mathbf {e}_\rho$ is in the horizontal plane and perpendicular to the reference trajectory and   the unit vector $\mathbf{e}_{y}$ is directed towards the reader. The line perpendicular to the reference trajectory at $s$  represents the {\it transverse plane}     extending in the vertical direction.   The dotted line  represents the projection in the horizontal plane of an arbitrary particle, intersecting the transverse plane at $s$ at $(x,y)$. 

%~\cite{}
We now make two canonical transformations~\cite{landau-1969-a, goldstein-2002-a} to change the variables to ones convenient for studying beams in accelerators:  First, we adopt $s$ as the independent variable and use $(x,y,-t)$ as the coordinates ~\cite{courant-1958-a}, where $t$ is the time the particle arrives at the transverse plane at $s$. The new canonically conjugate variables are:
 \begin{equation}\label{new can vab}
 \mathbf {x}_{new}= (x, y, -t) ,\; \mathbf{p}_{new} =\left( p_x, p_y, U \right),
 \end{equation}
 where $U$ is the energy of the particle
 %\footnote{In most references, including ~\cite{courant-1958-a}, the longitudinal conjugate pair is chosen to be $ (t ,-U )$. However, our choice leads to the sign convention for the $z$ variable below.}. 
The new Hamiltonian is ;  
\begin{multline}\label{new H}
{\mathcal H}\left( \mathbf {x}_{new} , \mathbf {p}_{new} ; s\right) =-eA_s - \left( 1+\frac{x}{\rho(s)} \right) \times \\ \sqrt{\left(\frac{U-e\phi}{c}\right)^2 - m^2 c^2 - \left( p_x-e A_x \right)^2-\left( p_y-e A_y \right)^2}.
\end{multline}
Here $\rho(s)$ is the radius of curvature,  $A_s=\left( 1 + x/\rho(s) \right)\mathbf{A}\cdot \mathbf{e}_s,$ where $\mathbf{e}_s$ is the unit vector in the tangential direction. Since the transformation is canonical, the equation of motion is the same as in Eq.(\ref{H eq m}) with the replacement $t \rightarrow s$, $\mathbf x \rightarrow \mathbf{x}_{new}$, $\mathbf p \rightarrow \mathbf{p}_{new}$, and $H \rightarrow \mathcal H$.

The second transformation is to  \textit{deviation variables}  relative to the trajectory of the reference particle $\left(x_0, y_0, -t_0;p_{0x},p_{0y},U_0\right)=\left(0,0, -t_0;0,0,U_0\right)$. The deviation variables are therefore $(x,y,-t+t_0;p_x,p_y,U-U_0)$. The change to  deviation variables is also a canonical transformation, in which the Hamiltonian $\mathcal{H}_D$ is the same as Eq.~\eqref{new H} except the zeroth and the linear terms in the power series expansion of $\mathcal H$ are excluded~\cite{duffy-2016-a}.

When the electric field is absent and the magnetic fields are static and perpendicular to $\mathbf{e}_s$, we can choose $A_x = A_y= \phi=0$.  Also, the energy $U$ and the curvature $\rho_0$ are constant. In this case, $\mathcal H_{D}/p_0$, where $p_0$ is the momentum of the reference particle, can be chosen as a scaled Hamiltonian with canonically conjugate variables $\left(x,y,-v_0 (t-t_0); p_x/p_0,p_y/p_0,(U-U_0)/(v_0 p_0) \right)$~\cite{ruth-1985-a, conte-2008-a}. Note we introduced an additional scale factor $v_0$, the velocity of the reference particle, for the longitudinal variables. For the highly relativistic case, $v \approx c $ if we ignore the correction factor of $1/2\gamma^2$ , then we can approximate
\begin{equation}\label{rela approx}
\frac{p_x}{p_0} \approx \frac{dx}{ds}= x',\; \frac{p_y}{p_0} \approx\frac{dy}{ds}= y',\;\frac{U-U_0}{v_0 p_0} \approx \frac{p_s-p_0}{p_0}=\delta .
\end{equation}
We will introduce the following notation for the longitudinal deviation variable:
\begin{equation}\label{long devi}
-c (t(s)-t_0(s))= z (s)
\end{equation}
The quantity $z(s)$ is  the distance (in general the arc distance) \textit {ahead} of the reference particle along $s$, thus arriving there at an \textit {earlier} time. Thus, the canonical variables in the present case are
\begin{equation}\label{can var noacc}
\left(x,y,z;p_x,p_y,\delta\right).
\end{equation}
The corresponding scaled Hamiltonian is
\begin{multline}\label{scaled H}
\mathcal{H}_S (x,y,z;x', y',\delta;s)\\ \approx\ \left[ -e\frac{A_s}{p_0}-\left(1+\frac{x}{\rho}\right)\sqrt{(1+\delta)^2-x'^2-y'^2}\right]\\ \approx 
-e\frac{\left[A_s \right]}{p_0}-\frac{x}{\rho}\delta++\frac{1}{2}\left(x'^2+y'^2\right)+... .
\end{multline}
In the above $\left[..\right]$ implies removing the zeroth and the first order terms in the power series expansion in the scaled variables of the expression inside the square brackets.   

When acceleration is present, then we need to go back to the original Hamiltonian equation of motion Eq.(\ref{H eq m} ) or the Lorentz force equation Eq.(\ref{Lorentz}), as will be done in Section \ref{acc cavity}. The presence of longitudinal magnetic field can be treated by working in a rotating frame, as discussed Section \ref{sec2:solenoid}.

\subsubsection{Symplecticity}\label{symplecticity } 
Our goal in this section is to illustrate the special property enjoyed by a mechanical system that can be described by the  Hamiltonian equation of motion. We will mostly consider the cases in which the variables Eq.\eqref{can var noacc} and the Hamiltonian Eq.\eqref{scaled H} are applicable. This is not valid when acceleration is present as in Section \ref{acc cavity}, where we will revert to the variables $(\mathbf x, \mathbf p)$. 

We rearrange the 6D canonical variables into a column vector $\pmb{\mathcal Z}$ as follows:
\begin{equation} \label{variables} 
\pmb{\mathcal Z}=\left(\begin{array}{c} {x} \\ {x'} \\ {y} \\ {y'} \\ {z} \\ {\delta } \end{array}\right)=\left(\begin{array}{c} {\zeta _{1} } \\ {\zeta _{2} } \\ {\zeta _{3} } \\ {\zeta _{4} } \\ {\zeta _{5} } \\ {\zeta _{6} } \end{array}\right) . 
\end{equation} 
It is often useful to introduce the subspace as follows:
\begin{equation} \label{XYZ space} 
\pmb{\mathcal Z}=\left(\begin{array}{c} {\mathbf X} \\ {\mathbf Y} \\ {\mathbf Z} \end{array}\right); \mathbf X=\left(\begin{array}{c} {x} \\ {x'} \end{array}\right),\mathbf Y=\left(\begin{array}{c} {y} \\ {y'} \end{array}\right), \mathbf Z=\left(\begin{array}{c} {z} \\ {\delta } \end{array}\right) . 
\end{equation} 
Note these are the canonical deviation variables in the curvilinear coordinates introduced in Section \ref{sec2a2 }, \textit{not} the usual Cartesian variables of the laboratory frame. 

Introducing the gradient vector  $\nabla$ in 6D phase space
\begin{equation}\label{6D grad}
\nabla_j=\frac{\partial}{\partial \zeta_j},
\end{equation}
the equation of motion with the scaled Hamiltonian can be written as
\begin{equation}\label{scaled H eq}
\frac{d}{ds}\pmb{\mathcal Z} =J \nabla \mathcal{H_S}(\pmb{\mathcal Z};s).
\end{equation}
Here, we have introduced the unit symplectic matrix:
\begin{equation} \label{unit symp} 
J=\left(\begin{array}{ccc} {J_{2D} } & {0} & {0} \\ {0} & {J_{2D} } & {0} \\ {0} & {0} & {J_{2D} } \end{array}\right),\; J_{2D} =\left(\begin{array}{cc} {0} & {1} \\ {-1} & {0} \end{array}\right). 
\end{equation} 

By solving Eq.(\ref{scaled H eq}), the map \textit{M }  $\pmb{\mathcal Z}\to \bar{\pmb{\mathcal Z}}$ corresponding to a section of the accelerator from $s$ to $\bar{s}$ can be found  :
\begin{equation} \label{H map} 
\bar{\pmb{\mathcal Z}}=M\left(\pmb{\mathcal Z}\right);\quad \; \bar{\zeta }_{i} =M_{i} \left(\pmb{\mathcal Z}\right). 
\end{equation} 
The inverse map is
\begin{equation} \label{5)} 
{\pmb{\mathcal Z}}=M^{-1} \left(\bar{\pmb{\mathcal Z}}\right). 
\end{equation} 

We introduce the Jacobian matrix \textit{R} whose components are:
\begin{equation} \label{Jacobian} 
R_{ij} \left(\pmb{\mathcal Z}\right)=\frac{\partial \bar{\zeta }_{i} }{\partial \zeta _{j} }.  
\end{equation} 
For Hamiltonian dynamics, the Jacobian matrix is symplectic:
\begin{equation} \label{ZEqnNum905720} 
R^{T} JR=RJR^{T} =J. 
\end{equation} 

It follows from Eq.~\eqref{ZEqnNum905720} and the continuity  of \textit{R }as $\bar{s} \to s $ that its determinant is unity:
\begin{equation} \label{unitjacob} 
\det(R)=1. 
\end{equation} 
All 2$\times $2 matrices with unit determinant are symplectic. For higher dimensions, Eq.~\eqref{ZEqnNum905720} imposes significant restrictions on the matrix. 

In the following we will mostly consider the case where the transformation in Eq.~\eqref{H map} is linear:

\begin{equation} \label{ZEqnNum451641} 
\bar{\pmb{\mathcal Z}}=R{\pmb{\mathcal Z}}. 
\end{equation} 
Here \textit{R }is the Jacobian matrix given by Eq.~\eqref{Jacobian} whose elements are independent of ${\pmb{\mathcal Z}}$.

\subsection{Single particle motion in external field}\label{sec2b}

In this sub-section, we present some important examples of the transformation matrix $R$ relevant for beam shaping.

%%%%%%%%%%%%%%%%%%%%%%
\subsubsection{Free space, bending magnets, and quadrupole magnets}\label{sec2a1}

With no electric fields, and the components of the static magnetic fields corresponding to a bending magnet and quadrupole are given by
\begin{equation}\label{stat mag}
B_x=B_1(s)y,\; B_y=-B_0 (s)+B_1(s)x .
\end{equation}
Here $B_0=p_0/\left(e\rho \right)$ is the strength of the dipole magnet bending the particle horizontally. In  computing the vector potential $A_s$ one finds the scaled Hamiltonian Eq.(\ref{scaled H}) up to the quadratic terms:
\begin{equation}\label{quad H}
{\mathcal H}_S=-\delta \frac{x}{\rho}+\frac{x^2}{2 \rho^2}+ \frac{K_q}{2}\left( -x^2 +  y^2 \right)+\frac{1}{2}\left(x'^2+y'^2\right)
\end{equation}
Here $ K_q=eB_1/\left(p_0c\right)$ is the quadrupole strength.  When $K_q=1/\rho=0$, this will be free space. The equation of motion is obtained from the Hamiltonian equation Eq.(\ref{scaled H eq}). The equations in the transverse directions, after reducing first order differential equations to second order ones,  become:
\begin{equation}\label{eq x} 
\frac{d^{2} x}{ds^{2}} +\left(-K_q (s)+\frac{1}{\rho (s)^{2} } \right)x = \frac{1}{\rho \left(s\right)} \delta,
\end{equation}
\begin{equation}\label{eq y} 
\frac{d^{2} y}{ds^{2}} + K_q (s) x = 0.
\end{equation}
In deriving these equations, we are using the symbol $x'$ both as the canonical momentum and  as the slope $dx/ds$.  We indicated that $\rho$ and $K_q$ are functions of $s$. However, we assume the functions are piece-wise constant and neglect the transition effects.  Equation(\ref{eq x}) shows that the motion in the $x$ direction is influenced both the quadrupole force as well as the centripetal force due to the curvature. For historical reason, the motion described by Eqs.\eqref{eq x} and \eqref{eq y} is known as the betatron motion. Although the sign of the quadrupole focusing strength $K_q$ in the $x$ direction is opposite to that in the y-direction, focusing in both direction can be achieved either by the centripetal focusing (weak focusing) or by arranging the adjacent quadrupoles to have opposite sign (strong focusing). See text books for details, for example~\cite{wiedemann-1999-a}.

%
% \begin{figure}
% \includegraphics[width=.55\linewidth]{./Sec2-KJK/fig2-twoCircles.eps}
%\caption{\label{fig:sec2:twoCircles} Focusing in a bending magnet. The solid (dotted) %circle represents the trajectory of the reference (displaced) particle in a bending %magnet. If the solid circle is straightened out, the displaced trajectory will look %similar to a sinusoidal curve with a period of $2\pi \rho$, explaining the focusing term %$1/\rho^2$ in Eq.~\eqref{eq x} }
%\end{figure}
%

Consider the motion in the $y$-direction determined by Eq.~\eqref{eq y}. Its solution can be written in the following form~\cite{courant-1958-a}:
\begin{eqnarray}\label{beta y}
y=\sqrt{2J_{y} \beta_y}\cos{(\psi_y)},\nonumber\\
y'=\frac{\beta_y'}{\beta_y}y - \sqrt{\frac{2J_{y}}{\beta_y}}\sin{(\psi_y)}
\end{eqnarray}
Here $J_{y}$ is a constant, and
\begin{equation}
\psi_y=\int_0 ^s \frac{d\bar{s}}{\beta_y(\bar{s})} +\psi_{y0}  ,
\end{equation}
where $\psi_{y0}$ is another constant. The function $\beta_y(s)$ is one of the Courant-Snyder amplitudes, commonly referred to as the beta function. The other two are:
\begin{equation}\label{albegam}
\alpha_y=-\frac{1}{2} \beta_y' , \quad  \gamma_y=\frac{1+\alpha_y^2}{\beta_y}.
\end{equation}
The oscillatory motion described by Eq.\eqref{beta y} is known as the betatron motion, which has an invariant known as the Courant-Snyder invariant given by:
\begin{equation}\label{CS Inv}
\gamma_y y^2 + 2\alpha_y y y' +\beta_y y'^2 = J_y .
\end{equation}
The pair $(\psi_y, J_y)$ are known as the angle-action variables in classical mechanics~\cite{ goldstein-2002-a, ruth-1985-a}. The beta function is determined by the following nonlinear, second order differential equation and appropriate boundary conditions:
\begin{equation}\label{beta eq}
2 \beta_y \beta_y''-{\beta_y'}^2 + 4 \beta_y ^2 K_q=4 .
\end{equation}
The advantage of writing the solution in the form of Eq.\eqref{beta y} is that the betatron motion is specified by two distinct characteristics, the initial conditions associated with each particle via $J_y$ and $\psi_{y0}$ and  the magnet arrangement of the beamline via the beta function.  The homogeneous part of the  Eq.~\eqref{eq x} has the same structure as that of Eq.\eqref{eq y}. Thus, we have the betatron motion in the $x$-direction and its associated Courant-Snyder invariant as well.

We now focus on Eq.~\eqref{eq x}. By forming linear combinations of solutions of the form given in Eq.\eqref{beta y} with appropriate constants $J_x$ and $\psi_{x0}$, we can construct two independent solutions of the homogeneous equation, $C_{s\tau}$ (cosine-like) satisfying $C_{\tau\tau}$=1, $C_{\tau\tau}'$=0 and  $S_{s\tau}$ (sine-like) with $S_{\tau\tau}$=0, $S_{\tau\tau}'$=1 . Then the solution of the inhomogenious equation can be written as follows ~\cite{brown-1982-a, wiedemann-1999-a}:
\begin{eqnarray}\label{solve x}
x_{s}=C_{s\tau}x_{\tau}+S_{s\tau}x'_{\tau}+\eta_{s\tau}\delta , \nonumber \\
x'_{s}=C'_{s\tau}x_{\tau}+S'_{s\tau}x'_{\tau}+\eta'_{s\tau}\delta ,
\end{eqnarray}
The last terms are special solution to the inhomogeneous equation with
\begin{equation} \label{eta} 
\eta _{s\tau}  =S_{s\tau } \int _{\tau }^{s}d\zeta \; C_{\zeta \tau } /\rho  \left(\zeta \right)-C_{s\tau } \int _{\tau }^{s}d\zeta \; S_{\zeta \tau } /\rho \left(\zeta \right). 
\end{equation}

The longitudinal motion is given by
 \begin{eqnarray}\label{eq z}
 \frac{dz}{ds}&=&-\frac{x}{\rho (s)},\nonumber \\ \frac{d\delta}{ds}&=&0,
 \end{eqnarray}
with the solution
\begin{equation}\label{solve z}
z_{s}= z_{\tau}-\int_{\tau }^{s}d\zeta \frac{ x_{\zeta}}{\rho(\zeta)},\quad \delta_s=\delta_{\tau}.
\end{equation}
Inserting Eqs.\eqref{solve x} into Eq.\eqref{solve z} and collecting results so far, we obtain the \textit{R} matrix for transformation from $\tau$ to $s$ in \textbf{X}, \textbf{Z} space: 
\begin{equation} \label{R matrix} 
R_{s\tau } =\left(\begin{array}{cccc} {C_{s\tau } } & {S_{s\tau } } & {0} & {\eta _{s\tau } } \\ {C'_{s\tau } } & {S'_{s\tau } } & {0} & {\eta '_{s\tau } } \\ {R_{51} {}_{s\tau } } & {R_{52} {}_{s\tau } } & {1} & {R_{56} {}_{s\tau } } \\ {0} & {0} & {0} & {1} \end{array}\right). 
\end{equation} 
The $R_{5,j}$ elements the above are
\begin{equation}\label{R elements}
\left(R_{51s\tau}, R_{52s\tau},R_{56s\tau} \right)=-\int_{\tau }^{s}d\zeta \; \left(C_{\zeta \tau },S_{\zeta \tau },\eta_{\zeta \tau } \right) /\rho \left(\zeta \right),
\end{equation}
The matrix given by Eq.~\eqref{R matrix} satisfies the symplectic condition, Eq.~\eqref{ZEqnNum905720}. 

For free space (or drift space), we have $\rho \rightarrow \infty$, $\eta=\eta'=R_{5j}=0$ and the $2\times 2$ upper-left  bloc of $R_{s\tau}$ becomes
\begin{equation}
D_{\ell}=\left(\begin{array}{cc}{1}&{\ell}\\{0}& {1}\end{array}\right).
\end{equation}
Here $\ell=s-\tau$ is the length of the free space. If we have a thin quarupole of focal length $f_C$, then the bock becomes
\begin{equation}
F_{f_C}=\left(\begin{array}{cc}{1}&{0}\\{-1/f_C}& {1}\end{array}\right).
\end{equation}
Using Eq.~\eqref{R elements}, the matrix for a sector bending magnet of constant $\rho$ and deflection angle $\theta $ can be found, see Eq.(77) in~\cite{brown-1982-a}.
%\begin{multline}
 %\label{17)} 
%R_{B} =\left(\begin{array}{cccc} {\cos (\theta) } & {\rho \sin (\theta) } & {0} & {\rho \left(1-\cos (\theta) \right)} \\ {-\sin (\theta) /\rho } & {\cos (\theta) } & {0} & {\sin (\theta) } \\ {-\sin (\theta) } & {-\rho \left(1-\cos (\theta) \right)} & {1} & {-\rho \left(\theta -\sin (\theta) \right)} \\ {0} & {0} & {0} & {1} \end{array}\right). 
%\end{multline}
The matrix for the corresponding rectangular magnet is obtained by multiplying a defocusing lens of focal length $\rho /\tan \left(\theta /2\right)$ on both sides of the sector magnet matrix with the result :
% ~\cite{chautard-1996-a}
%\begin{eqnarray} \label{18)} 
%\begin{split}
\begin{multline} \label{18}
R_{Brect} \left(\rho \right) =  \\ 
\left(\begin{array}{cccc} {1} & {\rho \sin (\theta) } & {0} & {\rho \left(1-\cos (\theta) \right)} \\ {0} & {1} & {0} & {2\tan \left(\theta /2\right)} \\ {-2\tan \left(\theta /2\right)} & {-\rho \left(1-\cos(\theta) \right)} & {1} & {-\rho \left(\theta -\sin (\theta) \right)} \\ {0} & {0} & {0} & {1} \end{array}\right). 
%\end{split}
\end{multline}
This matrix is also given in~\cite{brown-1999-a}.\footnote{ However, the sign convention in this reference such that the $z$ coordinate has the opposite sign to ours. Thus, the $(5,j)$ and $(j,5)$ elements there should be multiplied by $-1$. Otherwise, the matrix is  not symplectic.}
%\end{eqnarray} 
%%%%
%\begin{split}
%    \tilde{m}&= P\tilde{M} \\
%             &= K(R^1R^2R^3t) \begin{pmatrix} X \\ Y \\ 0 \\ 1 \end{pmatrix} \\
%             &= K(R^1R^2t)\begin{pmatrix} X \\ Y \\ 1 \end{pmatrix} \\
%             &= H \begin{pmatrix} X \\ Y \\ 1 \end{pmatrix} 
%\end{split}

A \textit{dogleg} consisting of a rectangular bending magnet, a drift of length $\ell$ , and a reverse rectangular bending magnet has the following matrix:
\begin{equation} \label{sec2:eq:Dogleg} 
R_{DL} =\left(\begin{array}{cccc} {1} & {\ell} & {0} & {\eta } \\{0} & {1} & {0} & {0} \\ {0} & {\eta } & {1} & {\xi } \\ {0} & {0} & {0} & {1} \end{array}\right) .
\end{equation} 
Here
\begin{eqnarray}\label{chiparam}
\ell &=& d+2\rho \sin (\theta),\nonumber \\
\eta &=& 2\left(d+\rho \sin (\theta) \right)\tan \left(\frac{\theta }{2} \right), \\
\xi &=& -2\rho \left(\theta +\sin (\theta) \right)+8\rho \tan \left(\frac{\theta }{2} \right)+4d\tan ^{2} \left(\frac{\theta }{2} \right).\nonumber
\end{eqnarray}
The dogleg is often employed in beam shaping, such as for bunch compression and emittance exchange ( Sections~\ref{sec2c3} and \ref{sec6: EEX}).

\subsubsection{RF photo-cathode cavity}\label{acc cavity}

\begin{figure}[hhhhh!]
 \includegraphics[width=.95\linewidth]{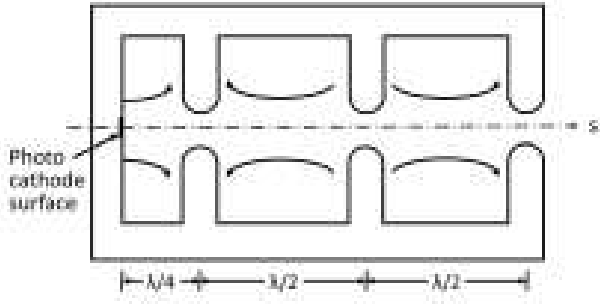}
\caption{\label{fig:sec2:photoinjector} A schematics of an RF photo-cathode cavity. The first cell where photocathode is located is a half cell. From Ref.~\cite{kim-1989-a}. }
\end{figure}

Here, we consider particle motion in an RF photo-cathode cavity ~\cite{fraser-1985-a}; an important device for generating bright beams. The cavity will be assumed to have cylindrical symmetry and shown schematically in Fig.~\ref{fig:sec2:photoinjector} .  Since acceleration is involved, we need to use the canonical variables introduced in Eq.\eqref{new can vab} in Section~\ref{Ham form} with the independent coordinate as the distance along the axis of the cavity $s$. The discussion here follows closely that in reference ~\cite{kim-1989-a}.

The non-vanishing components of the EM field , keeping terms quadratic in radial coordinate $r$, take the following form:   
\begin{eqnarray} \label{acc fields}
E_{s} &=& \left(\mathcal {E}(s)-\frac{r^2}{4}\left(\mathcal {E}''(s)-k^2\mathcal {E}(s)\right) \right)\sin (\omega t) ,\nonumber \\ 
E_{r } &=& -\frac{1}{2} r \mathcal {E}'(s) \sin (\omega t) ,\nonumber \\ 
B_{\phi } &=& \frac{k}{2c} r \mathcal {E}(s) \cos (\omega t) .
\end{eqnarray}
Here $k=2\pi /\lambda=\omega/c$, $\lambda$ is the RF wavelength, and 
\begin{equation}
\mathcal {E} (s)=E_0 \Theta (s) \cos( kz ).
\end{equation}
Here, $E_0$ is the peak on-axis electric field and $'=d/ds$.  The function $\Theta \left(s\right)$ accounts for the cavity exit at $s=s_f =(n+1/2 )\lambda/2$, close to a step function, unity inside the cavity and decreases rapidly to zero as $s$ leaves the cavity at $s_f$. Therefore $\Theta '$ and $\Theta ''$ are similar to the delta function and its derivative respectively, with appropriate sign. Inside the cavity $0 \le s < s_f$, the transverse field $E_r$ and $B_{\phi}$ are linear in $r$ and while $E_s$ is independent of $r$. These properties are crucial in minimizing the emittance growth, especially near the cathode located at $s=0$. The profile of the cavity boundary supporting these field 
is not realizable since it has logarithmic divergence at $s=m\lambda/2$. However, the parameters of the finite profile as shown in Fig.~\ref{fig:sec2:photoinjector} can be chosen to produce fields close to those in Eq. \eqref{acc fields}~\cite{mcdonald-1988-a}.\\

\paragraph{Longitudinal motion}

First consider the longitudinal motion on the axis $r=0$. The canonical variables can be taken as $(\phi,\gamma)$, where
\begin{equation}\label{phi}
\phi=\omega t - ks =k\left(ct-s \right)=\phi_0 - k z.
\end{equation}
Here $\phi_0$ is the reference phase. Note $z$ is the same as that in Eq.\eqref{long devi} for particles moving with relativistic velocity $c$.
The equations of motion are
\begin{eqnarray}\label{long motion}
    \frac{d\phi}{ds}&=&k \left( \frac{\gamma}{\sqrt{\gamma^2 -1}}-1\right) \nonumber\\
    \frac{d\gamma}{ds}&=&\frac{eE_0}{2mc^2}\left[\sin(\phi)+\sin(\phi+2ks) \right].
\end{eqnarray}
 For photo-cathode cavity operation, we are interested in solving Eq.\eqref{long motion} with the initial condition $(\phi, \gamma)=(\phi_i, 1)$. Noting that the RHS of the phase equation is appreciable only near the cathode, $0 \le ks\ll 1$, an approximate solution for the entire range $0 \le s\le n\lambda/2$ can be found:
\begin{eqnarray}\label{appr sol}
\phi&=&\frac{1}{2\alpha \sin{(\phi_i)}}\left[\sqrt{\tilde{\gamma}^2-1}-\left(\tilde{\gamma}-1\right)\right]+\phi_i, \nonumber \\
\gamma&=& 1+\alpha\left[ks \sin{(\phi_i)}+\frac{1}{2}\left(\cos{(\phi)}-\cos{(\phi+2ks)}\right)\right].
\end{eqnarray}
Here $\phi_i$ is the initial phase and
\begin{equation}\label{rough gamma}
\tilde{\gamma}=1+2\alpha \sin{(\phi_i)} ks
\end{equation}
and we introduced the acceleration strength parameter
\begin{equation}\label{alpha}
\alpha=\frac{e E_0}{2mc^2 k}.
\end{equation}
For operating RF photo-cathode cavities, the value of $\alpha$ is between 1 and 2. The approximate solution is fairly accurate for $3+1/2$ cell cavity with $\alpha=1$ and $30^{\circ} \le \phi_i \le 70^{\circ}$. Improvement of the approximation is discussed in reference~\cite{floettmann-2015-a}. At the exit, the particle would have accelerated to a high energy $\gamma_f \gg 1$. The final phase $\phi_f $ becomes
\begin{equation}\label{fin phase}
\phi_{f}=\frac{1}{2\alpha \sin{(\phi_i)}} +\sin{(\phi_i)}.
\end{equation}
Thus the ratio of the final to initial phase spread, or the the bunch length is
\begin{equation}\label{bunch comp}
\frac{\Delta \phi_{f}}{\Delta \phi_i}=1-\frac{\cos{(\phi_i)}}{2\alpha \sin^2{(\phi_i)}}.
\end{equation}
The bunch at the photo-cathode cavity exit will be compressed if $0 < \phi_0 < \pi/2 $.  Equation~ \eqref{long motion} can be cast in Hamiltonian form by using the longitudinal Hamiltonian
\begin{eqnarray}\label{long H}
\mathcal H_L (\phi,\gamma;s)&=&\frac{eE_0}{2mc^2}\left[\cos(\phi)+\cos (\phi+2ks)\right]\nonumber\\
&+&\sqrt{\gamma^2-1}+\arctan{\left(\frac{1}{\sqrt{\gamma^2-1}}\right)} .
\end{eqnarray}
Therefore the map connecting the initial to the final points in the $(\phi, \gamma)$ phase space is symplectic and area-preserving. However, the transformation is nonlinear and will lead to an increase in the longitudinal emittance.\\

\paragraph{Transverse motion}
Let's now turn to the transverse motion. The Lorentz force  given by Eq.\eqref{Lorentz} is purely radial with the magnitude

\begin{equation}\label{r force}
F_{r } =e\left( E_{r } -v_s B_{\phi } \right).
\end{equation}
Here $v_s=ds/(dt)$.
Using Eq.\eqref{acc fields}, the force can be shown to be written in the following form:
\begin{eqnarray}\label{r force diff}
F_r &=&eE_{0} \frac{1}{2} \rho \left[ -\frac{1}{c} \frac{d}{dt} \left(\Theta \sin (ks)\; \cos (\omega t)\right)\right. \nonumber \\&-&\left. \Theta '\sin (\phi) +(1-\beta )\Theta '\sin (ks)\; \cos (\omega t )\right]
\end{eqnarray} 
The change of the radial momentum from the cathode to exit is the time integral of the above equation. The contribution of the last term is negligible for relativistic velocities and the first term integrates to zero. The contribution of the middle term comes only from the exit region with the result 
\begin{equation} \label{exit t mom}
\Delta p_r = eE_{0} \frac{r }{2c} \sin (\phi _{f}).
\end{equation}
 Equation\eqref{exit t mom} gives rise to a bow-tie shaped $(r, p_r)$ phase space at the cavity exit due to the different exit phase $\phi_f $ of each particle, that is, each particle receives a different kick that depends on when it exits the cavity .   This transverse longitudinal coupling leads to an increase in the projected emittance in the $(r, p_r)$ space. 
 
  The radial  force due to the space charge fields, Eq.\eqref{t space charge F}, leads to a similar result since it also depends on the longitudinal position $z$. The space charge effect actually dominates the transverse emittance growth since the force is not localized at the cavity exit but persistent from the beam creation at the cathode. Fortunately, the increase can be corrected by the emittance compensation technique~\cite{serafini-1997-a,miginsky-2009-a,floettmann-2017-a,wang-2006-a,carlsten-1989-a,ferrario-2007-a,ferrario-2000-a}.
 %Another type of longditudinal-transverse coupling was pointed out in~\cite{flottmann-2003-a}.

\subsubsection{Transverse deflecting cavity}\label{sec2:deflectingcavity}
%Here we consider a cavity consisting of several pill-box cells in $\pi$-mode configuration. For %deflection, the mode in each cell is TM${}_{011}$. 
%Particle motion in a transverse deflecting cavity was analyzed in ~\cite{cornacchia-1995-a}in %thin cavity approximation. The analysis for a $\pi$ mode cavity was given in %e~\cite{cornacchia-2002-a} but without derivation. A full derivation was given by Edward in an %unpublished note~\cite{edwards-2007-a} .

\begin{figure}[t]
 \includegraphics[width=.75\linewidth]{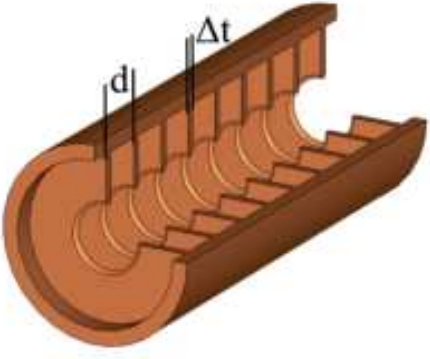}
\caption{\label{fig:sec2:cavityIrisLoaded} Iris-loaded structure producing HEM modes for transverse deflection.  $\Delta t$ is the thickness of the discs and $d$ the structure periodicity. From Ref~\cite{paramonov-2019-a}.}
\end{figure}

In the RF photo-cathode cavity considered in previous section, the transverse deflection occur only at the cavity exit. It has been known since the celebrated paper by Panofsky and Wenzel~\cite{panofsky-1956-a} that neither TM or TE mode of a smooth waveguide cannot provide a sustained transverse kick to a relativistic particle moving along the waveguide axis. To see this, note that the transverse part  of the third and fourth of Maxwell equations, Eq.\eqref{Maxwell}, can be written as follows:
\begin{eqnarray}\label{Maxwell perp}
\frac{\partial}{\partial s} \mathbf{E}_{\perp}-\mathbf{e}_s\times\frac{\partial}{\partial t} \mathbf{B}_{\perp}&=&-\mathbf{e}_s\times \nabla_{\perp}E_s ,\nonumber \\
\frac{\partial}{c^2 \partial t} \mathbf{E}_{\perp}-\mathbf{e}_s\times\frac{\partial}{\partial s} \mathbf{B}_{\perp}&=&-\mathbf{e}_s\times \nabla_{\perp}B_s .
\end{eqnarray}
The first of the above is the differential form of Panofsky-Wenzel theorem~\cite{panofsky-1956-a} and the second was noted by ~\cite{paramonov-2019-a}. Let's consider a traveling wave with fields  in the following form:
\begin{eqnarray}
\left(\mathbf{E}_{\perp},\mathbf{B}_{\perp}\right)&=&\left(\mathbf{\tilde{E}}_{\perp}(x,y),\mathbf{\tilde{B}}_{\perp}(x,y) \right)\sin{(\phi)},\nonumber\\ E_s&=&\tilde{E}_s (x,y) \cos{(\phi)} . 
\end{eqnarray}
Assuming that the phase velocity of the wave and the particle velocity are both equal to $c$, we can show that the LHS of both equations are proportional to the transverse force. From the expressions on the RHS, it then follows that  the transverse gradient of both $E_s$ {\it{and}} $B_s$ should not vanish for a sustainable transverse force . 
An iris-loaded structure, such as shown in Fig.~\ref{fig:sec2:cavityIrisLoaded}, supports hybrid electromagnetic (HEM) modes, that can provide a sustained transverse force. The mode was studied by many authors since 1963~\cite{hahn-1963-a,garault-1964-a,paramonov-2019-a,floettmann-2014-a}. 

When the iris thickness $\Delta t$ is much smaller than the structure period $d$, which in turn is much smaller than the mode wavelength $\lambda$, the electric field should be perpendicular to the iris radius at $r=a$. The transverse Laplacian $ {\nabla_{\perp}}^2 =\partial^2/\partial x ^2+\partial^2/\partial y $ of the fields $\mathbf{\tilde{E}}$ and $\mathbf{\tilde{B}}$ must vanish.  Considering polynomials in $x$ and $y$ of order up to 2, they are linear combinations of $1, x, y, xy$, and  $x^2-y^2$. From symmetry consideration, we may choose  $\tilde{E}_x=\alpha +\beta \left(x^2-y^2 \right)$ and $\tilde{E}_y=\gamma x y$.  Then, all electric field components are determined up to an overall constant by the condition that the electric field should be perpendicular to the iris at $r=a$ and  the first of Maxwell equations Eq.\eqref{Maxwell} (setting  $\rho_S=0 $) . The magnetic field components are then determined from  Maxwell's third equation . The results are
\begin{eqnarray} \label{HEM}
E_{x} &=&-\hat{E} \frac{k^2}{4}  \left(a ^{2}  + x^{2} -y^{2} \right)\sin{(\phi)}, \nonumber\\ 
E_{y} &=&-\hat{E} \frac{k^2}{2}  x y  \sin{(\phi)} ,\nonumber \\ 
E_{s} &=& \hat{E} k  x \cos{(\phi)},\nonumber \\ 
cB_{x} &=&\hat{E} \frac{k^2}{2} xy \sin{(\phi)} \nonumber, \\ 
cB_{y} &=&  -\hat{E} \frac{k^2}{4} \left(a ^{2}-  4/k^2  + x^{2} -y^{2} \right)\sin (\phi) \nonumber, \\
cB_{s}  &=&  -\hat{E} k y \cos{(\phi)}.  
\end{eqnarray}
The Lorentz force is then
\begin{equation}\label{HEM force}
\mathbf F = (F_x,F_y,F_s)=e\hat{E} \left( -\sin{(\phi)}, 0, k x \cos{(\phi)}\right)
\end{equation}
For beam shaping applications,  we wish the transverse deflection changes sign as $z$ varies across a bunch so that the head and tail receive an opposite kick. Thus, we choose $\phi_0=0 $ in Eq.\eqref{phi}. Assuming $ k z \ll 1$, the force to first order in the deviation variables $(x,z)$ becomes
\begin{equation}\label{approx HEM force}
\mathbf F \approx e\hat{E} \left( k z, 0, k x \right)
\end{equation}
The equation of motion inside the HEM structure is then
\begin{eqnarray}\label{EOM HEM}
\frac{d x}{ds} &=& x',\quad \frac{d x'}{ds}= \frac{\kappa_T}{\ell}  z , \nonumber \\
\frac{dz}{ds} &=& 0,\quad \frac{d\delta}{ds}= \frac{\kappa_T}{\ell}  x .
\end{eqnarray}
Here $\ell$ is the length of the cavity and
\begin{equation}
\kappa_T =\frac{e\hat{E} \ell k}{mc^2 \gamma}
\end{equation}
is a parameter characterizing the deflection strength.
In Eq.\eqref{EOM HEM},  $dz/ds$ vanishes since particles are highly relativistic.  The solution of Eq.\eqref{EOM HEM} can be written by the following matrix connecting the entrance and exit values of the vector $\left(\mathbf X , \mathbf Z\right)=\left(x, x', z, \delta \right)$~\cite{cornacchia-2002-a}: 
\begin{eqnarray} \label{TDC} 
R_{TDC}= \left(\begin{array}{cccc} {1} & {\ell} & {\kappa_T \ell /2} & {0} \\ {0} & {1} & {\kappa_T } & {0} \\ {0} & {0} & {1} & {0} \\ {\kappa_T } & {\kappa_T \ell/2 } & {\kappa_T ^{2} \ell/6} & {1} \end{array}\right) .
\end{eqnarray} 
The matrix $R_{TDC}$ has the desirable property that $\left(R_{TDC}\right)^n$ is the same as the matrix for a cavity $n$ times longer, corresponding to the matrix obtained  by substituting $\ell \rightarrow n\ell$ and $\kappa_T \rightarrow n \kappa_T$ in Eq.\eqref{TDC}.

The $(6,5)$ element in $R_{TDC}$ wreaks havoc in the transverse-to-longitudinal emittance exchange  discussed in Section~\ref{sec6}. However, it can be removed by suitable accelerating cavities~\cite{zholents-2011-a}.

\subsubsection{Axial magnetic field }\label{sec2:solenoid}

Axial magnetic fields produced inside solenoidal coils are used for guiding and focusing electron beams~\cite{reiser-1994-a}. The cathode may be immersed in an axial magnetic field. We assume, as is usually the case, that because the magnetic field is cylindrically symmetric around the \textit{z}-axis perpendicular to the cathode surface, it can be derived from the vector potential
\begin{equation} \label{31)} 
\mathbf A(r ,s)=\frac{1}{2} r B\left(s\right)\mathbf e_{\phi } . 
\end{equation} 
Here we are using the cylindrical coordinate $\left(r ,\phi ,s\right)$ with origin at the cathode center, and \textit{z} is the distance away from the cathode. The magnetic field is
\begin{equation} \label{32)} 
\mathbf{B}(r ,s)=\mathbf{\nabla} \times \mathbf{A}=B(s)\mathbf{e}_{s} -\frac{1}{2} r B'\mathbf{e}_{\rho } . 
\end{equation} 
The second term involving $B'=dB/ds$ gives the magnetic flux spreading out radially outside the solenoidal coil.  We are assuming the variation in \textit{s} is slow, $B''\approx 0$, so that $\mathbf{\nabla} \times \mathbf{B}$ vanishes.

The transverse motion is given by
\begin{equation} \label{ZEqnNum793453} 
\frac{d^{2} \mathbf{x}}{ds^{2} } =-\frac{e B\left(s\right)}{p_{s}} \mathbf{e}_{s} \times \frac{d\mathbf{x}}{ds} -\frac{e B'\left(s\right)}{2p_{s}} \mathbf{e}_{s} \times \mathbf{x} .
\end{equation} 

The motion looks simpler in a special rotating frame. To see this, consider the relation between the coordinate vector \textbf{x} in the laboratory frame to the coordinate vector $\mathbf{x}_{R} $ in a frame rotating at a rate $\kappa $~\cite{kim-2000-a}:
\begin{eqnarray} \label{ZEqnNum263851} 
\mathbf{x}&=&M_{\phi}\mathbf{x}_{R} ,\quad M_{\phi}=\left(\begin{array}{cc} {\cos (\phi) } & {\sin (\phi) } \\ {-\sin (\phi) } & {\cos (\phi) } \end{array}\right) , \nonumber \\ 
&&\mbox{with~} \phi \left(s\right)=\int _{0}^{s}\kappa \left(\tau \right) \, d\tau . 
\end{eqnarray} 
By differentiating, we obtain
\begin{equation} \label{36)} 
\frac{d\mathbf{x}}{ds} =\left(\frac{d}{ds} M_{\phi}\right)\mathbf{x}_{R} +M_{\phi}\frac{d}{ds} \mathbf{x}_{R}  .
\end{equation} 
The first term is
\begin{eqnarray}
\label{ZEqnNum122450} 
\left(\frac{d}{ds} M_{\phi}\right)\mathbf{x}_{R} &=& -\kappa\left(\begin{array}{c} {\sin (\phi) \, x_{R} -\cos (\phi) \, y_{R} } \\ {\cos (\phi) \, x_{R} +\sin (\phi) \, y_{R} } \end{array}\right) \nonumber \\
 &=&-\kappa M_{\phi}\left(\mathbf{e}_{s} \times \mathbf{x}_{R} \right) .
\end{eqnarray} 
We thus obtain the well-known relationship between the space-frame (laboratory frame) and body-frame (moving with the body) differentiation~\cite{goldstein-2002-a}:
\begin{equation} \label{labvec} 
\frac{d}{ds} \mathbf{x}=M_{\phi}\left(\frac{d}{ds} -\kappa \mathbf{e}_{s} \times \right)\mathbf{x}_{R} .
\end{equation} 
The second derivative is
\begin{equation} \label{ZEqnNum955853} 
\frac{d^{2} \mathbf{x}}{ds^{2} } =M_{\phi}\left(\frac{d}{ds} -\kappa \mathbf{e}_{s} \times \right)\left(\frac{d}{ds} -\kappa \mathbf{e}_{s} \times \right)\mathbf{x}_{R} .
\end{equation} 
Using Eq. \eqref{ZEqnNum955853} in Eq. \eqref{ZEqnNum793453}, and choosing
\begin{equation} \label{eq:kappaL} 
\kappa \left(s\right) \rightarrow \kappa_L\left(s\right) \equiv \frac{eB\left(s\right)}{2p_{s} }  ,
\end{equation} 
we find that the equation of motion in the rotating frame becomes very simple:
\begin{equation} \label{40)} 
\frac{d^{2} \mathbf{x}_{R} }{ds^{2} } =-\kappa_{L}^{2}\left(s\right)\mathbf{x}_{R}.
\end{equation} 
The spatial frequency  $\kappa_L$ given by Eq.~\eqref{eq:kappaL}, is one half of the cyclotron frequency, and is known as the \textit{Larmor} frequency, because he was the first to note such simplification as explained  by~\cite{brillouin-1957-a}.  In the Larmor frame the particle motion is the same as in a focusing channel, thus with two Courant-Snyder invariants. 
Equation~\eqref{labvec} can also be written as
\begin{equation} \label{rotvec1} 
\frac{d}{ds} \mathbf{x}_R=R^{-1}\left(\frac{d}{ds} + \kappa_L \mathbf{e}_{s} \times \right)\mathbf{x}  .
\end{equation} 
Noting that $\mathbf{e}_s\times\mathbf{x}=\rho \mathbf{e}_{\phi}$ and in view of Eq.~\eqref{31)}, we have
\begin{equation} \label{43} 
\mathbf{p}_R \equiv p_s\mathbf{x'}=R^{-1}\left(\mathbf{p} + e \mathbf{A} \right)  .
\end{equation} 
Therefore, the transverse momentum in the instantaneous rotating frame is simply the canonical momentum. The canonical angular momentum is
\begin{equation}
{\pmb{\mathcal L}}=\mathbf{x}_R\times \mathbf{x'}_R ={\mathcal L} \mathbf{e}_s  .
\end{equation}
The canonical angular momentum is conserved. Its magnitude 
 in laboratory frame quantities is 
\begin{equation} \label{canAMM} 
{\mathcal L}=x y' -x' y+\kappa_L \left(x^2  +y^2 \right)  .
\end{equation} 
The canonical angular momentum consists of a field part $\kappa \left(x^2+y^2 \right)$ and a kinetic part $x y'-y x'$. The conversion of field angular momentum to kinetic angular momentum occurs as the particle exits the solenoid when it receives an azimuthal kick from the radial magnetic field in the transition region.

The relation between the 4D phase-space vector in the rotating frame and that in the laboratory frame is
\begin{equation} \label{rotvec2} 
\left(\begin{array}{c} {\mathbf{X}_R}\\ {\mathbf{Y}_R } \end{array}\right)=\left(\begin{array}{c}{x_R}\\{x'_R}\\{y_R}\\{y'_R} \end{array}\right)=R_B\left(\begin{array}{c}{x}\\{x'}\\{y}\\{y'} \end{array}\right).
\end{equation}
Here
\begin{equation}\label{45b)}
R_B = \left(\begin{array} {c c c c} {1} &{0} &{0} & {0}\\{0}& {1} &{-\kappa_L} & {0} \\{0}& {0} &{1} & {0} \\{\kappa_L}& {0}& {0} & {1} \end{array}\right)   .
\end{equation}
Note $M_B$ is {\it{not}} simplectic. However, its determinant is unity:
\begin{equation}\label{46x)}
    \det[R_B]=1.
\end{equation}

\subsection{Beam transformation under external field}\label{sec2c}

\subsubsection{Liouville's theorem and manipulation of phase-space distribution}\label{sec2c1Liouville}

A beam consists of many particles and can be specified by its phase-space distribution function $f(\pmb{\mathcal Z};s)$.  Let the phase-space volume $\Omega $ in $\pmb{\mathcal Z}$ at $s_1$ transform to $\bar{\Omega }$ in $\bar{\pmb{\mathcal Z}}$ at $s_2$. Here, we are going back to the general nonlinear map $M$ , Eq.\eqref{H map}. Assuming that there are no obstructions leading to particle loss, the distribution function can be normalized to unity as the beam goes through an accelerator beamline: 
\begin{equation} \label{ZEqnNum800732} 
\int _{\Omega }d^{6} \pmb{\mathcal Z} f\left({\mathcal Z},s_{1} \right)=\int _{\bar{\Omega }}d^{6} \bar{\pmb{\mathcal Z}} f\left(\bar{\pmb{\mathcal Z}},s_{2} \right)=1.
\end{equation} 
The left-hand side of Eq.~\eqref{ZEqnNum800732} can, noting Eq.~\eqref{unitjacob}, be written as
\begin{eqnarray} \label{ZEqnNum164592} 
&&\int _{\bar{\Omega }}d^{6} \bar{\pmb{\mathcal Z}} \; \det\left(R^{-1} \right)f\left(M^{-1} \left(\bar{\pmb{\mathcal Z}}\right),s_{1} \right) \nonumber \\
&=&\int _{\bar{\Omega }}d^{6} \bar{{\mathcal Z}} \; f\left(M^{-1} \left(\bar{\pmb{\mathcal Z}}\right),s_{1} \right) .
\end{eqnarray} 
Here  $R$ is the Jacobian matrix given by Eq.\eqref{Jacobian}. Comparing this with RHS of Eq.~\eqref{ZEqnNum800732},  we find that the transformation of the distribution function corresponding to the transform of the phase-space variable in Eq.~\eqref{H map}:
\begin{equation} \label{ZEqnNum508283} 
f\left(\pmb{\mathcal Z},s_{2}\right ) = f \left(M^{-1} \left(\pmb{\mathcal Z}\right),s_{1} \right) .
\end{equation} 
This is the celebrated Liouville's theorem~\cite{liouville-1838-a}, fundamental to many beam shaping schemes involving external fields.

Section \ref{sec3} discusses how  $f(\pmb{\mathcal Z};s)$ at the cathode leads to different distribution subsequently.  If the 6D distribution function $f\left(\pmb{\mathcal Z}, s\right)$ is integrated over all components of $\pmb{\mathcal Z}$ except the variable $\zeta_j$, we obtain the 1D density distribution in  $\zeta_j$.  Section \ref{sec4} discusses various ways of obtaining interesting 1D distributions by suitable map $M$.

We now return to the case where the map is linear and can be represented by matrix $R$.

 To give a simple example, consider the distribution in the $\left(z,\delta \right)$ subspace:
\begin{equation} \label{1DGaussian} 
f\left(z,\delta ;s_{1} \right)=\frac{1}{\ell \sqrt{2\pi } \sigma _{\delta } } \exp \left[-\frac{\delta ^{2} }{2\sigma _{\delta } {}^{2} } \right] .
\end{equation} 
Here $\ell$ is the bunch length, over which the distribution is regarded as constant.

  The corresponding line density in \textit{z} is constant:
\begin{equation} \label{lambda1D} 
\lambda \left(z;s_{1} \right)=\int d\delta f\left(z,\delta ;s_{1} \right) =\frac{1}{\ell} .
\end{equation} 
The matrix for the bunch compression section consisting of chirping (energy change linear in $z$) followed by an $R_{56} $ is
\begin{equation} \label{47)} 
R_{C} =\left(\begin{array}{cc} {1} & {R_{56} } \\ {0} & {1} \end{array}\right)\left(\begin{array}{cc} {1} & {0} \\ {h} & {1} \end{array}\right)=\left(\begin{array}{cc} {1+hR_{56} } & {R_{56} } \\ {h} & {1} \end{array}\right) .
\end{equation} 
Its inverse is
\begin{multline} \label{48)} 
R_{C} {}^{-1} \left(\begin{array}{c} {z} \\ {\delta } \end{array}\right)=\left(\begin{array}{cc} {1} & {-R_{56} } \\ {-h} & {1+h R_{56} } \end{array}\right)\left(\begin{array}{c} {z} \\ {\delta } \end{array}\right)\\
=\left(\begin{array}{c} {z-R_{56} \delta } \\ {-hz+\delta /C} \end{array}\right).
\end{multline} 
Here we introduced the compression factor $C$:
\begin{equation}\label{compf}
C=1/\left(1+hR_{56} \right).
\end{equation}

  Therefore, the distribution after the beamline, applying Eq.~\eqref{ZEqnNum508283}, becomes
\begin{equation} \label{49)} 
f\left(z,\delta ;s_{2} \right)=\frac{1}{\ell \sqrt{2\pi } \sigma _{\delta } } \exp \left[-\frac{\left(-hz+\delta /C\right)^2}{2\sigma_{\delta }^{2} } \right]. 
\end{equation} 
The corresponding line density is
\begin{equation} \label{50)} 
\lambda \left(z;s_{2} \right)=\int d\delta f\left(z,\delta ;s_{2} \right) =\frac{C}{\ell} .
\end{equation} 
The line density is increased by \textit{C} since the bunch length is compressed by the same factor.

\subsubsection{Beam matrix and emittance}\label{sec2:beamemittance}

A beam can also be completely specified by all of its beam moments, which are:
\begin{equation} \label{ZEqnNum595960} 
\left\langle \zeta _{i} \zeta _{j} \zeta _{k} ...\right\rangle _{s} = \int d^{6}  \pmb{\mathcal Z}\; f\left({\mathcal Z};s\right)\zeta _{i} \zeta _{j} \zeta _{k} ...,
\end{equation} 
with $i,j,k,..=1,2,..6$.  The moments transform accordingly to 
\begin{equation} \label{ZEqnNum863542} 
\left\langle \zeta _{i2} \zeta _{j2} \zeta _{k2} ..\right\rangle _{s2}   =R_{i2,i1} R_{j2,j1} R_{k2,k1} ..\left\langle \zeta _{i1} \zeta _{j1} \zeta _{k1} ..\right\rangle _{s1}  .
\end{equation} 
The first order moments can be made to vanish noting that $\zeta$'s are deviation variable, that is variables relative to the trajectory of the reference particle. The reference particle is at the ``center" of the beam in the sense that the first order moments of the deviation variables vanish:  
\begin{equation} \label{53)} 
\left\langle \zeta _{i} \right\rangle =\int d^{6}  {\pmb{\mathcal Z}}\; \zeta _{i} f(\pmb{\mathcal Z},s)=0.
\end{equation} 
The second order moments are 
\begin{equation} \label{54)} 
\Sigma =\left\langle \pmb{\mathcal Z} \pmb{\mathcal Z}^{T} \right\rangle ;\; \Sigma _{ij} =\int d^{6} \pmb{\mathcal Z}\; f\left({\mathcal Z}\right)\; \zeta _{i} \zeta _{j}   .
\end{equation} 
Since $f\left(\pmb{\mathcal Z}\right)\ge 0$ and cannot vanish identically, the \textit{beam matrix} $\Sigma $ is symmetric and positive definite. 
%If $f\left(\pmb{\mathcal Z}\right)$ follows a Gaussian distribution, the first two orders of beam moments completely determine the beam shape in phase space.

Note that the \textit{R} matrix is symplectic, satisfying Eq.~\eqref{ZEqnNum905720}.  The first moments vanish by suitably choosing the coordinate frame. The second moments are elements of the beam matrix, which can be written as 
\begin{equation} \label{ZEqnNum901896} 
\Sigma =\left\langle \pmb{\mathcal Z} \pmb{\mathcal Z}^{T} \right\rangle =\left(\begin{array}{ccc} {\left\langle \mathbf{XX}^{T} \right\rangle } & {\left\langle \mathbf{XY}^{T} \right\rangle } & {\left\langle \mathbf{XZ}^{T} \right\rangle } \\ {\left\langle \mathbf{YX}^{T} \right\rangle } & {\left\langle \mathbf{YY}^{T} \right\rangle } & {\left\langle \mathbf{YZ}^{T} \right\rangle } \\ {\left\langle \mathbf{ZX}^{T} \right\rangle } & {\left\langle \mathbf{ZY}^{T} \right\rangle } & {\left\langle \mathbf{ZZ}^{T} \right\rangle } \end{array}\right) .
\end{equation} 
Here $\left\langle XX^{T} \right\rangle $, etc., are 2$\times $2 matrices and $\left\langle \right\rangle $ is the averaging operation as in Eq.~\eqref{ZEqnNum595960}.  The beam matrix transforms as
\begin{equation} \label{ZEqnNum554620} 
\Sigma (s_{2} )=R\Sigma (s_{1} )R^{T}  .
\end{equation} 
We introduce the quantities called  \textit{projected emittances }in each \textbf{X},\textbf{Y},\textbf{Z} subspace:
\begin{eqnarray} \label{57)} 
\varepsilon _{x}^{proj} &=& \sqrt{\det\left(\left\langle \mathbf{XX}^{T} \right\rangle \right)} =\sqrt{\left\langle x^{2} \right\rangle \left\langle x'^{2} \right\rangle -\left\langle xx'\right\rangle ^{2} }, \nonumber \\ 
\varepsilon _{y}^{proj} &=& \sqrt{\left\langle y^{2} \right\rangle \left\langle y'^{2} \right\rangle -\left\langle yy'\right\rangle ^{2} }, \\
\varepsilon _{z}^{proj} &=&\sqrt{\left\langle z^{2} \right\rangle \left\langle \delta ^{2} \right\rangle -\left\langle z\delta \right\rangle ^{2} } \nonumber.  
\end{eqnarray} 
The projected emittances are in general not invariant. 

Williamson~\cite{williamson-1936-a} has proved that a positive definite, symmetric matrix such as $\Sigma $ can be transformed to a 2$\times $2 block diagonal form by a \textit {symplectic} matrix \textit{A}:
\begin{equation} \label{ZEqnNum483392} 
        A\Sigma A^{T} =\left(\begin{array}{ccc} {\left\langle \mathbf{X}_1\mathbf{X}_1 {}^{T} \right\rangle } & {0} & {0} \\ {0} & {\left\langle \mathbf{X}_2\mathbf{X}_2{}^{T} \right\rangle } & {0} \\ {0} & {0} & {\left\langle \mathbf{X}_3 \mathbf{X}_3{}^{T} \right\rangle } \end{array}\right) .
\end{equation} 
Here the 2-vectors $\mathbf{X}_j$, $j$=1,2,3 represent a new partition of the 6D space into three, decoupled, 2D phase spaces. The emittances in each decoupled space
\begin{equation} \label{unnormalized} 
\varepsilon_j = \sqrt{\det \mean{\mathbf{X}_j\mathbf{X}_j{}^T }} .
\end{equation}
will be referred to as the \textit{principal emittance}\footnote{In reference~\cite{dragt-2011-a}, this emittance is referred to as {\em eigen-emittance}.}.
The elements of the 2$\times $2 beam matrix can be parametrized as follows:
\begin{equation}\label{beam para}
\mean{\mathbf{X}_j \mathbf{X}_j{}^{T}}\equiv \left(\begin{array}{cc} { \mean{x_j^2} } & {\mean{x_j x'_j}} \\ {\mean{x_j x'_j}} & \mean{{x'_j}^2 }\end{array}\right) =\varepsilon_j\left(\begin{array}{cc} { \beta_j } & {-\alpha_j} \\ {-\alpha_j} & {\gamma _j }\end{array}\right).
\end{equation}
From Eq.\eqref{unnormalized}, it follows that the parameters are related by $\beta_j \gamma_j - \alpha_j ^2 =1$. Note that this relation is the same as the second of Eq.\eqref{albegam}.
The parametrization here is directly connected to that introduced in the  Courant-Snyder form of betatron motion \eqref{beta y} (identifying the $y$-subspace with the $j$-subspace); If we compute the beam matrix using Eq.\eqref{beta y} as the beam trajectory, assuming that the betatron phase $\psi_{j0}$ is uniformly distributed, then we find that the principal emittance in jth subspace is the statistical average of particles' action variable in that subspace:
\begin{equation}
\varepsilon_j=\mean{J_j}.
\end{equation}
Therefore, the parameters $\alpha_j$ ,$\beta_j$  and $\gamma_j$ here are in fact identical to those introduced in Eq.\eqref{beta y}. 

The beam matrix Eq.\eqref{beam para} can be diagonalized with the matrix
\begin{equation} \label{59)} 
B^j =\left(\begin{array}{cc} {1} & - {\alpha_j /\gamma_j} \\ {0} & {1} \end{array}\right).
\end{equation} 
Indeed
\begin{equation} \label{60)} 
B_j\mean{\mathbf{X}_j \mathbf{X}_j{}^{T}} B_j {}^{T} =\left(\begin{array}{cc} {\varepsilon_j \beta_j{}^* } & {0} \\ {0} & {\varepsilon_j /\beta _j{}^* } \end{array}\right) .
\end{equation} 
Here we used the notation
\begin{equation}\label{waist beta}
\beta_j{}^* \equiv 1/\gamma_j .
\end{equation}
The transformation represented by matrix $B^j$ is a free-space translation going back a distance $\ell=\alpha_j/\gamma_j$ to the \textit{waist} location of the beam where the correlation $\alpha_j$ vanishes. In this sense, $\beta_j{}^*$ is known as the beta function at the waist. Since angular divergence does not change under a free-space translation, the value of $\gamma_j$ does not change while translating to the waist.  Note the second of Eq.\eqref{albegam} can be written as
\begin{equation}\label{beta trans}
 \beta_j=\left(1+\alpha_j{}^2 \right)/\gamma_j=\beta_j{}^*\left(1+(\ell/\beta_j{}^*)^2\right)
 \end{equation}
This is the well-known and useful equation describing how the beta function changes away from the waist.

Performing  diagonalization  in other dimensions, we  obtain
\begin{multline} \label{ZEqnNum421539} 
BA\Sigma A^{T} B^{T} = \\ \mbox{diag}\left(\varepsilon_1 \beta_1{}^* ,\varepsilon_1 /\beta_1{}^* ,\varepsilon_2 \beta _2 {}^* ,\varepsilon_2 /\beta_2 {}^* ,\varepsilon _3 \beta_3{}^* ,\varepsilon_3 /\beta_3{}^* \right) .
\end{multline} 
 The beta functions at the waist,  $\beta_j{}^* $s, are also positive but not invariant.   With an additional transformation, Eq.~\eqref{ZEqnNum421539} can be reduced  to Williamson's normal form $\text{diag}\left(\varepsilon_1 ,\varepsilon _1 ,\varepsilon _2 ,\varepsilon _2 ,\varepsilon _3 ,\varepsilon _3 \right)$.  However, this last transformation is not physical since it changes the dimensions of the elements. 

The principal emittance $\varepsilon_j$ is invariant under any transformation that leaves  the subspace $j$ intact, \textit{if there is no acceleration}. If $j=x$, a translation along z-axis is an example of such transformation. If particles are accelerated in the $z$ direction, for example, then $x'=dx/ds$ is no longer a canonical variable, and we need to start from the correct canonical variable $p_x=m\gamma dx/dt=m c\gamma \beta x'$ and its conjugate $x$. When the particle velocities are nearly the same, the emittance that is invariant  under acceleration also is 
\begin{equation}\label{normalized}
\varepsilon_j{}^n=\beta\gamma\varepsilon_j .
\end{equation}
Note that $\beta$ and $\gamma$ in the above equation are the velocity in the $z$-direction divided by $c$ and the particle energy divide by $mc^2$, \textit {not} the Courant-Snyder amplitudes.  The emittance defined by Eq.\eqref{normalized}  is referred to as {\it normalized} emittance, while that defined by Eq.~\eqref{unnormalized} is known as the {\it un-normalized} emittance.  The emittances in this review are un-normalized emittances unless specified otherwise.

The principal emittances can be found by adopting the standard eigenvalue problem~\cite{dragt-2011-a}.  They can also be obtained as follows~\cite{courant-1966-a,neri-1990-a}.  First, note that the quantities
\begin{equation} \label{63)} 
\Gamma ^{n} =\text{Tr}[\left(J\Sigma \right)^{2n} ],\quad n=1,2,3.. 
\end{equation} 
are, in view of Eq.~\eqref{ZEqnNum905720}, invariant under symplectic transformation. Note also that  $\Sigma$ inside the trace of Eq.\eqref{63)} can be replaced by the diagonal matrix Eq.\eqref{ZEqnNum421539}. Thus, we obtain
\begin{equation} \label{ZEqnNum914550}
\Gamma ^{n}=  2(-1)^n\left(\left(\varepsilon _1 \right)^{2n} +\left(\varepsilon _2 \right)^{2n} +\left(\varepsilon _3 \right)^{2n} \right).
\end{equation} 
The above gives three equations for three principal emittances. We also have
\begin{equation} \label{ZEqnNum935832} 
\det\left(\Sigma \right)=\left(\varepsilon _1 \varepsilon _2 \varepsilon _3 \right)^{2}  .
\end{equation} 
Any of the three equations from Eqs.~\eqref{ZEqnNum914550} and \eqref{ZEqnNum935832} can be solved for the three principal emittances.

The square root of Eq.\eqref{ZEqnNum935832}, or the product of three principal emittances, are known as the 3D emittance.

%%%%%%
\subsubsection{Emittance exchange  (EEX) and phase-space exchange (PSE) }\label{sec2c3}

An important corollary of Williamson's theorem is that \textit{an arbitrary re-partition of the emittances is not possible by means of symplectic transformation}. Thus, for example, we exchange $\left(\varepsilon _{x} ,\varepsilon _{y} ,\varepsilon _{z} \right)$ to $\left(\varepsilon _{z} ,\varepsilon _{y} ,\varepsilon _{x} \right)$ but cannot re-partition it to $\left(\varepsilon _{x}/4 ,4\varepsilon _{y} ,\varepsilon _{z} \right)$. This fact appears to be first noted in the accelerator physics context by Courant~\cite{courant-1966-a}.  

Although the set of three principal emittances does not change, the ordering in the set could be changed. For example, the emittance in $\mathbf{X}$ space can be exchanged to the emittance in  $\mathbf{Z}$ space. To explain the meaning of the emittance exchange (EEX), consider a beam matrix of the form:
\begin{multline}\label{SigI}
\Sigma_I=\left(\begin{array}{cc}{\Sigma_x}& {0}\\{0}& {\Sigma_z}\end{array}\right);\\
\Sigma_x=\left(\begin{array}{cc}{\mean{x^2}}& \mean{xx'}\\{\mean{xx'}}& {\mean{x'^2}}\end{array}\right), \Sigma_z=\left(\begin{array}{cc}{\mean{z^2}}& \mean{z \delta}\\{\mean{z \delta}}& \mean{\delta^2}\end{array}\right).
\end{multline}
The matrix $\Sigma_I$ is uncoupled, that is, the off-diagonal blocks vanish. Therefore, the determinants of the 2$\times$2 matrices $\Sigma_x$ and $\Sigma_z$ are, respectively, $\varepsilon_x{}^2$ and $\varepsilon_z{}^2$, where $\varepsilon_x$ and $\varepsilon_z$ are the principal emittances . Suppose a beamline gives rise to the following transformation:
\begin{equation}\label{EEXbl}
R \left(\begin{array}{cc}{\Sigma_x}& {0}\\{0}& {\Sigma_z}\end{array}\right) R^T =\left(\begin{array}{cc}{\Sigma_z}{}{'} & {0}\\{0}& {\Sigma_x}{}{'}\end{array}\right).
\end{equation}
Here, $\Sigma_x{}{'}$ is a 2$\times$2 beam matrix symplectically connected to $\Sigma_x$ and similarly $\Sigma_z{}{'}$ to $\Sigma_z$. $R$ is a symplectic matrix that can be written in 2$\times$2 block matrices:
\begin{equation}\label{transM}
    R=\left(\begin{array}{cc}{A}& {B}\\{C}& {D}\end{array}\right).
\end{equation}
 Equation~\eqref{EEXbl} holds if
\begin{equation}\label{EEXblc}
A\Sigma_x C^T +B \Sigma_z D^T=0.
\end{equation}
A beamline  is referred to as an EEX beamline if Eq.~\eqref{EEXbl} holds for any \textit{arbitrary} pair of beam matrices $\Sigma_x$ and $\Sigma_z$. Thus Eq.~\eqref{EEXblc} must hold for arbitrary $\Sigma_x$ and $\Sigma_z$. There are four possibilities:  ($i$) $A=0, D=0$, ($ii$) $C=0,B=0$, ($iii$) $A=0,B=0$, and ($iv$) $C=0,D=0$. In case (ii) emittances are not exchanged, and in cases (iii) and (iv) $R$ is not symplectic. Thus, a matrix for emittance exchange will be of the form 
\begin{equation}\label{ABnull}
A=0, D=0.
\end{equation}
Then
\begin{equation}\label{Phaseswap}
R \left(\begin{array}{c} {\mathbf{X}}\\{\mathbf{Z}}\end{array} \right)=\left(\begin{array} {c} {B \mathbf{Z}} \\{C\mathbf{X}}\end{array}\right).
\end{equation}
Therefore, the EEX transformation is much more than an  exchange of the magnitude of the sub-space area---it exchanges the whole sub-space $\mathbf{X}$ with the whole sub-space $\mathbf{Z}$. A phase-space shape in $\mathbf{X}$ will be transformed to a corresponding shape in $\mathbf{Z}$, making the exchange transformation useful for beam shaping purposes. Although the transformation was named EEX since it was considered in the context of the emittance exchange~\cite{cornacchia-2002-a,kim-2006-a,emma-2006-a}, it would be more proper to refer to it as \textit{phase-space exchange}(PSE).

Examples of emittance and phase-space exchange methods are discussed in Sections~\ref{sec6: EEX} and~\ref{sec6c2} , respectively.

\subsubsection{Emittance repartitioning}\label{sec2c4}
 Although emittances can only be exchanged wholly in a symplectic transformation, an emittance \textit{re}-\textit{partitioning} is possible if a non-symplectic step occurs at some point during the transformation. An example is provided by a beam produced from a photocathode immersed in an axial magnetic field $B$, as discussed in Section~\ref{sec2:solenoid}. We construct the beam matrix  of the rotating frame-vector Eq.~\eqref{rotvec2}: 
\begin{equation} \label{80)} 
\Sigma_B  =\left(\begin{array}{cc} {\left\langle \mathbf{X}_R \mathbf{ X}_R {}^{T} \right\rangle } & {\left\langle \mathbf{X}_R \mathbf{ Y}_R {}^{T} \right\rangle }  \\ {\left\langle \mathbf{Y}_R \mathbf{X}_R {}^{T} \right\rangle } & {\left\langle \mathbf{ Y}_R \mathbf{Y}_R {}^{T} \right\rangle }  \end{array}\right) .
\end{equation} 
The subscript B indicates the presence of the magnetic field. Let $\varepsilon _{R1}$ and $\varepsilon _{R2}$ be the two principal emittances associated with the beam matrix $\Sigma_B$, which will in general be different from the principal emittance of the beam in the absence of the magnetic field, $\varepsilon _{01}$ and $\varepsilon _{02}$. In view of Eq.~\eqref{46x)}, we have
\begin{equation}\label{75x)}
    \det\left[\Sigma_B\right]=\det\left[R_B \Sigma_0 R_B {}^T\right]=\det\left[\Sigma_0 \right].
\end{equation}
Here, $\Sigma_0$ is the beam matrix for $B=0$.  Thus,
\begin{equation}\label{emm product}
    \varepsilon_{B1} \varepsilon_{B2} =\varepsilon_{01} \varepsilon_{02}.
\end{equation}
The canonical angular momentum associated with the beam is
\begin{equation} \label{beamCAM} 
\langle {\mathcal L} \rangle = \langle x y' -x' y \rangle +\kappa_L \left\langle x^2  +y^2 \right\rangle ,
\end{equation} 
where $\kappa_L$ was defined in Eq.\eqref{eq:kappaL}.
The conservation of the canonical angular momentum of a beam is known as Busch's theorem~\cite{bush-1926-a}. A generalization of Busch's theorem to a non-symmetric system was discussed by~\cite{groening-2018-a}.

Suppose a cathode produces  a round beam, $\varepsilon_{01}=\varepsilon_{02}$. If we now immerse the cathode in a magnetic field, the two principal emittances can become different. It is shown in Section~\ref{sec6} that an arbitrary value of the emittance ratio $\varepsilon_{B1} / \varepsilon_{B2}$ can be achieved by varying the magnetic field and hence the canonical angular momentum Eq.~\eqref{beamCAM}. Is this a violation of the Williamson-Courant theorem?  No, since the matrix $M_B$ in Eq.~\eqref{45b)} is not symplectic. Since the beam is really born in a magnetic field with the beam matrix $M_B \Sigma_0 M_B {}^T$, one may object calling this example an emittance re-partition. Indeed, if the beam is first produced from a cathode in a field-free region and then encounters an axial magnetic field, its principal emittances will not change. 
Emittance re-partitioning schemes \textit{always} involve non-symplectic elements, such as beam masks or tapered absorbing blocks. A general emittance re-partitioning was investigated in Ref.~\cite{carlsten-2011-a}.    

Details of experimental setup and results are discussed later in Sections~\ref{sec6b1} and~\ref{sec6b2}. 

\subsubsection{Nonlinear case}\label{sec2:sec2c5nonlinear}
If the Hamiltonian contains polynomials of order higher than quadratic in the scaled deviation variables, the variables in Eq.~\eqref{variables} are not canonical and  the  map $M$ in Eq.~\eqref{H map} becomes nonlinear. If the nonlinearity is small, the canonical variables Eq.~\eqref{new can vab} can be expressed in Taylor series in $\pmb{\mathcal Z}$. The map from $\pmb{\mathcal Z}\to \bar{\pmb{\mathcal Z}}$ can be found by solving the equation of motion using the original Hamiltonian Eq.~\eqref{new H} in the following form: 
\begin{equation} \label{81-nonlinear} 
\bar{\zeta _{i} }=R_{ij} \zeta _{j} +T_{ijk} \zeta _{j} \zeta _{k}+... 
\end{equation} 
The coefficients $T_{ijk} ,..$ were worked out in detail in~\cite{brown-1982-a} ( However, see footnotes 1 for the sign convention). The constraints on these coefficients from symplectic property was discussed in~\cite{wollnik-1985-a}.

Higher order solutions can be obtained in classical mechanics by canonical perturbation theory, which provides a procedure for finding canonical transformations  in which the new canonical momenta become constants of motion~\cite{arnold-1978-a}. The procedure was applied to accelerator beam dynamics, see for example~\cite{ruth-1985-a}.  A powerful method using Lie canonical transformation with polynomial generators has been developed that can handle very high order polynomial terms by numerical computation~\cite{dragt-2011-a}.

 If the transformation is nonlinear, Liouville's theorem still applies microscopically. However, there can be an apparent increase in the macroscopic phase-space volume due to filamentation~\cite{sorensen-1988-a}. For a weakly nonlinear system, one can still introduce adiabatic invariants, which is the phase-space area following the physical orbit~\cite{landau-1969-a}.

\subsection{Beam-generated fields }\label{sec2d}
The systematic study of the effects of beam-generated fields on the operation of intense, high-brightness accelerators was begun by accelerator physicists in the Mid-Western Universities Association (MURA)~\cite{jones-2010-a}.  Vlasov's equation was applied to the study of the beam instability problems in~\cite{nielsen-1959-a} (What is referred to as ``Boltzmann's equation" in this paper is actually identical to Vlasov's equation since the collision terms are neglected) . The impedance concept was found to be an efficient tool for expressing the force in the frequency domain for long beam bunches in storage rings~\cite{faltens-1975-a}. Later, the concept of wakefields as a force in the time domain was found to be more appropriate in studying short bunch phenomena in linacs~\cite{wilson-1989-a}. Here we summarize these concepts relevant to the beam shaping topics in electron linacs, limiting ourselves to the longitudinal interaction.

\subsubsection{ Wakefield and impedance}\label{sec2d1}

A particle  moving uniformly at a speed $\beta c,\; \beta \approx 1$ carries a Coulomb field with it, squeezed to an angular width of $\sim 1/\gamma =\sqrt{1-\beta ^{2} } $ due to Lorentz contraction. If there is a surface parallel to the particle trajectory at a distance \textit{b}, a moving area of longitudinal length $b/\gamma $ is under the influence of the Coulomb field.  If the surface is perfectly conducting and smooth, the boundary condition at the surface will be maintained. If there are interruptions in the surface , however, then the field interacts with the surface and produces fields behind the particle ( thus, the term wake) that can influence the motion of the trailing  particles.  

The longitudinal wakefield $w(z)$ is defined as the EM field $E_{s} $ on a test particle trailing a fixed distance \textit{z} behind the drive particle ~\cite{wilson-1989-a,heifets-1991-a,chao-1993-a,stupakov-2001-a}:
\begin{equation} \label{82)} 
w\left(z\right)=-\frac{1}{e} \left. E_{s} \right|_{z=s-ct} . 
\end{equation} 
The minus sign in the above is to make a positive wake that corresponds to the test particle losing its energy. The sign of $z$ is that it is ahead if positive. The dependence on the transverse coordinates is not important in most of the following and is thus neglected. The unit of the wake field is V per C per m. In the above, we are considering the case where the wake is uniform along the beam chamber. When the wake is localized, the wake function is defined by the integral over the passage of the local structure.

The energy \textit{loss} per unit distance of a particle at position \textit{z} due to other particles is\footnote{Note that $w (z-z')$ is written as   $w(z'-z)$ in some references, e.g., ~\cite{chao-1993-a,stupakov-2001-a}.  Our choice is convenient since it  ensures the same Fourier-transform convention for impedance as well as current.}
\begin{equation} \label{ZEqnNum411253} 
\Delta E\left(z\right)=eQ\int _{-\infty }^{\infty }dz'\; w\left(z-z'\right)\lambda \left(z'\right)  .
\end{equation} 
Here \textit{Q} is the total charge and $\lambda (z)$ is the line-charge density of the particles in the beam normalized as $\int dz\lambda  \left(z\right)=1$.  

Now we introduce the \textit{impedance per unit length} and Fourier transform of the electric field and current profile as follows~\cite{nielsen-1959-a,chao-1993-a}: 
\begin{eqnarray}  
Z(k) &=& \frac{1}{c} \int _{-\infty }^{\infty }ds\;  w\left(z\right)e^{-ikz} ,\label{Zk}\\
\tilde{E}(k)&=&\int _{-\infty }^{\infty }dz\; e^{-ikz}  E\left(z\right),\label{Ek}\\
I(k) &=& cQ\int _{-\infty }^{\infty }dz\; e^{-ikz}  \lambda \left(z\right) \label{Ik} .
\end{eqnarray} 
From Eq.~\eqref{ZEqnNum411253} these quantities are related via 
\begin{equation} \label{wakevolt} 
 \tilde{E}\left(k\right)\equiv V\left(k\right)=-Z\left(k\right)I\left(k\right),  
\end{equation} 
where $V\left(k\right)$ is the (negative) voltage applied per unit distance. In Eq.~\eqref{lambda1D} the line density $\lambda (z)$ was given as an integral of the phase space distribution keeping only the $\mathbf Z$. In the more general case, we should write
\begin{equation}\label{lambda3D}
\lambda(z;s)=\int d\delta d^2{\mathbf X} d^2{\mathbf Y} f\left(\pmb{\mathcal Z};s\right) .
\end{equation}
In discussing coherent instabilities including free-electron lasers,  one often uses the term  \textit {bunching factor} $b(k;s)=I(k)/cQ$. Note that $I(k)$ or $b(k)$ can be expressed as an integral in 6D phase space $\pmb{\mathcal Z}$:
\begin{equation}\label{bunchingf}
b\left(k;s\right)=\int dz e^{-ikz}\lambda(z;s)=\int d^6\pmb{\mathcal Z}e^{-ikz} f\left(\pmb{\mathcal Z};s\right) .
\end{equation}
The collective force due to a beam-generated field on a particle at $z$ will change the longitudinal momentum and hence $\delta$:
\begin{equation}\label{collforce}
\frac{d \delta_{coll}}{ds}=eE(z)=-\frac{r_e}{\gamma}N \int dk e^{ikz}Z(k)b(z;s) .
\end{equation}
Here $r_e$ is the the classical electron radius, and $N$ is the total number of particles. The subscript $coll$ is to emphasize that this is the beam-generated, collective force.

The meaning of equations\eqref{wakevolt} and \eqref{collforce} is that a current modulation at spatial frequency $k$ impresses an energy modulation via the impedance $Z(k)$. Passing through the subsequent beamline, energy modulation can cause beam instabilities, as discussed in Section~\ref{sec2d5} .

Next, we discuss three representative cases of impedance.

\subsubsection{Fields due to boundary perturbation}

 % A particle moving near the velocity of light through a cylindrical pipe in the \textit{z}-direction carries a space-charge field concentrated in a thin disc perpendicular to the motion.  Wakefields are produced when the resistivity of the pipe is finite or if the shape of the boundary cross section is not constant. 
 Since wakefields originate from the EM field in the moving disc scattered off by the surface interruptions, a wakefield satisfies the causality condition~\cite{wilson-1989-a}
\begin{equation} \label{ZEqnNum634563} 
w\left(z\right)=0;\quad z>0 .
\end{equation} 

A wakefield device, referred to as a \textit{dechirper}, with corrugated walls can be useful in correcting the energy chirp that may arise while compressing a bunch for high peak current~\cite{bane-2012-a,bane-2016-a,emma-2014-a,deng-2014-a}.  A flat dechirper consisting of two opposing corrugated plates separated by a half gap \textit{a} has been shown to be effective in removing the chirp in the bunches driving an X-ray FEL oscillator~\cite{qin-2016-a}. The wakefield of such a device was computed and can be represented approximately by the following form:
\begin{equation} \label{ZEqnNum441598} 
w\left(z\right)\approx \frac{\pi cZ_{0} }{16a^{2} } \quad {\rm for}\; z<0;\quad =0\quad {\rm for}\; z>0, 
\end{equation} 
with $Z_{0}=377$~$\Omega$ being the free-space impedance. For a flat-top charge density, we see readily from Eq.~\eqref{ZEqnNum411253} that the energy loss is linear in \textit{z}, and the particles in the tail losing more energy than those at the head.  The difference in energy correction from head to tail is found to be
\begin{equation} \label{88)} 
\Delta E\left(z\right)=-\frac{\pi cZ_{0} QL}{16a^{2} } .
\end{equation} 
The impedance per unit length corresponding to Eq.~\eqref{ZEqnNum441598} can be computed from Eq. \eqref{ZEqnNum735989}:
\begin{equation} \label{89)} 
Z_{flat} \left(k\right)=i\frac{\pi Z_{0} }{16a^{2} k} . 
\end{equation} 

We note here that the impedance of a round pipe of radius \textit{a} with random or periodic corrugations in the high frequency limit is given by~\cite{gluckstern-1989-a}
\begin{equation} \label{90)} 
Z_{round}(k)=i\frac{Z_{0} }{\pi a^{2} k} . 
\end{equation} 
Impedances due to the interruption of perfectly conducting walls are sometimes referred to as \textit{geometric} impedances.

\subsubsection{Space-charge force}\label{sec2d3}

The longitudinal space-charge wake was computed for a bunch moving between two parallel conductors~\cite{nielsen-1959-b} and moving inside a pipe~\cite{neil-1965-a} with  the following approximate method: 

Recall the variable $z=s-c\beta t\approx s-ct$.  Consider a beam of uniform charge density $\lambda (z)= $ constant and a uniform cross section, of radius \textit{b}, travelling at  velocity $\beta c,\; \beta \approx 1$ along the axis of a circular, perfectly conducting pipe of radius \textit{a}. The non-vanishing EM field components in cylindrical coordinates are the radial electric field $E_{r } $ and the azimuthal magnetic field \textbf{$B_{\phi } $} given by
\begin{eqnarray} \label{t space charge F} 
r \le b: E_{r } (r,z) &=&\frac{Q\lambda(z) r }{2\varepsilon _{0} b^{2} } ,B_{\phi }(r,z) =\frac{\beta Q\lambda(z) r }{2\varepsilon _{0} b^{2} c} ,\nonumber \\ b<r <a: E_{r }(r,z) &=&\frac{Q\lambda(z) }{2\varepsilon _{0} r }, B_{\phi }(r,z) =\frac{\beta Q\lambda(z) }{2\varepsilon _{0} r c}.   
\end{eqnarray} 
Now consider the density has a variation so that the $z$-dependence of $\lambda \left(z\right) $ needs to be taken account. If the dependence on $z$ is slow, the main fields are still given by Eq.~\eqref{t space charge F} , but , in addition, there will be  a longitudinal electric field on axis $E_{z} \left(0, z\right)$ .  This can be determined from Faraday's law: The line integral of electric field along the loop consists of straight lines connecting the points in cylindrical coordinates, $(0,z)\to (0,z+\Delta z)\to (a,z+\Delta z)\to (a,z)\to (0,z)$ should be equal to the time derivative of the magnetic flux into the loop. The electric field for the segments $(0,z+\Delta z)\to (a,z+\Delta z)$ and $(a,z)\to (0,z)$ are $E_r$'s in the above, and vanishes for the segment $(a,z+\Delta z)\to (a,z)$. The e-field $E_z (0,z)$ for the segment $(0,z)\to (0,z+\Delta z)$ is to be determined. The magnetic flux into this loop can be computed from $B_{\phi}$. 
%\begin{multline}
%E_{z} \left(z\right)\Delta z+\frac{1}{2\varepsilon _{0} } \left(1+2\log %\frac{a}{b} \right)Q\lambda '\Delta z \\
%=\frac{\beta }{c} \frac{1}{2\varepsilon _{0} } \left(1+2\log \frac{a}{b} %\right)Q\lambda 'c\beta \Delta z\ .
%\end{multline}
%Here $\lambda '=d\lambda /dz$.  We then obtain:
We obtain in this way
\begin{equation} \label{ZEqnNum728960} 
E_{z} \left(0,z\right)  = -\frac{Q}{2\varepsilon _{0} \gamma ^{2} } \left(1+2\log \frac{a}{b} \right)\frac{d \lambda}{dz} .
\end{equation} 

According to Eq.~\eqref{ZEqnNum728960}, the space-charge field tends to smooth away a density bump , as the particles repel each other. This is dramatically illustrated by Fig.~\ref{sec5_fig_scshaping} (b) in Section \ref{sec5_SC1}. If there is a periodic modulation in the density, the other hand, then the modulation could become enhanced, as discussed in Section\ref{sec5_SC3}.

Equation~\eqref{ZEqnNum728960} diverges logarithmically as $a\to \infty $. However, the derivation is valid only if the density bump is not too steep, implying [see problem of 1.5 of~\cite{chao-1993-a}]
\begin{equation} \label{ZEqnNum316207} 
a<\gamma \Delta z ,  
\end{equation} 
where $\Delta z$ is the extent of the density variation. The corresponding impedance is
\begin{equation} \label{94)} 
Z_{SC} =i\frac{Z_{0} k}{4\pi \gamma ^{2} } \left(1+2\log \left[\frac{a}{b} \right]\right) .
\end{equation} 

In the absence of a vacuum chamber pipe, the inequality Eq.~\eqref{ZEqnNum316207} is violated. In this case, one can use the squeezed electric field due to a single particle moving with  uniform velocity $\beta c\approx c$ parallel to the $z$-axis ~\cite{jackson-1998-a} as the Green's function, perform the Fourier transform in $z$ variable, and observe that only the $0$-th component in the azimuthal series will contribute.  
%\begin{multline} \label{ZEqnNum225309} 
%\left. E_{z} \right|_{z=s-ct} =-\frac{Q}{4\pi \varepsilon _{0} \gamma ^{2} %} \frac{\partial }{\partial z} \\
%\int d{\mathbf x}'_{\bot } dz\frac{1}{\sqrt{\left({\mathbf x }_{\bot } -{\mathbf x}'_{\bot } \right)^{2} /\gamma ^{2} +\left(z-z'\right)^{2} } }  \lambda \left(z'\right)\lambda _{\bot } \left({\mathbf x }'_{\bot } \right) .
%\end{multline} 
%Here, $\lambda _{\bot } \left({\mathbf x}_{\bot } \right)$ is the transverse density normalized as $\int d{\mathbf  x}_{\bot } \lambda _{\bot }  \left({\mathbf x}_{\bot } \right)=1$. In the cylindrical coordinates ${\mathbf x}_{\bot } =\left(r ,\phi \right)\; {\rm and}\; {\mathbf x}'_{\bot } =\left(r ',\phi '\right)$, we have the following expansion~\cite{jackson-1998-a}:
%\begin{multline} \label{96)} 
%\frac{1}{\sqrt{\left({\mathbf x }_{\bot } -{\mathbf x}'_{\bot } \right)^{2} /\gamma ^{2} +\left(z-z'\right)^{2} } } \\ 
%= \frac{2}{\pi } \sum _{m=-\infty }^{\infty }\int _{0}^{\infty %}dke^{im\left(\phi -\phi '\right)}   \cos \left[k\left(z-z'\right)\right] %\\ 
%\times I_{m} \left(\frac{k r _{<} }{\gamma } \right)K_{m} \left(\frac{k r _{>} }{\gamma } \right) 
%\end{multline} 
The impedance per unit length can then be computed with the result  ~\cite{rosenzweig-1996-a,venturini-2008-a}:
\begin{equation} \label{97)} 
Z_{SC}(k)=i\frac{Z_{0} }{\pi b^{2} k} \left(1-\frac{kb}{\gamma } K_{1} \left(\frac{kb}{\gamma } \right)\right) .
\end{equation} 
Expanding $K_{1} \left(kb/\gamma \right)$ for small $kb/\gamma $~\cite{venturini-2008-a,huang-2000-a}:
\begin{eqnarray} \label{98)} 
Z_{SC} \left(k\right)&=&i\frac{Z_{0} k}{4\pi \gamma ^{2} } \left(2\log \left[\frac{\gamma }{kb} \right]+2\log 2-2\gamma _{E} +1\right) \nonumber \\ 
&\approx& i\frac{Z_{0} k}{4\pi \gamma ^{2} } \left(2\log \left[\frac{\gamma }{kb} \right]+1.23\right) ,
\end{eqnarray} 
where $\gamma _{E} =0.577$ is the Euler number. 

\subsubsection{ Coherent synchrotron radiation}\label{sec2d4}

A particle on a circular path emits synchrotron radiation. The radiation from the tail of a bunch, proceeding in a path tangential to the circular path, can exert a force to the head of the bunch moving on an arc. Making use of the results known in 1912~\cite{schott-1912-a}, the wakefield at a distance \textit{s} \textit{ahead} of the emitting particle (note \textit{s} is negative) can be written as follows~\cite{derbenev-1995-a,murphy-1997-a,saldin-1997-a}:
\begin{eqnarray} \label{99)} 
E_{\parallel } (z)&=&-\frac{e}{4\pi \varepsilon _{0} } \frac{2}{3^{1/3} \rho ^{2/3} } \frac{d}{ds} G\left(z\right), \\
&& \mbox{with~ }  G(z)=\frac{1}{\left(z\right)^{1/3} }, {\mbox {for}~}
 z\gg\frac{2\rho }{3\gamma ^{3} } . \nonumber  
\end{eqnarray} 
In the above equation $\rho$ is the radius of curvature. The function $G\left(s\right)$ vanishes for $s\ge 0$. The force due to density distribution $\lambda \left(s\right)$ will then be
\begin{equation} \label{100)} 
E_{\parallel ,\lambda } \left(z\right)=\frac{e}{2\pi \varepsilon _{0} } \frac{1}{3^{1/3} \rho ^{2/3} } \int _{-\infty }^{s}dz' \frac{1}{\left(z-z'\right)^{1/3} } \frac{d}{dz'} \lambda \left(z'\right) . 
\end{equation} 
In the above equation we used integration by parts and also assumed that $\left|s-s'\right|$ is about the bunch length, which is much greater than $\rho /\gamma ^{3} $.  The impedance associated with this force is referred to as \textit{coherent synchrotron radiation} (CSR) impedance:
\begin{eqnarray} \label{101)} 
Z_{CSR} \left(k\right)&=&\frac{Z_{0} }{2\pi 3^{1/3} \rho ^{2/3} } \int _{0}^{\infty }\frac{1}{\left(z\right)^{1/3} } {\kern 1pt}  e^{-ikz} dz \nonumber \\
&=& \frac{Z_{0} k^{1/3} }{2\pi 3^{1/3} \rho ^{2/3} } e^{i\pi /6} \Gamma \left(2/3\right) \nonumber \\
&\approx&  \left(1.63+0.94i\right)\frac{Z_{0} k^{1/3} }{4\pi \rho ^{2/3}} .
\end{eqnarray} 

The CSR impedance was first identified in connection with storage ring physics, in which the mode number $n=k\rho $, where $\rho $ is the radius of the curvature~\cite{iogansen-1959-a,faltens-1975-a}.  The CSR impedance can be significant when bending magnets are used in linacs, e.g., in the final chicane magnet of a bunch compressor where the current is high~\cite{heifets-2002-a,huang-2002-a}.

In addition to CSR, synchrotron radiation has high frequency incoherent part, which can lead to diffusion in energy and also transverse phase space dilution through dispersion. This was studied by M. Sands~\cite{sands-1955-a,sands-1969-a} and revisited in context of X-ray free-electron laser operation~\cite{saldin-1996-a}. However, the ISR (incoherent Synchrotron Radiation) cannot be used for shaping due to its incoherence. Also, the ISR  effects that may reduce shaping accuracy are not significant for the typical electron beams for shaping. 

\subsubsection{Collective motion}\label{sec2d5}

  We now discuss how the force due to the beam-generated fields acting back on the beam,  limiting to the 4D phase space $\pmb{\mathcal Z}=\left(\mathbf{X},\mathbf{Z}\right)$ for simplicity. The first question to ask is whether the Liouville's theorem, Eq.\eqref{ZEqnNum508283}, is still valid in the presence of the beam-generated field. The answer is yes, as long as  we can neglect the discrete particle aspects and regard the beam as a continuous fluid. The fluid description is valid if the Debye length $\lambda_D$ is much shorter than the length scale of the collective disturbance: 
\begin{equation}\label{fluidc}
    k \lambda_D \ll 1 ,
\end{equation}
where $k$ is the wave number of the disturbance. The criteria in Eq.~\eqref{fluidc} were first derived for non-relativistic plasma~\cite{pines-1952-a} and extended to relativistic beams~\cite{sorensen-1988-a,kim-2011-a,rosenzweig-1996-a}. For beam physics, we define the Debye length as the transverse spread of the particles during one plasma oscillation: 
\begin{equation}\label{b105)}
\lambda_D =\frac{c\sigma_{\Delta \beta}}{\omega_p} ,
\end{equation}
where $c\sigma_{\Delta_{\beta}}$ is the RMS velocity spread and 
\begin{equation} \label{omegap}
\omega_p=\sqrt{\frac{e^2 n }{\varepsilon_0 m \beta\gamma^3}} .
\end{equation}
is the plasma frequency for a relativistic beam.
The basic process behind collective motion is as follows:  A region in phase space may develop higher spatial density when the beam goes through some part of the beamline, such as a compressor. This part of the phase space then exerts a collective force via Eq.~\eqref{collforce}. This force is added to the external force to modify the beam evolution. A classic example is the plasma oscillation when the density of a part of the beam is increased at the expense of another part. These two parts then oscillate against each other with the plasma frequency given by Eq.~\eqref{omegap}. Collective motion is often detrimental for beam shaping, but it can be useful in some particular cases, such as generation of ultrashort bunch trains via nonlinear plasma oscillation, as will be discussed in Section~\ref{sec5_SC3}.

Let's now discuss how a beam-generated field affects the evolution of the phase-space distribution $f\left(\pmb{\mathcal Z};s\right)$ in a beamline beginning at $s=0$ ~\cite{heifets-2002-a,huang-2002-a}. In the absence of a beam-generated field, the beam distribution functions $f_0$ at two different locations are related by the Liouville's theorem
\begin{equation}
f_0\left(\pmb{\mathcal Z}_s ;s\right)=f_0\left(\pmb{\mathcal Z}_\tau ;\tau \right);\pmb{\mathcal Z}_\tau=R_{s\tau}{}^{-1} \pmb{\mathcal Z}_s .
\end{equation}
This is the same as Eq.~\eqref{ZEqnNum508283}, $R_{\tau s}$ being a combination of any transformation matrix discussed in Section \ref{sec2b}. For notational clarity, a variable at location $\tau$ was given the same subscript, e.g., $\pmb{\mathcal Z}_\tau$.
 Let $d \delta_{coll}\left(z_\tau;\tau\right) $ be the increase in the electron's relative  energy   due to the beam-generated field in a small interval  between $\tau$ and $\tau+d \tau$.  We obtain
\begin{equation} \label{distchange} 
f(\pmb{\mathcal Z}_s,s)=f_0(\pmb{\mathcal Z}_s;s)-\int _{0}^{s}d\tau \frac{\partial f_0(\pmb{\mathcal Z}_{\tau } ;\tau)}{\partial \delta_\tau }  \frac{d\delta _{coll}\left(z_\tau;\tau\right) }{d\tau }  .
\end{equation} 
From this equation the bunching factor at $s$, Eq.\eqref{bunchingf}, can be obtained after some mathematical manipulation~\cite{huang-2002-a}: 
%\begin{eqnarray} \label{bunchf-2} 
%b\left(k;s\right)&=& b_{0} \left(k;s\right)-ik \int d\tau R_{56,s\tau }\int d\pmb{\mathcal Z}_{\tau }  \nonumber \\ 
%&& \times   e^{-ikz\left(\pmb{\mathcal Z}_{\tau } \right)} f_0\left(\pmb{\mathcal Z}_{\tau } ;\tau \right)\frac{d\delta _{coll}\left(z_\tau;\tau\right) }{d\tau } . 
%\end{eqnarray} 
\begin{eqnarray} \label{bunchfeq} 
b\left(k,s\right)&=& b_{0} \left(k,s\right)+\nonumber \\ &&\frac{ik r_e}{\gamma } \int d\tau R_{56,s\tau }  \int \frac{dk_{1} }{2\pi }  Z\left(k_1;\tau \right)b\left(k_{1} ;\tau \right) 
\nonumber \\
&& \times \int d\pmb{\mathcal Z}_{0} e^{-ikz_s\left(\pmb{\mathcal Z}_{0} \right)+ik_{1}z_\tau \left(\pmb{\mathcal Z}_{0} \right)}  f_{0} \left(\pmb{\mathcal Z}_{0} \right).  
\end{eqnarray}
%In the second integral, we changed the integration variable from %$\pmb{\mathcal Z}_s$ to $\pmb{\mathcal Z}_\tau$, based on the fact that $ %d\pmb{\mathcal Z}_s=d\pmb{\mathcal Z}_\tau$. We also performed integration %by parts in the integral over $\delta_\tau$, making use of the relation
Here
\begin{eqnarray} \label{zst} 
z_s\left(\pmb{\mathcal Z}_{\tau } \right)&=&\sum_{j=1}^{6} R_{5j,s\tau}\zeta_{j,\tau} \nonumber \\
&=&z_{\tau } +R_{51,s\tau } x_{\tau } +R_{52,s\tau } x'_{\tau } +R_{56,s\tau } \delta _{\tau }  .
\end{eqnarray} 
In the first part, $\zeta_{j,\tau}$ denote the components of $\pmb{\mathcal Z}_\tau$, see Eq. \eqref{variables}.
The second part of the above equation is for the case $\pmb{\mathcal Z}=\left(\mathbf{X},\mathbf{Z}\right)$.
%Here $z_\tau\left(\pmb{\mathcal Z}_0\right)$ and $z_s\left(\pmb{\mathcal Z}_0\right)$ are obtained from Eq.~\eqref{zst} by appropriate replacement of the subscripts.
 Equation~\eqref{bunchfeq} is an integral equation for the evolution of the bunching factor. We consider the initial distribution in the following form:
 \begin{equation}\label{initdist}
  f_{0} \left(\pmb{\mathcal Z}_{0} \right)=\bar{ f}_{0} \left(\pmb{\mathcal Z}_{0} \right)+\hat{ f}_{0} \left(\pmb{\mathcal Z}_{0} \right) .
 \end{equation}
 The first part $\bar{f}_0$ is smooth  and the second part $\hat{f}_0$ contains high-frequency modulation giving rise
 to the initial bunching factor $b_0$. This term is regarded to be small. Therefore, $f_0$ in Eq.~\eqref{bunchfeq} can be replaced by $\bar{f}_0$. Equation~\eqref{bunchfeq} can be solved iteratively.

To see the physical meaning of Eq.~\eqref{bunchfeq}, we neglect the $\left(\mathbf{X}\right)$ and $\left(\mathbf{Y}\right)$ phase space, that is, $\pmb{\mathcal Z}=\mathbf{Z}=(z,\delta)$. We assume the initial distribution  given by 
 \begin{equation}\label{chirpedG}
 f_0(z,\delta;0)=n_0\frac{1}{\sqrt{2\pi}\sigma_\delta} e^{-\left(\delta-hz\right)^2/2\sigma_\delta{}^2} .
 \end{equation}
 This is a Gaussian distribution, as in Eq.~\eqref{1DGaussian}, but chirped with chirping coefficient $h$, with $n_0=1/\ell$ being the line density. The quantities in the exponent of Eq.~\eqref{bunchfeq} are
 \begin{eqnarray}
z_s\left(\pmb{\mathcal Z}_{0} \right)&=& z+R_{56}(s)\delta, \nonumber \\  z_\tau\left(\pmb{\mathcal Z}_{0} \right)&=& z+R_{56}(\tau)\delta . \end{eqnarray}
Here we simplified the notation by $R_{s0,56}\rightarrow R_{56}(s)$ and $\left(z_0,\delta_0\right)\rightarrow (x,\delta)$. The integral over $\delta$ and $z$ can be performed to obtain
\begin{eqnarray}\label{1Dbunchf}
    b(k;s)&=&b_0(k,s)+\int_{0}^{s} d\tau R_{56,s\tau}\frac{I(\tau)}{\gamma I_A} \nonumber \\
    &&\times Z\left[k(\tau);\tau\right]b\left[k(\tau);\tau\right]\xi(s,\tau) .
\end{eqnarray}
Here, $I_A=ec/r_e$ is the Alfven current, $I=ecn_0 C(\tau)$ is the peak current at  $\tau$, and
\begin{equation}\label{k&k1}
k(\tau)=\frac{C(\tau)}{C(s)}k; C(\tau)=\frac{1}{1+hR_{56}(\tau)} .
\end{equation}
The compression factor $C$ was introduced in Eq.~\eqref{compf} and  
\begin{equation}\label{constxi}
\xi(s,\tau)=e^{-k^2{\sigma_\delta}^2\left[C(s)R_{56}(s)-C(\tau)R_{56}(\tau)\right]^2/2} .
\end{equation}
The meaning of Eq.~\eqref{1Dbunchf} is clear: The density modulation at $\tau$ becomes an energy modulation through the impedance (that can be enhanced through compression), which in turn becomes density modulation through $R_{56}$. The factor $\xi$ gives the                                  degradation due to the energy spread. This equation was first derived by ~\cite{saldin-2002-a}.

%%%%%%%%%%%%%%%%%%%%%%%%%%%%%%%%%

\section{Bunch control via the electron gun~\label{sec3}}

In this Section, we present a variety of gun-based methods for controlling the electron bunch distribution. We classify these methods according to the distribution of the generated bunch: either \textit{mesoscopically} shaped or \textit{macroscopically} smooth bunches. (see Fig.~\ref{fig:sec1:mesoscopic} and the related discussion in Section~\ref{sec1}). In the shaped case, the electron gun is used to directly generate the desired shaped bunch.  In the smooth case, the gun generates a smooth bunch which is subsequently shaped by the methods presented in the later Sections,~\ref{sec4} and~\ref{sec5}, of this review. Gun-based methods presented will include those that have achieved control over the 2D transverse or 1D longitudinal distribution, as well as recent progress in controlling the complete 3D distribution.

\subsection{Introduction}

The electron gun consists of a cathode surface in a region of accelerating and focusing fields and is used to generate the initial electron bunch which is injected into the linac.  The emission of the initial electron bunch distribution from the cathode surface and its evolution through the gun is complex and varied. The reader interested in understanding electron gun physics in-depth is pointed to~\cite{dowell-2008-a,dowell-2010-a,rao-2015-a,dowell-2016-a} and the references therein.  In the introduction to this Section, we only present high-level details needed for understanding gun-based shaping.

Electron bunch generation begins at the cathode surface and continues until the bunch exits the external fields (accelerating and focusing) of the gun. The initial electron distribution emitted from the cathode surface is affected by both the properties of the cathode material and the fields at the cathode surface while the evolution of the bunch is affected by the self-generated fields of the bunch (e.g. space charge) as well as the external fields. However, in this Section, we focus on an idealized regime where the self-generated fields are negligible, and the external fields of the gun merely accelerate and guide the bunch to high energy while preserving the shape (although not the size) of initial cathode distribution. The later Sections of this review include the other effects.

Electron guns used in e- linacs can be classified by their electric fields (DC and RF) or by their cathode type. There are a wide variety of cathodes in use, but they can conveniently be classified by their emission mechanism: field, thermionic, and photo emission (FE, TE and PE, respectively) ~\cite{dowell-2008-a,dowell-2010-a}. When these cathodes are operated in an electron gun, they go by the names of (DC or RF) field emission gun, (DC or RF) thermionic gun, and (DC or RF) photoemission cathode, which is usually shorted to "photocathode" gun.  However, as stated above, the impact of the gun fields on shaping are being ignored (in this Section) so we will drop the DC and RF labels, unless noted otherwise.  On the other hand, the classification according to the emission mechanism is crucial to understanding gun-based shaping methods since the methods presented in this section differ substantially in their capabilities for controlling the initial cathode distribution emitted from each cathode type.

Historically, all three cathode types (FE, TE and PE) have been used to generate smooth beams, but the same cannot be said of shaped beams. TE cathode guns are robust electron sources and they are the workhorse of storage rings for light sources around the world~\cite{lightsources-a}. They are widely used in e- linacs for producing smooth bunches but have not been used in applications requiring either transverse or longitudinal shaping. FE cathode guns are still in the R\&D phase and are not yet used in any e- linacs, at least not the ones we are concerned with in this paper. Nonetheless, the FE cathode gun has great potential and it has demonstrated transverse shaping, although not longitudinal, in proof of principle experiments.  The PE cathode gun is the workhorse for the linacs in SASE FEL facilties and they are being used for both transverse and longitudinal shaping applications. The reason FE and TE cathodes are not used for longitudinal shaping is because the electron emission from these cathodes follows the time structure of the applied electric field (ranging from continuous in a DC gun to ns-scale in an RF gun) which is too long for the time scales of interest (fs-ps) in this review. In principle, TE cathodes could be used for generating transversely shaped bunches, but this avenue has not been pursued to the knowledge of the authors.  For a thorough discussion of the various cathode types, we refer the interested reader to the tutorial paper from~\cite{jensen-2018-a}.

The remainder of this Section is organized into three parts. In~\ref{sec3:cathReview}, we present a selected review of cathodes that focuses on cathode properties relevant to bunch shaping.  Second, we present demonstrated methods for generating smooth bunches (~\ref{sec3:smooth}) and end with methods for generating shaped bunches (~\ref{sec3:shaped}).

\subsection{Cathode review~\label{sec3:cathReview}} 
In this Section, we will review the different cathodes types and summarize their properties relevant to shaping (see~Table \ref{sec3:tab:cathode-characterisitics}).  These properties include the magnitude of the electric at the cathode ($E_{c}$) and the work function ($W$) of the cathode, which is the minimum energy needed to extract an electron from the cathode surface. There are also several properties of the emitted electron bunch of relevance, its: current density ($J$), average kinetic energy emitted from the cathode ($E_{k}$), transverse and longitudinal rms spot sizes ($\sigma_x$, $\sigma_z$) and dimensionless rms momentum ($\sigma_{p_x}$, $\sigma_{p_z}$). Finally, we also include some parameters that are specific to the particular cathode type but those will be introduced in the corresponding Sections below.  The interested reader can find more complete details on cathodes and guns in~\cite{dowell-2008-a,dowell-2010-a} and the references therein.

\subsubsection{Current density}~\label{sec3:cathReview:curDen} 
 
There is great variation in the magnitude of the current density, $J$, generated by the three cathode types.  Typical values of $J$ for the three cathode types are shown in row~\textit{Longitudinal Parameters} of~Table \ref{sec3:tab:cathode-characterisitics}. Note that while the current density does not play a direct role in bunch shaping (it determines the number of electrons) we include this Section since it provides the necessary background to understand how we classify cathodes in this review.    

A FE cathode consists of an arrangement of one or more sharply pointed tips, or emitters, located on a cathode immersed in an applied "macroscopic" electric field, $E_{c}$. The geometry of the tip is characterized by the ﬁeld enhancement factor $\beta$ which enhances $E$ ($\mathrm{\sim}$100 MV/m) to generate extremely high local fields, $F=\beta E$ ($\mathrm{\sim}$10 GV/m) on the tip, from where the electrons are extracted. The local FE  current density is given by the Fowler-Nordheim equation~\cite{fowler-1928-a}
\begin{equation} \label{eq:sec3-FNequation}
j_{FE}\left(F\right)=A_{FE} F^2e^{-\frac{{w}^{{3}/{2}}}{F}} ,
\end{equation}
with material constant $A_{FE}$. (Note, that lower case letters $j_{FE}$ and $w$ are used to differentiate the local values at the tip from the macroscopic parameters (upper case letters, $J_{FE}$ and $W$) averaged across the cathode surface.) The well-known field emission electron microscope (FEM) is an example of a DC FE gun based on a single FE tip.  In a typical FEM, the tip is made of tungsten ($W=4.5$~eV) with radius ranging from 100 nm to 1 mm. Small tips are capable of extremely high local current densities, $j_{FE}$ $\mathrm{\sim}$10${}^{6}$A/mm$^{2}$, with correspondingly small emission areas $\mathrm{\sim}$100 nm$^{2}$.

The challenge of operating an electron gun with a single FE tip is that the current emitted is only $\mathrm{\sim}$0.1 mA per tip, which is inadequate for operation of the linacs we consider in this paper since they require larger current (e.g. photoinjector current $I_{PE}=1nC/10ps=100A$ but at least $1A$ is needed) from the cathode.  The solution is to use large surface area FE cathodes engineered to hold many microscopic emitters immersed in the applied "macroscopic" electric field on the cathode, $E_{c}$. Three such engineered FE cathodes have been developed and used in FE electron guns: field emission arrays (FEAs)~\cite{jarvis-2010-a}, carbon nanotubes (CNTs)~\cite{laszczyk-2020-a}, and ultra-nanocrystalline diamonds (UNCDs)~\cite{baryshev-2014-a}. These surfaces are composed of a quasi-continuous distribution of  electron emitters of macroscopic area $\mathrm{\sim}$1 mm$^{2}$ and achieve macroscopic current densities of $J_{FE}$ $\mathrm{\sim}$1 A/mm$^{2}$, sufficient for electron linacs. The FE parameters shown in~Table \ref{sec3:tab:cathode-characterisitics} are for large area engineered FE cathodes, not single tips.

TE is the liberation of electrons from a heated surface and are operated in guns with applied electric field on the cathode $E_{c}$ ($\mathrm{\sim}$100 MV/m).  Dispenser cathodes have work functions of $W$ ($\mathrm{\sim}$1.6 eV) and are representative of TE cathode and are operated with a temperature of $T$ ($\mathrm{\sim}$1400 K) to allow electron emission. The TE macroscopic current density is given by the Richardson's Law~\cite{richardson-1913-a},

\begin{equation}  \label{eq:sec3-RLDequation}
J_{TE}\left(T\right)=A_{TE}T^2e^{-\frac{W}{k_bT}} 
\end{equation}
at temperature $T(K)$, where $A_{TE}$ is a material constant, and $k_b$ is the Boltzmann constant.  Dispenser cathodes operated in electron guns with high electric fields, $E_{c}$, ($\mathrm{\sim}$100 MV/m) can achieve current densities of $J_{TE}$ $\mathrm{\sim}$1 A/mm$^{2}$ and are therefore sufficient for electron linacs. 

PE occurs when the photons illuminating the photocathode surface have an energy in excess of the work function of the cathode material. The current density emitted from a PE cathode~\cite{dowell-2009-a} is 
\begin{equation} \label{eq:sec3-PCcurrentequation}
J_{PE}\left(h\nu \right)=n_{ph}A_{PE}{\left(h\nu -W \right)}^2 ,
\end{equation}
where $h\nu$ is the photon energy, $h$ is Plank's constant, $A_{PE}$ is a material constant, $n_{ph}$ is the number of photons per unit area, and $\left(h\nu -W \right)\ $is known as the excess energy and is the kinetic energy of the emitted electrons. PE cathodes are operated in electron guns with high electric fields, $E_{c}$, ($\mathrm{\sim}$100 MV/m) and achieve the highest macroscopic current density of the three cathode types, $J_{PE}$ $\mathrm{\sim}$1000 A/mm$^{2}$.

\subsubsection{Kinetic energy}~\label{sec3:cathReview:kineticE} 
$E_{k}$ is the average, total kinetic energy of the electrons emitted from the cathode which are emitted isotropically into the half-sphere over the cathode.  In this Section, we give the expression for $E_{k}$ for each of the different cathode types. As will be seen in the next section, the intrinsic emittance is determined by $E_{k}$. Note that the total kinetic energy, $E_{k}$, is equipartitioned into each degree of freedom, $x,y,z$.

The average kinetic energy of electrons emitted from a FE cathodes~\cite{forbes-2016-a} is ,
\begin{equation} \label{eq:sec3-FE-kineticEnergy}
E_{k}=\frac{e \hbar F}{\sqrt{8 m_e W}},
\end{equation}
where $m_e$ is the electron mass and typical $E_k$ = $0.3$ eV for FE cathode tips. For electrons emitted from a TE cathode, their average kinetic energy is,
\begin{equation} \label{eq:sec3-TE-kineticEnergy}
E_{k}=\frac{3}{2} {k}_bT,
\end{equation}
where, for example, $E_k$ = 0.12 eV for a dispenser cathode operating at $T$=$1400$~K. Finally, electrons emitted from PE cathodes have average kinetic energy given by,
\begin{equation} \label{eq:sec3-PE-kineticEnergy}
E_{k}=\frac{h\nu -W }{2},
\end{equation}
where, for example, a copper cathode (work function of 4.65 eV) is illuminated by a laser of wavelength $\lambda$=248nm, has $E_k$ = 0.18 eV.

An examination of the three expression for $E_k$ reveal two things worth pointing out. First, the expressions for FE and PE both depend on the work function $W$. In practice, the work function $W$ is usually replaced by the effective work function $\phi_{eff}$ to account for the Schottky effect~\cite{dowell-2009-a} which is the reduction of the work function by the applied field and it plays a role in all emission processes, especially field emission. However, we keep $W$ in all expressions for simplicity. Second, note that a parameter associated with the emission mechanism appears in each expression for $E_k$: it depends on the local field $F$ (or equivalently $\beta$) for FE cathodes, on temperature $T$ for TE cathodes, and on photon energy $h\nu$ for PE cathodes.. Typical values of $E_k$ are shown in~\textit{Excess Kinetic Energy} of~Table \ref{sec3:tab:cathode-characterisitics}. 
 
\subsubsection{Emittance}~\label{sec3:cathReview:Emit} 

In the first half of this Section, we explain the fundamental role that emittance plays in limiting the resolution of the shape.  Once this is established, we then discuss the intrinsic emittance generated by the three cathode types and end with a discussion on how to choose a cathode for an application.

\paragraph{Shaping resolution~\label{sec3:cathReview:shapRes}}

We define the shaping resolution to be the smallest spot size that can be obtained at the end of a beamline, of transfer matrix $R$, for a fixed beam spot size at the beginning of the beamline.  If we place a lens (i.e. a quadrupole magnet) at the beginning of this beamline~\cite{carey-1987-a}, then the lens can be varied to minimize the transverse spot size at end,
\begin{equation}\label{carey book eqn}
{\sigma }_{x,min}=\frac{R_{12}{\varepsilon }_x}{{\sigma }_{x,0}} ,
\end{equation}
where $R_{12}$ is the $(1,2)$ element of $R$, $\sigma_{x,0}$ is the beam size at the beginning of the beamline (which is also at the lens) and ${\varepsilon }_x$ is the horizontal beam emittance (assumed to be constant). The reader should recall that we are ignoring collective effects of the beam and higher-order magnetic optics of the beamline. (For the derivation of Eq.~\eqref{carey book eqn} see Eq. 6.45 in~\cite{carey-1987-a} and the discussion therein. The book's notation is slightly different than what we use in this paper. Also note that smallest spot size is not necessarily the waist of Eq.~\eqref{waist beta}.) A similar expression for the minimum longitudinal bunch length ($\sigma_{z,min}$) at the end of the beamline can be derived and is given by,

\begin{equation}
{\sigma }_{z,min}=\frac{R_{56}{\varepsilon }_z}{{\sigma }_{z,0}} ,
\end{equation}
where ${\sigma }_{z,0}$ is the longitudinal bunch length at the beginning of the beamline, ${\varepsilon }_z$ is the longitudinal beam emittance, and $R_{56}$ is an element of the transfer matrix. Note that in this case, we use a \textit{longitudinal lens} (i.e. an RF cavity) at the beginning of the beamline to minimize the bunch length at the end of the beamline

The significance of the preceding two equations is that they give us three ways to minimize the spot size (i.e. increase the resolution of the shape). Two of these ways can be easily controlled, either increase the initial spot size $\left({\sigma }_{x,0},{\sigma }_{z,0}\right)$ or change the beamline to decrease elements ($R_{12}$ and $R_{56}$).  The third way of increasing the resolution is to decrease the emittance (${\varepsilon }_x$, ${\varepsilon }_z$), however, this is not possible once the beam leaves the cathode since the emittance is an invariant. Therefore, the intrinsic emittance of the beam emitted from the cathode is the ultimate limit on the shaping resolution (assuming no collimation or cooling of the beam) since it cannot be reduced once it leaves the cathode.  We now turn to the intrinsic emittance generated at the cathode.

\paragraph{Intrinsic emittance~\label{sec3:cathodeEmit}}

Having established the fundamental importance of emittance in shaping, we now discuss the emittance of the electron bunch emitted from the cathode, i.e. the \textit{intrinsic emittance} (also called thermal emittance in the literature). Assuming there is no correlation between the three phase space planes of the electrons emitted from the cathode, the normalized intrinsic emittance (see Eq.~\eqref{normalized}) in 6D is given by the product of the three normalized intrinsic 2D emittances,
\begin{equation}
{\varepsilon }_{6D}^{n}={\varepsilon }_{x}^{n}{\varepsilon }_{y}^{n}{\varepsilon }_{z}^{n} ,
\end{equation}

Assuming there is no correlation between the position of an emitted electron and its transverse momentum, then  intrinsic rms normalized horizontal emittance at the cathode is,
\begin{equation} \label{eq:sec3:2Demit}
{\varepsilon }_{x}^{n}={{\sigma }_x}{{\sigma}_{p_x}},
\end{equation}
where ${\sigma }_x$=${\sqrt{<x^2>}}$ and ${\sigma}_{p_x}$=${\sqrt{<p_{x}^{2}>}/{m_{0}c}}$. Corresponding expressions apply for each coordinate so we can rewrite, Eq.~\eqref{eq:sec3:2Demit} to give the \textit{intrinisic rms normalized emittance at the cathode} for each of the 2D phase space planes as,
\begin{equation} \label{eq:sec3:2D-xemit}
{\varepsilon }_{c}^{n}={{\sigma }_c}{{\sigma}_{p_c}},
\end{equation}
where ${\sigma }_c$ and ${\sigma}_{p_c}$ are the initial rms bunch size and the dimensionless rms momentum at the cathode, respectively, in the horizontal/vertical/longitudinal plane. The accelerator designer can easily control ${\sigma }_c$, but ${\sigma}_{p_c}$ is an intrinsic property of the cathode.  Developing cathodes with low initial ${\sigma}_{p_c}$ is a challenging and active area of research for each of the emission mechanisms. The intrinsic rms normalized transverse emittance of all three cathode types (e.g., using the $x$-direction to be definite) ~\cite{flottmann-1997-a} can be written in terms of the kinetic energy of the electrons emitted from the cathode, $E_{k}$, and is given by,
\begin{equation} \label{eq:sec3-cathodeemittance}
{\varepsilon }_{x}^{n}={\sigma }_x\sqrt{\frac{{2E}_k}{3m_ec^2}} ,
\end{equation}
where $c$ is the speed of light, ${\sigma }_x$ is the rms horizontal spot radius on the cathode. The dimensionless rms horizontal momentum at the cathode, ${\sigma}_{p_x}$, is given by the \textit{square root term} in Eq.~\eqref{eq:sec3-cathodeemittance} and can also be thought of as the \textit{intrinisic rms normalized emittance at the cathode per rms spot size} (i.e.~${\sigma}_{p_x}$ = ${\varepsilon }_{x}^{n} / {\sigma }_x$) by virtue of Eq.~\eqref{eq:sec3:2Demit} and has units of $\upmu$m/mm. It is worth pointing out that this equation shows that the intrinsic emittance of the cathode depends on $E_k$ (of the emitted electrons) and implies that low emittance can be achieved if the electrons are emitted with small $E_k$.  Unfortunately, as we can see in the PE cathode case, as $E_{k}$ approaches 0, so does the emitted current $J_{PE}$~\eqref{eq:sec3-PCcurrentequation}. Thus there is no way to have both low intrinsic emittance and high current density.

The expression for the intrinisic rms normalized horizontal emittance at the cathode per rms spot size of any of the cathode types can be found by substituting the appropriate expression for $E_k$ from~\ref{sec3:cathReview:kineticE} into Eq.~\eqref{eq:sec3-cathodeemittance}. As an example, the expression for ${\sigma}_{p_x}$ for PE is found to be, 
\begin{equation} \label{eq:sec3-PE-transP}
{\varepsilon }_{x}^{n} / {\sigma}_x=\sqrt{\frac{h\nu -W }{3{m_e}{c}^2}},
\end{equation}
with typical values shown in the row \textit{Transverse Parameters} of~Table \ref{sec3:tab:cathode-characterisitics} along with typical spot sizes on the cathode, ${\sigma }_x$.  Note that it is becoming increasingly common to talk about the mean transverse energy ($MTE$ = $2 / 3$ $E_k$) of the electrons emitted from the cathode. By equipartition of energy, we know the mean longitudinal energy is $1/3$ $E_k$ and we can find an expression the  intrinisic rms normalized longitudinal emittance at the cathode per rms bunch length as,
\begin{equation} \label{eq:sec3-PE-transP}
{\varepsilon }_{z}^{n} / {\sigma}_z=\sqrt{\frac{{h\nu -W}}{6{m_e}{c}^2}},
\end{equation}

\subsubsection{Response time~\label{sec3:cathReview:respTime}} 
The response time of a cathode is defined as the temporal lag between the excitation of the cathode and emission of electrons.  While all types of cathodes have intrinsically short response times, in practice, the duration of the electron bunch emitted by FE and TE cathodes is controlled by the duration of the electric field, not the cathode response time. On the other hand, the emission of electrons from PE cathodes is gated by a laser pulse which can be much shorter than the electric field duration of the gun.  Therefore, the PE cathode is the only one used for generating longitudinally shaped bunches. 

The response time of PE cathodes varies from the fs-scale, for metallic photocathode, to 10’s ps-scale, for certain semiconductor photocathodes, ~\cite{smedley-2012-a}.  Row \textit{Longitudinal Parameters} of~Table \ref{sec3:tab:cathode-characterisitics} shows response times of the cathodes.  However, we do not list these times for FE and TE cathodes since they are not used for longitudinal shaping in electron guns.

%%%%%%%%%%%%%%%%%%%%%%%%%%%%%%%% CATHODE CHARACTERISTICS TABLE BELOW 
%\begin{table*}[] 
%    \caption{\label{sec3:tab:cathode-characterisitics}
%    Cathode characteristics}
%    \centering
%    \begin{tabular}{l |c | c | c |  c}
	
%    \toprule
%    & (large area) Field Emission & Thermionic Emission (TE) & Photoemission (PE) & units\\

%    \hline
%    {\bf Operating Conditions}
%	& $E_c\sim 100$~MV/m & $E_c\sim 100$~MV/m & $E_c\sim 100$~MV/m  \\
%    & $F\sim 10$~GV/m & $T=1400$~K & $\lambda\simeq$ 260~nm &  \\
%    & $W=4.5$~eV (tungsten) &  $W=1.6$~eV (dispenser) &  $W=4.6$~eV (copper) \\
	
%    \hline 
%    {\bf Excess kinetic energy} \\
%	typical $E_k$  & 0.3  & 0.12  & 0.2   & eV \\

%    \hline 
%    \multicolumn{5}{l}{\bf Transverse Parameters}  \\
    
%	\hline
%    typical spot size $\sigma_x$ & 1-3  & 1-3  & 0.1-10 & mm  \\
%    typical ${\varepsilon }_{x,y}$  & 0.3-1.0  & 0.12-0.36  & 0.02-2   &  $\upmu$m  \\

%    \hline
%    \multicolumn{5}{l}{\bf Longitudinal Parameters}  \\
    
%	\hline
%    emission time & --- & --- & $10^{-3}-10$ & ps \\ 
%    current density & $J_{FE}\le 1$ & $J_{TE}\le 1 $ & $J_{PE}\le 10^3$ & A/mm$^2$ \\
	
%	\hline
%    \end{tabular}
%\end{table*}

\begin{table*}[] 
    \caption{\label{sec3:tab:cathode-characterisitics}
    Cathode characteristics}
    \begin{ruledtabular}
    \begin{tabular}{l c  c  c   c}
	
    & (large area) Field Emission & Thermionic Emission (TE) & Photoemission (PE) & units\\

    \hline
    typical field at cathode ($E_c$)
	& $100$ & $100$ & $100$ & MV/m  \\
    Local field, temperature, wavelength & $F\sim 10$~GV/m & $T=1400$~K & $\lambda\simeq$ 260~nm &  \\
    typical work function ($W$) & $4.5$ (tungsten) &  $1.6$ (dispenser) &  $4.6$ (copper) & eV\\
	
    typical excess kinetic energy ($E_k$)  & 0.3  & 0.12  & 0.2   & eV \\
    typical spot size $\sigma_x$ & 1-3  & 1-3  & 0.1-10 & mm  \\
    typical ${\varepsilon }_{x,y}$  & 0.3-1.0  & 0.12-0.36  & 0.02-2   &  $\upmu$m  \\

    emission time & --- & --- & $10^{-3}-10$ & ps \\ 
    current density ($J$) & $\le 1$ & $\le 1 $ & $\le 10^3$ & A/mm$^2$ \\
	
    \end{tabular}
    \end{ruledtabular}
\end{table*}

%%%%%%%%%%%%%%%%%%%%%%%%%%%%%%%% END TABLE 

\subsubsection{Cathode selection guidance~\label{sec3:cathReview:select}} 

Based on the above considerations, the source designer must choose one of the three electron cathodes for their shaping application.  In general, the optimization process is complicated and this short Section only aims to give a flavor of the decision making process. The choice between cathodes is made by considering the trade-offs between the various factors shown in~Table \ref{sec3:tab:cathode-characterisitics}. The following factors must be considered by the electron source designer in choosing a cathode that meets the requirements of the shaping application.

\begin{enumerate}
\item  \textit{Longitudinal shaping.} If the application requires longitudinal shaping at the source then there is only one choice, choose PE. This is because PE cathodes are gated by the exciting photocathode laser pulse as has already been described. 

\item  \textit{Charge per bunch required.} This is a product of the transverse spot size and current density. The larger the spot size on the cathode the more charge.

\item  \textit{ Transverse and longitudinal emittance requirements}. These will determine the bunch shaping resolution.The smaller the spot size on the cathode the lower the emittance.
\end{enumerate}

As an example, consider an application that requires transverse shaping and low transverse emittance using an RF gun. At first it would appear that TE is the best choice based on it having the lowest intrinsic transverse emittance.  However, if the application requires 1000~pC, then TE's low current desnity $J_{TE}$ would require a large emission radius compared to PE cathode thus driving up the transverse emittance.  Therefore, a PE cathode would be a better choice. Further, even after the PE cathode gun is chosen, the choice of laser spot size on the cathode entails a tradeoff between charge and emittance. 

Based on the preceding analysis, it is clear that PE cathodes have the most shaping capabilities so the reader may wonder why FE and TE cathodes are pursued.  The reason is that PE cathodes come with the most operational complexity as they require large laser systems that must be actively synchronized the linac. TE cathodes do not require laser systems but have the added complexity of operating at $T>1000^\circ C$. While FE cathodes do not require laser systems or heating so are the simplest of all.  However, if longitudinal shaping is required for FE or TE cathodes, then a gating system will be required which will increase the operational complexity.

We end this Section with the caveat that we left out two sub-classes of cathodes, namely, hybrid cathodes that combine two of the three basic emission mechanisms.  The two hybrid cathodes are: photo-assisted field emission cathode~\cite{swanwick-2014-a, mustonen-2011-a} and photo-assited thermionic emission cathode~\cite{sun-2006-a}.  These are potentially important cathodes, since they open up the possibility of longitudinal shaping for FE and TE cathodes, however, we do not cover them here due to space limitations.

\subsection{Smooth distributions~\label{sec3:smooth}}

We now turn to the most common initial electron distribution emitted from an electron gun, a smooth bunch.  The methods introduced in this section focus on generating smooth bunch profiles, and therefore, bunches with low intrinsic emittance, Eq.\eqref{eq:sec3-cathodeemittance}, in order to provide high shaping resolution. In this case, shaping is accomplished by shaping systems located downstream of the electron gun described in Sections~\ref{sec4}, \ref{sec5}, and~\ref{sec6} of this review.  In this Section, we describe how the smooth distributions are produce for each of the three cathode types~\cite{jensen-2018-a}.

\subsubsection{FE based smooth transverse distributions~\label{sec3c1}} 

\begin{figure} 
\includegraphics[width=0.48\textwidth,keepaspectratio=true]{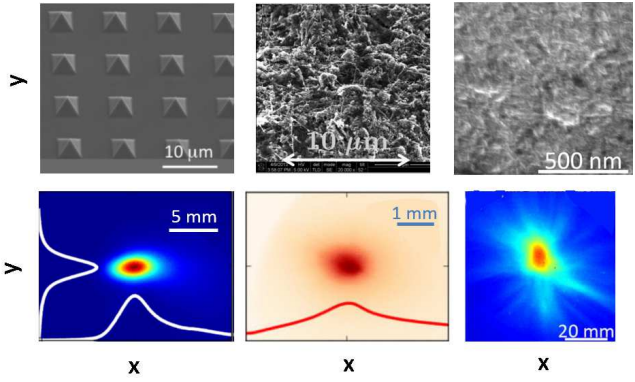}
\caption{ Three types of large area FE cathodes.  Surface images (top row, left-right) of FEA, CNT, and UNCD cathodes. The distance between the emitters is largest for FEA (10~$\upmu$m), smaller for CNT (0.1-1~$\upmu$m) and smallest for UNCD (10-100~nm). Smooth beam images (bottom row, left-right) generated by FEA, CNT, and UNCD cathodes. Figure adapted from \cite{piot-2014-a,mihalcea-2015-a,baryshev-2014-a}.\label{sec3:fig5:FEA-default-trans-shapes} }
\end{figure}

FE cathodes are thought to have the potential of providing a robust source of low emittance electrons. As discussed in~\ref{sec3:cathReview:curDen}, large area, engineered FE cathodes, can be based on field emission arrays (FEAs), carbon nanotubes (CNTs) or ultra-nanocrystalline diamonds (UNCDs). All of their surfaces are composed of a quasi-continuous distribution of electron emitters (Fig.~\ref{sec3:fig5:FEA-default-trans-shapes}). In this Section, we describe demonstrated methods of producing smooth transverse distributions for each of the engineered FE cathode types.

The FEA is an arrangement of a large number of discrete tips on the cathode surface. The emission sites of the FEA, shown in Fig.~\ref{sec3:fig5:FEA-default-trans-shapes}(a), come from diamond tips on pyramid bases~\cite{piot-2014-a} separated by 10~$\upmu$m.  The CNTs can be deposited into regular arrays, like the FEA, or randomly oriented on the cathode surface. In the later case, emission sites of the CNT, shown in Fig.~\ref{sec3:fig5:FEA-default-trans-shapes} (top, center), comes from the randomly oriented fibers covering the CNT surface~\cite{mihalcea-2015-a}. The emission site of the ultra-nanocrystalline diamond (UNCD) cathode, shown in Fig.~\ref{sec3:fig5:FEA-default-trans-shapes}(top, right), is believed to come from the grain boundaries of the UNCD thin film deposited on the surface ~\cite{baryshev-2014-a}. The emitter separation of the FEA, UNCD, and CNT cathodes are approximately 1-10~$\upmu$m, 0.1-1~$\upmu$m, and 10-100 nm, respectively (Fig.~\ref{sec3:fig5:FEA-default-trans-shapes}, top row). All engineered FE cathode types have experimentally produced smooth electron distributions (Fig.~\ref{sec3:fig5:FEA-default-trans-shapes}, bottom row).  The smoothness of the transverse distribution from large-area FE cathodes is limited by two factors: the need to merge emission from discrete emission sites and non-uniformity of the emitters across the cathode.   The discreteness of the emitters becomes less noticeable as the separation between sites decreases.  Further, another limitation of operating a large area FE cathode in an RF gun is the large energy spread and long duration of the electron bunches due to emission taking place over a large range of RF phases. Even though these are longitudinal parameters, they adversely affect the transverse properties of the bunch. Most importantly, the large energy spread leads to strong chromatic aberrations.

\subsubsection{TE based smooth transverse distributions~\label{sec3c2}} 
The TE cathode gun is one of the simplest and most robust electron sources; it is the workhorse of light source facilities around the world~\cite{lightsources-a}.  Historically, the TE cathode gun has not been used for the more demanding application of driving FELs, due to its lower transverse beam brightness and difficulty with achieving a short pulse.  However, researchers at SACLA~\cite{asaka-2017-a} have developed a low-emittance thermionic-gun-based injector. In the injector, electron beams are emitted from a CeB$_6$ thermionic cathode of 3-mm diameter located in a DC 500-kV gun followed by a beam chopper and a bunch compresso to produce an electron beam with high peak current (3–4 kA) and low transverse normalized-slice emittance (below 1 $\upmu$m) sufficient to drive a compact free-electron laser. 

\subsubsection{PE based smooth transverse distributions~\label{sec3c3}} 

\begin{figure} 
\includegraphics[width=0.48\textwidth]{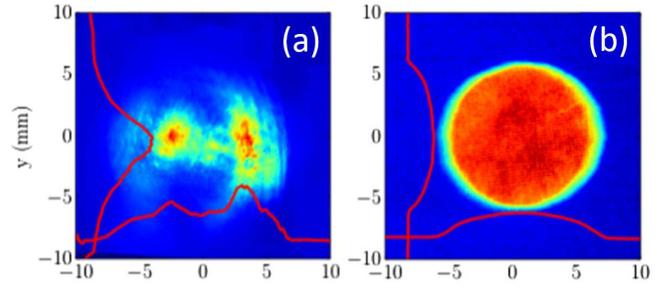}
\caption{Comparison of inhomogeneous (a) and homogeneous (b) laser profiles at the plane of the PE cathode.\label{sec3:fig1:homogen-v-nonhomogen} From ~\cite{halavanau-2017-a}.}
\end{figure}

Many applications of the electron source require smooth transverse distributions, such as the uniform flattop or the Gaussian profile.  While the typical IR output from the photocathode laser (TEM$_{00}$) is very nearly Gaussian at the output, its shape can become distorted during the frequency upconversion process to generate UV. In the typical case, smoothing is usually needed due to inhomogeneities in the far-field image of the UV beam at the PE cathode plane (see ~Fig.~\ref{sec3:fig1:homogen-v-nonhomogen}(a)) to produce the desired homogeneous profile (see Fig.~\ref{sec3:fig1:homogen-v-nonhomogen}(b)). Methods to achieve smooth transverse electron distributions with PE cathodes use optical elements, inserted into the laser path, of which there are two types: passive (e.g., micro-lens array) and active (e.g., deformable mirror). In general, passive elements are simpler but active elements have greater capabilities for obtaining complicated transverse distributions as we describe below.

\begin{figure} 
\includegraphics[width=0.48\textwidth,keepaspectratio=true]{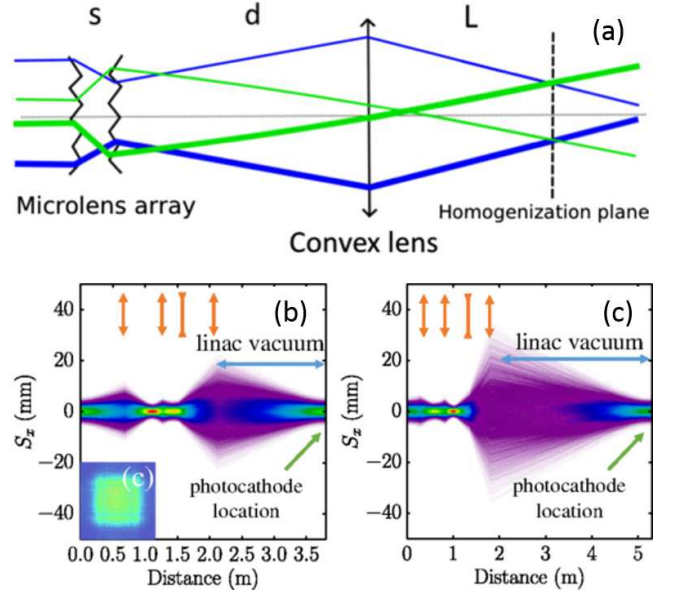}
\caption{Microlens array (MLA) based homogenization system. (a) Diagram of several rays passing through the MLA pair onto the homogenization plane. (b,c) ray tracing from the homogenization plane, through imaging optics and to the photocathode plane for two different systems. From~\cite{halavanau-2017-a}. \label{sec3:fig2:MLA}}
\end{figure}

Microlens arrays (MLAs) are passive optical elements used to homogenize the laser's transverse profile~\cite{bich-2008-a}. ~Fig.~\ref{sec3:fig2:MLA}(a) shows a schematic of homogenizing optics where a UV laser beam ($248$~nm) passes through a pair of MLAs followed by a convex lens resulting in a continuous and homogenized laser profile at the homogenization plane (see Fig.~\ref{sec3:fig1:homogen-v-nonhomogen}(b)). The MLA system is located outside the beam vacuum, and the laser emerging from the MLA system (at the homogenization plane) has large beam divergence, which makes transport to the cathode difficult.  A solution was found in Ref.~\cite{halavanau-2017-a} where the laser was imaged from the homogenization plane to the PE cathode plane located approximately 3.5 m away with an imaging system Fig.~\ref{sec3:fig2:MLA}(b,c).  Another advantage of passive homogenization systems (e.g. MLA) have over active ones is that they will homogenize a laser profile even if its profile fluctuates from shot to shot.  Active homogenization systems cannot do this because they use feedback loops to transform an incoming inhomogeneous laser profile into a homogeneous one which requires the incoming profile to be stable on the time-scale of the feedback loop. On the other hand, the passive system will instantaneously transform an arbitrary inhomogeneous laser profile into a homogeneous.

\begin{figure} 
\includegraphics[width=0.48\textwidth,keepaspectratio=true]{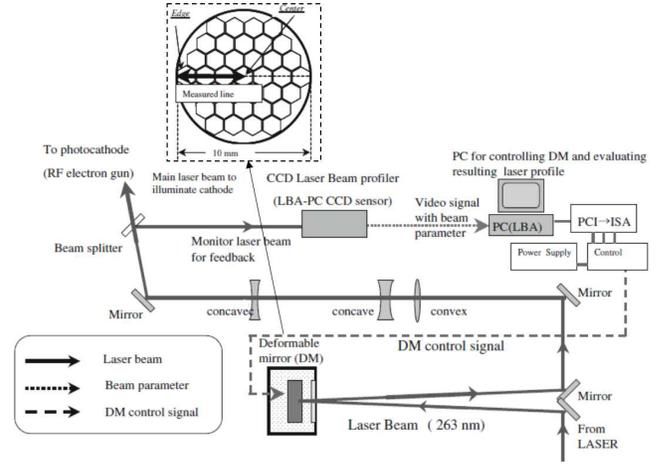}
\caption{Deformable mirror (DM) based homogenization system. A deformable mirror (inset) is actively adjusted via a computer [PC(LBA)] in a feedback loop to homogenize the laser profile on the photocathode by monitoring the profile at the virtual cathode location with a camera (LBA-PC CCD sensor). From~\cite{matsui-2008-a}.\label{sec3:fig3:deformable-mirror} }
\end{figure}

\begin{figure}
\includegraphics[width=0.48\textwidth,keepaspectratio=true]{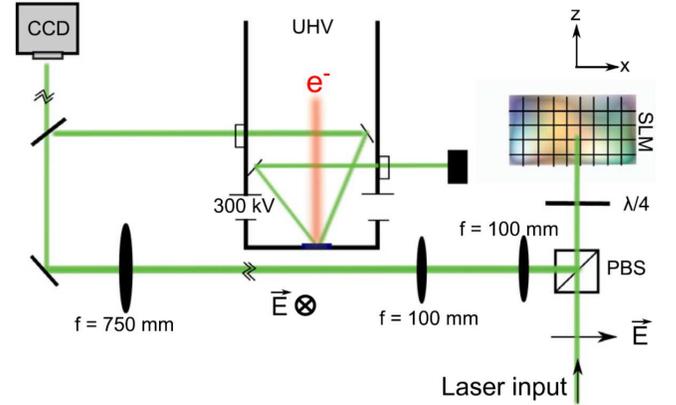}
\caption{Spatial light modulator (SLM) homogenization system. An SLM is actively adjusted via a computer (not shown) in a feedback loop to homogenize the laser profile on the photocathode by monitoring the profile at the virtual cathode location with a camera (CCD). From~\cite{maxson-2015-a} \label{sec3:fig4:SLM} }.
\end{figure}

Active optical elements provide a flexible, yet more complicated system for obtaining a homogenized transverse distribution at the phototcathode. The first systems were based on deformable mirrors (DMs) consisting of an array of electrically adjustable small mirrors, see Fig.~\ref{sec3:fig3:deformable-mirror}. The intensity profile of the laser pulse is controlled by adjusting the angles of the small mirrors which are under the control of a computer. After the laser beam reflects off the DM, a beam splitter sends a small fraction of the beam to the CCD camera located at the \textit{virtual cathode} while the majority continues to the PE cathode.  Numerical optimization algorithms, such as the genetic algorithm (GA), are run on a computer to adjust the angles of the small mirrors to optimize the profile~\cite{matsui-2008-a}. The DM method is still actively underdevelopment by researchers at~\cite{li-2017-a}. A second active approach is based on spatial light modulators (SLMs) as shown in Fig.~\ref{sec3:fig4:SLM}. While SLMs work on a different optical principal (birefringence) than DMs (reflection), their shaping functionality is the same. The SLM is placed in a feedback loop which monitors the transverse profile of the laser (again, a small fraction at the virtual cathode) while a computer running a GA is used to control the SLM element. In Fig.~\ref{sec3:fig4:SLM}, laser light enters from the bottom (in the z direction) and is polarized along x. A quarter-wave plate and SLM act as a polarization rotator with spatial dependence, which shapes the light when used with a polarizing beam splitter (PBS). The surface of the SLM is then 4-f imaged ($f$~=100mm lens pair) onto an intermediate plane to preserve the beam divergence, and then this intermediate plane is imaged with a single long focal length lens ($f$~=750mm) onto either the photocathode or a CCD. An ultrahigh vacuum (UHV) mirror reflects light to the center of the photocathode. The SLM-based system was found to have greater capacity in handling poor input laser quality~\cite{maxson-2015-a} than the DM-based system.  On the other hand, the SLM only works with IR and visible light while DM-based methods can work in the UV~\cite{li-2017-a} thus avoiding the distortions in the up-conversion process.  In other words, this continues to be a lively area of research.

\subsection{Shaped distributions~\label{sec3:shaped}}

Transverse bunch shaping methods have been demonstrated for both FE and PE cathodes.  In this subSection, we present shaping methods developed for these two emission mechanisms. 

\subsubsection{FE based shaped transverse distributions~\label{sec3:shaped:trans}} 

Transverse shaping of the electron distribution generated by large area FE cathodes is controlled by engineering the emitting surface.  This is an active research area but is not yet capable of generating high quality electron bunches suitable for the modern electron linacs we consider here.  None the less, due to the recent activity in this area coupled with its great potential to be used in electron linacs we present it here.  In particular, FEA-based cathodes have been used to generate both continuous (e.g., triangular) and modulated (e.g., array of beamlets) transverse distributions.

\begin{figure} 
\includegraphics[width=0.45\textwidth,keepaspectratio=true]{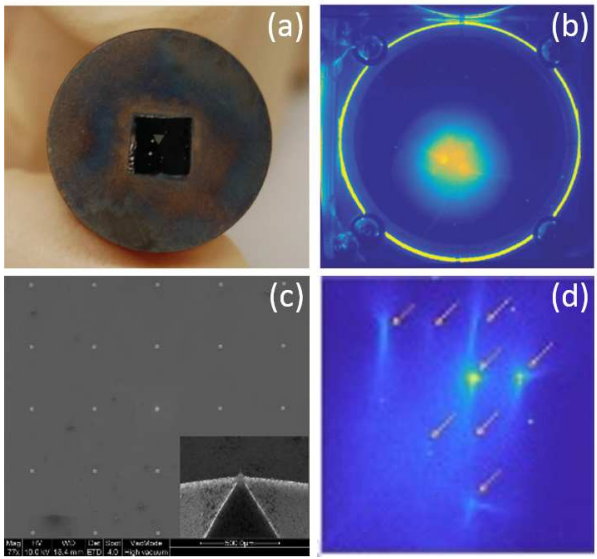}
\caption{FEA cathodes for transverse shaping.  Photograph of an FEA cathode with a triangular array inside a central square area of a round cathode plug (a) and its corresponding electron bunch image on a downstream screen showing a slight triangular shape (b). SEM image of rectangular array on a cathode plug (c) with diamond tips (inset), and its corresponding electron bunch image on a downstream screen showing a grid-shaped pattern (d). 
From \cite{andrews-2020-b,Nichols-2020-a}.
\label{sec3:fig7:FEA-nonuniform-trans-shapes} }
\end{figure}

As an example of a transversely shaped distribution, a triangular array of pyramid emitters was deposited into a 1-mm equilateral triangle with $\sim 10$~$\upmu$m spacing on a cathode plug ~\cite{andrews-2020-b}, as shown in ~Fig.~\ref{sec3:fig7:FEA-nonuniform-trans-shapes}(a). The downstream electron bunch image captured on an electron imaging screen, as shown in ~Fig.~\ref{sec3:fig7:FEA-nonuniform-trans-shapes}(b),  suggests a triangular shape, but space-charge effects and the long phase emission period are suspected to have blurred the image.  The emission period can be shortened with gated FEAs ~\cite{jarvis-2009-a}. Modulated transverse distributions (e.g., an array of spots) have been generated at the source with FEA cathodes, see ~Fig.~\ref{sec3:fig7:FEA-nonuniform-trans-shapes}(c) ~\cite{Nichols-2020-a}.  The downstream electron bunch image ~Fig.~\ref{sec3:fig7:FEA-nonuniform-trans-shapes}(d) shows that original modulation was maintained but degraded due to the non-uniformity of the emitters.  Note that as spacing of the emitters gets closer (e.g., a nanoengineered FEA ~\cite{graves-2012-a}, it becomes more difficult to maintain the modulation. In order to reproduce the initial source modulation out of the gun, the charge must be kept very low, to avoid space charge dilution, and the FEA must be gated to keep energy spread low.

\subsubsection{PE based shaped transverse distributions~\label{sec3:shaped:PEtrans}} 

In this Section, we present methods for shaping the transverse distribution of the electron bunch generated by a PE cathode in an electron gun.  This distribution is controlled by controlling the transverse profile of the photocathode laser beam.  

\begin{figure} 
\includegraphics[width=0.45\textwidth,keepaspectratio=true]{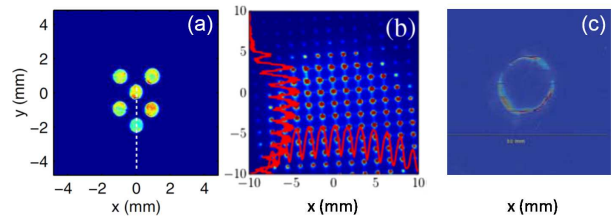}
\caption{Transversely shaped electron bunch distributions from a PE cathode. An array of spots generated with an optical mask (a) and an MLA (b). A hollow beam generated with an optical mask (c). Images from \cite{rihaoui-2009-a,wisniewski-2012-a,halavanau-2017-a}.\label{sec3:fig6:PE-nonuniform-trans-shapes} }
\end{figure}

Certain accelerator applications require shaped electron bunches with modulated transverse distributions, such as an array of beamlets [(~Fig.~\ref{sec3:fig6:PE-nonuniform-trans-shapes}(a,b)] or hollow beams [~Fig.~\ref{sec3:fig6:PE-nonuniform-trans-shapes}(c)]. To date, these patterns have only been generated with passive optical systems in PE cathode guns based on an optical mask~\cite{rihaoui-2009-a,wisniewski-2012-a} or an MLA~\cite{halavanau-2017-a}. In the former case, an optical mask is inserted into the laser path to block the unwanted part of the laser beam to create the desired laser pattern, which is then imaged onto the PE cathode to create the electron distribution on the cathode. Examples include: an aluminum plate with six holes was used to generate a \textit{positive} array of laser spots [(Fig.~\ref{sec3:fig6:PE-nonuniform-trans-shapes}(a)], and a painted quartz plate was used to create a \textit{negative} hollow laser ring [(Fig.~\ref{sec3:fig6:PE-nonuniform-trans-shapes}(c)].  In the latter case, an MLA system was used to create a large array of spots [(Fig.~\ref{sec3:fig6:PE-nonuniform-trans-shapes}(b)] by changing the location of the convex lens (Fig.~\ref{sec3:fig2:MLA}(top)) as described in the reference.

\subsubsection{PE based shaped longitudinal distributions~\label{sec3:shaped:PElong}}  

As stated above, due to the difficulties in gating FE and TE cathodes at the picosecond (and shorter) timescale, longitudinal bunch shaping is the domain of PE cathode guns. The longitudinal bunch shape of the electron distribution generated by a PE cathode gun depends on both the temporal laser pulse shape and the response time of the PE material. Photocathode laser systems can generate laser pulse duration ranging from 10's fs to 10's of ps while PE cathode response times range from the fs-scale, for metallic cathodes, to 10's ps for some semiconductor cathodes~\cite{dowell-2008-a}. For longitudinal shaping applications via the electron gun, one chooses a PE cathode with a response time much less that the duration of the laser pulse. In this way, the electron bunch temporal shape will simply follow the laser temporal shape. Laser pulse shaping methods have been used to generate both single bunches (smooth and shaped) and bunch trains. Various methods used to control the longitudinal distribution are presented below and are categorized as either frequency domain or time domain. 

\paragraph{Frequency domain laser shaping~\label{sec3:shaped:PElong:freq}}

Frequency domain methods manipulate the frequency spectrum of the input laser pulse to control the time profile of the output laser pulse. There are two approaches to frequency domain-based laser shaping that have been applied to PE cathode guns: acousto-optic programmable dispersive filter and Fourier Transform pulse shaping.

\begin{figure} 
\includegraphics[width=0.48\textwidth,keepaspectratio=true]{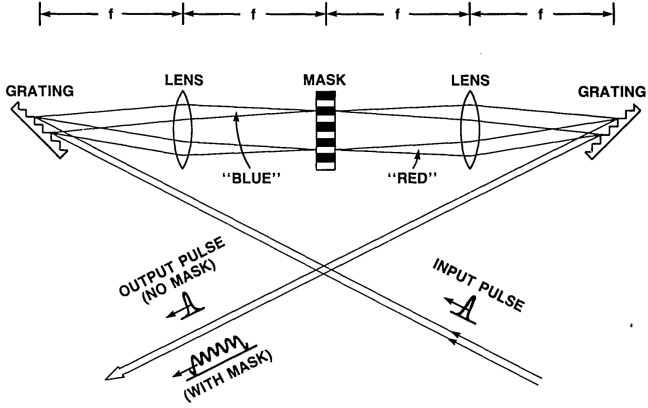}
\caption{Fourier Transform laser pulse shaping system.  A laser pulse (INPUT PULSE) enters a 4-f grating stretcher and its frequency spectrum is exposed at the Fourier Transform Plane. A mask (MASK) described by a frequency domain transfer function, $H(\omega)$, is placed at this plane to modify the spectrum of the input laser pulse resulting in a modified time structure of the output laser (OUTPUT PULSE). From~\cite{weiner-1998-a}. \label{sec3:fig8:long-freq-domain-4f-fourier-plane-mask} }
\end{figure} 

The acousto-optic programmable dispersive filter (AOPDF) relies on a longitudinal interaction between a polychromatic acoustic wave and a polychromatic optical wave in the bulk of a birefringent crystal.  It controls the IR spectrum by controlling the group delay versus wavelength with a programmable acoustic wave in the birefringent crystal.  Optical signals in the hundreds of Terahertz range are controlled with RF signals in the tens of MHz range.  This is a compact IR device, installed between the photocathode laser oscillator and amplifier, with high temporal resolution, which makes it suitable for the shaping of femtosecond IR pulses. A widely utilized commercial AOPDF---the {\sc dazzler}\texttrademark---can shape pulse length over a 6-ps duration at a maximum repetition rate close to a MHz~\cite{tournois-1997-a, verluise-2000-a}.  

In Fourier Transform pulse shaping, a 4-f grating stretcher is used to expose the spectrum of the input laser pulse in the spectral Fourier Transform plane of the stretcher (see~Fig.~\ref{sec3:fig8:long-freq-domain-4f-fourier-plane-mask}).  The spectrum can be controlled with an optical element placed in this plane. Let the spectrum of the input laser pulse be given by $X(\omega)$ and the frequency-domain transfer function of the optical element at the Fourier plane be represented by, $H(\omega)$, then the spectrum of the output laser pulse is given by their convolution,
\begin{equation} \label{eq:sec3-mask}
Y(\omega)=X(\omega)H(\omega),
\end{equation}
so that the output laser pulse in the time-domain is just the inverse Fourier Transform of this convolution, 
\begin{equation} \label{eq:sec3-mask}
y(t)=F^{-1}[Y(\omega)],
\end{equation}

The first systems used for shaping with an RF photocathode gun~\cite{neumann-2003-a, neumann-2009-a} used a fixed mask, described by $H(\omega)$, located in the Fourier Transform plane to modify the amplitude of the input spectrum. However, this method can be based on the modification of any of the amplitude, phase, or polarization of the input spectrum. described by the appropriate $H(\omega)$.  After the spectrum is modified, the second grating is used to bring the spectrum back to a line in the time domain to generate the desired temporal pulse shape of the output pulse. The laser community has achieved both laser pulse trains and temporal flattop laser pulses (Fig.~\ref{sec3:fig8:long-freq-domain-4f-fourier-plane-mask}) using masks to modify the amplitude of the spectrum~\cite{weiner-1998-a}.  Recent Fourier Transform pulse shaping approaches are based on programmable spatial light modulators (SLMs) due to their superior resolution and flexibility.  The first beam physics applications used an SLM system to convert a Gaussian pulse of 9 ps FWHM to a flattop pulse of the same length to drive a photocathode RF gun~\cite{yang-2002-a}.   Gaussian pulses were transformed into both triangular and super-triangular laser pulses with an SLM-based system~\cite{kuzmin-2018-a} as shown in~Fig.~\ref{sec3:fig10:SLM-long-shapes}.

\begin{figure} 
\includegraphics[width=0.48\textwidth,keepaspectratio=true]{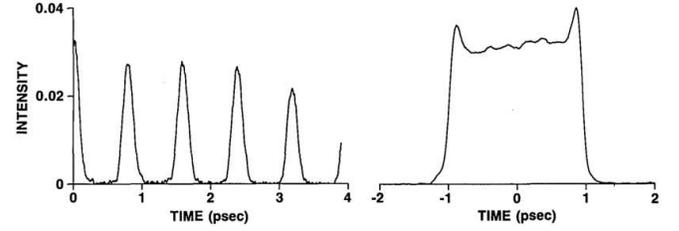}
\caption{Demonstrated laser pulse temporal shapes generated with the Fourier Transform laser pulse shaping system of~Fig.~\ref{sec3:fig8:long-freq-domain-4f-fourier-plane-mask} using an amplitude mask. Laser pulse train (left) and quasi-flattop laser pulse (right). Courtesy ~\cite{weiner-1998-a}. \label{sec3:fig10:SLM-long-shapes} }
\end{figure}

\begin{figure} 
\includegraphics[width=0.4\textwidth,keepaspectratio=true]{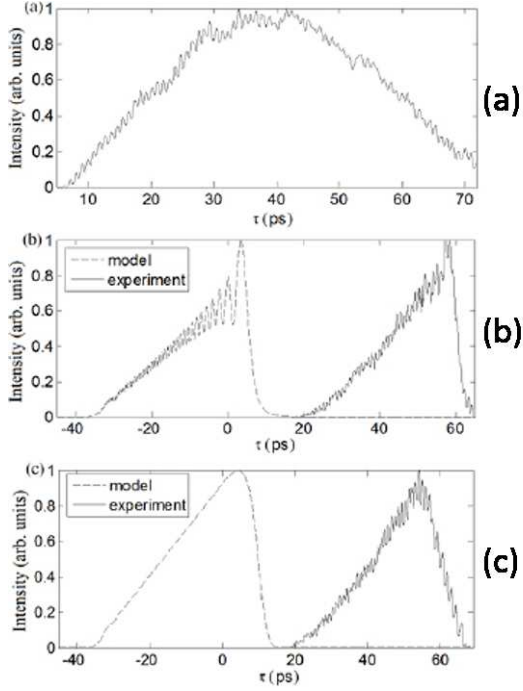}
\caption{Comparison of simulated and demonstrated laser pulse temporal shapes achieved with a Fourier Transform shaping system using SLMs. (a) unmodified laser pulse shape; (b) spectrum optimized for a right triangle; (c) spectrum optimized for a right triangle with the leading edge replaced by a super-Gaussian section. Courtesy ~\cite{kuzmin-2018-a}. \label{sec3:fig10:SLM-long-shapes} }
\end{figure}

The advantage of the frequency domain method is its high resolution and flexibility to produce various pulse shapes while the downside is its complexity and stability.  In addition, the need to transmit the IR laser pulse shape through the laser amplifier and harmonic conversion crystals introduces non-linearities causing differences between the IR and UV laser pulse shapes. Recent trends point toward further development of programmable Fourier Transform pulse shaping applied directly in the UV.  Another solution would be the development of PE cathodes that can respond to IR; this is another active area of research. 

\paragraph{Time domain laser shaping~\label{sec3:shaped:PElong:time}} 

\begin{figure}
\includegraphics[width=0.48\textwidth,keepaspectratio=true]{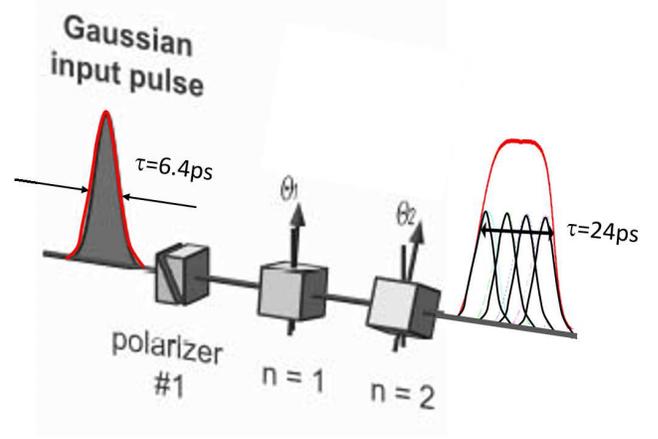}
\caption{Flattop laser pulse generated with $\alpha$~-BBO laser pulse stacking. From~\cite{will-2008-a} \label{sec3:fig11:alphaBBO-long-shapes} }
\end{figure}

\begin{figure} 
\includegraphics[width=0.425\textwidth,keepaspectratio=true]{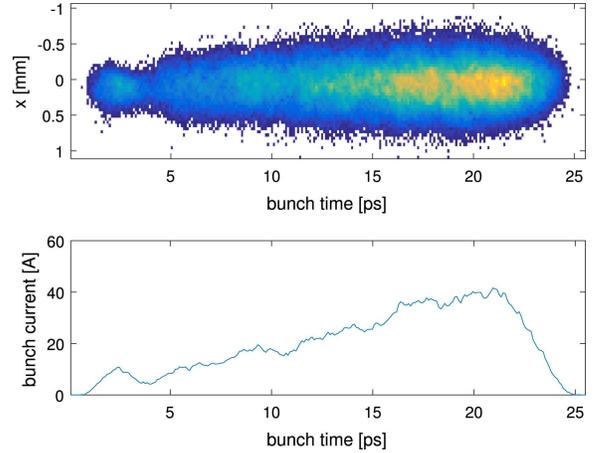}
\caption{Measured double-triangular electron bunch generated with $\alpha$~-BBO laser pulse stacking. (top) x-z projection of the bunch on a screen. (bottom) corresponding current profile. From~\cite{loisch-2018-a}.  \label{sec3:fig12:PITZ_alphaBBO_triangle} }
\end{figure}

Time domain methods are based on laser pulse stacking. This is where a series of short laser pulses are longitudinally combined (i.e., stacked) to form the desired longitudinal profile.  There are a number of methods that have been tried in the past. An early method used for PE cathode guns~\cite{siders-1998-a} was based on a Michelson interferometer in which $n$ 50/50 beam splitters, embedded in a set of optical delay legs, were used to split a single input laser pulse into a series of $2^{n}$ output laser pulses. Their device was used to generate a train of 16 pulses at THz spacing with very low loss. More recently, pulse stacking with birefringent crystals is almost exclusively used.  It exploits the group velocity mismatch between the ordinary and extraordinary axes of a birefringent crystal (e.g. $\alpha$~-BBO)~\cite{zhou-2007-a,will-2008-a,power-2009-a}. When a single Gaussian input pulse passes through a uniaxial birefringent crystal rotated at an angle relative to the crystal axis, two output pulses will emerge projected onto the ordinary and extraordinary axes; see Fig.~\ref{sec3:fig11:alphaBBO-long-shapes}. The time delay difference between these two components is due to the differing group velocities along the ordinary and extraordinary axes
\begin{equation}
\Delta t=L_{Xtal}\left(\frac{1}{v_{g,e}}-\frac{1}{v_{g,o}}\right),
\end{equation}
where $L_{Xtal}$ is the length of the crystal, and $v_{g,e}$ and $v_{g,o}$ are the group velocities associated with the extraordinary and ordinary optical axes, respectively. The crystal length controls the delay between the different polarizations while crystal angle controls the relative intensity of the different polarizations. If the angles are set to $45^{\circ}$, a flattop laser pulse is obtained~\cite{will-2008-a}, as shown in~Fig.~\ref{sec3:fig11:alphaBBO-long-shapes}.  However, it is also possible to obtain more complicated shapes (e.g., a double triangle) by passing the input laser pulse through a stack of crystals rotated at optimized angles relative to the incident laser pulse. In ~\cite{loisch-2018-a,liu-2019-a}, they used this method to generate a double-triangular laser pulse which was then used in PE cathode gun to generate the corresponding double-triangular electron bunch as shown in~Fig.\ref{sec3:fig12:PITZ_alphaBBO_triangle}.

The advantage of the time domain method lies in its simplicity.  Moreover, if the pulse stacker is implemented in the UV~\cite{power-2009-a} the distortion of the pulse due to nonlinearities arising during the frequency-conversion process is completely averted. Its downside is its limited flexibility since there is no way to change the crystal length on the fly though some control over the pulse shape can be achieved by rotating the crystals.

In addition to producing shaped the ``continuous" distributions shown above (i.e. triangle and flattop), birefringent crystals have also been used to produce bunch trains. Reference~\cite{li-2008-c} proposes a simple and more compact system specifically adapted to bunch train generation. The setup consists of a concentric stair-step echelon combined with a focusing lens. The echelon consists of a series of concentric flat zones with different thicknesses, so as to introduce a discrete delay correlated to the transverse radius. At the focal point of a downstream lens, such a configuration produces pulses that are delayed in time. The nature of the echelon design introduces small delays so that the method is well adapted to the generation of pulses with ps-scale temporal separation. In~\cite{li-2008-c} the technique is numerically investigated for the formation of a train comprising 20- to  100-fs bunches with an associated bunching factor peaked at $\sim 0.5$~THz for application to a coherent Smith-Purcell THz source based on a 50-kV electron beam.  

\subsubsection{PE based spatio-temporal (3D) shaping}

A natural continuation of the progress with programmable 2D transverse and 1D longitudinal laser shaping are methods for shaping the complete 3D spatio-temporal distribution of the laser. This method gives the electron source designer complete control over the initial electron distribution in a PE cathode gun. Recall, however, that this does not mean complete control over electron distribution in general, since, once it leaves the cathode it is subject to nonlinear space-charge forces that will distort the initial distribution. Interestingly, one of the main motivations for this line of R\&D is to generate electron bunches with uniform 3D ellipsoid distributions since these are the only distributions whose space-charge fields (i.e., internal force fields) are linear functions of position ~\cite{luiten-2004-a}. This gives rise to particularly simple dynamical behavior: a uniform ellipsoid under the influence of its self-fields (electrostatic) will change its size but retain its shape; a uniform ellipsoid with linear internal fields. In addition to the 3D ellipsoid laser pulse shapes, other distributions of interest is the 3D cylindrical distribution and distributions with 2D circular transverse distributions but combined with triangular 1D longitudinal distributions.

\begin{figure} 
\includegraphics[width=0.4\textwidth,keepaspectratio=true]{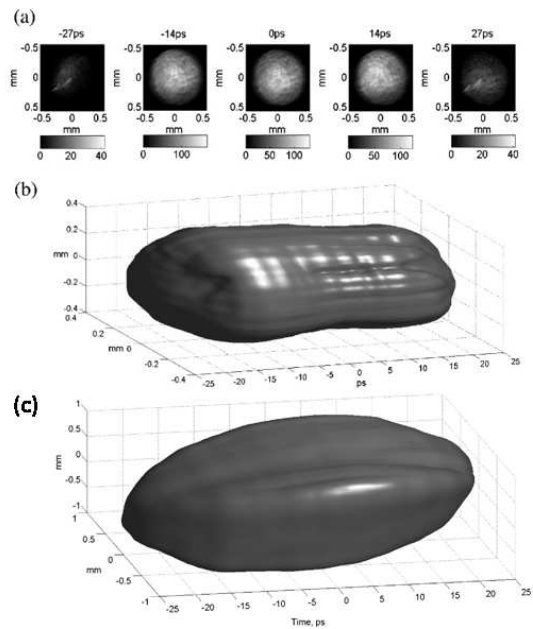}
\caption{Demonstrated 3D laser distributions based on SLMs. (a) measured transverse intensity distributions of quasi-cylindrical pulses; (b) reconstructed 3D distribution of quasi-cylindrical; (c) reconstructed 3D distribution of quasi-ellipsoidal. From~\cite{mironov-2016-a}.  \label{sec3:fig13:3D-quasi-shapes} }
\end{figure}

Recent progress in 3D laser shaping with IR laser pulses has been achieved by several groups~\cite{mironov-2016-a,kuzmin-2019-a,kuzmin-2020-a}.  For example, researchers used multiple programmable SLMs (Fig.~\ref{sec3:fig13:3D-quasi-shapes}) to generate both a 3D quasi-cylinder and 3D quasi-ellipsoidal IR laser distribution~\cite{mironov-2016-a}. In addition to SLM-based techniques, 3D shaping can also be accomplished via control of chromatic aberration. as discussed by~\cite{li-2008-a}, where a {\sc dazzler}\texttrademark~system is used to introduce a complex spectral structure that, in combination with a highly dispersive section, resulted in the generation of a 3D ellipsoidal bunch~\cite{li-2008-b}. 

Despite the successful demonstration of these 3D IR laser distributions, these have not yet been used to extract electrons from a photocathode gun.  The next steps require converting the IR pulse shapes to the UV where they can be used to exicte PE cathodes.  Nonetheless, this is an exciting and important area of research.  Further, it is interesting to note that the demand for 3D laser pulse shaping, and its subsequent progress, has been lead by the PE cathode gun community, not the laser community.

\section{Beam  control  within  one  degree  of freedom using external fields \label{sec4}}

This section discusses the use of externally applied  fields to shape the beam distribution. We introduce the general formalism and discuss various methods commonly employed or recently proposed  to control the beam distribution in longitudinal or transverse directions and some of the associated projections. 

\subsection{General considerations}
\begin{figure}
 \includegraphics[width=0.95\linewidth]{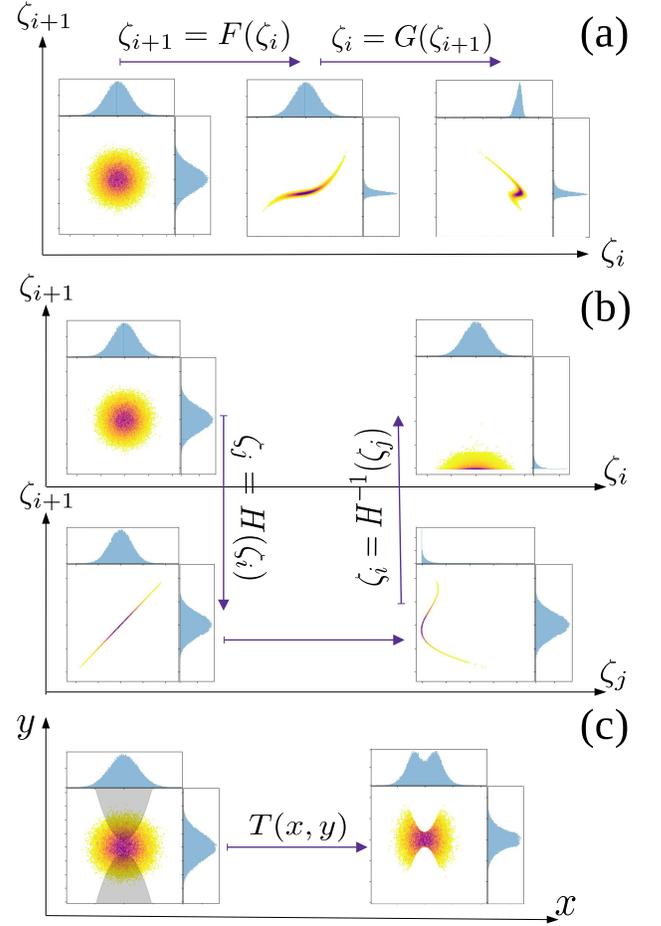}
\caption{\label{fig:sec4:overview}Overview of phase-space shaping method discussed in Section~\ref{sec4} with the labels ``initial" and "final" referring to the initial and final distributions. Shaping via non-linear fields and transport (a): a momentum dependence on the coordinate is introduced via the function $F$ and combined with a downstream beamline introducing a momentum-dependent position via $G$ so that $\zeta_i \mapsto G\circ F (\zeta_i)$. Shaping using a local correlation between two DOFs (b): one of the coordinates $\zeta_i$ is coupled to another DOF coordinates $\zeta_j$, via the transformation $\zeta_j=H(\zeta_i)$, $\zeta_j$ and the coupling removed via $H^{-1}$. Shaping through collimation (c): an intercepting mask in the $(x,y)$ plane shown as a shaded area applies a transmission function $T(x,y)$ resulting in a shaped transverse distribution.}
\end{figure}

We categorize beam-shaping methods using external fields into three main approaches as depicted in  Figure~\ref{fig:sec4:overview}. The first approach consists of finding a phase-space transformation that maps a given initial distribution to a final target distribution; see Fig.~\ref{fig:sec4:overview}(a). A second approach introduces a local coupling between two variables so that the overall transformation is still within one degree of freedom (DOF) but the local coupling enables access to another variable Fig.~\ref{fig:sec4:overview}(b). A common example is the use of dispersive collimation where particles with given energies are removed via the introduction of a local correlation between energy and position. Finally, another technique employs an interceptive mask that modifies the momentum or affects the particles' transmission according to their transverse position Fig.~\ref{fig:sec4:overview}(c).  

The use of an external electromagnetic field to impart correlation within one DOF is commonly employed to control the beam properties along any of the DOFs. For instance time-dependent fields are commonly used to introduce, e.g., a linear correlation in the longitudinal phase space (LPS) such as $\delta =hz$, that can later be exploited using a longitudinally dispersive beamline to alter the final bunch length via $ \sigma_{z,f}= C^{-1}\sigma_{z,0}$, where the compression factor is defined in Eq.~\eqref{compf} and  $h$ and $R_{56}$ are, respectively, the chirp in the incoming ($z,\delta$) LPS and the longitudinal dispersion associated with the dispersive section. It should be pointed out that the same mechanism applies to the other DOFs, e.g., in the transverse planes to focus the beam. 

Generally, this type of manipulation can be split into two stages referred to as a  modulator and a  convertor. The modulator introduces a position-dependent momentum whereas the convertor consists of a beamline providing a momentum-dependent change in the position.  For the sake of simplicity, we consider the case of uncoupled motion between the three DOFs, and focus on one of the DOFs, with particle conjugated variables $(\zeta_i,\zeta_{i+1})$ [where $i \in[1,3,5]$]. We consider the effect of the modulator to impact a position-dependent external force along $\hat \zeta_{i+1}$ of the form ${F}_{i+1}(\zeta_i)$, resulting in the coordinate transformation
\begin{eqnarray}
\zeta_{i,0}  \rightarrow \zeta_{i,m} &=&\zeta_{i,0} , \nonumber \\
\zeta_{i+1,0}  \rightarrow \zeta_{i+1,m} &=&\zeta_{i+1,0} +{F}_{i+1}(\zeta_{i,0}) ,
\end{eqnarray}
under an impulse approximation (where we ignore change in the particle position). In the latter equation,  the subscripts $_0$ and $_m$ respectively refer to the values before and after the modulator section. The downstream converter beamline introduces a momentum-dependent  change in position so that the final position $\zeta_{i,f}$ is related to the upstream coordinate as 
\begin{eqnarray}
\zeta_{i,m}  \rightarrow \zeta_{i,f} &=&\zeta_{i,m} +{G}_i (\zeta_{i+1,m}) , \nonumber \\
\zeta_{i+1,m}  \rightarrow \zeta_{i+1,f} &=&\zeta_{i+1,m} .
\end{eqnarray}

Overall, the  transformation associated with the beamline takes the coordinate $(\zeta_{i,0},\zeta_{i+1,0})$ and transforms it to the final coordinate 
\begin{eqnarray} \label{eq:sec4-tranformation}
\begin{pmatrix}
\zeta_{i,f}   \\
\zeta_{i+1,f}
\end{pmatrix}
=
\begin{pmatrix}
\zeta_{i,0} +{G}_i [\zeta_{i+1,0} +{F}_{i+1}(\zeta_{i,0})] \\
\zeta_{i+1,0} +{F}_{i+1}(\zeta_{i,0})
\end{pmatrix} ,
\end{eqnarray} 
and a transfer map defined such that $(\zeta_{i,f}, \zeta_{i+1,f})={\cal M}(\zeta_{i,0}, \zeta_{i+1,0})$ can be formally associated with the transformation. 

Considering an initial phase-space-density distribution $\Phi_0(\zeta_{i,0},\zeta_{i+1,0} )$  and invoking Liouville's theorem~\ref{sec2c1Liouville} $\Phi_f(\zeta_{i,f} ,\zeta_{i+1,f} ) d \zeta_{i,f} d \zeta_{i+1,f}= \Phi_0(\zeta_{i,0},\zeta_{i+1,0} ) d\zeta_{i,0}d \zeta_{i+1,0}$, we can write the final phase-space distribution as
\begin{eqnarray}
\Phi_f(\zeta_{i,f} ,\zeta_{i+1,f} ) = \Phi_0[{\cal M}^{-1}(\zeta_{i,0},\zeta_{i+1,0} )] , 
\end{eqnarray}
since the Jacobian of the transformation is unity. We now consider the projection along the position direction $\widehat{\zeta_i}$ defined as 
\begin{eqnarray}
P_i(\zeta_i)=\int d\zeta_{i+1} \Phi (\zeta_{i} ,\zeta_{i+1} ).
\end{eqnarray}
By virtue of the charge-conservation we now have
\begin{eqnarray}
P_f(\zeta_{i,f}  )  = P_0[\zeta_{i,0}(\zeta_{i,f})] \frac{\partial \zeta_{i,0} }{\partial  \zeta_{i,f}}, 
\end{eqnarray}
where the RHS can be written solely in term of $\zeta_{i,f}$ via inversion of the map described in Eq.~\eqref{eq:sec4-tranformation}. Therefore, by properly tailoring the transformation ${\cal M}$, one can modify the shape of the projection along any direction. To illustrate the set of derived equations we first consider the simple example of the bunch compression discussed in Section~\ref{sec2c1Liouville}.  Given the incoming-LPS coordinate $(z_0,\delta_0)$, the correlation introduced by the linear accelerator and the energy-dependent path length from the compression can be described, respectively, by  $F(z)=h  z$ and $G(\delta)=R_{56}\delta$. Here for sake of simplicity we take both $F$ and $G$ to be linear functions of $s$ such that the overall transformation is described by 
\begin{eqnarray}
\begin{pmatrix}
z_{f}   \\
\delta_{f}
\end{pmatrix}
=
\begin{pmatrix}
z_{0} +R_{56} [\delta_{0} +h(z_{0})] \\
\delta_{0} +h(z_{0})
\end{pmatrix} ,
\end{eqnarray}
so that the final longitudinal charge distribution is 
\begin{eqnarray}
\lambda_f(z_{f}  )  = C \lambda_0\left( \frac{z_{f}-R_{56}\delta_{0}}{{ C}} \right) , 
\end{eqnarray}
where ${ C}$ represents the compression factor defined in Eq.~\eqref{compf}. Taking $\lambda_0(z_0)=1/\sqrt{2\pi\sigma_{z,0}^2}\exp[-z_{0}^2/(2\sigma_{z,0}^2)]$ and assuming that $\delta_{0}$ represents a random uncorrelated fraction energy spread, we obtain 
\begin{multline}
\lambda_f(z_{f}  )  = \frac{ C}{\sqrt{2\pi}\sigma_{z,0}}  \exp \left( -\frac{ C^2 z^2_f}{2\sigma_{z,0}^2} \right) 
\\
\times \exp \left(- \frac{C^2R^2_{56}\delta_{0}}{2\sigma_{z,0}^2} \right) , 
\end{multline}
which showcases the well-known results associated with linearized bunch compression previously mentioned in Section~\ref{sec2}. A simple extension is to consider the case where nonlinearities play a role in the compression. For instance, owing to RF curvature in the linac employed to impart the chirp, a quadratic dependence on $\zeta_{5,0}$ is also introduced when the bunch length does not strictly verify the condition $\sigma_{z,0} \ll \lambda$ (where $\lambda$ is the wavelength of the accelerating mode in the linac) so that $F(z)=h  z + h_2z^2$. Likewise, standard four-bend bunch compressors are known to introduce a second-order longitudinal dispersion $T_{566}$, and consequently $G(\delta) = R_{56}\delta + T_{566}\delta^2$. Following the same approach as before yields a final distribution along the longitudinal axis to be of the form~\cite{li-2001-a}
\begin{eqnarray} \label{eq4-nonlinearcompression}
\lambda_f(z_{f}  )  = \frac{1}{\sqrt{2\pi}\sigma_{z,f}} \frac{\exp({z_{f}/(\sqrt{2}\sigma_{z,f}})} {[-z_{f}/(\sqrt{2}\sigma_{z,f})]^{1/2}}\Theta(-z_f), 
\end{eqnarray}
where $\Theta()$ is the Heavide function. 
Thus, the nonlinear transformation introduced by the functions $F$ and $G$ now results in a change of the bunch current profile commonly encountered in, e.g., magnetic compression, subjected to strong nonlinearities; see for instance ~\cite{dohlus-2004-a}. 

The example considered so far can be generalized via the introduction of arbitrary nonlinear functions $F$ and $G$ to control the final phase space $(\zeta_i, \zeta_{i+1})$ correlations. By controlling the degree of the non-linearity introduced, one can, in principle, tailor the correlations within the phase space to produce the desired profile along each of the phase-space directions. 

In beam physics, it is customary to expand the transport map via a truncated Taylor series and consider each nonlinear order separately; see Section~\ref{sec2:sec2c5nonlinear}. This is a typical process in optics where, e.g., the energy-dependent path length introduced by a dispersive section, e.g., a bunch compressor as discussed above, is now written as $\zeta_{5,f}= \sum_{i=1}^6 R_{5i} \zeta_{i,0}+\sum_{i=1}^6\sum_{j\ge 1}^6 T_{5ij} \zeta_{i,0} \zeta_{j,0} + {\cal O}(\zeta_0^3)$ and likewise for any beamline; see Eq.~\eqref{81-nonlinear}.  For sake of simplicity we assume a pencil bunch so that high-order coupling between the longitudinal and transverse phase spaces can be neglected in the expansion of the transfer map.

So far we have considered the modulator function to be a continuous function of the coordinate over a finite interval. Another approach is to consider a periodic function of the form $F(u)=F_0\cos (k u +\phi)$.  In such a case, a series of concatenated modulators described with a function $F_i(u)=F_{0,i}\cos(k_i u + \phi_i)$ could be used to synthesize the desired final distribution by controlling the term of the Fourier series associated with the final distribution. Such a description can be employed, for example, to describe a chain of linacs operating at different frequencies. By shaping the energy spectrum and using a transformation in $z = {\cal M}(\delta) $ one could tailor the energy spread or current profile. In practice introducing an arbitrarily high harmonic may be challenging, especially for the time-dependent field given, e.g., the limited set of klystron frequencies.

\subsubsection{Interceptive beam shaping}
The shaping techniques described so far combine nonlinear external fields with the beamline providing nonlinear correlation between the position and momentum of particles. A straightforward shaping technique  consists of intercepting the beam with a mask that has a  given ``transmission" function $T(x,y)$ so that the mask can only affect the distribution in the transverse spatial coordinates; see Fig.~\ref{fig:sec4:overview}(c). The incoming phase-space distribution $\Phi_i(\zeta_1...\zeta_4)$ is then simply transformed as 
\begin{eqnarray}
\Phi_f(x, y) = \Phi_i(x,y) T(x,y), 
\end{eqnarray}
where we assume the mask to be thin and only affecting the beam according to the transverse coordinate (and not the momentum). Consequently, the transverse profile along, e.g., $\zeta_1$ can be found from 
 \begin{eqnarray}
P_f(x) &=& \int_{-\infty}^{+\infty} \Phi_i(x,y) T(x,y) dy \nonumber \\
&=&  \int_{\Upsilon_-(x}^{\Upsilon_+(x)}  \Phi_i(x,y)  d y, 
\end{eqnarray}
where we have taken the mask to be a binary function with unity value in the domain $y\in[\Upsilon_-(x), \Upsilon_+(x)]$. 

A drawback of such a masking technique is its intrusive nature, which may hinder its application to high-power or high-repetition-rate beams, as the beam loss associated with the shaping process could result in radiological activation or hardware damage. 

\subsubsection{Manipulation with local coupling}
Another class of transformation involves local correlations between coordinates within two DOFs; see Fig.~\ref{fig:sec4:overview}(b).  In such a transformation,  an external field introduces the required correlations $\zeta_j=H(\zeta_i)$ and the coordinate $\zeta_j$ is manipulated (e.g., via a function similar to $F$ or $G$). Finally, the inverse transformation $H^{-1}$ to remove the correlation between $\zeta_j$ and $\zeta_i$. In the process the shaping imparted to $\zeta_i$ via shaping of $\zeta_j$ is preserved; see Fig.~\ref{fig:sec4:overview}(b). A simple example of implementation of a manipulation based on local coupling regards dispersive collimation where a local dispersion bump locally introduces a correlation between transverse position and energy where a collimator is used to tailor the energy distribution (e.g. remove energy tail). One advantage of local-coupling methods combined with a mask is its simple implementation while providing a high degree of control over the beam shape (via precise shaping of the intercepting mask~\cite{majernik-2021-a}). However, the mask can result in significant particle losses.  

%%%%%%%%%%%%%%%%%%%%%

\subsection{Generation of shaped current distributions\label{sec4:subsec:currentshaping}}
One important aspect of LPS control resides in the ability to control the beam current profiles for application in beam-driven wakefield accelerators and light sources. Historically, current beam shaping has been an integral part of electron injectors based on continuous wave (CW) electron sources where a combination of masks and RF cavities $-$ often dubbed ``chopping" systems $-$ are commonly employed to form bunches  for injection in the subsequent linear accelerators~\cite{smith-1986-a,tiefenback-1993-a}.  
\subsubsection{Local coupling combined with transverse masking}\label{sec4b1}
Current shaping techniques were initially discussed as a means to prebunch the beam for free-electron laser (FEL)~\cite{nguyen-1996-a} applications. The method was eventually demonstrated at the ATF facility~\cite{muggli-2008-a} where it was also extended to shaping beyond microbunch generation~\cite{shchegolkov-2015-a}. In brief, the method combined local coupling between the transverse (horizontal) and longitudinal phase spaces with masking. To explain the technique we consider an incoming bunch with an LPS chirp $h_0$ sent to a dispersive section with a transfer matrix producing the horizontal and longitudinal dispersions $\eta$ and $\xi_-$, respectively.  Downstream of such a dispersive section the final horizontal and longitudinal positions of an electron is 
\begin{eqnarray}\label{eq:sec6:maskModulation}
x_m&=& x_{\beta}+\eta \delta_0, \nonumber \\
z_m&=& z_0 +\xi_- \delta_0, 
\end{eqnarray}
where $x_{\beta}$ is the geometric contribution to the beam spot. An intercepting mask in $(x_m,y_m)$ with transmission function dependent on the horizontal coordinate $f(x_m)$ will tailor the transverse profile. For a beam with correlated momentum spread $\delta_0=h_0z_0$ the mask will alter the shape of the longitudinal distribution following   $f[x_m/(\eta h_0)]$. A downstream dispersive section with longitudinal dispersion $\xi_+$ and designed to suppress the dispersion introduced by the section upstream of the mask results in the overall shaping function $f[x_m/(\eta h_0)]$. This technique bears similarities with the frequency-domain temporal shaping method commonly encountered in ultrafast laser shaping, where a frequency-chirped laser is dispersed, its spectrum modified with a mask, and subsequently recombined~\cite{kuzmin-2018-a,weiner-1998-a}; see Section~\ref{sec3:shaped:PElong:freq}. 

The method was implemented in~\cite{muggli-2008-a} to form train of sub-ps bunch. In such a case the mask consists of a set of $N$ slits so that the transmission can be approximated as $f(x_m)=\sum_{\ell=1}^N \delta(x_m-X_\ell)$ where $X_\ell$ are the horizontal positions of the vertical slits. Considering $X_\ell=\ell D$, with  $D$ being the inter-slit spacing on the mask, the final modulation period is shown to be~\cite{muggli-2008-a, hyun-2019-a}
\begin{eqnarray}
\Delta z &\simeq& D \frac{1 + \xi_+ h_0}{\eta h_0}. 
\end{eqnarray}

\begin{figure}
 \includegraphics[width=0.85\linewidth]{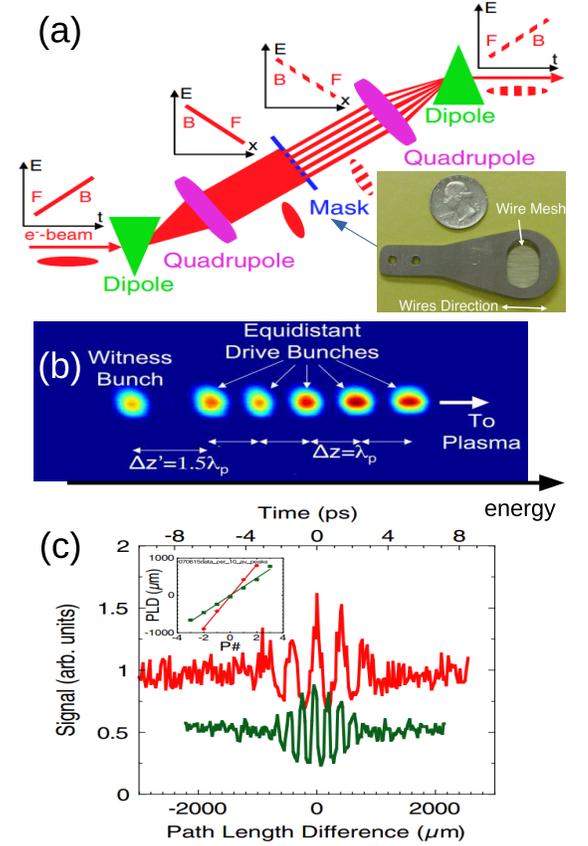}
\caption{\label{fig:sec4:atfexperiment}  Experimental generation of a sub-picosecond bunch train at the ATF using a dogleg beamline (a) with final beam distribution in a downstream spectrometer (b) and measured temporal distribution via interferometry of coherent transition radiation (c). Figure adapted from~\cite{muggli-2008-a} and ~\cite{muggli-2010-b}. }
\end{figure}

A practical implementation of the method used for a proof-of-principle experiment at the Accelerator Test Facility (ATF) appears in Fig.~\ref{fig:sec4:atfexperiment}(a). The beam was locally dispersed in a dogleg beamline with vanishing net dispersion. The dogleg included an optical lattice providing an antisymmetric dispersion function with maximum value attaining $\eta\simeq 1.5$~m in close proximity to the mask location. The mask consisted of sub-mm tungsten wire; see inset photograph in  Fig.~\ref{fig:sec4:atfexperiment}(a). The produced bunch was modulated in energy and time ($z$) owing to the final LPS chip.  Given the fixed dispersion, the mask sets the energy modulation period and ultimately the final current modulation, which can be controlled via the incoming chirp and $\xi_+$ function of the beamline.  It was  demonstrated that the technique enables some control over the final temporal period downstream of the dogleg beamline: for $\sim 10$~pC modulations with sub-mm periods were produced, consistent with the resonant excitation of wakefield in plasmas~\cite{muggli-2010-a}. Likewise, this method was extended to produce bunches with triangular beam distribution at the ATF~\cite{shchegolkov-2015-a} for wakefield excitation in dielectric-lined waveguides~\cite{antipov-2012-a}. 

More recently, a current-shaping method based on directly introducing a spatio-temporal coupling was proposed using a pair of transverse-deflecting cavities~\cite{kur-2009-a,zholents-1999-a}. Such a concept is similar to the one commonly employed in DC electron injectors as part of the low-energy ``chopping" system required to form the bunched beam before injection in an RF linac~\cite{wilson-1985-a}. 

\begin{figure}
 \includegraphics[width=\linewidth]{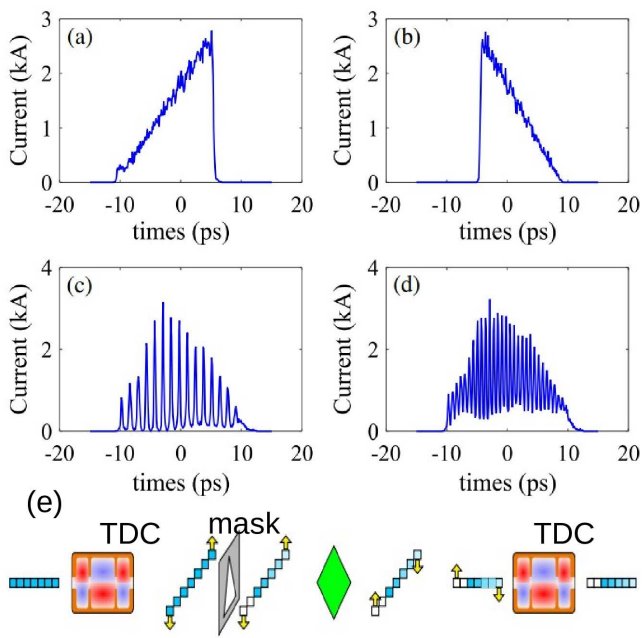}
\caption{\label{fig:sec4tdcsimulationshaper}  Numerical modeling of current-profile shaping showcasing the generation of ramped (a), reverse-triangle (b) and modulated (c,d) bunch distributions using local correlations imparted by a pair of transverse-deflecting cavities (TDC) combined with a mask (e) (adapted from ~\cite{ha-2020-a}).}
\end{figure}

The main advantage of this shaping method compared to the one based on dispersive coupling stems from its ability to directly couple $z$ to one of the transverse coordinates. In the method does not suffer from CSR-induced phase-space degradation. It can also provides control over the LPS chirp. In its simplest implementation, the beamline consists of two transverse deflecting cavities (TDCs) separated by a beamline described by a transfer matrix $M$. The overall matrix of the system in $(x,x',z,\delta)$ is 
\begin{eqnarray}
S = R_{TDC}(\kappa_d)M R_{TDC}(\kappa_u)\equiv
\begin{pmatrix}
A & B\\
C & D \\
\end{pmatrix},
\end{eqnarray}
where $R_{TDC}(\kappa)$ represents the matrix of a TDC with strength $\kappa$, see Section~\eqref{sec2:deflectingcavity}. In order for the beamline to be globally uncoupled, the $2\times2$ block matrix should vanish: $B=C=0$. 
Solving for $B=0$ yields the condition on the elements of $M$ and cavity strength to 
\begin{eqnarray}
m_{12} &=&\displaystyle - \frac{L \left(\kappa_{d} + \kappa_{u} \left(L m_{21} + m_{11} + 2 m_{22}\right)\right)}{2 \kappa_{u}}, \nonumber \\
\kappa_d &=&\displaystyle - \frac{\kappa_{u} \left(L m_{21} + 2 m_{22}\right)}{2}, \mbox{and} \\
m_{11}&=&\displaystyle - \frac{L m_{21}}{2} - \frac{\kappa_{u}}{\kappa_{d}}. \nonumber
\end{eqnarray}
These matching conditions ensure $S$ becomes block diagonal. The $(x,x')$ block matrix is independent of the TDC strengths, and the overall  effect of the combined TDCs is to modify the LPS chirp $h\equiv \mean{\zeta \delta}/\mean{\zeta^2}^{1/2}$ as $h \rightarrow h=h_0-S_{65}$ with
\begin{eqnarray}
S_{65}=\displaystyle \frac{L \kappa_{u}^{2} \left[( L m_{21} +2m_{22} )^2 + 4\right]}{16}.
\end{eqnarray} We have $S_{65}>0$ for a cavity length $L\ne 0$.
It is interesting to note that the chirp vanishes in the thin-lens model of the TDC ($L=0$). Likewise the block matrix becomes
\begin{eqnarray}
M_{\perp} =
\begin{pmatrix}
m_{22}^{-1} & 0 \\
m_{21} & m_{22} \\
\end{pmatrix},
\end{eqnarray}
consistent with Ref.~\cite{kur-2009-a} where the choice $ m_{11}=m_{22}=-1$ was made so that $\kappa_u=\kappa_d$ (given that decoupling forces $m_{12}=0$). The shaping mask is located between the two TDCs and with sufficient phase advance from the upstream TDC to ensure the beam has a significant correlation $\mean{xz}$ at its location. The profile of the mask $T(x,y)$ is then tailored to provide the desired temporal shape. The application of this method to produce a modulated relativistic electron bunch was first discussed in~\cite{du-2012-a} in the context of superradiant THz radiation, and further investigated numerically in~\cite{ha-2020-a} to support the formation of various current shapes to support a beam-driven wakefield accelerator with enhanced transformer ratio. Figure~\ref{fig:sec4tdcsimulationshaper} presents numerical simulation results for a minimal system composed of two TDCs and a quadrupole magnet with a different type of mask to produce doorstep distributions [Fig.~\ref{fig:sec4tdcsimulationshaper}(a,b)] for beam-driven acceleration and sub-ps bunch trains with variable microbunch spacing [Fig.~\ref{fig:sec4tdcsimulationshaper}(c,d)].

The masks described so far stop a fraction of the incident beam. An alternative solution is to use a spoiling mask made of a thin foil to alter the beam properties associated with a fraction of the beam. An example of implementation uses a slotted foil at the Linac Coherent Light Source (LCLS)~\cite{emma-2004-a} to selectively spoil the transverse emittance of an electron beam (the fraction of the beam intercepted by the foil undergoes multiple scattering and suffer an emittance growth). The method was employed to ultimately control the duration of X-ray pulses in FELs, as only the unscattered beam population contributed to the lasing mechanism and supported the generation of isolated femtosecond X-ray pulses along with twin variable-delay X-ray pulses using a dual-slot foil~\cite{ding-2015-a}.

\subsubsection{Modulators combined with longitudinally dispersive sections\label{sec4:subsec:modulatorsWlongDisp}}

A simple implementation of a current-shaping technique consists of introducing an energy modulation using RF accelerating cavities to properly control the longitudinal dispersion in the downstream beamline. The energy modulation is introduced as a time-varying field with accelerating voltage of the form $V(z)=\widehat{V}\cos(kz+\varphi)$, where $\widehat{V}$, $\varphi$, and $k$ are, respectively, the peak accelerating voltage and phase, and the accelerating mode wavevector. The LPS transformation associated with the downstream dispersive sections is described via the Taylor expansion of the longitudinal map with, e.g., first- and second-order the coefficients $R_{56}\equiv \frac{\partial z}{\partial \delta}$, $T_{566}\equiv \frac{\partial^2 z}{\partial \delta^2}$ as detailed in Section~\ref{sec2:sec2c5nonlinear}. The two widely-used approaches are to either ($i$) introduce a linear energy modulation and controls the nonlinear dispersion in the dispersive section, or to ($ii$) control the nonlinear modulation using accelerating sections operating at multiple frequencies.

 \begin{figure}
 \includegraphics[width=1\linewidth]{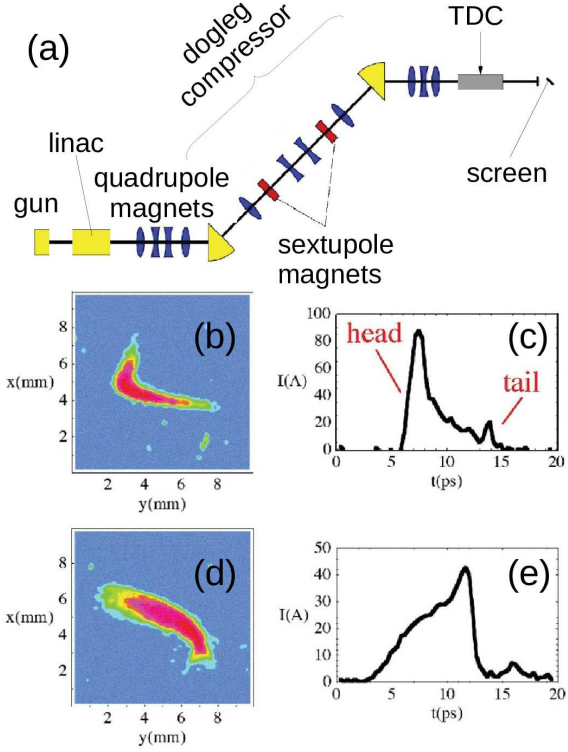}
\caption{\label{fig:sec4:shapingNeptune} Experimental setup for phase-space shaping via introduction of nonlinear longitudinal dispersion at the NEPTUNE facility (a) and resulting LPS (b,d) with associated current distribution (c,e) experimentally generated. Figure adapted from~\cite{england-2008-a}. }
\end{figure}
A technique to control the nonlinear longitudinal dispersion to shape the beam current distribution was introduced in ~\cite{england-2005-a}. It was observed that compression of a ``long" incoming bunch accelerated in an S-band linac such that $kz\ll1$ is not strictly valid. As a result, the final fractional energy offset is
\begin{eqnarray}
\delta\simeq \frac{e\widehat{V}}{ \bar{E}} \left[-kz \sin\varphi+ \frac{(kz)^2}{2} \cos \varphi\right],
\end{eqnarray}
(where $ \bar{E}$ is the final beam energy) resulting in significant quadratic correlation in the LPS with associated current profile described by an equation similar to Eq.~\eqref{eq4-nonlinearcompression}. It was recognized that the addition of sextupole magnets in the dispersive section (arranged as a dogleg beamline) provided a sufficient handle over the $T_{566}$ coefficient to ultimately control the current distribution. The experimental implementation of this technique at the NEPTUNE facility demonstrated the generation of $\sim 12$-MeV bunches with linearly ramped current profiles~\cite{england-2008-a}; see Fig.~\ref{fig:sec4:shapingNeptune}. The technique was also shown to provide some control over the ramp shape and direction. \\

The introduction of higher-order longitudinal dispersion (up to third order) was also proposed to suppress current spikes arising from collective effects in the LCLS accelerator~\cite{charles-2017-a}. At LCLS,  extreme current values at the head and tail of the electron bunch are responsible for substantial CSR-induced projected emittance growth. The nominal operating mode of LCLS consits of truncating the head and tail to suppress the current spikes to improve the FEL lasing  performances. It was shown that current spikes can be strongly suppressed via control of the  $U_{5666}\equiv \frac{1}{6} \frac{\partial^3 z}{\partial \delta^3}$ longitudinal-dispersion term using an octupole magnet without a significant increase of the horizontal slice emittance.\\

Further improving the precision of shaping methods discussed in this section can be accomplished using multiple energy modulators operating at different frequencies. Such an approach allows for the incoming LPS to be an arbitrary polynomial function $\delta(z)$ with its coefficients controlled by the settings of the modulator accelerating-voltage amplitude and phase. The solution was investigated using a dual frequency modulator  operating at the frequencies $f_1$ and $f_n\equiv n f_1$ with total accelerating voltage $V(z)=V_1\cos(k_1z+\varphi_1) + V_n \cos(k_n z+\varphi_n) $, where $\widehat{V_{1,n}}$ and  $\varphi_{1,n}$ are, respectively, the accelerating voltages and operating phases of the two linac sections, and $k_{1,n}\equiv2\pi f_{1,n}/c$. Assuming $k_{1,n}z_0 \ll 1$,  the electron's LPS coordinates downstream of the linac are $(z_l=z_0, \delta_l=a_l z_0+b_lz_0^2)$, where  $a_l\equiv a_0-e(k_1 V_1 \sin\varphi_1 + k_n V_n \sin\varphi_n )/\bar{E}_l$, $b_l\equiv b_0-e(k_1^2 V_1 \cos\varphi_1 + k_n^2 V_n \cos\varphi_n )/(2 \bar{E}_l)$ with $e$ being the elementary charge and  $\bar{E}$ the beam's average energy downstream of the linac. The passage of the bunch through a  longitudinally dispersive section results in an electron final coordinate to  be given as a function of the initial coordinates following $z_f = a_f z_0 + b_f z_0^2 $ with $a_f \equiv 1 + a_l R_{56}$ and $b_f\equiv b_l R_{56}+ a_l^2 T_{566}$. Taking the initial current  to follow the  Gaussian distribution  $I_0(z_0)=\hat{I_0} \exp [-z_0^2/(2\sigma_{z,0}^2)]$ (where $\hat{I_0}$ is the initial peak current), and invoking the charge conservation gives the final current distribution 
\begin{eqnarray}
I^u_f(z_f)&=&\frac{\hat{I}_0}{\Delta^{1/2}(z_f)} \exp\left[-\frac{(a_f + \Delta^{1/2}(z_f))^2}{8b_f^2 \sigma_{z,0}^2}\right], \nonumber \\
  && \times \Theta [\Delta(z_f)], 
\end{eqnarray}
where $\Delta(z_f) \equiv a_f^2+4b_f z_f $ and $\Theta()$ is the Heaviside function. 

\begin{figure}
 \includegraphics[width=1\linewidth]{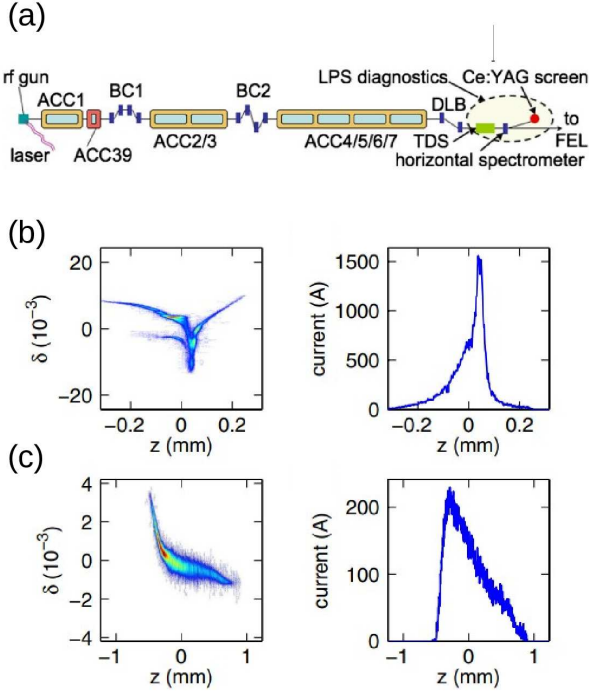}
\caption{\label{fig:sec4:shapingFLASH} Experimental results on phase-space shaping using a dual-frequency linac at the FLASH facility at DESY (a) with corresponding nominally compressed beam (b) and linearly-ramped bunch generation (c) obtained via proper control of the accelerating voltage phase and amplitude of the linac sections ACC1 (1.3 GHz) and ACC39 (3.9 GHz) prior to the bunch compressor 1 (BC1) (adapted from Ref.~\cite{piot-2012-a}).  }
\end{figure}

An example of the method discussed above was experimentally implemented in the FLASH facility at DESY~\cite{piot-2012-a}, where a linac composed of  1.3-GHz and 3.9-GHz superconducting-linac modules and two magnetic bunch compressors was used; see Fig.~\ref{fig:sec4:shapingFLASH}(a). Specifically, the accelerator settings were optimized to form a linear-ramped current profile; see Fig.~\ref{fig:sec4:shapingFLASH}(b,c). The technique provided ample control over the shape owing to the number of variables available and produced a shaped beam at 300 MeV for future injection in the FLASH-Forward facility~\cite{aschikhin-2016-a}.

This method could be generalized in principle by introducing an arbitrary number of accelerating sections operating at different frequencies to ultimately synthesize any correlation in the LPS via a Fourier series.  The generalized scheme is challenging in practice due to the limited number of RF sources and associated prohibitive cost. However, it could be implemented passively as discussed in Section~\ref{sec5}. \\

\begin{figure}
 \includegraphics[width=1\linewidth]{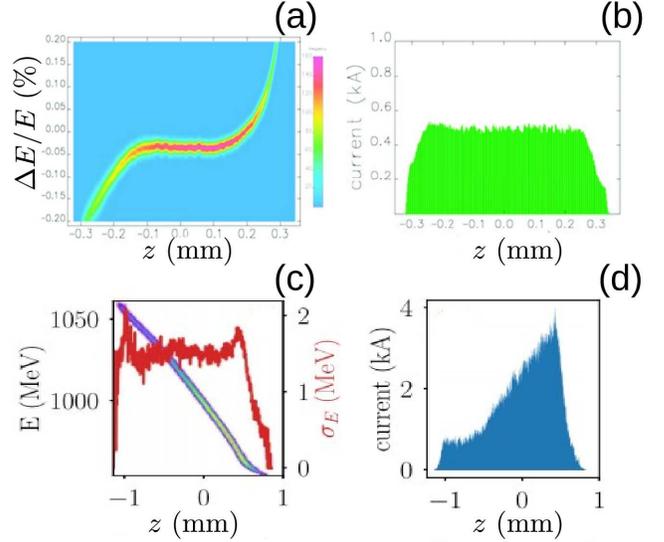}
\caption{\label{fig:sec4:integratedexample} Examples of LPS (a,c) and associated current distribution(b,d) obtained via a fully integrated combining the photocathode-laser shaping method discussed in Section~\ref{sec3} with multi-frequency linacs and nonlinear dispersive sections. Plots (a,b) are reproduced from ~\cite{cornacchia-2006-a}, which simulated a possible operation mode of the FERMI@ELETTRA FEL. Plots (c,d) are extracted from \cite{tan-2021-a}, which investigated the generation of shaped bunches for efficient acceleration in a corrugated-waveguide wakefield accelerator. In plot (c), the superimposed red trace corresponds to the slide RMS energy spread (right label).}
\end{figure}

As a final note, we should point out that most of the time, current shaping relies on an integration of different shaping techniques as explored in ~\cite{cornacchia-2006-a,lemery-2015-a}. For instance, the generation of uniform beams required for cascaded harmonic lasing in X-ray FELs combines photocathode laser shaping to precisely precompensate for nonlinearities introduced in the compression process or arising from collective effects~\cite{cornacchia-2006-a}. This work especially demonstrated the generation of uniform current distribution based on numerical simulations; see  Fig.~\ref{fig:sec4:integratedexample}(a,b). Likewise, numerical simulations indicate that a properly shaped photocathode laser pulse combined with a multi-frequency linac and a two-stage nonlinear compression process could provide precise control over the LPS and current distribution~\cite{tan-2021-a}, ultimately realizing the sought-after door-step distribution without the requirement for any collimation; see Fig.~\ref{fig:sec4:integratedexample}(c,d).
 
%%%
%
%
%%%

%
\subsection{Realizing ultra-low energy spread}
A recurrent topic associated with the use of bright electron beams regards the generation of ultra-low longitudinal emittance. One application of such a capability is the production of a low-energy-spread beam, ultra-short electron bunch for use in ultra-fast electron scattering experiments. There has been a renewed interest in meV energy-spread, sub-MeV electron bunches for applications in electron energy-loss spectroscopy~\cite{egerton-2008-a}.

In principle, the method described above and detailed in ~\cite{zeitler-2015-a} could be tailored to produce low energy spread. However, most of the low-energy-spread instruments  have been based on dispersive collimation implementing an ‘omega filter’~\cite{tsuno-1997-a}. One issue associated with this class of monochromator commonly used in electron-scattering instrument is the limited beam transmission. To palliate this limitation, a lossless monochromator was  proposed~\cite{duncan-2020-a} and demonstrated via numerical simulation the ability to produce 200-keV electron bunch with meV-level energy spread. The setup combines a photoemission electron source with a pair of accelerating cavities. The bunch-length lengthening ultimately set the final fraction energy spread $\sigma_{\delta,f}$ under Liouville's theorem as $\sigma_{\delta,f}=\sigma_{\delta,0}\frac{\sigma_{z,0}}{\sigma_{z,f}}\ge \frac{\hbar}{2}$ where $_0$ refers to the initial bunch length and fractional energy spread.~\cite{duncan-2020-a} note that the final kinetic energy $E_K$ distribution along the bunch is quadratically dependent on the particles' time of arrival and radial position within the bunch $E_K(t,r)=a t^2+ b r^2$ such a dependence can be compensated by the pair of accelerating cavities. 
% ; see Fig.~\ref{fig:sec4:monochromation}(b)
% \begin{figure}
% \includegraphics[width=.75\linewidth]{./Sec4-PP/fig4-monochromation.eps}
%\caption{\label{fig:sec4:monochromation} Monochromation process in a low-energy beam using an $\Omega$ filter (a) and a lossless cavity-based system (b). Adapted from ~\cite{duncan-2020-a}. }
%\end{figure}
 %
%

%%%%
\subsection{Controlling longitudinal-phase-space (LPS) correlations~\label{sec4:subsec:controLPS}}

As discussed so far, the shaping of the bunch's current profile is often implemented by controlling correlation in the LPS. In this section, we focus on methods that have been implemented with the primary purpose of controlling correlation in the LPS. An example is the removal of correlated energy spread downstream of an accelerator to produce, e.g., a more monoenergetic bunch.

 \begin{figure}
 \includegraphics[width=1\linewidth]{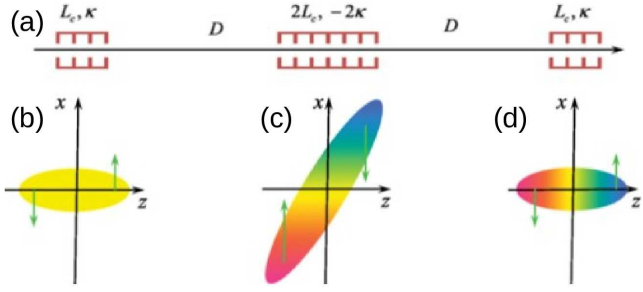}
\caption{\label{fig:sec4:TDCLPScontrol} A TDC-based beamline for the generation of strong LPS chirps (a). The diagrams (b-d) illustrate the $x-y$ electron-bunch distribution inside each TDC with the green arrows gives the direction of the transverse kick $x'$. The color map indicates the energy distribution with larger and lower energies respectively appearing as red and blue colors. Adapted from~\cite{yampolsky-2020-a}. }
\end{figure}
Although linear energy-spread control is conventionally achieved using an accelerating cavity operated off the crest, the method introduced in ~\cite{yampolsky-2020-a} combines several deflecting cavities to provide a power-efficient  control of the LPS chirp. Specifically, the technique considers three horizontally deflecting cavities separated by a distance $D$ and with respective deflection strengths $\kappa$, $-2\kappa$, and $\kappa$ to produce an uncoupled transport matrix in $(x,x',z,\delta)$ with longitudinal $2\times 2$ block given by
\begin{eqnarray}
R_{z|\delta} =
\begin{pmatrix}
1 & 0 \\
-\frac{2}{3}\kappa^2(3D+2L_c) & 1
\end{pmatrix} ,
\end{eqnarray}
where $L_c$ is the transverse-deflecting cavity length; see Fig.~\ref{fig:sec4:TDCLPScontrol}. The latter equation reveals the main advantage associated with the proposed TDC-based scheme over conventional off-crest acceleration. The introduced chirp $R_{65}$ scales quadratically with the cavity’s strength $\kappa$ while the off-crest acceleration scales linearly with the accelerating field. Although this simple model is not capable of altering the polarity of the chirp, introducing focusing elements between the TDC does enable such control. Additionally, the focusing elements can also be optimized to mitigate transverse-emittance degradation throughout the beamline~\cite{yampolsky-2020-a}. 

 \begin{figure}[hhhh!!!]
 \includegraphics[width=0.80\linewidth]{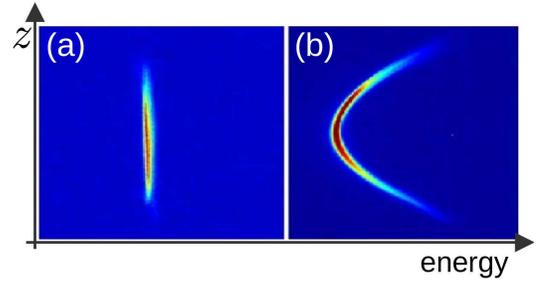}
\caption{\label{fig:sec4:39GHzXFEL} Example of longitudinal phases-space linearization at the LCLS facility (adapted from ~\cite{akre-2008-a}. Shown is the LPS measurement without (a) and with (b) operation of the fourth-harmonic linearizing cavity (operating at 11.424~GHz).  }
\end{figure}

Accessing higher-order correlation in the LPS is more challenging especially when dealing with short (picosecond-scale) bunches, as the required RF wavelength needs to be comparable to the bunch length. However, some correction can be achieved with proper choice and interplay between RF systems operating at different frequencies or through non-relativistic effects. For example, correcting or imposing quadratic nonlinearities in the LPS can be achieved by combining a harmonic RF field to the fundamental mode, as introduced in ~\cite{smith-1986-a} and later adapted for the LCLS and FLASH FELs~\cite{flottmann-2001-a}. The method relies on introducing a harmonic frequency so that the final fractional energy offset associated with an electron located at $z$ takes the form
\begin{eqnarray}
\delta(z)&=&\delta_0 \frac{E_0}{E_f}+\frac{eV_1}{E_f}[\cos(k_1 z + \varphi_1)-\cos\varphi_1 ] \nonumber \\
&& +\frac{eV_n}{E_f}[\cos(k_n z + \varphi_n)-\cos\varphi_n ] \nonumber \\
&\simeq & \delta_0 \frac{E_0}{E_f}+\left(\frac{eV_1}{E_f}k_1\sin\varphi_1 + \frac{eV_n}{E_f}k_n\sin\varphi_n\right) z \nonumber \\
&& - \left(\frac{eV_1}{2E_f}k_1^2\cos \varphi_1 + \frac{eV_n}{2E_f}k^2_n\cos\varphi_n\right) z^2 ,
\end{eqnarray}
where $\delta_0$ is the initial relative energy offset, $\varphi_1$ and $V_1$ (resp. $\varphi_n$ and $V_n$) are, respectively, the phase and amplitude of the fundamental (resp. harmonic) section of the linac, and $E_f$ the final energy. Considering the simple case of on-crest operation $\varphi_1=\varphi_n=0$, we note that cancellation of the second-order term can be accomplished when the harmonic section is operating on the deceleration phase with voltage amplitude $V_n=V_1\left(\frac{k_1}{k_n}\right)^2$. Such an approach was first demonstrated at the Free Electron LASer in Hamburg (FLASH) facility in DESY~\cite{harms-2011-a} by adding a third-harmonic linac ($f_n=3.9$-GHz) to the fundamental-frequency linac ($f_1=1.3$~GHz). The same scheme was implemented at the European X-ray FEL~\cite{decking-2020-a}. Figure~\ref{fig:sec4:39GHzXFEL} showcases a measurement of the LPS-linearization process at the LCLS where a fourth-harmonic cavity ($f_n=11.424\mbox{~GHz}=4\times 2.856$~GHz) is employed~\cite{akre-2008-a}. Similar, methods were investigated in the context of ultrafast electron-diffraction setups~\cite{floettmann-2014-b}.

It has also been recognized that this type of LPS linearization can be implemented without the need for a harmonic field when dealing with non-ultrarelativistic bunches. Such a scheme was introduced in~\cite{krafft-1996-a}  were it was shown via numerical simulations and experimentally confirmed that a proper control of the phase and amplitude of a buncher and capture cavities could control the $R_{56}$ and $T_{566}$ in the  CEBAF injector~\cite{wang-1998-a}. A similar technique was explored theoretically and via simulation for an ultra-fast electron source using an RF photoinjector; see~\cite{zeitler-2015-a}. Specifically, the authors show that expanding the beam after the electron source enabled a higher-order correction of the longitudinal focus by a subsequent accelerating cavity that is operated at the same frequency as the electron gun. Although the method was implemented as part of a ballistic-compression scheme to demonstrate the generation of sub-fs bunches at low energy, it could in principle be extended to higher energy when combined with standard compression beamlines.

\begin{figure}[hhhh!!!]
 \includegraphics[width=0.95\linewidth]{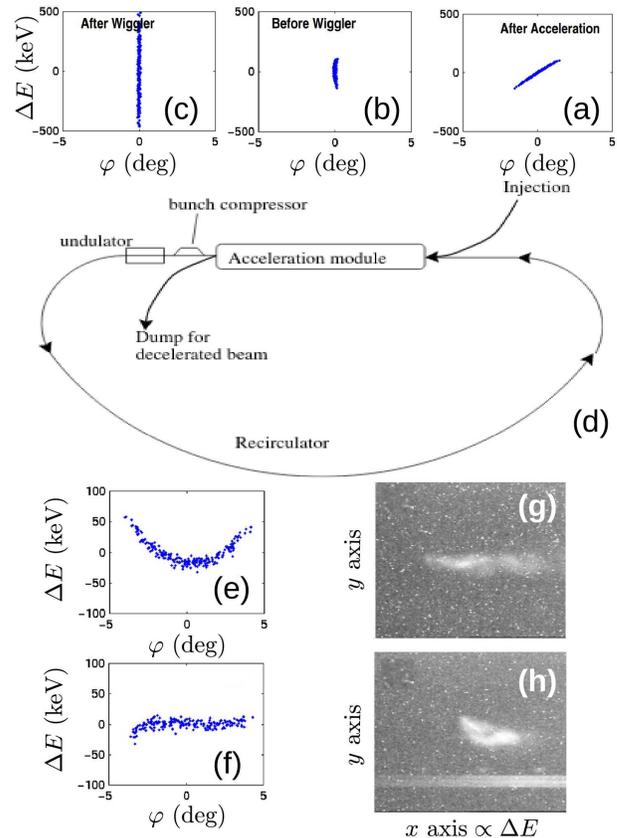}
\caption{\label{fig:sec4:ERLsextupole} Nonlinear control of LPS  with high-order longitudinal dispersion in JLab IR-FEL Demo for two tuning of the longitudinal lattice along with resulting energy spread measured after energy recovery [images (g) and (h)]. The simulated LPS are displayed at acceleration-module exit (a), before(b) and after (c) the FEL iteration, and after energy recovery for the cases of linear (e) and nonlinear (f) correction (via independent control of the $R_{56}$ and $T_{566}$ of the recirculator) of the LPS. . The measured $(\delta E, y)$ distributions shown in images (g) and (h) correspond respectively to simulated LPS presented in (e) and (f). The images are recorded on the final beam dump window and the horizontal axis is representative of the energy spread after deceleration in the acceleration module. Adapted from Ref.~\cite{piot-2003-b}.  }
\end{figure}

Additionally, introducing high-order longitudinal dispersion, as discussed previously in Section~\ref{sec4:subsec:currentshaping} can also be used to control LPS nonlinearity. Such a method was successfully implemented in a high-power energy-recovery linac~\cite{piot-2003-b} and supported the first demonstration of same-cell energy recovery of a high-power electron beam with increased energy-spread after its interaction in an FEL oscillator~\cite{neil-2000-a}; see Fig.~\ref{fig:sec4:ERLsextupole}. Higher-order correction using octupole magnets was also implemented in the JLab 10-kW FEL facility~\cite{neil-2006-a}.

\begin{figure}[hhhh!!!]
 \includegraphics[width=0.85\linewidth]{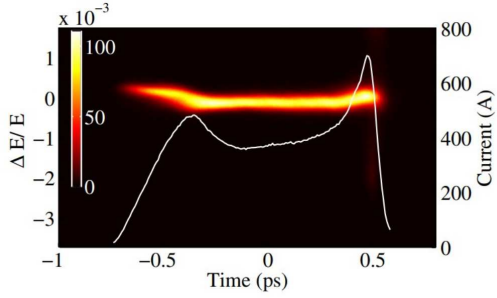}
 \caption{\label{fig:sec4:LaserPenco} Experimental demonstration of control of LPS correlation at at the FERMI@ELETTRA accelerator using a 1.4-GeV beam. Electron-bunch LPS (image) with superimposed current profile (white line, right axis). These data were obtained using a shaped photocathode laser with ramped temporal profile configuration. The 300-fs bunch core has a nearly constant incoherent energy spread of about $\sigma_{E} \simeq 150$~keV. From~\cite{penco-2014-a}. }
\end{figure}

Finally, a laser-pulse shaping similar to the one discussed in Sections~\ref{sec2} and~\ref{sec4:subsec:currentshaping} initially proposed in~\cite{cornacchia-2006-a} can also be combined with a beamline with nonlinear longitudinal dispersion to control the final LPS of a relativistic bunch. An experimental demonstration of the concept was performed at the FERMI@ELETTRA accelerator using a 1.4-GeV beam~\cite{penco-2014-a}.  The photocathode laser was temporally shaped to precompensate nonlinear correlations nominally accumulated during the acceleration and two-stage bunch compression processes. The measured LPS upstream of the FEL beamline is uncorrelated and linearized; see Fig.~\ref{fig:sec4:LaserPenco}. In the process, the authors demonstrated that the beam dynamics of the ramped bunch did not significantly affect the beam transverse emittance compared to their nominal operating point.

\subsection{Transverse phase-space control}
We now examine techniques to transversely shape the beam. This type of shaping has commonly been implemented via transverse shaping of the photocathode-laser pulse in a photoemission source or by proper shaping of the potential in a thermionic-cathode-based electron source as detailed in Section~\ref{sec2}. This type of shaping is ultimately impacted by collective effects so that devising shaping methods implementable at high energy is beneficial. For instance, producing uniform electron beams with uniform transverse distribution  is critical to a broad range of applications including beam irradiation of a target, e.g., to produce X rays, as it provides a uniform dose and mitigates thermo-mechanical stress on the target~\cite{pasquali-2019-a}. Likewise, using bright electron beams with uniform transverse distribution offers an 
effective way of improving the performance and output power of tapered X-ray FELs~\cite{emma-2014-b}. Finally, transversely shaped beams combined with the mask-based shaping techniques discussed earlier can lower beam losses on the mask. These transverse-shaping methods have gained considerable interest, as their possible combination with the phase-space-exchanging techniques discussed in Section~\ref{sec6} could enable control of the current distribution with unprecedented precision and versatility.

\subsubsection{Nonlinear transformations~\label{sec4:sec:nonlinearbeammapping}}
\begin{figure}[hhhh!!!!]
 \includegraphics[width=.85\linewidth]{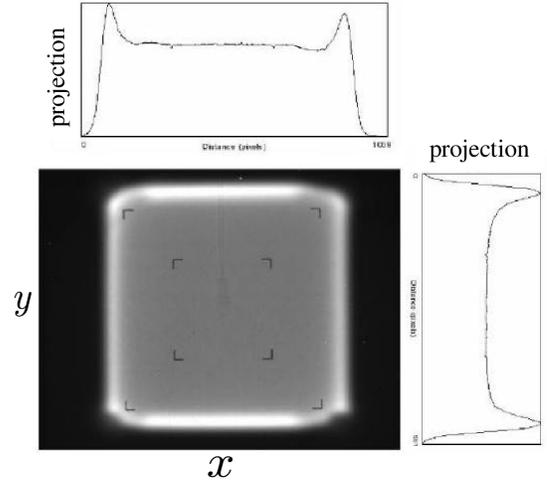}
\caption{\label{fig:sec4:octupole_uniform} Measured beam distribution obtained downstream of a uniformizing beamline composed of octupole magnets. Adapted from ~\cite{tsoupas-2007-a}. }
\end{figure}

The idea of using nonlinear transformations parallels the early discussion related to LPS manipulation. Such an approach was first explored in ~\cite{merminga-1991-a} to remove the beam tail via nonlinear focusing (implemented with sextupole magnets) as part of a ``nonlinear collimation" scheme proposed for future linear colliders. Likewise, the use of nonlinear focusing to form uniform beam distributions was discussed in ~\cite{meads-1983-a}. Numerical simulations ~\cite{kashy-1987-a} demonstrated the formation of uniform distribution using an octupole magnet implemented in practical beamline. Likewise, ~\cite{batygin-1993-a} derives the nonlinear force required to redistribute an incoming beam non-uniform distribution into a uniform distribution using a beamline composed of a multipole lens followed by a drift space. Further work discussed in ~\cite{meot-1996-a} analytically shows that the odd-order multipole fields, such  octupole and dodecapole components, are required for transverse uniformization. Efforts to produce transversely uniform distributions over appreciable distances along an accelerator beamline has been investigated for possible use in combination with undulator tapering in X-ray FELs~\cite{jiao-2015-a}. Over the years several experiments have been conducted. An early demonstration experiment using a 200-MeV $H^-$ beam was performed at BNL~\cite{tsoupas-1991-a}, and an example of distribution measured at the NASA Space Radiation Laboratory (NSRL), using an electron beam at BNL~\cite{tsoupas-2007-a}, appears in Fig.~\ref{fig:sec4:octupole_uniform}. Similarly, uniform beam distributions are sometime used in medical accelerators. 

\begin{figure}
 \includegraphics[width=.975\linewidth]{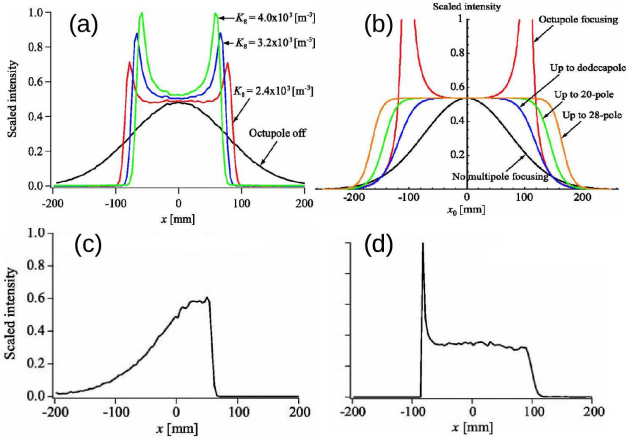}
\caption{\label{fig:sec4:yuritheoryuniform} Example of computed transverse profiles for a nonlinear-transport beamline composed of odd-order fields (a,b) and two sextupole magnets (c,d). Adapted from~\cite{yuri-2007-a}. }
\end{figure}

Reference~\cite{yuri-2007-a} develops a theoretical framework to understand the formation of shaped beam profiles using a nonlinear focusing beamline. Following this work, the phase-space coordinates downstream of the beamline under consideration can be expanded as a truncated power series of the initial phase-space coordinate upstream of the beamline ($x_0,x_0'$) via a transformation of the form
\begin{eqnarray}
\begin{pmatrix}
x_t \\
x'_t
\end{pmatrix}
=
\begin{pmatrix}
(M_{11}-\frac{\alpha_0}{\beta_0}M_{12})x_0 - M_{12} \sum_{n=3}^{\infty} \frac{K_{2n}}{(n-1)!}x_0^{n-1}\\
(M_{21}-\frac{\alpha_0}{\beta_0}M_{22})x_0 - M_{22} \sum_{n=3}^{\infty} \frac{K_{2n}}{(n-1)!}x_0^{n-1}
\end{pmatrix} . \nonumber
\end{eqnarray}
Imposing Liouville's theorem as discussed in Section~\ref{sec2c1Liouville} indicates that the final distribution along one of the position coordinates can be written as
\begin{eqnarray}
\rho_t=\frac{\rho_0}{M_{11}-\frac{\alpha_0}{\beta_0}M_{12} -\sum_{n=3}^{\infty} \frac{K_{2n}}{(n-2)!}x_0^{n-1} } .
\end{eqnarray}
This equation provides the values of the magnetic multipole strength $K_{2n}$ required to transform the initial distribution into the desired one. The work also extends previous methods by investigating the effect of the even-order multipole fields on beam uniformization and demonstrating the feasibility of uniformization of a Gaussian beam for even-order fields, or considering incoming transversely asymmetric beams. The authors conclude that using a combination of even-order multipoles may ultimately provide uniform beam over a larger transverse section; see Fig.~\ref{fig:sec4:yuritheoryuniform}(b). 

 Finally, although the emphasis has often been on the generation of uniform beams, the technique can also produce beams with transversely shaped distribution. For instance, Fig.~\ref{fig:sec4:yuritheoryuniform} demonstrates the generation of a ramped horizontal distribution obtained as an intermediary step to generate a uniform distribution. Furth development demonstrated the generation of patterned hallow beams~\cite{yuri-2019-a}. This latter capabilities could be taken advantage of and combined with phase-space-exchange methods discussed in Section~\ref{sec6}.

 \begin{figure}[hhhh!!!!]
 \includegraphics[width=.95\linewidth]{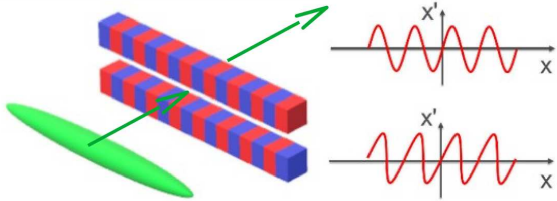}
\caption{\label{fig:sec4:twiggler} Configuration for transverse beam shaping using a ``transverse"-wiggler magnet (left) and horizontal phase-space (right) immediately downstream (upper plot) the wiggler and after a subsequent drift (lower plot).  Adapted from ~\cite{ha-2019-a}. }
\end{figure}

Another method toward shaping the transverse phase space is to use a transversely periodic magnetic field to modulate the electron divergence across the transverse beam distribution as proposed in ~\cite{ha-2019-b} and illustrated in Fig.~\ref{fig:sec4:twiggler}. Taking the example of the horizontal phase space $(x,x')$ passing through a set of  transverse wigglers $-$ a wiggler oriented transverse to the beam direction $-$ with wiggler parameter $K_{w,i}$, the divergence of an electron at initial position $x_0$ will be given by 
\begin{eqnarray}
x'=x_0'+ \sum_{i=1}^n K_{w,i} (\sin k_i x), 
\end{eqnarray}
where $k_i\equiv2\pi/\lambda_i$ is the transverse wavevector with $\lambda_i$ being the $i^{th}$ wiggler period.  A proper choice of $k_i$ and $K_{w,i}$ allows the synthesis of any correlation in the $(x,x')$ phase space that could produce a tunable profile along the horizontal spatial or divergence direction. 

\subsubsection{Interceptive beam shaping: beyond binary masks\label{sec4:beyondBinMask}}
As discussed in Section~\ref{sec4:subsec:currentshaping}, an interceptive ``binary" mask with optimized contour provides a versatile and simple tool to shape the beam  transverse distribution but is rarely used owing to the accompanying losses. These losses are especially problematic when the shaping technique is implemented in a high-current accelerator as they can lead to, e.g., radiological activation and damage of beamline hardware. 

For ultra-low emittance beams where quantum coherence is achieved and the beam can be described by its wavefunction, optical (photon) techniques have been adapted to shape the beam with a high degree of control, e.g., in electron microscopes~\cite{nagayama-2011-a,shiloh-2019-a}. An example of such a manipulation is the generation of ``vortex" electron beams carrying orbital angular momentum~\cite{uchida-2010-a} using a nano-engineered spiral-like phase plate made of stacked graphite thin film.

\begin{figure}[t]
\includegraphics[width=1\linewidth]{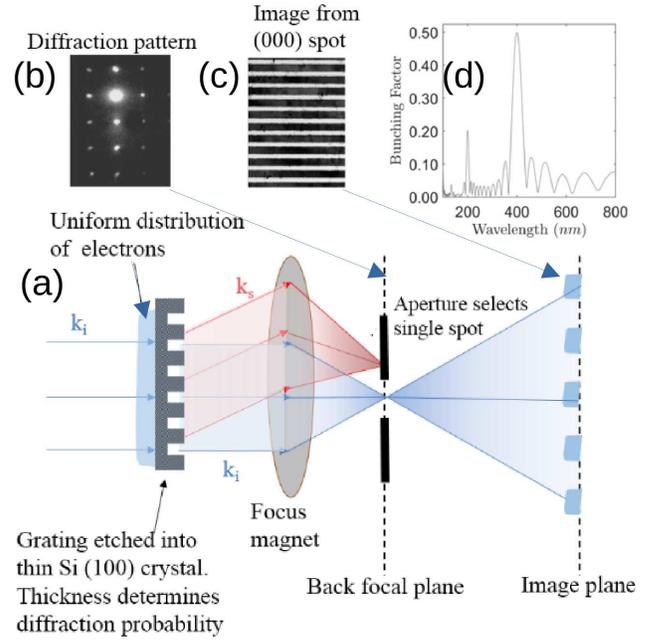}
\caption{\label{fig:sec4:diffractionmodulation} Principle of diffraction modulation to produce a nano-patterned beam (a) with diffraction pattern downstream of the grating (b) and final transverse distribution of the transmitted beam (c) with associated spatial spectrum along the vertical coordinate (d). Figure adapted from~\cite{graves-2019-a}. }
\end{figure}

Although electron beams produced in conventional accelerators are incoherent, ultralow-emittance relativistic electron beams can be manipulated with variable-transmission masks, as discussed in ~\cite{nanni-2018-a}, provided their transverse coherent $L_{\perp}=\frac{\hbar}{mc}\frac{\sigma_{\perp}}{\varepsilon_{\perp}}$~\cite{vanoudheusden-2007-a} is comparable or larger than the intercepting-mask crystalline structure (where $ \sigma_{\perp}$ and $\varepsilon_{\perp}$ respectively refer to the transverse beam size and emittance). In this reference, a diffraction contrast modulation technique was proposed to realize nano-patterned electron beams. The method relies on electron diffraction in a transmission grating with periodically variable thickness. The grating structure spatially modulates the fraction of electrons diffracted into a particular Bragg peak. An aperture selecting the transmitted beam or one of the Bragg peaks then results in a final modulated beam. The method was demonstrated at SLAC using a 2.3-MeV 1-pC electron beam from an ultrafast electron-diffraction setup~\cite{graves-2019-a}. The beam was diffracted through a thin lithographically patterned Si membrane [Fig.~\ref{fig:sec4:diffractionmodulation}(a)] to produce a vertically modulated beam distribution [Fig.~\ref{fig:sec4:diffractionmodulation}(d)] with a spatial period of $\sim 400$~nm and an associated bunching factor of 0.5 [see spatial spectrum in Fig.~\ref{fig:sec4:diffractionmodulation}(b)]. In this experiment, the unwanted electrons were scattered into the (220) Bragg peak, and the transmitted electron associated with the (000) peak was used to produce the final 300-fC modulated beam  [Fig.~\ref{fig:sec4:diffractionmodulation}(c,d)]. This masking technique has some limited tunability accomplished by varying the tilt angle of the grating and selecting the Bragg peak to diffract the beam where desired. This method ultimately suffers from the mask's limited lifetime due to atomic displacement over long periods of time. The achieved sub-micrometric spatial period combined with demagnifying optics could produce beams with nanometric modulations.

\section{Longitudinal shaping with beam self-generated field\label{sec5}}

As described in the previous section, the rule of thumb in beam shaping is to introduce controlled electromagnetic fields, and external fields are easy-to-control sources for beam shaping. However, we can also use the beam's self-generated fields to shape the beam. 

Self-generated fields are usually considered to be an obstacle to improving the beam quality. While fully decoupled Gaussian beam or beams having linear correlations are desirable, most self-generated fields introduce non-linear correlations on the beam. The non-linearity on the beam's phase space not only increases the emittance, but it introduce intrinsic limitations on the manipulation of the beam. Overcoming the degradation of the beam quality caused by the self-fields is one of the major research topics in accelerator physics. In the best case, self-fields might enable manipulation of the beam  and avoid its degradation simultaneously. 

This section focuses on shaping mechanisms that are based on beam-generated fields. Most of the methods introduced in this section are for longitudinal shaping because the use of self-generated fields for transverse shaping is scarce.  We first discuss longitudinal profile shaping using the space-charge field and CSR in Sec.~\ref{sec5_SC} and Sec.~\ref{sec5_CSR}. Then we describe the mechanism using wakefields in Sec.~\ref{sec5_wake}. Each subsection provides a short description of the results and their applications.

%%%%%%%%%%%%%%%%%%%%%%%%%%%%%%%%%%%%%%%%%%%%%%%%%%%%%%%%%%%%%%%%%%%%%%%%%%%%%%%%%%%%%%%%%%%%%%%%%%%%%%%%

\subsection{Shaping profiles using space-charge field} \label{sec5_SC}
The space-charge field has two characteristics. The first is that it depends on the beam's spatial distribution. If the beam has a symmetric profile, the field strength is symmetric to the beam center.  The second is that it acts in a way to defocus or lengthen the bunch when the beam is not density modulated because electrons repel each other. These two characteristics give us several methods for using the space-charge field for shaping. The following subsections describe these shaping methods and give relevant results.

\subsubsection{Space-charge field with a single bunch} \label{sec5_SC1}
Due to the field's dependence on the beam's spatial distribution, the usefulness of the space-charge field from a single bunch is limited. However, we can imagine two different uses of the space-charge field from a single bunch.

\begin{figure}
\centerline{\includegraphics[width=0.5\textwidth, keepaspectratio=true]{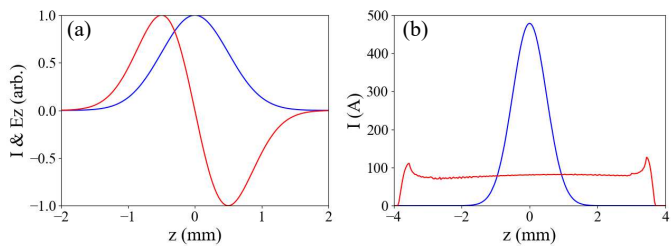}}
\caption{Current profile shaping using space-charge force.  (a) Gaussian longitudinal profile (blue) and corresponding longitudinal space-charge field from Eq.~\eqref{sec5_eq_sc3}. (b) Longitudinal profile of the initial particle distribution (blue), and the longitudinal profile after a 3-m-long drift (red). (b) is a simulation result of a beam with 1-nC charge and 3-MeV energy.}\label{sec5_fig_scshaping}
\end{figure}

First, the space-charge force applied to the bunch is not linear except for a few profiles. Thus, similar to the transverse shaping using an octupole magnet \cite{yuri-2007-a} (see Sec.~\ref{sec4:sec:nonlinearbeammapping} for further details), a symmetric-nonlinear space-charge field can induce a symmetric change in the profile. For example, in a cylindrical beam with a uniform transverse distribution  and a  longitudinal distribution of $\lambda(z)$, the space-charge field can be expressed as in Eq.~\eqref{ZEqnNum728960}. If the initial longitudinal distribution is Gaussian, then the axial space-charge field will be
\begin{equation}\label{sec5_eq_sc3}
E_z(z) = {e\over4\pi\epsilon_0 \gamma^2} {N_b z\over \sigma_z^3}\left(1+2\ln{b\over a}\right)\exp\left[-{z^2\over 2\sigma_z^2}\right].
\end{equation}
The Gaussian profile and corresponding space-charge field in Eq.~\eqref{sec5_eq_sc3} are visualized in Fig.~\ref{sec5_fig_scshaping}(a). In the case of the example shown in Fig.~\ref{sec5_fig_scshaping}, all particles move outward due to the space-charge field as the beam drifts, and the bunch length will increase. The space-charge field in this example is nonlinear and has its maximum strength at around $\pm0.5$ mm. Thus, particles initially located at $\pm$0.5 mm gain more momentum than particles in $|z|>0.5$ mm, and eventually overtake these particles in $|z|>0.5$ mm. This situation is very similar to octupole-based transverse shaping, see Fig.~\ref{fig:sec4:octupole_uniform}. Thus, if the beam traverses a long enough distance (i.e., large enough $R_{56}$), then this Gaussian distribution would evolve into a uniform distribution. Figure~\ref{sec5_fig_scshaping.eps}(b) shows a particle tracking simulation result, which demonstrates the idea. A Gaussian profile with a charge of 1 nC and energy of 3 MeV evolves into a uniform profile due only to its space-charge field as it drifts for 3 m.  As we can see from Eq.~\eqref{sec5_eq_sc3}, the space-charge field gets weaker as the beam energy increases. Thus, the change arising from the space-charge field requires a large R$_{56}$ to convert momentum change to position change. This requirement limits the use of the space-charge field to near the electron gun; thus, so far this type of shaping is not demonstrated experimentally.

We can consider a similar type of shaping for generating asymmetric bunches, but it is necessary to introduce an originally asymmetric profile. Thus, using the space-charge field for asymmetric control is not an attractive option, but the understanding of this mechanism may help to improve the shaping quality of emission-based shaping methods introduced in Section~\ref{sec3}. 

\begin{figure}
\centerline{\includegraphics[width=0.5\textwidth, keepaspectratio=true]{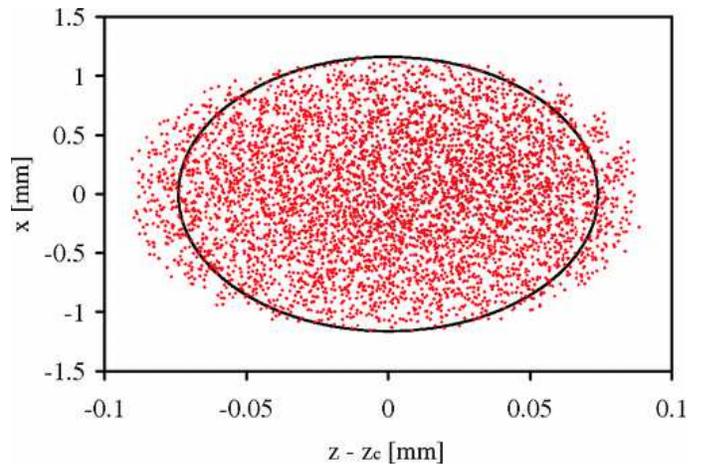}}
\caption{Blow-out regime concept and its demonstration using simulation. The figure shows simulated particle distribution in x-z. The image is taken 50 ps after the laser illuminates the photocathode. The black circle represents the theoretically ideal result. From \cite{luiten-2004-a}.}\label{sec5_fig_3dellipsoid}
\end{figure}

The other usage is exploiting the symmetric lengthening feature of the space-charge field. In \cite{luiten-2004-a} they suggested a new concept, the so-called blow-out regime, to generate a 3D ellipsoidal distribution with uniform density.  In this concept, an ultra-short laser pulse with an appropriate radial distribution ($\lambda_{\perp}(r)\sim\sqrt{1-(r/A)^2}$ where A is the maximum radius) shines on a photocathode. This generates an ultra-thin sheet of electrons that evolves into a uniform ellipsoidal shape. Here, the ultra-short pulse length plays a key role because it allows that all particles in the bunch to experience a similar space-charge field strength. This allows the beam to evolve into a uniform ellipsoid regardless of the original pulse shape. The imperfection here will appear as a soft edge of the ellipsoid. In \cite{luiten-2004-a} they ran simulations using a 100-pC beam with 1-mm radius and 0.4-eV energy at the cathode surface. The electric field at the cathode surface is assumed to be 100 MV/m, and the incident laser pulse is assumed to have a Gaussian profile with 30-fs FWHM. Figure~\ref{sec5_fig_3dellipsoid} shows a simulated particle distribution 50 ps after the laser illuminates the cathode. The original disk-shaped distribution has evolved into a nearly elliptical one that is slightly less than ideal (black curve) due to the imperfection.

\begin{figure}
\centerline{\includegraphics[width=0.45\textwidth, keepaspectratio=true]{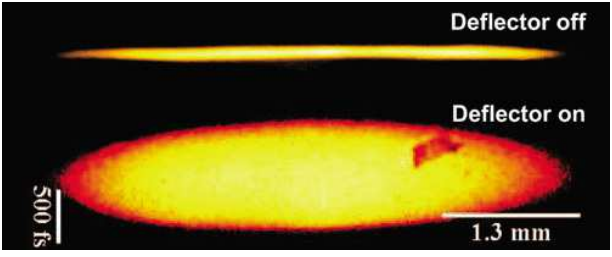}}
\caption{Experimental demonstration of ellipsoidal beam generation using blow-out method. The beam images were measured after a deflecting cavity. The beam was focused vertically, and the deflecting cavity kicked the beam vertically. x and y axes correspond to the beam's horizontal and longitudinal distributions, respectively. The beam's energy was 3.75 MeV and the charge was 15 pC. From \cite{musumeci-2008-a}.}\label{sec5_fig_3dellipsoidexp}
\end{figure}

In~\cite{rosenzweig-2006-a}, further theoretical work was performed to combine the blow-out regime and emittance compensation. Later, this blow-out regime was experimentally demonstrated by \cite{musumeci-2008-a,oshea-2011-a,piot-2013-a}. The first experiment in \cite{musumeci-2008-a} demonstrated the blow-out regime using 15-pC beam; see Fig.~\ref{sec5_fig_3dellipsoidexp}. Later, the experiment in \cite{piot-2013-a} demonstrated the blow-out using a higher charge level (0.5-nC) with a Cs$_2$Te cathode, and it showed well-shaped bunches. However,  the method required a disk-shaped beam, making it difficult to reduce the laser spot size at the cathode to obtain target charge. This limits the intrinsic emittance, which scales with laser spot size,  Eq.~\ref{eq:sec3-cathodeemittance}. The emittance demonstrated so far does not show a clear advantage of this method. There is another emission scheme, based on a cigar-beam, introduced in \cite{rosenzweig-2019-a} that achieved lower emittance. While the blow-out regime uses a short but large radius laser profile, which produces a longitudinal expansion of the beam, the cigar-beam regime uses a long but small radius laser profile which produces radial expansion of the beam. Simulations showed a noticeable improvement in beam brightness of the cigar compared to the blow-out regime. Due to these limitations, research efforts were moved to using ellipsoidal laser profiles instead of the blow-out regime to generate an ellipsoidal beam; see \cite{khojoyan-2014-a}. See Section~\ref{sec3} for further details of laser shaping.

\subsubsection{Space-charge field with a few bunches} \label{sec5_SC2}

When more than a single bunch exists, the bunches mutually repel each other via the space-charge field. If there are extra bunches in front of and behind the target bunch, the electromagnetic field felt by the target bunch will depend on these extra bunches and the target bunch's self-field. Here, the charge ratio of each bunch and the spacing between the bunches can be used as knobs to control the space-charge field applied to the target bunch. 

In \cite{lu-2018-a}, this concept was used to compress an electron bunch. During the experiment, they generated a low-charge main bunch and two extra bunches for space-charge field shaping. The bunches were generated by a photocathode gun and using a laser beam splitter and $\alpha$-BBO crystals to generate a laser pulse train. The main bunch was located  between two field-shaping bunches. These two field-shaping bunches push the target bunch, which generates a negative longitudinal chirp of the target bunch’s longitudinal phase space (LPS). Note that the negative chirp means the head of the bunch has lower energy. Because the space-charge field outside of the bunch shows the usual 1/$r^2$ tendency, they located the target bunch at the middle so that the nonlinearity of the space-charge field coming from the field-shaping bunches almost cancels out. This provided a nearly linear longitudinal chirp to the target bunch; the relative separation was adjusted by using a laser delay line. The laser was injected into the RF photocathode gun that accelerated the beam to about 3.4 MeV. Due to the beam's low energy, a few meters of drift provided enough $R_{56}$ for ballistic bunching. Ballistic bunching means that the bunch compression is accomplished by the velocity difference of the particles in a drift. Velocity bunching is a more general term, but the ballistic bunching is used when acceleration is not included in the compression.

\begin{figure}
\centerline{\includegraphics[width=0.5\textwidth, keepaspectratio=true]{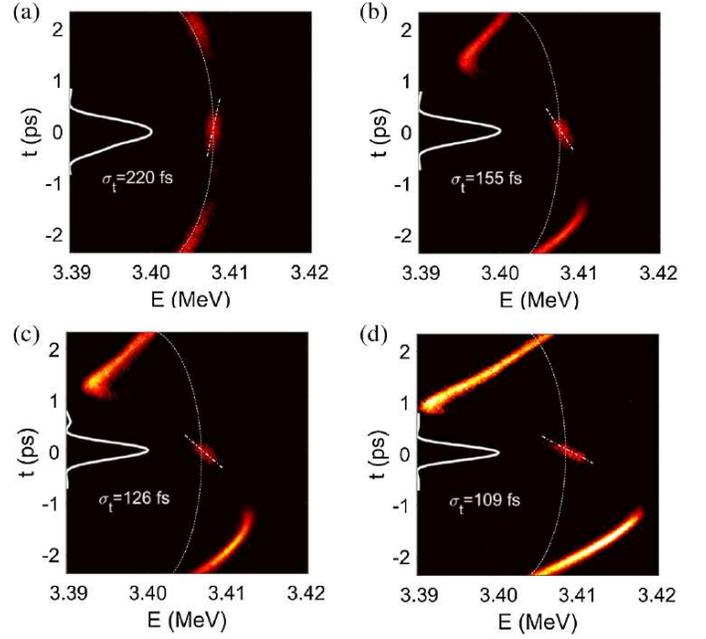}}
\caption{Space-charge field shaping for bunch compression. Measured longitudinal phase spaces are displayed. Each panel corresponds to different charge level of field-shaping bunches. (a-d) correspond to  charges of 0.2, 3.2, 4.5, and 6.7 pC, respectively. $t>0$ corresponds to the head. From \cite{lu-2018-a}.}\label{sec5_fig_sccompression_exp}
\end{figure}

During the experiment, they fixed the target bunch's charge level to 50 fC while varying the field-shaping bunches' charges from 0.2 to 6.7 pC. Here, the charges of the field-shaping bunches were used to control the field strength applied to the target bunch. They observed bunch compression with an LPS measurement as shown in Fig.~\ref{sec5_fig_sccompression_exp}. Also, the main bunch length was controlled from 220 fs to 109 fs by varying the charge. As we can see from Fig.~\ref{sec5_fig_sccompression_exp}(d), there is still room for further compression.

Another interesting point of this work is stability. In the case of chicane compression, the RF jitter of the cavity controlling the chirp affects the longitudinal chirp or the beam energy. This RF jitter, in turn, result in an increase of the bunch length jitter and arrival time jitter. On the other hand, all three bunches originated from a single laser pulse, and there are no RF cavities other than the gun in this scheme. Thus, it can provide better stability than other methods. This is an interesting method for applications requiring a short bunch with good stability such as ultra-fast electron diffraction (UED), ultra-fast electron microscopy (UEM), or FELs.

A similar method was also experimentally demonstrated for transverse phase space. In \cite{rihaoui-2009-a}, they introduced extra bunches surrounding a central main bunch in the transverse space; see Sec.~\ref{sec3:shaped:PEtrans}. Due to the interaction via the space-charge force and the beam's distribution change by solenoid focusing, these extra bunches either introduced extra focusing to the main beam or changed the beam's distribution to a shape that was other than a Gaussian distribution. This could be an interesting option for a system having difficulty using magnets. For example, locating a solenoid near the cathode of superconducting guns is difficult, but transverse control is necessary to preserve the beam's quality. These extra bunches may provide all-beam-focusing to the target beam, so it may be able to locate a solenoid further downstream while preserving the beam quality.

Although the experimental demonstrations discussed in this section are limited to one-dimensional density control, we can imagine appling this method to more general bunch shaping applications. For example, nonlinearity is a key for asymmetric bunch shaping, as we saw in Section~\ref{sec4}. We can  imagine exploiting nonlinearities that space-charge force introduces. Bunch charges and spacing can be used to control the shape of LPS.  Combining this method with laser shaping, may be an opportunity to overcome the high-charge limitation of the laser shaping method. Laser shaping methods are limited to low-charge bunches because of the strong space-charge force of the main bunch. However, the space-charge field of the main bunch can be controlled by using additional bunches which may provide a new way to apply a laser-shaping technique to high-charge bunches.

We should note that beam shaping with extra bunches always has the problem of eliminating the extra bunches after they are no longer needed for shaping. The elimination process may introduce additional disadvantages that make the method less beneficial than described so far. A beam cutting process may happen using a mask and an additional beamline element such as chicane or RF deflecting cavities; \cite{ha-2020-a}. However, these beamlines can easily add extra timing jitter to the beam, negating one of the benefits of this method. Thus, it is necessary to consider this elimination process during the design.

%%%%%%%%%%%%%%%%%%%%%%%%%%%%%%%%%%%%%%%%%%%%%%%%%%%%%%%%%%%%%%%%%%%%%%%%%%%%%%%%%%%%%%%%%%%%%%%%%%%%%%%
%%%%%%%%%%%%%%%%%%%%%%%%%%%%%%%%%%%%%%%%%%%%%%%%%%%%%%%%%%%%%%%%%%%%%%%%%%%%%%%%%%%%%%%%%%%%%%%%%%%%%%%

\subsubsection{Space-charge field with multiple bunches: space-charge oscillation}\label{sec5_SC3}
We now consider more than three bunches by providing a density modulation at the beginning. This initial density modulation introduces oscillatory behavior in the longitudinal phase space, which is similar to the plasma oscillation that ions and electrons form. In the electron bunch case, the initial density modulation and bunch-to-bunch interaction via the space-charge force introduces the same oscillation.

\begin{figure}
\centerline{\includegraphics[width=0.5\textwidth, keepaspectratio=true]{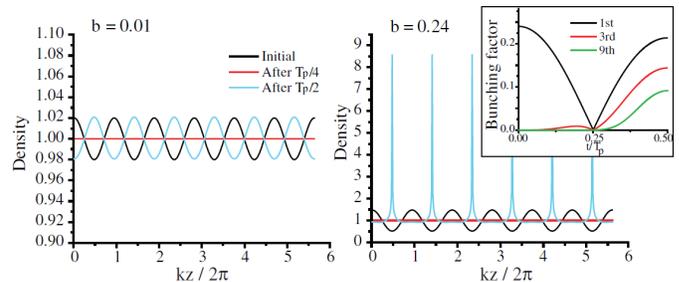}}
\caption{Nonlinear plasma oscillation of the beam to generate a bunch train with a high bunching factor. Beam’s longitudinal density profiles and their evolution are displayed. The left panel started with a bunching factor of 0.01, and the right panel started with a bunching factor of 0.24. These are numerical solutions of the system of equations describing a coasting beam's density profile evolution. From \cite{musumeci-2011-a}.}\label{sec5_fig_scoscillation_theory}
\end{figure}

This behavior is shown in the left panel of Fig.~\ref{sec5_fig_scoscillation_theory}. The initial density modulation (black) lengthens each micro-bunch, due to the space-charge force, so that the modulation becomes weaker as the beam drifts until it finally disappears at a certain distance (red). Once the beam passes this point, the particles in each micro-bunch continue to move in the same direction, due to their momentum. These particles build another density modulation whose phase is 180$^\circ$ off from the original modulation, thus the direction of the space-charge force now is reversed. At a certain point (blue), the momentum direction changes, due to the reversed space-charge force, and this density modulation starts to oscillate. The oscillation frequency of this oscillation is equal to the plasma oscillation's frequency in Eq.~\eqref{omegap}.  The change of the relative energy spread induced by the longitudinal space-charge force can be written as
\begin{equation}\label{sec5_eq_momentumkick}
    \Delta\delta_{LSC} = 4{\gamma^3\omega_p b\over ck},
\end{equation}
where k is wave number ($={2\pi\over\lambda}$), and $b$ is bunching factor $=|\int \lambda(z,0)e^{ikz}dz|$; see \cite{musumeci-2013-a}.

%%% nonlinear oscillation
In the linear oscillation situation described above, a sine-like shape on the phase space does not break and keeps oscillating because the energy modulation induced by the initial density modulation is small so that all particles cannot arrive at the density peak area before the momentum is reversed (i.e., $R_{56}\delta\ll \lambda/4$). However, when the modulation becomes big enough, the space-charge force can induce wave-breaking where the momentum modulation is large enough to cause all particles to reach the density peak area (i.e., $R_{56}\delta\approx \lambda/4$).  

It is also possible to understand this process as the summation of harmonics. In the case of linear oscillation, only the fundamental mode governs the oscillation.  However, as the initial density modulation gets larger, the oscillation process picks up more harmonic components of the fundamental wavelength.  After half of the plasma period, these harmonic components sum in-phase and generate current spikes.  This viewpoint can be seen from the equation below showing the electron density. More details are described in \cite{musumeci-2013-a}.
\begin{equation}
\lambda(z,t) = \left[1+\cos\left(\omega_p t\right) \sum_{m=1}^{\infty} m c_{m}(t) \cos \left(mkz\right)\right],
\end{equation}
where $c_m(t)={(-1)^{m+1}\over m}{2\over\alpha(t)}J_m\left[m\alpha(t)b\right]$, and $\alpha(t)=2\sin^2(\omega_pt/2)$.

Although this concept can be applied to any coasting beam, it is impractical for a high-energy beam because of its long plasma wavelength.  Thus, it was applied to an electron gun by several groups \cite{neumann-2003-a,harris-2007-a,neumann-2009-a}. Here we describe more details of the experiment done by \cite{musumeci-2011-a}, which successfully demonstrated the generation of current spikes.

\begin{figure}
\centerline{\includegraphics[width=0.45\textwidth, keepaspectratio=true]{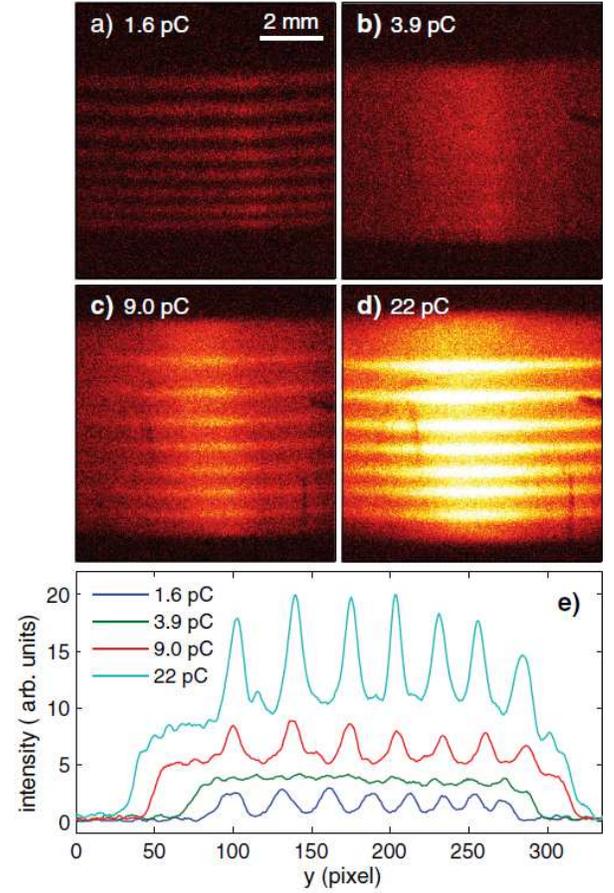}}
\caption{Experimental demonstration of bunch train generation using nonlinear plasma osciallation. (a-d) shows temporally streaked beam images, and corresponding longitudinal profiles are shown in (e). Streaking direction for beam images is vertical. From \cite{musumeci-2011-a}.}\label{sec5_fig_scoscillation_exp}
\end{figure}

They used three $\alpha$-BBO crystals to generate eight equally spaced laser pulses to form a 1-THz ($\lambda=300$ $\upmu$m) pulse train; see Sec.~\ref{sec3:shaped:PElong}. They tested the bunch charge up to 40 pC, and the gun solenoid was used to focus the beam. To demonstrate the method, they varied the laser intensity, which controls the charge, to change the phase advance of the oscillation. The phase advance determines the final modulation density. The modulation was measured with a deflecting cavity at the end of an experimental beamline. The solenoid field was fixed, and the gun phase was set to 35$^\circ$ to preserve the original modulation period. The result is summarized in Fig.~\ref{sec5_fig_scoscillation_exp}.

According to their simulation, the phase advance of the lowest charge case (1.6 pC) was about 0.18$\pi$; see Fig.~\ref{sec5_fig_scoscillation_exp}. Thus, it still showed the original modulation and all eight peaks appeared at the end of the beamline. When the charge was increased to 3.9 pC, the modulation was washed out; this charge level corresponds to the phase advance of 0.25$\pi$ in their simulation. As the charge increased further, the phase advance became slightly lower or higher than 0.5$\pi$, so a density modulation appeared again, but now there were only seven peaks because of the $\pi$ phase shift of density peaks in a given total bunch length. 

A few years later, this method was revisited and experimented at a higher charge level to consider an application to terahertz (THz) radiation. In \cite{zhang-2016-a} they set up an experimental beamline similar to \cite{musumeci-2011-a} and added a chicane to broaden the frequency tuning range. They used a charge up to 1 nC, and results were very similar to the ones from  \cite{musumeci-2011-a}. They controlled both the charge level and the solenoid focusing strength to control the phase advance. The initial modulation was washed out with a low phase advance while the modulation was shifted by 180$^\circ$ when the phase advance was 0.25$\pi$.

In addition to confirming the oscillation at a higher charge level, they also tried to generate THz coherent transition radiation (CTR) using a foil. They adjusted two tuning parameters to control the modulation frequency and measured the spectrum of CTR using a Michelson interferometer equipped with a Golay cell \cite{golay-1947-a}. The first parameter was the launching phase of the gun. Depending on the launching phase, the bunch had different longitudinal chirps. This eventually introduced different compression ratios for ballistic bunching in the low energy area. They varied the phase from 25$^\circ$ to 50$^\circ$, and the frequency of CTR was varied from ~1 THz to ~0.7 THz, correspondingly. The second parameter was the chicane's bending angle, which controlled $R_{56}$. Compared to the launching phase, this second parameter provided a much wider tuning range. They changed the current to the dipole magnet from 0 A to 30 A, which corresponded to $R_{56}$ values from 0 to ~18 mm, and it changed the frequency from ~0.8 THz to 1.6 THz.

Because of the high charge level, they achieved ~2 $\upmu$J of THz energy from CTR.  They expected that a 30-mm-long quartz tube with 0.3- and 0.4-mm inner and outer diameters, respectively, would provide ~8-MW, ~1.4-mJ THz radiation at 0.7-THz frequency. This intense and tunable THz radiation from a compact beamline can be useful for a typical spectroscopy-type equipment, pump-probe measurements in XFELs, or THz wakefield accelerators.

\subsubsection{Space-charge field with multiple bunches: longitudinal cascade amplifier}\label{sec5_SC4}
\begin{figure}
\centerline{\includegraphics[width=0.5\textwidth, keepaspectratio=true]{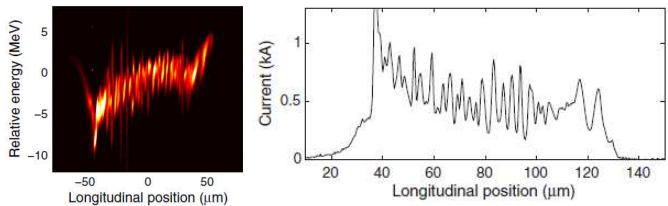}}
\caption{Measurement of microbunching instability. The left panel shows the measured longitudinal phase space without a laser heater. The strong modulation originated from the shot noise and its amplification. The right panel shows the corresponding current profile. Adapted From \cite{ratner-2015-a}.}\label{sec5_fig_mbi}
\end{figure}

%%% cascade amplification
When a momentum modulation from an initial density modulation is strong enough, the initial density modulation can be amplified with an appropriate $R_{56}$. This phenomenon was studied extensively in the early 2000s to understand the significant beam quality degradation and radiation generated from the beam. This is called the microbunching instability and many theories and experimental papers were published to explain and suppress this phenomenon; for example, \cite{saldin-2002-b,saldin-2002-a,saldin-2004-a,heifets-2002-a,huang-2002-a,wu-2008-a,lumpkin-2009-a,huang-2010-a,marinelli-2011-a,spampinati-2014-a,ratner-2015-a,prat-2017-a}. One of them, \cite{ratner-2015-a}, showed a picture of the longitudinal phase space with various conditions, and Fig.~\ref{sec5_fig_mbi} clearly shows the modulation amplified by the gain process; see Sec.\ref{sec2d5}.

As described in Sec.~\ref{sec2d5}, \cite{saldin-2002-a} derived a simple equation to estimate the gain ($G$) of the density modulation, which is the ratio of final to initial modulation amplitude. Longitudinal space-charge fields introduced energy modulation to the beam, as appeared in Eq.~\eqref{constxi} as an impedance term. Then, $R_{56}$ of the given beamline converted an energy modulation to a density modulation.  For the microbunching instability, the longitudinal space-charge force from the shot noise makes the gain higher than 1 and amplified this unwanted shot noise. Thus, an effort, such as using a laser heater, was made to reduce the gain [see the $\sigma_{\delta}$ term in the exponent of Eq.~\eqref{constxi}].

\begin{figure}
\centerline{\includegraphics[width=0.5\textwidth, keepaspectratio=true]{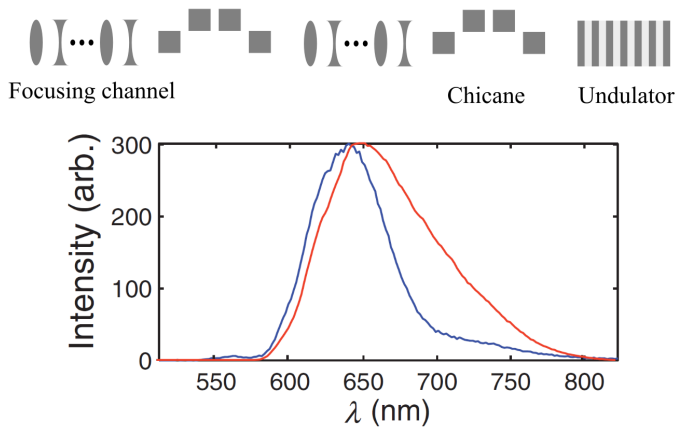}}
\caption{ Measurement of radiation spectrum from a bunch train generated by a longitudinal cascade amplifier. The top panel shows a configuration of a longitudinal cascade amplifier with two cells. The bottom panel shows a measured spectrum of undulator radiation with the cascade amplifier. The blue corresponds to the on-axis radiation spectrum while the red shows the integrated one. The bottom figure is adapted from \cite{marinelli-2013-a}.}\label{sec5_fig_sccascade}
\end{figure}

On the other hand, \cite{Schneidmiller-2010-a} introduced a concept to use this amplification of shot noise as a way for generating a high-frequency bunch train; it was called the longitudinal space-charge cascade amplifier (LSCA).  In this concept, they used a drift with a focusing channel to introduce a momentum modulation from the density modulation, then a chicane is used to introduce an $R_{56}$ optimized for the modulation conversion with maximum gain (see the top panel of Fig.~\ref{sec5_fig_sccascade}). This focusing channel and chicane form a single cell of the amplifier, and each cell provides gain to the density modulation. Thus, they used a cascade of cells to saturate the density modulation. The numerical example they provided shows that the gain per cell reaches more than 40 for a beam with an energy of 3 GeV, a peak current of 2 kA, a normalized emittance of 2 $\upmu$m, and an energy spread of 0.3 MeV, which is close to typical FEL parameters. Thus, two cascades would provide a $G>1,000$. This is high enough to saturate the shot noise for this example case. Here the amplitude of shot noise ($A_{sh}\simeq \sqrt{ec\over I\lambda}$) is around 0.001 for $\lambda=15$ nm getting optimal gain. 

 The LSCA method was experimentally demonstrated in 2013 by \cite{marinelli-2013-a}. The experiment amplified shot noise using three chicanes with drifts between them. The amplified modulation was used to generate radiation by using an undulator, with an undulator parameter of 0.58, a period of 1.9 cm, and a total of 11 periods. Each chicane provided $R_{56}$ of 4 mm, 2.5 mm, and 1.5 mm, in order, and the chicanes were 2 m apart. The initial momentum modulation was accumulated for a 10-m-long drift before the first chicane, and the beam energy was 72 MeV.  The spectrum of undulator radiation was measured.  They used a photodiode detector to measure the integrated intensity gain. With a 12-pC bunch, the average gain in the integrated intensity over the incoherent background was ~600. When they measured the gain of on-axis radiation and only considered  the bunch charge contributing to the coherent radiation, the gain went up to 1.5$\times 10^4$, which was in good agreement with their estimated gain of 2.5$\times 10^4$ from the linear theory.

This method is considered for FEL facilities to generate high-power coherent radiation, large radiation bandwidth, or attosecond pulses as described in \cite{Schneidmiller-2010-a} and \cite{dohlus-2011-a}. On the other hand, another use of this cascade method was proposed by \cite{ratner-2013-a} wherein they suggested using the cascade method to achieve a high cooling rate for coherent electron cooling. Here, the modulation imprinted by an hadron beam onto the electron beam is amplified with the cascade amplifier, and the field from the amplified modulation of electron beam is then used to cool the hadron beam.

%%% Plasma cascade amplifier
\subsubsection{Space-charge field with multiple bunches: plasma cascade amplifier}\label{sec5_SC5}
While the previous LSCA did not consider transverse behavior much, \cite{litvinenko-2018-a} introduced a new method called the plasma-cascade amplifier (PCA), which generates modulation at the plasma frequency using solenoid focusing. As previously mentioned, $G>1$ occurs when the space-charge induced momentum is strong enough, which is converted to the density by an appropriate $R_{56}$.  It is hard to expect a high gain in a drift because the $R_{56}$ of a drift ($L_d/\gamma_0^2$) is much lower than the one from a chicane ($2L_{dogleg}\theta^2$, $\theta\ll1$). However, the final density modulation is governed by how many particles gather together at one spot, so $R_{56}\delta$ is the critical term. Thus, even though a drift provides a low $R_{56}$, providing a stronger momentum modulation from the space-charge force can make a similar amplification. One of the clear ways to achieve it is to increase the density by focusing the beam, as we can infer from Eq.~\eqref{sec5_eq_momentumkick}.

\begin{figure}
\centerline{\includegraphics[width=0.4\textwidth, keepaspectratio=true]{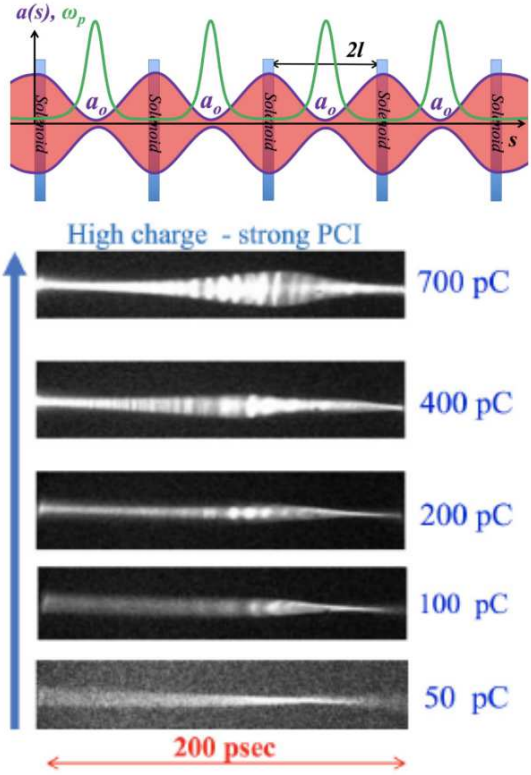}}
\caption{Experimental demonstration of plasma cascade amplification. The conceptual figure on top shows the configuration of a PCA with corresponding plasma wavelength modulation (top), and the measured longitudinal density distribution (bottom). Adapted from \cite{petrushina-2019-a}.  }\label{sec5_fig_pci}
\end{figure}

The plasma cascaded instability can also be understood as a parametric resonance of harmonic oscillators as described by \cite{litvinenko-2018-a}. In the harmonic oscillator
\begin{equation}
    \ddot{x} + \omega_0^2x=0,
\end{equation}
the oscillation can increase exponentially when the oscillation frequency is modulated at a certain frequency (i.e., $\omega_0(t) = \omega_0\left[ 1+A\cos(\omega t) \right]$).  This resonance happens when $\omega\simeq 2\omega_0$. In the case of the PCA, $\omega_0$ corresponds to the plasma oscillation frequency. Thus, if the plasma oscillation frequency can be modulated, then a parametric resonance will occur in the beam's longitudinal density modulation. Here, a series of focusing solenoids are used to modulate the plasma oscillation frequency, as shown in Fig.~\ref{sec5_fig_pci}.  Due to the required relationship between the plasma oscillation frequency and its modulation frequency, it is clear that system parameters such as the beam envelope, solenoid-to-solenoid distance, and charge level should be carefully designed.  Thorough theoretical work has been done by \cite{litvinenko-2018-a}, and further work is on-going [see \cite{blaskiewicz-2019-a}].

The method was experimentally demonstrated by \cite{litvinenko-2019-a,petrushina-2019-a}. A long bunch (400 ps) with low energy (1.76 MeV) was used for the experiment, and the charge was varied. They used 5 cells to amplify the shot noise, and the solenoids were not placed periodically, which works equally well as a periodic setup. From the simulation, they expected a gain of 400-500 for 0.4 THz, and a gain around 200 for 0.6 THz. Figure~\ref{sec5_fig_pci} shows measured longitudinal beam profiles. The density modulation is very clear, and it shows a charge dependency, which is expected because $G>1$ occurs with a strong enough momentum modulation from the space-charge force. The spectrum of these modulated bunches is broadband with a peak around 0.4 THz as \cite{litvinenko-2019-a,petrushina-2019-a} expected from their simulation. See \cite{litvinenko-2021-a} for further details. 

As LSCA had applications to radiation and coherent electron cooling, the PCA can be applied for the same purposes. While the LSCA requires a dispersive beamline such as a chicane, the PCA does not introduce any dispersion to the beam. This may provide a relative benefit in terms of beam quality preservation due to the lack of CSR and timing issues for the electron cooling.

%%%%%%%%%%%%%%%%%%%%%%%%%%%%%%%%%%%%%%%%%%%%%%%%%%%%%%%%%%%%%%%%%%%%%%%%%%%%%%%%%%%%%%%%%%%%%%%%%%%%%%%%

\subsection{Shaping profiles using coherent synchrotron radiation} \label{sec5_CSR}

CSR is another beam-generated field that changes a longitudinal beam momentum distribution along time.  Similar to the space-charge field, a CSR field is determined by the bunch's longitudinal profile, and the field strength is the only controllable knob, as we can see from Eq.~\eqref{100)}. CSR cannot be generated without a dispersive element such as a dipole magnet. This means that the transverse and longitudinal phase space will be correlated along the path, and it makes the analysis and control more difficult than the space-charge field of a single bunch.  Thus, using CSR directly for shaping is challenging, and no direct use of it has been proposed so far.

However, CSR is indeed a field that can change the longitudinal profile. It is normally considered an obstacle for longitudinal beam manipulation. There have been several research efforts that tried to preserve the shape through a beamline with CSR. In the rest of this section, we describe efforts to suppress CSR's impact on the profile.  Studying these methods will tell us how to use CSR as a shaping tool.

\begin{figure}
\centerline{\includegraphics[width=0.5\textwidth, keepaspectratio=true]{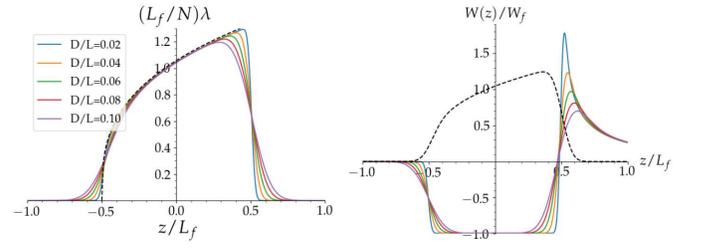}}
\caption{Theoretical estimation of aberrations of ideal current profile flattening CSR wake and aberration's impact on the CSR wake. Current profile with aberrations are displayed in the left panel. The right panel shows corresponding CSR wakes. The dashed curve in the left panel shows the ideal profile from Eq.~\eqref{sec5_eq_CSRSuppressingProf}. Solid curves show profiles aberrated by the chicane. Each curve corresponds to a different $\mathcal{D}$ in Eq.~\eqref{sec5_eq_CSRProfEvol}; $\mathcal{D}/L=$ 0.02, 0.04, 0.06, 0.08, and 0.10, where L is the initial bunch length. $L_f$ ($=L/C$) is the final bunch length after compression. Adapted from \cite{mitchell-2013-a}.  }\label{sec5_fig_CSRsuprresion}
\end{figure}

\cite{mitchell-2013-a} showed theoretically that a certain beam shape can flatten the CSR wake, and it can minimize the CSR's impact on transverse emittance growth through the chicane.  The beam profile is given by
\begin{equation}\label{sec5_eq_CSRSuppressingProf}
\lambda(z) = {4\over3}{(z-a)^{1/3}\over(b-a)^{4/3}},
\end{equation}
where $a$ and $b$ are the limits of the longitudinal bunch domain ($a\leq z \leq b$). This shape is given on the left of Fig.~\ref{sec5_fig_CSRsuprresion}, and the corresponding CSR field is on the right. 

Here, preserving the initial shape is critical to minimize CSR's impact on the emittance, so \cite{mitchell-2013-a} derived an equation for the longitudinal profile starting from a decoupled 6D distribution. The profile at distance $s$ can be written as
\begin{equation}\label{sec5_eq_CSRProfEvol}
    \lambda(z;s) = {C\over\sqrt{2\pi}\mathcal{D}} \int_{-\infty}^{\infty} \lambda(\zeta;0) \exp\left( - {(\zeta-zC)^2\over2\mathcal{D}^2} \right) d\zeta,
\end{equation}
where $C$ is the compression factor, $\mathcal{D}$ is given by
\begin{equation} \label{sec5_eq_CSRsuppression}
    \mathcal{D}^2(s) = C^2\left( R_{56}^2(s)\sigma_{\delta}^2 + \mathcal{H}(s) \right),
\end{equation}
and $\mathcal{H}$ is given by
\begin{equation}
    \mathcal{H}(s) = {\left[\sigma_x^2 R_{51}(s) - \sigma_{xx'} R_{52}(s)\right]^2 + \varepsilon_x^2 R_{52}^2(s) \over \sigma_x^2 }.
\end{equation}
The change of the profile depends on $C$ and $\mathcal{D}$.  When $\mathcal{D}$ is close to zero, $\lambda(z;s) = C\lambda(zC;0)$, which means the final profile has the same shape as the initial profile but is compressed. Thus, the profile would not change significantly when $\mathcal{D}$ is small.  Figure~\ref{sec5_fig_CSRsuprresion}, which shows the longitudinal profile for different values of $\mathcal{D}$ and their corresponding CSR wakes, proves that CSR has a small impact on the profile when $\mathcal{D}$ is small.  This tells us that a small $R_{56}$, initially low energy spread, and a small transverse emittance are preferred. Also, beam matching into the beamline is another important factor.

\cite{ha-2016-a} tracked CSR effects on the beam in an EEX beamline to find a way to preserve the shaping quality of EEX-based shaping (see Sec.~\ref{sec2c3} and Sec.~\ref{sec6: EEX}).  The longitudinal shift due to CSR can be written as
\begin{equation}
    \Delta z_f = \xi_2\Delta\delta_2^{\eta=const} + \xi_2^* \Delta\delta_2^{\eta\neq const} + \xi_2(1+\kappa\eta^*)\Delta\delta_1^{\eta\neq const},
\end{equation}
where the $\Delta\delta$ terms are integrated momentum changes made by CSR from each section of the EEX beamline, $\kappa$ is the deflecting cavity's kick strength, $\eta$ is the dispersion of a dogleg, and $\xi_2$ is the $R_{56}$ of the second dogleg. $\eta^*$ and $\xi_2^*$ are effective dispersion and $R_{56}$ terms to account for their effects when dispersion is not a constant. This simple equation shows that $R_{56}$ is the carrier of CSR-induced momentum change. Thus, reducing $R_{56}$ is clearly a good way to minimize CSR's impact on the profile.

\begin{figure}
\centerline{\includegraphics[width=0.5\textwidth, keepaspectratio=true]{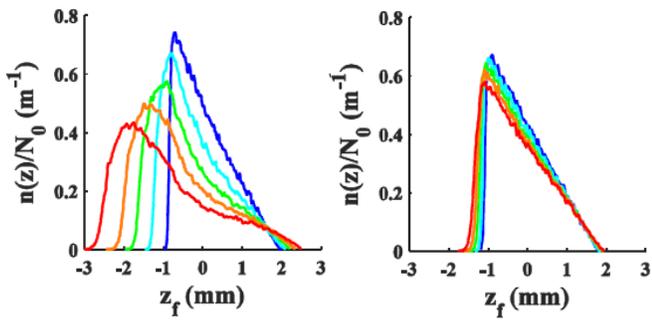}}
\caption{Correction of CSR-induced aberration in start-to-end simulation. The x-axis is the beam's longitudinal position after a EEX beamline, and the y-axis is the normalized longitudinal density. The colors represent charges of 1.1, 3.3, 5.5, 7.7, and 11 nC with blue corresponding to a low charge and red corresponding to a high charge.  The left panel was simulated with a 20$^\circ$ bend for the dogleg, and the right panel used a 12$^\circ$ bend.  Adapted from \cite{ha-2016-a}.  }\label{sec5_fig_CSRsuprresion2}
\end{figure}

Figure \ref{sec5_fig_CSRsuprresion2} shows current profiles at the exit of an EEX beamline for several different charge levels. The left figure corresponds to an EEX beamline with a 20$^\circ$ bend, and the right figure corresponds to a bending angle of 12$^\circ$. The $R_{56}$ of the doglegs were 0.29 m and 0.18 m, respectively. The profile of 11 nC case (red) has several clear differences from the 1.1 nC case (blue) when $R_{56}$ is high. For example, the tail is lengthened, the peak is rounded, the linear ramp is changed to a concave curve, and the bunch length is significantly lengthened.  On the other hand, the low $R_{56}$ case preserved all features (tail, peak, ramp, and bunch length) even for 11 nC.

 \cite{tan-2021-a} performed an optimization by using a reverse tracking simulation to find a beamline setting and an initial beam condition for providing the desired final profile. By optimizing the entire beamline, nonlinearities introduced by each part of the beamline could cancel each other out. The beamline they used in the simulation consisted of an SRF gun, 650-MHz linac cavities, a 3.9-GHz linac cavity, and two chicane compressors. Optimization variables included accelerating gradient, phase, frequencies of 650-MHz linac cavities, $R_{56}$ and $T_{566}$ of each chicane, and parameters defining the initial profile.

Even though the charge of the beam they used was 10 nC, they found that a reasonable beamline setting provided the desired final longitudinal profile. Even this setting included quite a high $R_{56}$ for both chicanes, 0.129 m and 0.131 m, which can introduce a strong CSR effect on the profile. However, the work was done with a simulation code with several simplified physics (e.g., beam propagation using $R_{56}$ and $T_{566}$, steady-state CSR only, etc.). Thus, more work is required to reach a definite conclusion.  However, it is promising that control of nonlinearities of the beamline may compensate a strong CSR's impact on the final profile.

From the study of this research on CSR suppression, we can infer how to make CSR's impact on the final profile stronger. Strong CSR, due to high charge or short bunch, coupled with high $R_{56}$ may provide the CSR dominant control over the final profile. This would induce a profile change along the beamline; thus the final profile could be made to have completely different features than the initial one. However, the final profile that the CSR can generate would be limited because the CSR field is determined by the beam's profile along the beamline.  This means that we need to either shape the initial profile or create other knobs to control the beam's longitudinal profile inside the beamline. $R_{51}$, $R_{52}$, and transverse phase space may be good candidates for the additional knob, and we may be able to use nonlinearities from other parts of the beamline to manipulate the final profile further.

%%%%%%%%%%%%%%%%%%%%%%%%%%%%%%%%%%%%%%%%%%%%%%%%%%%%%%%%%%%%%%%%%%%%%%%%%%%%%%%%%%%%%%%%%%%%%%%%%%%%%%%%

\subsection{Shaping profiles using wakefields}\label{sec5_wake}

The longitudinal wakefield inside a given structure can be expressed as the convolution of current distribution and wake function, which is the wakefield from a single particle; see Eq.~\eqref{ZEqnNum411253}.  Here, the longitudinal wake function can be written as the summation of each mode excited in the structure:
\begin{equation} \label{sec5_eq_wakefunction}
w_z(z)=\sum_{i} 2\mathcal{K}_i cos(k_i z),
\end{equation}
where $k_i$ is the wavenumber of the $i$th mode, and $\mathcal{K}_i$ is the loss factor of $i$th mode which is defined as the energy lost by the particle exciting the mode per unit charge squared. While the space-charge field and CSR did not provide an additional knob other than its strength, in the case of the wakefield, the material or geometry of the vacuum pipe determines the wake function (i.e., loss factor and fundamental frequency). Thus, manipulation using the wakefield provides higher degrees of freedom than other beam-generated fields.

In this section, we discuss the use of the wakefield in shaping the longitudinal density profile. Similar to methods described in Section~\ref{sec4}, wakefields control time-energy correlation and the subsequent beamline (e.g. drift and anisochronous beamline) make the correlation that determines the longitudinal density distribution. Thus, we can easily imagine applications such as bunch compression, bunch train generation, and single-bunch shaping. These applications will follow the same principle as the ones in Section~\ref{sec4}, but the wakefield-based techniques could be more compact and cost effective than external devices. 

In addition to profile shaping, the degrees of freedom that wakefields provide enable further control over the longitudinal phase space. Therefore, we also discuss manipulations of the energy distribution and longitudinal phase space using wakefields.

%%%%%%%%%%%%%%%%%%%%%%%%%%%%%%%%%%%%%%%%%%%%%%%%%%%%%%%%%%%%%%%%%%%%%%%%%%%%%%%%%%%%%%%%%%%%%%%%%%%%%%%%

\subsubsection{Bunch compression} \label{sec5_wake_compression}

A wakefield can work with two compression mechanisms; ballistic bunching and magnetic bunching. Both mechanisms require control over the longitudinal chirp, which can be provided by the wakefield. The chirp can be produced in two different ways.  If there is a wakefield-driving bunch, the target bunch can be placed at the zero-crossing phase of the wakefield, to provide either a negative or a positive chirp. If there is no driving bunch, then the self-wakefield inside the target bunch always makes the tail lose more energy than the head of the bunch (with an appropriate choice of frequency). Thus, the target bunch always has a positive chirp. The positive chirp can work with magnetic bunching, but it is not appropriate for ballistic bunching because a drift has $R_{56}>0$. 

\begin{figure}
\centerline{\includegraphics[width=0.5\textwidth, keepaspectratio=true]{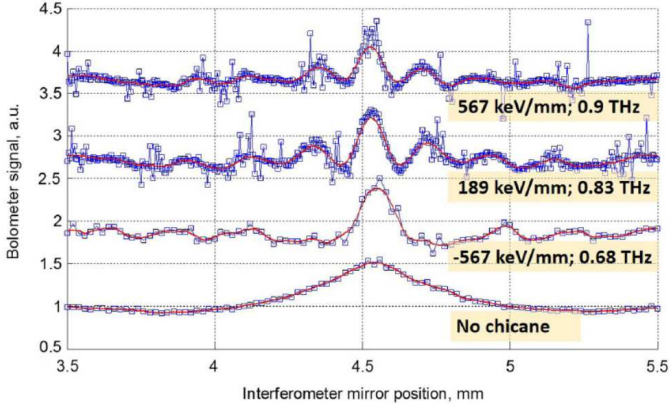}}
\caption{Experimental demonstration of wakefield-based bunch compression using THz structure with drive bunch. The figures show measured longitudinal phase space of the target bunch and corresponding projections. Each figures corresponds to different drive bunch charges. The charges were (a) 0, (b) 0.6, (c) 0.9, and (d) 1.3 pC. Adapted from \cite{zhao-2018-a}.  }\label{sec5_fig_wakecompression}
\end{figure}

Ballistic bunching using a wakefield was experimentally demonstrated by \cite{zhao-2018-a}. They split a Ti:sapphire laser at 800 nm into three pulses. Two pulses were used to generate the drive and target bunch while the third pulse was used for diagnostics purposes by converting it to THz radiation through optical rectification. A 5-cm-long quartz tube with an inner diameter of 400 $\upmu$m was used to generate a wakefield, and a drive bunch charge ranging from 0.6 to 1.3 pC was used to generate different longitudinal chirps. Thus, for a given $R_{56}\approx 3.07$ cm, they generated different compression ratios. The results are displayed in Fig.~\ref{sec5_fig_wakecompression}. An uncompressed bunch length of 150 fs rms was compressed to 2.8 fs rms in the case of (c).

While a similar compression can be easily accomplished with conventional RF cavities, the wakefield-based ballistic bunching offers a few advantages. First, this compression requires a driving bunch that can easily be generated by splitting UV laser pulses. Therefore, it does not require extra cavities for chirping and corresponding RF power sources. Second, RF sources always introduce jitter on the beam energy. Because the wakefield fully relies on the given structure and the beam, it can provide better energy stability. \cite{zhao-2018-a} measured energy stability of 2.4$\times10^{-4}$ rms for wakefield-based bunching, while it was 1.5$\times10^{-3}$ rms for bunching with a buncher cavity. Third, this bunching also reduces arrival time jitter. In \cite{zhao-2018-a} they measured arrival timing jitter using a THz streaking method proposed by \cite{zhao-2018-b}. The measured arrival time jitter for wakefield-based bunching was about 60 fs rms, while the buncher provided a jitter of 170 fs rms.

Similarly,  magnetic bunching with a wakefield can also have the benefit of reduced timing jitter. Small timing jitter is highly demanded in many modern accelerators. Especially, it is important for wakefield accelerators. Wakefield accelerators usually operate at very high frequency, so placing a short target bunch at the right time is critical because it determines the beam's energy spread and energy gain. However, controlling the separated bunch's timing in a few ps or lower is not straightforward. 

\cite{zhao-2018-a} proposed a scheme to use the wakefield to reduce the timing jitter between an externally injected drive bunch and main bunch. The proposed concept uses the wakefield from a drive bunch to produce a negative chirp on the main bunch's longitudinal phase space. This chirping process is followed by a chicane to compress the main bunch. If the main bunch is a little late compared to the ideal time, it will be accelerated by the driving bunch's wakefield. In the following chicane, the main bunch will take a shorter path than the path it would take when it arrives at the ideal time (i.e., zero-crossing of the wakefield). This path length difference compensates the original timing error after the compression. Similarly, if the main bunch arrives earlier than the ideal time, the main bunch will be decelerated by the wakefield and this bunch will take a longer path than the bunch at the right timing. Thus, the timing errors can be compensated.

%%%%%%%%%%%%%%%%%%%%%%%%%%%%%%%%%%%%%%%%%%%%%%%%%%%%%%%%%%%%%%%%%%%%%%%%%%%%%%%%%%%%%%%%%%%%%%%%%%%%%%%%
\subsubsection{Bunch train generation}
The bunch compression discussed in the previous section requires a condition ${\sigma_z\over \lambda_{wake}}\ll 1$.  If we go to the other regime (${\sigma_z\over \lambda_{wake}}\gg 1$), the wakefield can imprint a sinusoidal energy modulation on the beam, which can be used for bunch train generation. Once again, we can use either a drift or other anisochronous beamline as a momentum-to-density modulation converter.  A drift requires a low-energy beam to provide sufficient $R_{56}$, and the overall beamline setup can be simple and compact. On the other hand, an anisochronous beamline, such as a chicane, can generate a high $R_{56}$ that is compatible with any beam energy. This was experimentally demonstrated by \cite{lemery-2019-a} using a low-energy beam with drift and by \cite{antipov-2013-a} using a compact chicane.

In \cite{lemery-2019-a}, they used a beam with charge of ~1 nC and an energy of 6.2 MeV. Two structures were prepared (Inner diameter: 450, 750 $\upmu$m, 5 and 8 cm long), followed by an accelerating cavity providing ~14 MeV of energy gain. For both structures, downstream diagnostics measured the phase space, and the modulation was successfully imprinted in the LPS. The measured wavelength of the final modulation was ~1.01, and 1.81 mm for each structure. They showed good agreement with the expected fundamental wavelengths of each structure.

\begin{figure}
\centerline{\includegraphics[width=0.45\textwidth, keepaspectratio=true]{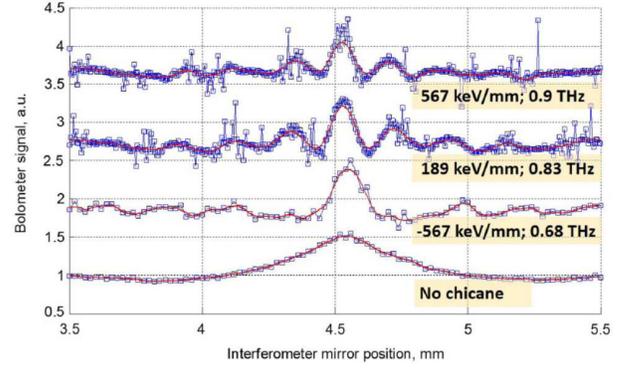}}
\caption{Experimental demonstration of bunch train generation using a dielectric structure followed by a chicane. The figure shows measured CTR signals, which are autocorrelation results. Each curve corresponds to a different initial energy chirp as described in the label next to the curve. The squares are the data points, and the red curves are the smoothed fits. From \cite{antipov-2013-a}.  }\label{sec5_fig_waketrain}
\end{figure}

On the other hand, \cite{antipov-2013-a} used a chicane consisting of four permanent magnets to provide a high $R_{56}$ ($\approx 4.9$ cm).  Here the beam's charge and energy were ~0.5 nC and 57 MeV, respectively. A 5-cm-long Kapton capillary with an inner diameter of 300 $\upmu$m was used. The corresponding fundamental frequency was around 0.8 THz. During the experiment, they controlled the incident longitudinal chirp by controlling the phase of the linac before the compressor. This macro-chirp on top of sinusoidal modulation, can provide either a frequency upshift or downshift. The modulation frequency was measured from the autocorrelation of the coherent transition radiation, and the result is shown in Fig.~\ref{sec5_fig_waketrain}. When there was no chicane, the modulation conversion to the density modulation was not observed. When a chicane was introduced, a density modulation appeared, and its frequency was close to the expected 0.8 THz (see the 189 keV/mm case). Also, depending on the linac phase, the chirp was changed from -567 keV/mm to +567 keV/mm. Although the initial modulation frequency was about 0.8 THz, the final density modulation frequency could be tuned from 0.68 to 0.9 THz.

Equation~\eqref{1Dbunchf} is applicable to both \cite{lemery-2019-a} and \cite{antipov-2013-a} methods.  In the case of the wakefield, the modulation amplitude induced by the initial energy modulation and $R_{56}$ was $A_{ind} = Ck|R_{56}|\Delta\delta\exp\left( -{1\over2}C^2k^2R_{56}^2\sigma_\delta^2 \right)$, where $\Delta\delta$ is the amplitude of the initial energy modulation [see \cite{saldin-2002-a}]. Note that the required $R_{56}\Delta\delta$ for the same final modulation amplitude decreases as modulation frequency increases. $R_{56}\Delta\delta$ is the particle's travel distance. To build a density spike from the sinusoidal energy modulation, particles need to travel a little less than $\lambda/4$. Thus, high-frequency modulation naturally requires either a small $R_{56}$ or a small initial modulation amplitude.  Although the beam energy was 57 MeV for \cite{antipov-2013-a}, the total length of the chicane was less than 50 cm, and $R_{56}$ was only around 0.049 m. Their high modulation frequency (0.8 THz) enabled the use of a compact chicane.  Also, as the modulation frequency increases, the initial uncorrelated energy spread becomes more important. The exponent of $A_{ind}$ includes a $k\sigma_\delta$ term. As the frequency increases, the modulation amplitude decays exponentially. Thus, a higher frequency requires a lower initial spread to keep the exponential decay at the same level. 

Finally, we note that modulation does not occur if the bunch’s longitudinal profile’s frequency spectrum does not overlap with the structure’s frequency spectrum. Thus, imprinting multi-period modulation is not possible for a Gaussian profile.  Both experiments described above used flat-top profiles.

%%%%%%%%%%%%%%%%%%%%%%%%%%%%%%%%%%%%%%%%%%%%%%%%%%%%%%%%%%%%%%%%%%%%%%%%%%%%%%%%%%%%%%%%%%%%%%%%%%%%%%%%
\subsubsection{Single-bunch profile shaping}
\begin{figure}
\centerline{\includegraphics[width=0.45\textwidth, keepaspectratio=true]{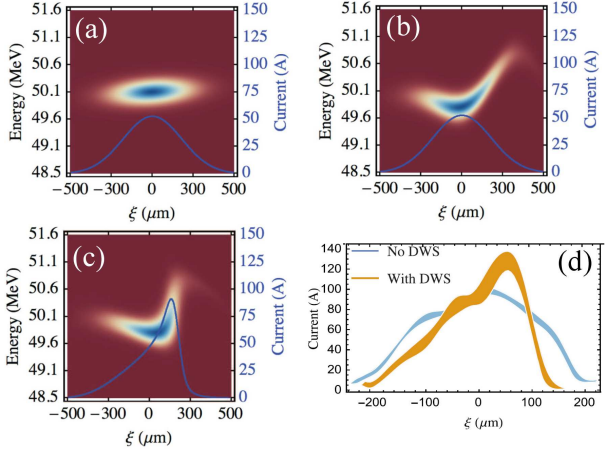}}
\caption{Experimental demonstration of triangular current profile generation using a dielectric structure followed by a chicane.(a-c) are simulation results, and (d) is reconstructed profile from interferometer measurements with and without the structure. (a-c) corresponds to longitudinal phase space before the structure, after the structure, and after the chicane, respectively. The initial phase space (a) is generated artificially using beam parameters used for the experiment. The modulation applied to this phase space is calculated from a formula using structure parameters used for the experiment. Adapted from \cite{andonian-2018-a}.}\label{sec5_fig_waketriangle}
\end{figure}

The current profile of a single bunch can be shaped by introducing an appropriate nonlinearity or by masking. As we saw from transverse manipulation by \cite{yuri-2007-a}, introducing an appropriate nonlinearity on the phase space can change the profile.  Figure~\ref{sec5_fig_waketriangle} shows an example for a triangular profile. A wakefield introduces a nonlinear curvature in the linear phase space. In this example, particles in the tail are accelerated due to the wakefield, while particles near the center lose their energy. Thus, when the beam enters a chicane, particles in the tail overtake the leading particles and build up a density spike. At the same time, due to the nonlinearity in the energy gain along the longitudinal position, the tail becomes sharp. The head of the current profile smears out for the same reason.  Note that this happens when the wavelength of the wakefield is comparable to the bunch length so that the appropriate nonlinearity is introduced to the phase space.

In \cite{andonian-2018-a} this method was experimentally demonstrated using an 80-pC, 50-MeV beam and a compact chicane with $R_{56}=9.2$ mm. They used a dielectric wakefield structure having an inner diameter of 200 $\upmu$m  and a length of 5 cm. The fundamental mode frequency was 0.39 THz, which is about a quarter of $1/\sigma_z$. The beam's profile was measured with and without the structure to compare the effect of the wakefield. The reconstructed profile from CTR interferometry is given in Fig.~\ref{sec5_fig_waketriangle}(d).

The method can be extended to generate arbitrary current profiles when we use a series of structures with appropriate frequencies. Similar to \cite{piot-2012-a} and \cite{ha-2019-a}, each structure represents a Fourier component of the target correlation. With an appropriate $R_{56}$, the correlation can change the initial profile to the desired profile. A scheme using transverse wigglers was proposed by \cite{ha-2019-a} and can be used for the longitudinal phase space directly by simply replacing transverse wigglers with wakefield structures.  Simulations in \cite{mayet-2020-a} showed the feasibility of using several structures with different frequencies. They also showed the feasibility of structure fabrication.

A masking-based technique has not been proposed yet. However, combining wakefield-based shaping with methods introduced this paper could provide advantages. For example, one can adopt a wakefield-based deflector (see \cite{novokhatski-2015-a}) instead of RF deflecting cavities for the method in \cite{ha-2020-a}; see Sec.~\ref{sec4b1}. The wakefield deflector uses a transverse wakefield that kicks the trailing particle transversely. Thus, it generates time and transverse correlation just like RF deflecting cavities. While this passive device may provide a big advantage on timing jitter, the nonlinearity of the transverse wake will require more extensive analysis to design the mask and eliminate the correlation after chopping \cite{bettoni-2016-a,craievich-2017-a,seok-2018-a}.

%%%%%%%%%%%%%%%%%%%%%%%%%%%%%%%%%%%%%%%%%%%%%%%%%%%%%%%%%%%%%%%%%%%%%%%%%%%%%%%%%%%%%%%%%%%%%%%%%%%%%%%%
\subsubsection{Control over the energy distribution} \label{sec5_wake2}
Because the wakefield controls a correlation between the longitudinal position and energy, it is also applicable for controlling the beam's energy distribution.  For example, most modern accelerators use bunch compression.  The bunch compression requires a longitudinal chirp, and it is a negative chirp in most cases.  Usually, the chirp remaining after the compression is eliminated by a combination of the linac's RF phase control and wakefields in the linac. However, it is a costly method since it requires operating the linac off-crest. Alternatively, methods using dielectric and corrugated metallic structures, which generate a strong wakefield, can remove the chirp inexpensively. In the case of profile shaping in Fig.~\ref{sec5_fig_waketriangle}, the wakefield frequency is chosen so that the particle in the tail is accelerated. If we make the frequency even lower, it is possible to make the beam only see a decelerating field. Because this wakefield always makes a positive chirp, the remaining negative chirp after the bunch compression can be  compensated. A device using the wakefield to compensate a positive longitudinal chirp is called dechirper.

\begin{figure}
\centerline{\includegraphics[width=0.5\textwidth, keepaspectratio=true]{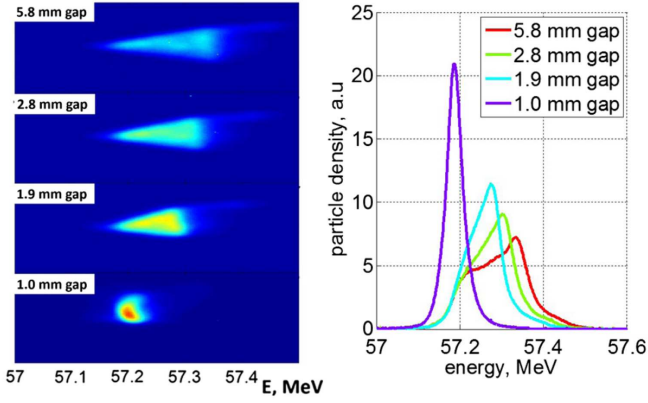}}
\caption{Experimental demonstration of energy spread reduction by wakefield from dielectric structure. The left panel shows beam images on the spectrometer with different structure gap sizes. Corresponding profiles are shown in the right panel. From \cite{antipov-2014-a}. }\label{sec5_fig_dechirper}
\end{figure}

Both \cite{antipov-2014-a} and \cite{emma-2014-a} experimentally demonstrated the effectiveness of the wakefield dechirper. Both experiments used a slab structure, which consists of two jaws. Thus, the wakefield strength was adjusted by the gap size. When the operating conditions, such as a beam's profile, are known, the structure can be optimized for the operating condition. However, these proof-of-principle experiments used the slab structure with an adjustable gap to provide flexibility to the experiment. Figure~\ref{sec5_fig_dechirper} shows measured beam images after the spectrometer and corresponding spectrum from \cite{antipov-2014-a}. As the gap size decreased, a quasi-linear wakefield amplitude increased, and it successfully reduced the energy spread.

The low correlated energy spread from a dechirper can benefit all modern accelerator applications. For example, the X-ray free-electron oscillator (XFELO) requires energy spread of $< 1\times 10^{-4}$ \cite{kim-2008-a}, which is challenging to achieve with existing methods. However, a recent study by \cite{qin-2016-a} showed its feasibility. Optimization of most of the linac parameters (phase, gradient, $R_{56}$, etc.) was performed, and the beamline had a dechirper at the end. This optimization in the simulation generated the core of the beam ($\sim$400 fs) having a relative energy spread of $\sim 2.6\times 10^{-5}$ (105 keV rms).

Although the methods just introduced demonstrate the compensation of the chirp by experiments or simulations, there was a common limitation. The methods introduce a nonlinearity in phase space that increases the correlated energy spread. The linearity in the core of the phase space can be preserved by designing a structure having a low-frequency single-mode, whose wavelength is much longer than the bunch length. However, preserving the linearity in the periphery is challenging.  When $k\sigma_z\ll1$, Eq.~\eqref{ZEqnNum411253} and Eq.~\eqref{sec5_eq_wakefunction} can be simplified to
\begin{equation}
    W_z(z) \approx 2\mathcal{K} Q_b \int_{-\infty}^{z} \lambda(\zeta) d\zeta,
\end{equation}
where $Q_b$ is the total charge, and $\lambda(z)$ is the normalized density profile. Thus, only a uniform profile provides $W_z(z)\sim z$, which preserves both the core and the periphery's linearity. All other profiles would would have a nonlinear periphery in their phase space. .

A simulation in \cite{antipov-2014-a} showed the feasibility of extending linear region of the wakefield by using a multi-mode structure. However, this approach does not provide any tunability after the fabrication of a structure. \cite{antipov-2014-a} also remarked about the use of two or more dechirpers to control nonlinearity issues. This multi-structure approach was recently simulated by \cite{mayet-2020-a}. It provided high-quality dechirping, but the dechirping for the periphery was still limited due to the limitation of Fourier synthesis known as the Gibbs phenomenon.

%%%%%%%%%%%%%%%%%%%%%%%%%%%%%%%%%%%%%%%%%%%%%%%%%%%%%%%%%%%%%%%%%%%%%%%%%%%%%%%%%%%%%%%%%%%%%%%%%%%%%%%%%%
\subsubsection{Control over longitudinal phase space}
As we notice with the previous single-bunch shaping and energy distribution control, a wakefield applied to the beam changes the longitudinal phase space, and it can be used to control the correlation of the phase space.  The most demanding control over the correlation is linearization, and it is experimentally demonstrated by \cite{deng-2014-a} and \cite{fu-2015-a}. For a simple description, if we assume that only the RF curvature from the linac generates nonlinearities and the beam's energy gain from the linac can be written as $E_{RF}\cos(k_{RF}z)$, this gain can be approximated to $E_{RF}\left[1-{(k_{RF}z)^2\over2}\right]$ for $k_{RF}z\ll1$. Similarly, the wakefield from a structure having only a single mode and a uniform profile can be written as ${n_0W_0\over k_{w}}\sin\left( k_{w}(z+\Delta z/2) \right)$, where $\Delta z$ is the width of the uniform profile, and $n_0$ is the density. Because $k_wt$ should be small while the wavelength of the wakefield should be shorter than the wavelength of the RF field in the linac, we can approximate it as $n_0W_0\left[ -{\Delta z k_w^2\over4}z^2 + \left(1-{k_w^2 \Delta z^2\over8}\right)z + {\Delta z\over2} \right]$. Thus, it is possible to eliminate the quadratic term that generates the curvature in the phase space by choosing the proper strength of the wakefield. This can be simply done by adjusting the gap size in the case of slab structures.

\begin{figure}
\centerline{\includegraphics[width=0.35\textwidth, keepaspectratio=true]{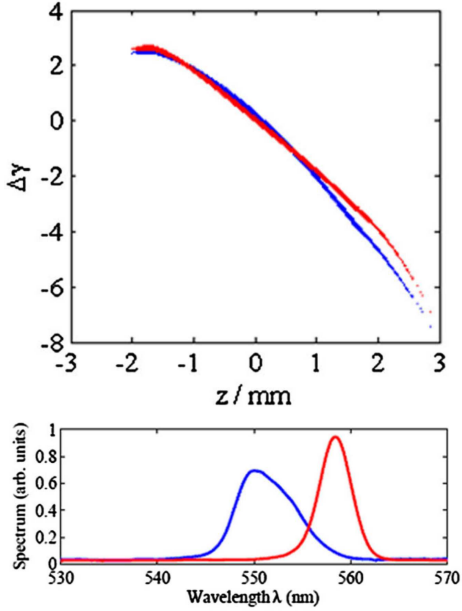}}
\caption{Demonstration of phase space linearization and its impact on the radiation spectrum. The top panel shows simulated longitudinal phase space before (red) and after (blue) a wakefield structure. The bottom panel shows the measured spectrum of undulator radiation. Here the red and the blue corresponds to results from different structure gap sizes. 2-mm gap used for the red, and 6-mm gap used for the blue.   Adapted from \cite{deng-2014-a}. }\label{sec5_fig_wakelinearization}
\end{figure}

The top panel of Fig.~\ref{sec5_fig_wakelinearization} shows the simulation by \cite{deng-2014-a}. The blue corresponds to the longitudinal phase space before a corrugated waveguide, and the red shows the corrected phase space after using the wakefield. It is clear that most of the region is linearized and the periphery is not affected due to a very strong initial nonlinearity. During the experiment, they generated undulator radiation to estimate the impact of the linearization on the radiation quality. They expected to see a significant reduction of the radiation bandwidth due to the linearization and a shift of the central wavelength. The bottom panel shows the measured spectrum of the radiation. Here, the blue spectrum corresponds to a gap size of 6 mm for the corrugated waveguide while the red corresponds to a 2-mm gap. As the wakefield strength increased, the bandwidth became narrower. The bandwidth from the 6-mm gap was measured to 7.8 nm. On the other hand, the 2-mm gap provided 3.7-nm bandwidth.

This experiment was designed to observe the effect of linearization on the radiation. However, we can imagine that the bunch compression will benefit from this linearization too. Currently, most of the linearization is done with harmonic cavities, which require additional RF power sources. However,the beam-generated field may be able to manage the nonlinearity with lower cost. \cite{penco-2017-a} used wakefields from a high-impedance linac and a dielectric waveguide structure to replace high-harmonic cavities. The result was similar to the linearization result by high-harmonic cavities.

\begin{figure}
\centerline{\includegraphics[width=0.5\textwidth, keepaspectratio=true]{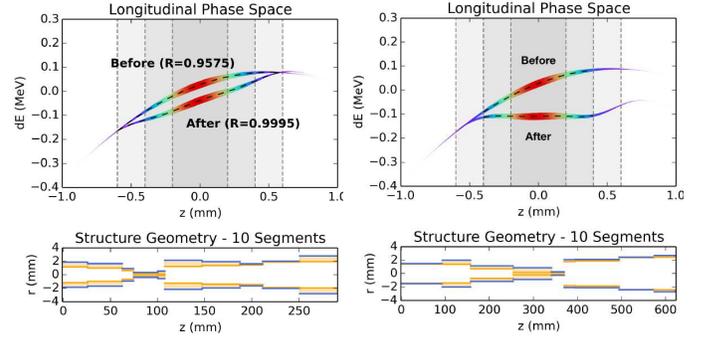}}
\caption{Longitudinal correlation control via a series of dielectric structures. The top panels show simulated longitudinal phase spaces before and after the correction using 10 dielectric structures, and the bottom panels show the corresponding structure geometries for simulations. The left column shows an example of linearization while the right column shows dechirping. Adapted from \cite{mayet-2020-a}. }\label{sec5_fig_fourier}
\end{figure}

Although Fig.~\ref{sec5_fig_wakelinearization} shows noticeable linearization accomplished with a single structure, a series of structures with different frequencies may enable further control of the phase space such as correlation control of the phase space for an arbitrary current profile or control of longitudinal chirp with the linearization. The wakefield from each structure will work as a Fourier component and can generate arbitrary correlation in the phase space; see \cite{ha-2019-a}. This concept was demonstrated in a simulation \cite{mayet-2020-a} using a series of dielectric structures. Because the choice of parameter set becomes too complex, they used an optimization algorithm to find the best result. Figure~\ref{sec5_fig_fourier} shows two of their simulation results. The top panel shows the longitudinal phase space with and without linearization by wakefields, and the bottom panel shows the corresponding structure geometries. They used a total of 10 structures to correct the shape and control the chirp. Although this approach requires experimental demonstration and further study on beam transport, instability, fabrication error, etc., this is indeed a powerful method that would enable us to optimize the longitudinal phase space for each application.

\section{Coupling between degrees of freedom for phase-space tailoring\label{sec6}}
\subsection{Introduction}
This section discusses shaping methods that rely on the coupling between phase spaces associated with different degrees of freedom (DOF). In Section~\ref{sec4}, the introduction of {\em local} coupling was shown to provide some control over a coordinate not usually accessible in an uncoupled beamline. In that case, shaping is performed within ``coupling bumps" and the correlation is removed downstream. Ultimately, the fact that the coupling is partial limits the precision of that shaping method. This section focuses on beamlines that have strong coupling or can swap phase-space coordinates. This class of phase-space manipulation enables precise shaping of the beam phase space and opens a path to emittance repartition among the different DOFs.  

\subsection{Coupling between the two transverse degrees of freedom}
\subsubsection{Producing beams with canonical angular momentum}\label{sec6b1}
Canonical angular momentum (CAM)-dominated, or ``magnetized", beams have important applications in electron cooling of heavy-ion beams~\cite{budker-1975-a,derbenev-1978-a,parkhomchuk-2000-a}. In such a scheme, a cold electron beam co-propagates with the ion beam at the same average velocity. Collisions between ions and electrons transfer thermal motion away from the ion to the electron beam. The cooling efficiency can be greatly improved by using a magnetized beam. More recently, the use of a CAM-dominated beam was also considered for mitigating a resonance-driven instability in long periodic focusing channels~\cite{cheon-2020-a}. A simple technique for forming CAM-dominated beam consists of immersing a cathode in an axial magnetic field $B_z$, i.e., inside a solenoid magnet. In such a case, the canonical momentum  $P_{\theta}=eA_{\theta}=\frac{eB_z}{2}r$ is proportional to $B_z$. In a rotationally invariant system, the conservation of angular momentum implies that the beam acquires a mechanical angular momentum (MAM) equal to the CAM once it exits the solenoidal field as $A_{\theta}$ vanishes; henceforth the beam's motion in the two transverse planes becomes coupled.

\begin{figure}[htb]
\centering
\includegraphics[width=0.49\textwidth]{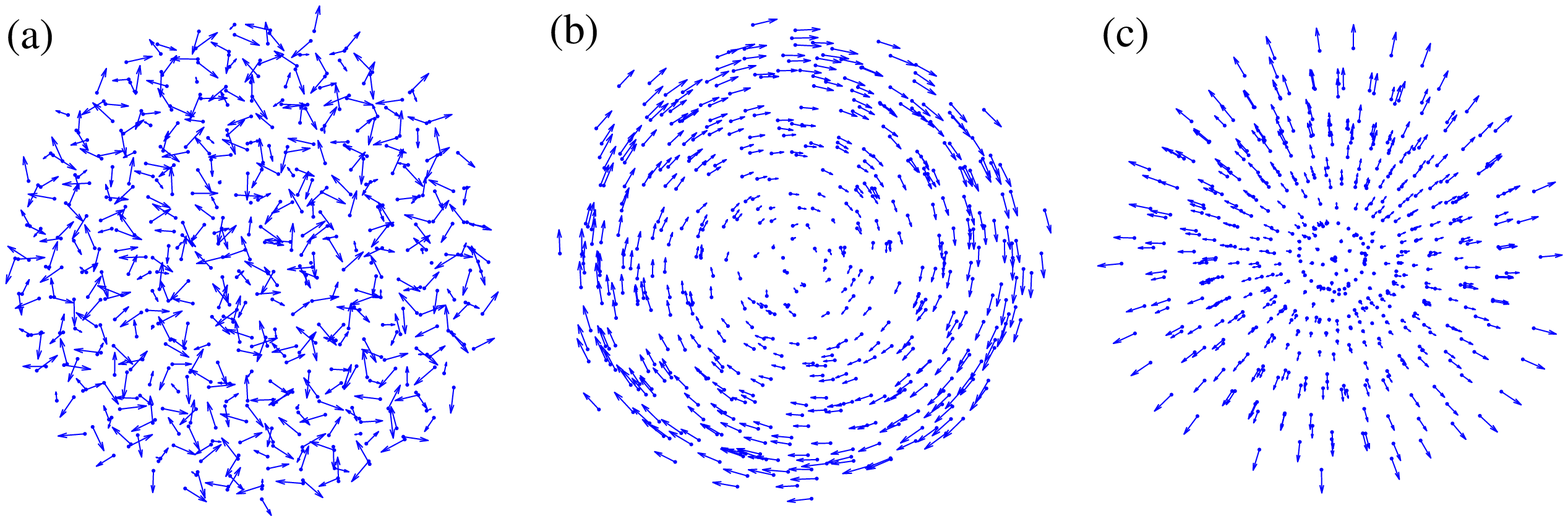}
\caption{Motions of the electrons when the beam is dominated by (a) emittance, (b) angular momentum or (c) space charge. Each dot
represents an electron in $(x,y)$ space, and the arrows show the magnitude and direction of the electrons' velocities. From ~\cite{sun-2005-a}.}
\label{fig:Sec6:camemitsc}
\end{figure}

As discussed in Section-~\ref{sec2}, beam dynamics is determined by several factors including space charge, thermal emittance, angular momentum, and external fields. However, beam dynamics differs drastically when one factor dominates over the others; see Fig.~\ref{fig:Sec6:camemitsc}. In Fig.~\ref{fig:Sec6:camemitsc}(a) the dynamics is dominated by the thermal emittance, and the electrons have random momentum direction; in (b) the beam is dominated by its angular momentum, and the
electrons shear in a vortex pattern; and in (c) the beam is dominated by space charge, and the electrons repel each other by the Coulomb force and move outwards.

Taking into account all if these contributions, the rms transverse envelope equation for an electron bunch propagating in a drift space is described by~\cite{reiser-1994-a} as
\begin{eqnarray} \label{eq:envelope}
\sigma'' - \frac{K_p}{4\sigma} -\frac{\varepsilon_u^2}{\sigma^3}
-\frac{ {\cal L}^2}{\sigma^3} =0,
\end{eqnarray}
where $\sigma$ is the transverse rms size, $K_p=\frac{2I}{I_0  \beta^3\gamma^3}$ is the generalized perveance, $I$ is the absolute value 
of the peak beam current, $I_0=4\pi\epsilon_0mc^3/e\approx 17$~kA is the Alfv\'en current, $\varepsilon_u$ is the uncorrelated transverse rms emittance, and $\mathcal L$ is related to the average canonical angular momentum $\mean{L}$~(see Section~\ref{sec2:solenoid}) and the longitudinal momentum $p_z$ of the beam via the magnetization
\begin{equation}
\mathcal L=\frac{\mean{L}}{2p_z}.
\end{equation}

The second, third, and fourth terms of Eq.~\eqref{eq:envelope} represent the effects due to space charge, emittance, and angular momentum, respectively. When the fourth term is much greater than the second and the third terms, the beam is {\em
angular-momentum-dominated}.  

If there is external electromagnetic linear focusing, an extra term in the form of $k_0 \sigma$ can be added to the envelope equation, where $k_0$ is related to the strength of the external focusing force.

The magnetic field on the photocathode is normally zeroed to minimize the projected emittances. This can be seen from Eq.~\eqref{eq:envelope}, where the canonical angular momentum term $\cal{L}$ plays the same role as the emittance term in the beam-envelope equation so that it can be introduced to an effective emittance  $\varepsilon_{eff}=\sqrt{\varepsilon_u^2+{\mathcal L}^2}$, as noted in Section~\ref{sec2}. However, a large magnetic field is required to produce an angular-momentum-dominated beam in order that the correlation between the two transverse degrees of freedom dominates, i.e., ${\cal L} \gg \varepsilon_u$. 

\begin{figure}[hbt]
\includegraphics[width=0.45\textwidth]{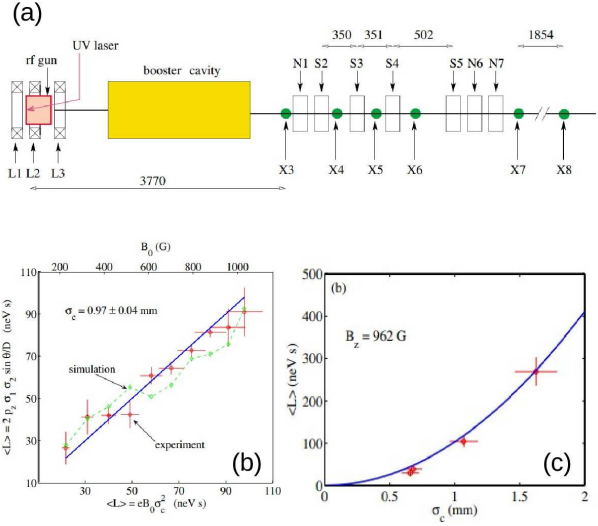}
\caption{Experimental setup used to produce a CAM-dominated beam at the Fermilab A0 facility (a), example measured conversion between CAM and MAM (b), and demonstration of the quadratic dependence on laser spot size on the cathode (c). In diagram (a), the letters represent solenoidal magnetic lenses (L), skew quadrupoles (Q), and
diagnostic stations (X, which means ``cross''). Dimensions are in mm. In (b) and (c), the red circles are measurements and the blue lines are fits. In plot (c), the green line shows the corresponding simulations. From~\cite{sun-2004-a}. \label{fig:sec6:magBeam}}
\end{figure}

In a rotationally invariant system, the conservation of canonical angular momentum $L$~\cite{noether-1971-a} states that
\begin{equation}\label{eq:buschtheorem}
L=\gamma m r^2 \dot{\phi}+ \frac{e}{2\pi} \Phi = const.,
\end{equation}
where $(r,\phi, z)$ are the cylindrical coordinates, and $\Phi$ is the magnetic flux enclosed inside a circle of radius $r$ at a given
location in $z$. Equation~\eqref{eq:buschtheorem} is also known as {\em Busch's theorem}~\cite{bush-1926-a,reiser-1994-a}. At the photocathode, the average of
the first term in Eq.~\eqref{eq:buschtheorem} is zero since $\mean{\dot{\phi}}=0$. The second term must not vanish in order to allow
the beam to acquire angular momentum. Therefore, an axial magnetic field on the cathode is required to generate an angular-momentum-dominated electron beam. The first photoinjector-based generation of a CAM-dominated beam was demonstrated at Fermilab's A0 photoinjector facility~\cite{sun-2004-a} using an L-band RF gun. The solenoidal lenses surrounding the RF gun were tuned to provide a variable magnetic field on the photocathode, and the beam's MAM was measured; see Fig.~\ref{fig:sec6:magBeam}. In the latter work, the beam was further accelerated using a superconducing cavity, and the MAM was measured at $\sim 15$~MeV. Similar investigations were most recently conducted at Jefferson Laboratory on a 300-keV electron beam produced from a DC gun~\cite{mamum-2018-a} and at the Fermilab FAST facility in a high-charge regime (3.2~nC) at 40~MeV~\cite{fetterman-2019-a}.  

\subsubsection{Decoupling of CAM-dominated beams and transverse emittance partitioning}\label{sec6b2}
A set of quadrupoles, with properly selected strength and separation, can apply a net torque to the CAM-dominated beam and remove its angular momentum. The result is an asymmetric beam with its two transverse degrees of freedom no longer coupled, i.e., a flat beam.	
A flat electron beam, i.e., a beam with high transverse emittance ratio, can be produced from an angular-momentum-dominated beam~\cite{brinkmann-2001-a}. The technique consists of manipulating an angular-momentum-dominated beam produced by a photoinjector using the linear transformation described in ~\cite{derbenev-1978-a}. A round-to-flat beam transformer, consisting of three skew quadrupoles and drift spaces, is discussed in ~\cite{burov-1998-a}. This technique was proposed as a way of producing a high-aspect-ratio beam to mitigate beamsstrahlung in future linear colliders while also circumventing the use of an electron damping ring conventionally used to reduce the vertical emittance~\cite{brinkmann-2001-a}. Likewise, the technique was also adapted for applications in of microwave- and THz-radiation generation~\cite{carlsten-2006-a,kumar-2007-a} using, e.g., the concept of Smith-Purcell backward oscillator~\cite{andrews-2005-a}. 

Finally, flat beams were also proposed as an intermediary way of transporting and accelerating magnetized beams by transforming them from magnetized beams into flat beams and back again~\cite{piot-2014-b,benson-2018-a}. 

%The transformation removes the angular momentum and results in a flat beam. A proof-of-principle flat-beam experiment was conducted at the Fermilab/NICADD Photoinjector Laboratory (FNPL)\footnote{NICADD isan acronym for Northern Illinois Center for Accelerator and Detector Development.} and an emittance ratio of $50$ was reported~\cite{edwards-2000-a,edwards-2001-a}.

The theory of generating a flat beam from an incoming angular-momentum-dominated beam is treated in several papers \cite{brinkmann-2001-a,derbenev-1998-a,burov-2002-a,kim-2003-a}. In this section, we follow the theoretical treatment based on the 4D beam matrix presented in ~\cite{kim-2003-a}, in which the round-to-flat beam transformation analysis was performed assuming that the beam and the transport channel upstream of the transformer are cylindrically symmetric and that the particle dynamics is Hamiltonian.

We will specify the coordinates of a particle in transverse trace space by two vectors:
\begin{equation} \label{e:vectordef}
 \bold{X}  =  \left(\begin{array}{c}x\\x'\end{array}\right)\;\mbox{and}\;\;
 \bold{Y}  =  \left(\begin{array}{c}y\\y'\end{array}\right).
\end{equation}

The corresponding $4\times 4$ beam matrix is
\begin{equation}\label{e:sigma}
\Sigma  =  \left(
\begin{array}{cccc}
\langle \bold{X{X}}^T\rangle&\langle \bold{X{Y}}^T\rangle\\
\langle \bold{Y{X}}^T\rangle&\langle \bold{Y{Y}}^T\rangle
\end{array}
\right).
\end{equation}

Let $R(\theta)$ be the $4\times 4$ rotation matrix of angle $\theta$:
\begin{equation}\label{e:rotation}
 R(\theta) = \left(
      \begin{array}{cc} I\cdot \cos{\theta} & I\cdot\sin{\theta} \\ -I\cdot\sin{\theta} & I\cdot\cos{\theta} \end{array}
      \right),
\end{equation}
where $I$ stands for the $2\times 2$ identity matrix. The beam matrix is rotationally invariant if
\begin{equation}\label{e:invar}
\Sigma = R (\theta)\cdot \Sigma \cdot R(\theta)^{-1}.
\end{equation}

From Eq.~\eqref{e:invar}, we obtain
\begin{eqnarray}\label{e:rotinvsigma11}
\langle \bold{X{X}}^T\rangle~\textrm{cos}^2\theta+\langle
\bold{Y{Y}}^T\rangle~\textrm{sin}^2\theta \nonumber \\ +(\langle
\bold{X{Y}}^T\rangle+\langle
\bold{Y{X}}^T\rangle)~\textrm{sin}\theta~\textrm{cos}\theta\nonumber \\ =\langle
\bold{X{X}}^T\rangle.
\end{eqnarray}
Since the rotation angle $\theta$ is arbitrary, Eq.~(\ref{e:rotinvsigma11}) leads to
\begin{eqnarray}
\langle \bold{X{X}}^T\rangle=\langle
\bold{Y{Y}}^T\rangle,\label{e:xxt}\\\langle
\bold{X{Y}}^T\rangle=-\langle \bold{Y{X}}^T\rangle.\label{e:xyt}
\end{eqnarray}

Taking the transpose of both sides of Eq.~\eqref{e:xyt}
\begin{equation}
{\langle \bold{X{Y}}^T\rangle}^T=-{\langle
\bold{Y{X}}^T\rangle}^T=-\langle \bold{X{Y}}^T\rangle,
\end{equation}
we find that $\langle \bold{X{Y}}^T\rangle$ is antisymmetric and can be written as
\begin{eqnarray}\label{e:xytJ}
\langle \bold{X{Y}}^T\rangle=\mathcal{L}J_{2D},
\end{eqnarray}
where $\mathcal{L}$ is a constant related to the angular momentum $L$ and longitudinal momentum $p_z$ by
\begin{equation}\label{e:calL}
\mathcal{L}=\langle xy'\rangle=-\langle x'y\rangle=\frac{L}{2p_z},
\end{equation}
and $J_{2D}$ is the $2\times2$ unit symplectic matrix given by Eq.~\eqref{unit symp}.

By expressing the beam matrix in terms of \emph{Courant-Snyder parameters} (also known as {\em Twiss parameters}) $\alpha$, $\beta$ ~(see, for example,~\cite{wiedemann-1999-a}), the general form of a round
beam matrix in $(x,x')$ or $(y,y')$ subspaces can be written as
\begin{eqnarray}
\langle \bold{X{X}}^T\rangle=\langle
\bold{Y{Y}}^T\rangle=\varepsilon T_0, \;\nonumber \\\mbox{with\;\;} T_0 =
\left(
\begin{array}{cc}
\beta&-\alpha\\
-\alpha&\frac{1+\alpha^2}{\beta}\\
\end{array}
\right)\label{e:T},
\end{eqnarray}
where $\varepsilon$ is the rms transverse emittance, and $|T_0|=1$.

Gathering Eq.~\eqref{e:xytJ} and Eq.~\eqref{e:T}, we may write the
general form of a cylindrically symmetric $4\times 4$ beam matrix in
the following convenient form:
\begin{eqnarray}\label{e:sigma0TJ}
\Sigma_0 =\left(
\begin{array}{cc}
\varepsilon T_0&\mathcal{L}J_{2D}\\
-\mathcal{L}J_{2D}&\varepsilon T_0\\
\end{array}
\right).
\end{eqnarray}

Let $M$ be the transfer matrix of the transformer, which is
symplectic. The beam matrix at the
exit of the transformer is
\begin{equation}\label{e:sigmaexit}
\Sigma=M\Sigma_0{M}^T.
\end{equation}

Kim noticed two invariants associated with the symplectic
transformation given by Eq.~\eqref{e:sigmaexit}~\cite{kim-2003-a}:
\begin{eqnarray}
&I_1=
\varepsilon_{4D}=\sqrt{\det(\Sigma)},\label{e:twoinv1}\\
&I_2(\Sigma)=  -\frac{1}{2}\mbox{Tr}(J_{4D}\Sigma
J_{4D}\Sigma).\label{e:twoinv2}
\end{eqnarray}
where ``Tr'' is the trace operator, and $J_{4D}$ is the $4\times4$ unit symplectic matrix .

Suppose a proper transfer matrix $M$ exists such that the beam matrix at the exit of the transformer is block diagonalized
\begin{equation}\label{e:sigma0diag}
\Sigma= \left(\begin{array}{cc}
\varepsilon_-T_- &0\\
0&\varepsilon_+T_+\\
\end{array}
\right),\;\;\mbox{with}\;\; T_{\pm}=\left(
\begin{array}{cc}
\beta_{\pm}&-\alpha_{\pm}\\
-\alpha_{\pm}&\frac{1+\alpha_{\pm}^2}{\beta_{\pm}}
\end{array}\right).
%|T_{\pm}|=1.f
\end{equation}

Applying Eq.~\eqref{e:twoinv1} to the symplectic transformation given by Eq.~\eqref{e:sigmaexit}, we have
\begin{equation}{\label{e:twoemits1}}
\sqrt{\det(\Sigma)}=\sqrt{\det(\Sigma_0)}\Rightarrow
\varepsilon_+\varepsilon_-=\varepsilon ^2-\mathcal{L}^2 .    \\
\end{equation}
It is easy to calculate the second invariant once we verify that
\begin{equation}
J_{2D}T_{0}J_{2D}T_{0}=-I,
\end{equation}
which leads to
\begin{eqnarray*}
\begin{aligned}
&J_{4D}\Sigma_0 J_{4D}\Sigma_0=\left(
\begin{array}{cc}
-(\varepsilon^2+\mathcal{L}^2) I&0\\
0&-(\varepsilon^2+\mathcal{L}^2) I
\end{array}\right),\quad\textrm{and}\\
&J_{4D}\Sigma J_{4D}\Sigma=\left(
\begin{array}{cc}
-{\varepsilon_-}^2 I&0\\
0&-{\varepsilon_+}^2 I
\end{array}\right).
\end{aligned}
\end{eqnarray*}
So, from Eq.~\eqref{e:twoinv2} we have
\begin{equation}\label{e:twoemits2}
I_2(\Sigma)=I_2(\Sigma_0)\Rightarrow
{\varepsilon_+}^2+{\varepsilon_-}^2=2(\varepsilon ^2+\mathcal{L}^2).
\end{equation}
Finally, the two transverse emittances can be derived from
Eq.~\eqref{e:twoemits1} and Eq.~\eqref{e:twoemits2}
\begin{equation}\label{e:twoemits}
\varepsilon_{\pm}=\varepsilon \pm \mathcal{L}.
\end{equation}
Equation~\eqref{e:twoemits} gives the two transverse emittances of a completely decoupled asymmetric beam. One
emittance~($\varepsilon_+$) can be orders of magnitude larger than the other~($\varepsilon_-$) given properly chosen initial conditions such as $\varepsilon$ and $\mathcal{L}$, which are related to the beam matrix at the cathode surface:
\begin{eqnarray}\label{sec7:cathodeBeamMatrix}
\Sigma_c= 
\left(\begin{matrix} 
\sigma^2_c & 0 & 0 & {\cal L}\\
0 & {\sigma'}^2_c &  -{\cal L} & 0\\
0 & -{\cal L} &  \sigma_c^2 & 0\\
{\cal L}  & 0 & 0 & {\sigma'}_c^2 \end{matrix}\right),
\end{eqnarray} 
where $\sigma_c$ and ${\sigma'}_c$ are the initial beam size and divergence spread at the cathode, respectively. The intrinsic  (or thermal) rms normalized emittance on the cathode (see Eq.~\eqref{eq:sec3:2D-xemit}) is $\varepsilon_{c}^{n}=\sigma_c\sigma_{pc}=\beta\gamma\sigma_c{\sigma'}_c=\beta\gamma\varepsilon$.

\begin{figure}[hhhhhhhh!!!!!!]\centering
\includegraphics[width=0.475\textwidth]{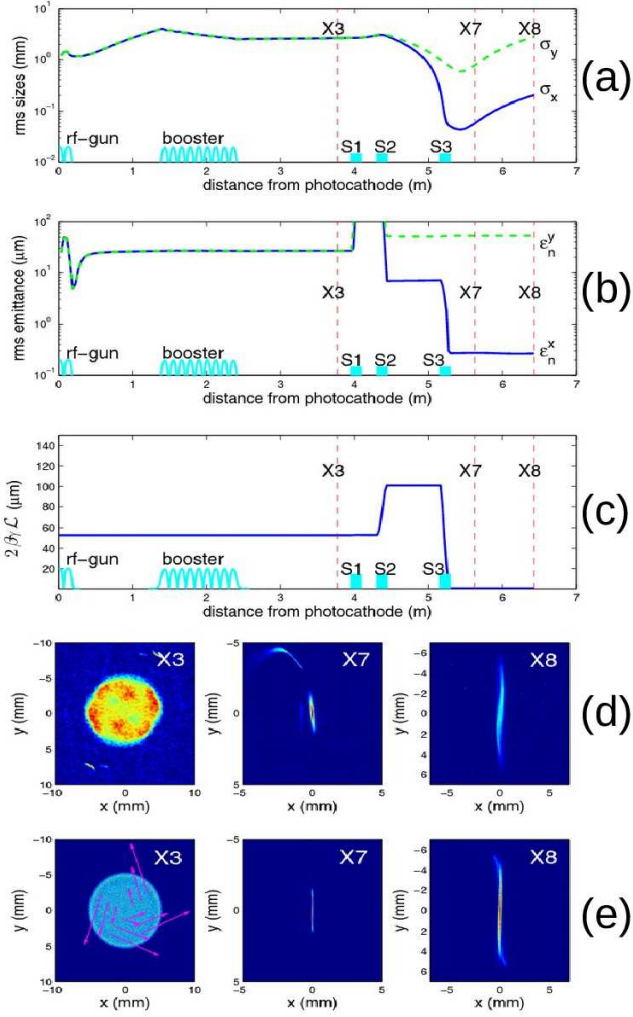}
\caption{Simulations (a-c,e) and measurements of flat-beam generation for the A0 photoinjector. Plots (a), (b), and (c) are \astra simulation of respectively the beam size, emittance and magnetization evolution along the beamline showcasing the removal of angular momentum using three skew-quadrupole magnets S1, S2 and S3 to transform the incoming magnetized beam into a flat beam. The  images displayed in (d) are measured beam distribution at X3, X7, and X8 screens while the distribution shown in (e) are the corresponding numerical simulations. The labels "X$i$" refer to the diagram appearing in Fig.~\ref{fig:sec6:magBeam}(a).  Adapted from \cite{piot-2006-a}\label{fig:sec6:A0flatbeam}}
\end{figure}

The experimental generation of a flat beam from a CAM-dominated beam was first demonstrated at the Fermilab A0 photoinjector~\cite{edwards-2000-a,edwards-2001-a}. The experimental setup was identical to the one displayed in Fig.~\ref{fig:sec6:magBeam} and three skew-quadrupole magnets located at $\sim 15$~MeV were employed to remove the angular momentum, as demonstrated via numerical simulation in Fig.~\ref{fig:sec6:A0flatbeam}(a-c). Ultimately, the experiment demonstrated the generation of a flat beam with a transverse emittance ratio of $\varepsilon_y/\varepsilon_x\simeq 100$~\cite{piot-2006-a}; see Fig.~\ref{fig:sec6:A0flatbeam}. The experiment confirmed the underlying physics and was validated against numerical simulation; see Fig.~\ref{fig:sec6:A0flatbeam}(d,e). Most notably, the smaller measured normalized emittance was $\varepsilon_{x}^{n}\simeq 0.4$~$\upmu${m}. This number, although limited by the diagnostics resolution, was a factor $\sim 2$ smaller than the thermal emittance estimated from the laser spot size on the cathode which was $\sim 1$~$\upmu${m}.

\begin{figure}[hbt]
\includegraphics[width=0.475\textwidth]{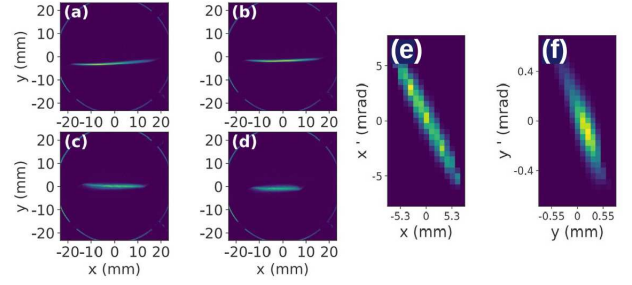}
\caption{Example of flat-beam generation at the Argonne Wakefield Accelerator. Images (a-d) showcase variable aspect ratio beams produced for $B_z=0.125$, 0.1,  0.074, and  0.049~T on the photocathode. Images (e) and (f) correspond to the reconstructed transverse phase spaces for a 1-nC bunch with $B_z=0.125$~T. Adapted from ~\cite{xu-2019-a}.  \label{fig:Sec6:awaflastbeam}}
\end{figure}

Further experiments, carried out at the AWA facility at Argonne, demonstrated the generation of a flat beam using a similar principle but with high charge up to 2 nC and an emittance ratio close to 200. The beamline also included a high-resolution phase-space measurement (using scanning slits) that permitted the reconstruction of the flat-beam phase space~\cite{xu-2019-a}; see Fig.~\ref{fig:Sec6:awaflastbeam}.

\subsubsection{Phase-space exchange between the two transverse planes~\label{sec6:XYexchange}}
In this section, we consider the phase-space coordinate exchange in the \textbf{X}-\textbf{Y} 4D phase space. Such a phase-space swapping was first proposed for spectrometer applications in 1972~\cite{kowalski-1972-a} and then for a M\"{o}bius accelerator~\cite{talman-1995-a}. If introduced in an electron storage ring for a light source, the beam emittance in the \textit{x}- and \textit{y}-directions become equal and one half of the natural emittance in the normal ring, thus mitigating the lifetime limitation for intra-beam scattering~\cite{aiba-2015-a}. The \textit{XY} exchange is also interesting for the next generation synchrotron radiation rings since it can lead to better horizontal injection efficiency~\cite{kuske-2016-a}. Likewise, such an exchange could be critical in lowering collective instabilities such as regenerative beam-break-up (BBU) instabilities in superconducting linacs~\cite{rand-1980-a,tennant-2005-a} and in beam-driven wakefield accelerators~\cite{gai-1997-a}. 

The transformation matrix of the form
\begin{equation} \label{sec6:eq:XZexchange} 
R_{XY} =\left(\begin{array}{cc} {0} & {A} \\ {B} & {0} \end{array}\right);\quad \left(\begin{array}{c} \bold{X} \\ \bold{Y} \end{array}\right)\to R_{XY} \left(\begin{array}{c} \bold{X} \\ \bold{Y} \end{array}\right)=\left(\begin{array}{c} {A\bold{Y}} \\ {B\bold{X}} \end{array}\right) 
\end{equation} 
can be constructed from an symmetric arrangement of five skew quadrupoles as follows:
\begin{equation} \label{sec6:eq:ZEqnNum720943} 
\; R_{XY} =Q_{S1} L_{1} Q_{S2} L_{2} Q_{S3} L_{2} Q_{S2} L_{1} Q_{S1} . 
\end{equation} 
Here $Q_{Si} $ and $L_{i} $ are the matrices for skew quadrupoles and drifts:
\begin{eqnarray} \label{68)} 
Q_{Si} =\left(\begin{array}{cccc} {1} & {0} & {0} & {0} \\ {0} & {1} & {q_{i} } & {0} \\ {0} & {0} & {1} & {0} \\ {q_{i} } & {0} & {0} & {1} \end{array}\right), \quad L_{i} =\left(\begin{array}{cccc} {1} & {d_{i} } & {0} & {0} \\ {0} & {1} & {0} & {0} \\ {0} & {0} & {1} & {d_{i} } \\ {0} & {0} & {0} & {1} \end{array}\right) .
\end{eqnarray} 
To simplify calculations, note that a skew quadrupole is obtained by a 45${}^\circ $ rotation of a normal quadrupole:
\begin{eqnarray} 
Q_{Si} &=&R_{\pi /4} {}^{-1} Q_{i} R_{\pi /4} ;\quad Q_{i} =\left(\begin{array}{cccc} {1} & {0} & {0} & {0} \\ {q_{i} } & {1} & {0} & {0} \\ {0} & {0} & {1} & {0} \\ {0} & {0} & {-q_{i} } & {1} \end{array}\right),  \nonumber \\ && 
\mbox{and~} R_{\pi /4} =\frac{1}{\sqrt{2} } \left(\begin{array}{cccc} {1} & {0} & {1} & {0} \\ {0} & {1} & {0} & {1} \\ {-1} & {0} & {1} & {0} \\ {0} & {-1} & {0} & {1} \end{array}\right) .
\end{eqnarray} 
Here we are using the 2$\times $2 block matrix notation and since drift matrices are rotationally invariant, we can write
\begin{eqnarray} 
R_{XY} &=&R_{\pi /4} {}^{-1} Q_{1} L_{1} Q_{2} L_{2} Q_{3} L_{2} Q_{2} L_{1} Q_{1} R_{\pi /4} . 
\end{eqnarray} 
Since the product of drifts and quadrupoles is a 2$\times $2 block diagonal, we have
\begin{eqnarray} 
R_{XY} &=&R_{\pi /4} {}^{-1} \left(\begin{array}{cc} {C} & {0} \\ {0} & {\bar{C}} \end{array}\right)R_{\pi /4} \nonumber \\
&=&\frac{1}{2} \left(\begin{array}{cc} {C+\bar{C}} & {C-\bar{C}} \\ {C-\bar{C}} & {C+\bar{C}} \end{array}\right) .
\end{eqnarray} 
The 2$\times $2 matrix $C$ is easy to calculate and $\bar{C}$ is obtained from $C$ by replacing $q_{i} \to -q_{i} $. The condition that $R_{XY} $ gives an emittance exchange is 
\begin{equation} 
\bar{C}=-C .
\end{equation} 
This gives
\begin{eqnarray} 
q_{1} &=&\frac{d_{2} q_{2} }{-d_{1} -d_{2} +d_{1} {}^{2} d_{2} q_{2} {}^{2} } ,\\ q_{3} &=&-\frac{d_{1} +d_{2} +d_{1} {}^{2} d_{2} q_{2} {}^{2} }{d_{1} d_{2} q_{2} \left(d_{1} +d_{2} \right)} . 
\end{eqnarray} 
The condition that $C$ is a pure drift gives
\begin{equation} \label{74)} 
d_{2} =\frac{-1+\eta ^{2} -\sqrt{1+2\eta ^{2} -4\eta ^{3} +\eta ^{4} } }{2q_{2} \eta \left(-1+\eta \right)} ;\quad \eta =d_{1} q_{2} . 
\end{equation} 
Lastly, the drift length becomes equal to the total drifts in the RHS of Eq.~\eqref{sec6:eq:ZEqnNum720943}, that is, $d_{T} =2\left(d_{1} +d_{2} \right)$ if
\begin{multline}
d_{1} =\frac{1}{9} \left[1-\frac{4\times 2^{2/3} }{\left(-67+9\sqrt{57} \right)^{1/3} } +\right. \\ \left. 2^{1/3} \left(-67+9\sqrt{57} \right)^{1/3} \right]\frac{1}{q_{2} } \approx \frac{-0.469}{q_{2} } . 
\end{multline} 
The system is therefore completely determined by specifying the value of $q_{2} $~\cite{kim-2020-a}. The final transform matrix is
\begin{equation} 
R_{XY} =\left(\begin{array}{cccc} {0} & {0} & {1} & {d_{T} } \\ {0} & {0} & {0} & {1} \\ {1} & {d_{T} } & {0} & {0} \\ {0} & {1} & {0} & {0} \end{array}\right). 
\end{equation} 

The exchange of transverse emittance via five skew quadrupoles in a transfer line prior to injection to the storage ring was numerically studied in~\cite{kuske-2016-a} and ~\cite{Armborst-2016-a} for the BESSY  II machine. The purpose of the manipulation was to exchange the smaller vertical emittance with the larger horizontal emittance prior to injection in order to stay within the acceptance of the ring. However, due the additional dispersion generated, the authors concluded that a different way of transverse emittance exchange, i.e., ``resonance crossing'' ~\cite{carli-2002-a, aiba-2020-a}, was more feasible. Indeed, the emittance exchange via resonance crossing in a booster synchrotron was demonstrated in ~\cite{Kallestrup-2020-a}. On the other hand, in the upgrade of the Advanced Photon Source at Argonne National Lab., a horizontal on-axis injection scheme from the booster to the storage ring has been adopted. The lattice of the booster to storage ring transport beamline was successfully designed to achieve transverse emittance exchange using a set of skew quadrupoles based ~\cite{kuske-2016-a}.    

%\subsubsection{beam conditioning Sessler}

%%% \subsubsection{Electron Cooling}

\subsection{Transverse-to-longitudinal phase-space exchangers}
\subsubsection{Emittance exchange} \label{sec6: EEX}
Transverse-to-longitudinal emittance exchange was first proposed by~\cite{cornacchia-2002-a} as a means to mitigate the microbunching instability in bright electron beams. The rationale there was that, generally, the transverse emittance produced by a state-of-the-art photoinjector is larger than desired for FEL generation, while the beam energy spread is smaller than necessary in order to avoid gain reduction in an x-ray FEL. As a result, they proposed a transverse-to-longitudinal phase-space exchange (i) to increase the slice energy spread, thereby reducing their sensitivity to the microbunching instability, while simultaneouly, to reduced the transverse emittance to improve the FEL performance (e.g., reduce the gain length). Utilization of emittance exchange in a compact XFEL facility was proposed in \cite{graves-2018-a}. It was also suggested as a way to mitigate BBU instabilities and to improve efficiency in energy-recovery linacs~\cite{piot-2009-a}. Mathematically, we follow a similar approach to the one described in Section~\ref{sec6:XYexchange} and seek a transformation that provides a matrix of the form 
\begin{eqnarray} \label{sec6:eq:XZexchange} 
&R_{XZ} =\left(\begin{array}{cc} {0} & {A} \\ {B} & {0} \end{array}\right);\nonumber \\
\quad 
&\left(\begin{array}{c} \bold{X} \\ \bold{Z} \end{array}\right)\mapsto R_{XZ} \left(\begin{array}{c} \bold{X} \\ \bold{Z} \end{array}\right)=\left(\begin{array}{c} {A\bold{Z}} \\ {B\bold{X}} \end{array}\right).
\end{eqnarray} 

\begin{figure}
\includegraphics[scale = 0.95]{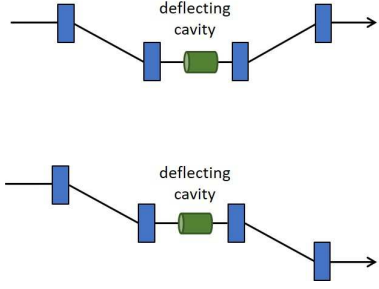}
\caption{Top: Partial emittance exchange using a deflecting cavity and and chicanes. Bottom: Complete emittance exchange using a deflecting cavity and double doglegs. \label{fig:sec6:chicDleg}}
\end{figure}

A possible beamline capable of providing such a transformation was first discussed in ~\cite{cornacchia-2002-a}. The beamline consists of a horizontal deflecting cavity located at the symmetry point of a chicane beamline; see Fig.~\ref{fig:sec6:chicDleg}(a). Using the transfer matrix of a deflecting cavity introduced in Section~\ref{sec2:deflectingcavity}, the overall transfer matrix of the system depicted in Fig.~\ref{fig:sec6:chicDleg}(a) is 
\begin{eqnarray}
R_{EEX}=R_{-DL}R_{TDC} R_{DL},
\end{eqnarray}
where $R_{DL}$ is the matrix of a dogleg under the small bending angle approximation; see Eq.~\eqref{sec2:eq:Dogleg} and  $R_{-DL}$ is the matrix of a reversed dogleg (i.e., $R_{-DL}$ is obtained from $R_{DL}$ via the substitution $\eta \mapsto -\eta$). 

In ~\cite{cornacchia-2002-a} they show that the condition $\kappa=1/\eta$ (here $\eta$ is the horizontal dispersion generated by one dogleg) causes most of the elements associated with the $2\times2$ anti-diagonal blocks of $R_{EEX}$ to vanish for a thin-lens TDC. From this, they show that the emittance is mapped as 
\begin{eqnarray}\label{sec6:eq:EEXDLmapping}
&(\varepsilon_{x,0},\varepsilon_{z,0})  \mapsto (\varepsilon_x, \varepsilon_z) \nonumber \\
=&(\sqrt{\varepsilon_{z,0}^2+\lambda^2\varepsilon_{x,0}\varepsilon_{z,0}},\sqrt{\varepsilon_{x,0}^2+\lambda^2\varepsilon_{x,0}\varepsilon_{z,0}}), 
\end{eqnarray}
where $\lambda$ is a coupling term that can be minimized via a proper choice of the incoming horizontal and longitudinal phase-space parameters. For $\lambda\ll 1$ we note that the beamline approximately exchanges the horizontal and longitudinal emittances [$(\varepsilon_{x,0},\varepsilon_{z,0}) \stackrel{R_{EEX}}{\longmapsto} (\varepsilon_{z,0},\varepsilon_{x,0}) $]. It was subsequently recognized by ~\cite{kim-2006-a} that flipping the second half of the chicane so that the TDC is flanked by two horizontally dispersive sections arranged as doglegs [see Fig.~\ref{fig:sec6:chicDleg}(b)] produces an ideal exchange (taking $L_c=0$). The emittance exchange condition is found to be
\begin{equation}\label{sec6:eqn:eexCond}
\kappa\eta+1=0,
\end{equation}
and the resulting transformation takes the form
\begin{multline}
 \label{sec6:Eq:DLEEXmatrix}
R_{EEX}=\left(\begin{matrix}0 & \frac{L_{c}}{4} & - \frac{L + \frac{L_{c}}{4}}{\eta} & \frac{\eta^{2} - \frac{\xi \left(4 L + L_{c}\right)}{4}}{\eta}\\0 & 0 & - \frac{1}{\eta} & - \frac{\xi}{\eta}\\- \frac{\xi}{\eta} & \frac{- L \xi - \frac{L_{c} \xi}{4} + \eta^{2}}{\eta} & \frac{L_{c} \xi}{4 \eta^{2}} & \frac{L_{c} \xi^{2}}{4 \eta^{2}}\\- \frac{1}{\eta} & - \frac{L + \frac{L_{c}}{4}}{\eta} & \frac{L_{c}}{4 \eta^{2}} & \frac{L_{c} \xi}{4 \eta^{2}}\end{matrix}\right),
\end{multline} 
where $L_c$ and $L$ are, respectively, the TDC length and the distances between the dogleg dipole magnets. Under the thin lens approximation ($L_c=0$), the latter matrix is $2\times2$-block anti-diagonal. In the most general case, it is possible to exchange the beam emittance by proper choice of the phase-space correlation on the incoming bunch. Specifically, the emittance mapping adopts the same form as Eq.~\eqref{sec6:eq:EEXDLmapping} with the coupling term  
\begin{eqnarray}\label{eq:sec6:lambda2}
{\lambda}^2=\frac{L_c^2(1+\alpha_{x,0}^2)[\xi^2+(\xi\alpha_{z,0}-\beta_{z,0})^2]}{16\eta^2\beta_{x,0}\beta_{z,0}}.
\end{eqnarray}
Here $\alpha_{i,0}$ and $\beta_{i,0}$ are the initial Courant-Snyder parameters associated with the horizontal ($i=x$) and longitudinal ($i=z$) degrees of freedom. The quantity $\lambda^2$ can be minimized by a proper choice of either the longitudinal or transverse C-S parameters. A possible solution consists of tuning the incoming LPS chirp to fullfill ${\alpha}_{z,0}=\,{\frac {{\it \beta_z}}{\xi}}$, corresponding to an incoming  longitudinal phase space chirp ${\cal C}\equiv\frac{d\delta}{dz}\big|_0=-\frac{1}{\xi}$ that produces a minimum bunch length at the cavity location~\cite{sun-2007-a}. Such a beamline was numerically investigated in~\cite{emma-2006-a} to confirm its ability to mitigate the microbunching instability in an X-ray FEL. 

\begin{figure}
\includegraphics[scale = 0.28]{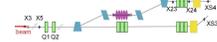}
\caption{Schematic of the A0 photoinjector beamline for bunch train generation using EEX. From \cite{ruan-2011-a}.  \label{sec6:fig:A0EEXschematics}}
\end{figure}

This double-dogleg configuration was used to experimentally demonstrate a near-ideal horizontal-to-longitudinal emittance exchange at the Fermilab A0 photoinjector~\cite{ruan-2011-a} with final results summarized in Table~\ref{sec6:tab:A0EEX}. See Fig.~\ref{sec6:fig:A0EEXschematics} for the experimental setup.

\begin{table}
\caption{\label{sec6:tab:A0EEX}
Direct measurements of horizontal transverse ($x$) to longitudinal ($z$) emittance exchange compared to simulation.  
Emittance measurements are in units of $\upmu${m} and are normalized. From~\cite{ruan-2011-a}; experiment performed using the configuration displayed in Fig.~\ref{sec6:fig:A0EEXschematics}.  
}
\setlength{\tabcolsep}{12pt}
\begin{center}
\begin{tabular}{rrrr@{$\,\pm\,$}lr@{$\,\pm\,$}lr@{$\,\pm\,$}l}
\toprule 
  & \multicolumn{2}{c}{Simulated} & \multicolumn{4}{c}{Measured} \\
  & \multicolumn{1}{c}{In} & \multicolumn{1}{c}{Out} & \multicolumn{2}{c}{In} & \multicolumn{2}{c}{Out} \\
\hline
$\varepsilon_\text{x}^{n}$  &  2.9  & 13.2  &  2.9  & 0.1 & 11.3 & 1.1   \\
$\varepsilon_\text{y}^{n}$  &  2.4  &  2.4  &  2.4  & 0.1 &  2.9 & 0.5   \\
$\varepsilon_\text{z}^{n}$  & 13.1  &  3.2  & 13.1  & 1.3 &  3.1 & 0.3   \\
\hline
\end{tabular}
\end{center}
\end{table}

Further work on improving the EEX beamline by~\cite{zholents-2011-a} proposed combining the TDC with two accelerating-mode cavities for providing a simple way of cancelling the thick-lens effect of the TDC, which induces beam energy gain, as detailed in Section~\ref{sec2:deflectingcavity}. The design also developed a chicane-based exhanger with quadrupole magnets to control the dispersion and circumvent the limitation associated with the early design~\cite{cornacchia-2002-a}. A similar design was later discussed in ~\cite{xiang-2011-a}. Ultimately, the capability to exchange emittance is limited by higher-order effects and requires the addition of higher-order multipole magnets, as discussed in ~\cite{nanni-2015-a}. Also, in the shaping process, collective effects such as CSR can significantly reduce the shaping quality \cite{zholents-2011-a, ha-2016-a, carlsten-2011-b}. Here, the exchange introduces extra difficulty than other well-studied beamlines such as chicane. Researches to control CSR's impact on the shaping are underway \cite{ha-2017-c,ha-2017-d,ha-2018-a,malyzhenkov-2018-a}.

\subsubsection{Current profile shaping}\label{sec6c2}
\begin{figure}[htb]
\includegraphics[width=0.45\textwidth]{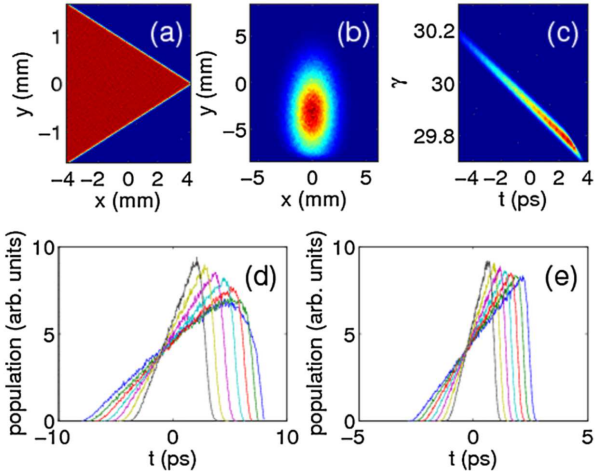}
\caption{Example of simulated generation of a linearly ramped current profile from an initial uniform triangular distribution in (a) shown with corresponding final transverse distribution (b) and longitudinal phase space (c) downstream of the EEX beamline. The current profiles after EEX are shown in plots (d) and (e) for initial horizontal and longitudinal profiles equal to (10,1)~$\upmu$m and (1,10)~$\upmu$m, respectively. From \cite{piot-2011-a}. \label{sec6:fig:arbCurrent}}
\end{figure}

An important application of emittance exchange is its potential capability for shaping the beam's temporal distribution with unprecedented versatility and precision~\cite{piot-2010-a}. Considering Eq.~\eqref{sec6:Eq:DLEEXmatrix} with $L_c=0$, we find that the initial phase space coordinates $(\mathbf X_0, \mathbf Y_0)^T$ of an electron will be mapped to final coordinates $(\mathbf X, \mathbf Y)^T=R_{EEX} (\mathbf X_0, \mathbf Y_0)^T$. In particular, the electron's final longitudinal coordinates $\mathbf Z=(z,\delta)$ are solely functions of its initial transverse coordinates $\mathbf X_0=(x_0,x'_0)$:
\begin{eqnarray}\label{eqn:phil}
\left\{ \begin {array}{ll} z =& -\frac{\xi}{\eta}x_0-\frac{L\xi-\eta^2}{\eta}x_0'\\
\delta =& -\frac{1}{\eta}x_0-\frac{L}{\eta}x_0'
\end{array}.\right.
\end{eqnarray}
Exploiting the mapping described by these equations, one can produce arbitrarily shaped current or energy profiles by controlling the incoming transverse phase space. Specifically, the incoming phase-space distribution in $\mathbf X$ is mapped into the $\mathbf Z$ space via 
\begin{eqnarray}
\Phi_z(\mathbf Z)= \Phi_x(B^{-1} \mathbf Z), 
\end{eqnarray}
where the subscript of the function $\Phi$ indicates to which of the two-dimensional phase spaces the function corresponds. 
Consequently, many of the techniques discussed in Section~\ref{sec4} can be readily applied. We can, for instance, use a mask to shape the incoming distribution in the $\mathbf X$ space and map this distribution into the $\mathbf Z$ plane as originally proposed in~\cite{piot-2010-a} and further discussed in~\cite{jiang-2011-a}. In principle, arbitrary bunch current distributions can be achieved using the EEX technique~\cite{piot-2011-a}, but ultimately the incoming emittance partition limits the shaping resolution, as discussed in Section~\ref{sec3}; see Fig.~\ref{sec6:fig:arbCurrent}.  

\begin{figure}[hhhhhh]
\includegraphics[width = 0.4750\textwidth]{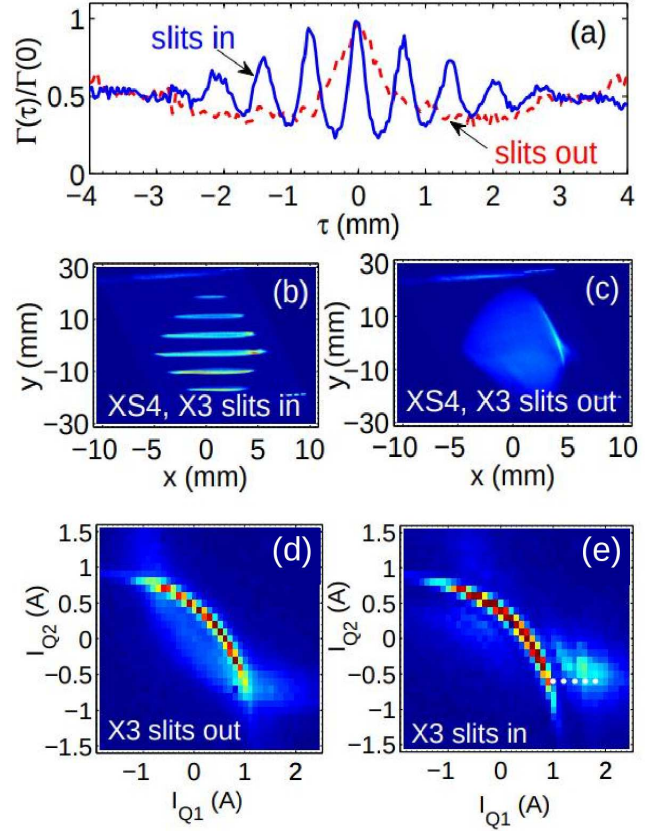}
\caption{Experimental demonstration of sub-picosecond bunch train generation with an EEX beamline. (a) Normalized autocorrelation function $\Gamma(\tau)/\Gamma(0)$ of the CTR signal recorded with (solid) and without (dashed) insertion of a multi-slit mask upstream of the EEX beamline ($\tau$ is the optical path difference). The corresponding beam transverse densities downstream of the EEX appear in (b) and (c).  The vertical axis on (b) and (c) is proportional to the beam's fractional momentum spread ($\delta$). The nominal bunch charge is $550\pm 30$~pC and reduces to $\sim 15 \pm 3$~pC when the slits are inserted. Total normalized CTR energy detected at X24 as a function of quadrupole magnet currents $I_{Q_1}$ and $I_{Q_2}$ with X3 slits out (d) and in (e) the beamline. The bolometer signal is representative of the inverse of the bunch duration $\sigma_t$. The intensity island appearing at $(I_{Q1},I_{Q2})\simeq$(1.5,-0.5) in (e) is indicative of a density-modulated bunch. From \cite{sun-2010-b}. \label{sec6:fig:totalCTR}}
\end{figure}

An experimental demonstration of the shaping capability of an EEX beamline was demonstrated in~\cite{sun-2010-a}, where the incoming beam was transversely sliced into multiple beamlets using a multislit plate and sent through an EEX beamline as diagrammed in Fig.~\ref{sec6:fig:A0EEXschematics}. Doing so produced a train of subpicosecond bunches with tunable separation as depicted in Fig.~\ref{sec6:fig:totalCTR}, where an autocorrelation of coherent transition radiation emitted by the beam downstream of the EEX beamline confirmed the beam was temporally modulated when the incoming beam was intercepted by the multislit mask located upstream of the EEX beamline. 
In addition, the experiment demonstrated that tuning the upstream quadrupole magnets Q1 and Q2 in Fig.~\ref{sec6:fig:A0EEXschematics} provided control over the final LPS correlation and could transfer the incoming transverse density modulation into an energy or temporal modulation as discussed in~\cite{sun-2010-b}; see also Fig.~\ref{sec6:fig:totalCTR}(b-e). This setup was also employed to investigate the generation of narrowband coherent transition radiation with tunable central wavelength~\cite{piot-2011-b}.

\begin{figure*}[ht]
\includegraphics[width=\textwidth]{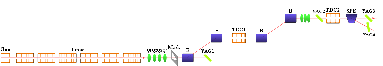}
\includegraphics[width=\textwidth]{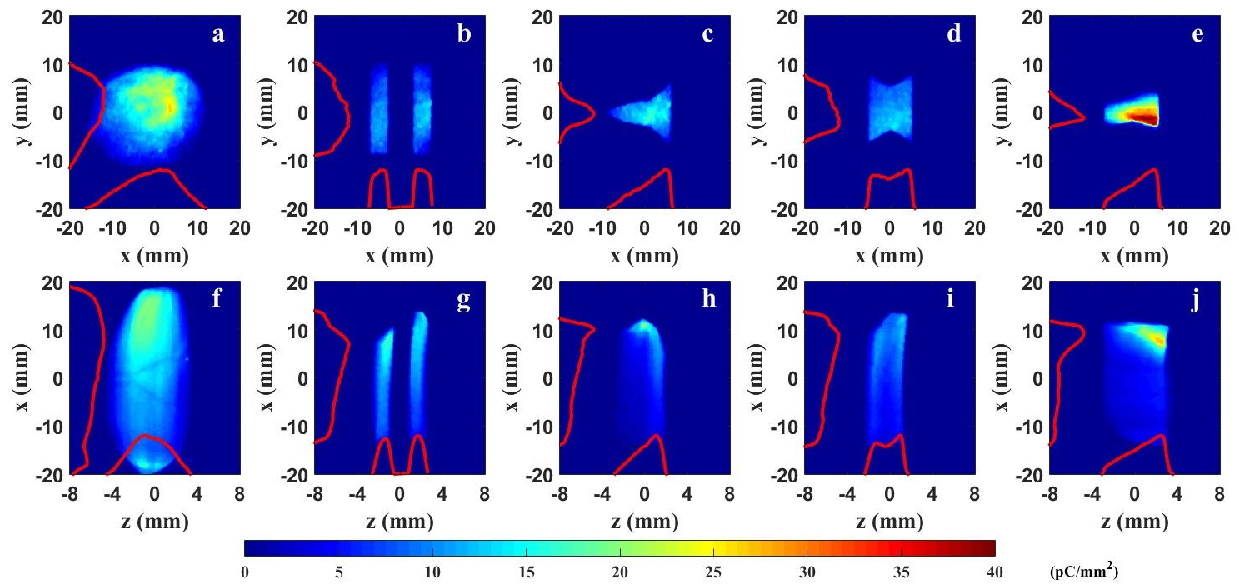}
\caption{Experimental demonstration of versatile bunch shaping using an EEX beamline at AWA. (Top) Schematic of the AWA EEX beamline for longitudinal beam shaping. (a-j) Beam shaping experiments using different transverse masks at the AWA EEX beamline. Images (a-e) are upstream of the EEX, and images (f-j) are downstream of the EEX. From \cite{ha-2016-a}.
\label{fig:awabeamine}}
\end{figure*}

LPS shaping was further investigated at the Argonne Wakefield Accelerator (AWA)~\cite{ha-2017-b} in the context of producing a beam suitable to improve the efficiency of a beam-driven wakefield accelerator. The EEX beamline used in the first generation of experiments at AWA adopted the double-dogleg configuration depicted in Fig.~\ref{fig:awabeamine} using a 48-MeV high-charge bunch. Four quadrupole magnets downstream of the linac were used to manipulate the transverse beam phase space prior to the exchanger. A set of 100-$\upmu$m-thick insertable tungsten masks of various shapes located $\sim$0.2~m upstream of the exchanging beamline were used to demonstrate high-precision control over the final temporal shape of the bunch. Each dogleg provided $\eta\simeq 0.9$~m and the TDC was a 1/2+1+1/2-cell L-band cavity. The experimental setup incorporated full, single-shot LPS diagnostics that were used to directly measure the produced temporal distribution downstream of the exchanger. Figure~\ref{fig:awabeamine} (g-j) showcases some examples of the final current distribution experimentally achieved with four different mask shapes. 
After these initial experimental demonstrations, AWA's EEX shaping beamline was then used set new records for transformer ratio.  The beamline produced ramped current profiles for driving the wakefield in a dielectric structure~\cite{gao-2018-a} and a plasma medium~\cite{roussel-2020-a} with transformer ratios of ${\cal R}\simeq 4.5$ and ${\cal R}\simeq 7.8$ respectively.

The most recent developments on EEX have focused on using this class of beamline to form nanometer-scale modulations for coherent-radiation emission at X-ray wavelengths either by combing structured beams formed using a structure, e.g., field-emission-array, photocathode~\cite{graves-2012-a}, or by impressing beam modulation in the transverse phase space upstream of the exchanging beamlines which is then converted in a modulation on the LPS (either along the energy or temporal direction). In~\cite{nanni-2018-a} a transmission mask such as described in Section~\ref{sec4:beyondBinMask} is employed to produce the initial transverse modulation while~\cite{ha-2019-b} explores the use of a transverse wiggler as shown in Fig.~\ref{fig:sec4:twiggler}. 
\subsubsection{Bunch compression}
The final bunch length downstream of the EEX is $\sigma_{z,f}=[R_{EEX,33}^2 \sigma_x^2 +R_{EEX,44}^2 \sigma_{x'}^2 +2R_{EEX,33}R_{EEX,44} \mean{xx'}]$. Substituting Eq.~\eqref{sec6:Eq:DLEEXmatrix}, with the assumption $L_c=0$ for simplicity, and expressing all beam quantities in term of $\varepsilon_x$ and C-S parameters $(\alpha_x,\beta_x)$ associated with the horizontal phase space upstream of the EEX beamline, we obtain
\begin{multline}
\sigma_{z,f}=\sqrt{\frac{\varepsilon_x}{\beta_x}}\left[\left(\frac{\xi}{\eta}\beta_x -\left(\eta-\frac{L\xi}{\eta}\right)\alpha_x\right)^2 +\right. \\
\left. \left(\eta-\frac{L\xi}{\eta}\right)^2 \right]^{1/2}. 
\end{multline}

The latter part of this equation indicates that upon choosing the proper incoming phase space correlation $\alpha_x/\beta_x= \xi/(\eta^2-L\xi)$, the final bunch length becomes $\sigma_{z,f}=|\eta-\frac{L\xi}{\eta}|\sqrt{\frac{\varepsilon_x}{\beta_x}}$, which depends on the EEX design and can be made very small. Based on this observation, ~\cite{carlsten-2011-b} investigated the performance of an EEX beamline for producing short bunches. The reference specifically discusses design choices that lead to extremely short final bunch duration $\sigma_{z,f}/c$ at the subfemtosecond time scale. The advantages of EEX beamlines employed for compression includes the reduced susceptibility to CSR-induced effects (including microbunching instability and bunch-length broadening), and elimination of the need for an initial longitudinal-phase-space chirp for compression using a chicane as well as any residual energy-phase correlation after compression. The drawback of an EEX-based bunch compressor is that the final horizontal phase space is determined by the incoming longitudinal phase space, and the final transverse partition is generally not symmetric, which may be problematic for some applications, such as FELs~\cite{emma-2006-a}. In addition, any longitudinal phase space jitter (e.g., timing, energy and energy spread) will be manifested in the transverse phase space, resulting in beam position and spot size jitter~\cite{ha-2017-c}. 
\subsubsection{Double phase-space exchangers}
A limitation of the EEX technique for shaping the beam profile stems from the associated emittance exchange between, e.g., the horizontal and longitudinal phase spaces. To circumvent this limit, a possible configuration consists of a beamline composed of two concatenated EEX beamlines providing the mapping $(\mathbf X_0,\mathbf Z_0)\mapsto (\mathbf Z_1, \mathbf X_1) \mapsto (\mathbf X_2,\mathbf Y_2)$. Consequently, inserting a mask at location ``1" produces a shaped beam downstream of the beamline while ideally leaving the incoming emittance unaltered. In practice such a method is not viable as collective effects such as space charge and CSR ultimately dilute the bending-plane emittance.   

Despite tihs limitation, another advantage of the double phase-space exchanger configuration was recognized in ~\cite{zholents-2011-a} as providing a way to transparently tune the final LPS parameters without requirements on the incoming LPS. To understand such an application, we can write the matrix of the beamline as a block diagonal matrix flanked by two matrices representing the EEX using Eq.~\eqref{sec6:Eq:DLEEXmatrix}:  
\begin{eqnarray}
R_{DEEX}=R_{EEX}
\left(\begin{array}{cc} A & 0 \\ 0 & B \end{array}\right) R_{EEX}, 
\end{eqnarray}
assuming $L_c=0$ so that $R_{EEX}\simeq \left(\begin{array}{cc} 0 & M \\ N & 0 \end{array}\right)$. Consequently, the total matrix of the double EEX (DEEX) simplifies as  
\begin{multline}~\label{sec6:eq:DEEXmatrix}
R_{DEEX}= \left(\begin{array}{cc}  MBN & 0  \\ 0 & NAM  \end{array}\right) \stackrel{\gamma\gg1}{\rightarrow} \left(\begin{array}{cc}  MN & 0  \\ 0 & NAM  \end{array}\right) ,
\end{multline}
which is a $2\times2$ block diagonal matrix and confirmed that the DEEX beamline does not provide any global coupling. The simplification $B=I$ in the ultra-relativistic limit $\gamma\gg1$ comes from the longitudinal dispersion associated with a longitudinal a drift and scales as $-L/\gamma^2$ where $L$ is the length of the telescope beamline. These results are generally available even when $L_c\ne 0$ as long as thick-lens effects are corrected via, e.g., the addition of accelerating-mode cavities~\cite{zholents-2011-a}. Equation~\eqref{sec6:eq:DEEXmatrix} suggests a simple way of tuning the final LPS correlation of the beam by properly designing the insertion beamline optical lattice between the two EEXs. The second EEX simply converts the transversely fine-tuned beam via the insertion beamline to the longitudinal phase space. Focusing on the LPS mapping and treating the insertion beamline as a telescope, i.e.  considering $A$ to be diagonal with element $A_{11}=1/A_{22}\equiv a$, we find
\begin{eqnarray} \label{sec6:eq:NAMmatrix0} 
NAM=\left(\begin{matrix}\frac{L \xi + a^{2} \left(L \xi - \eta^{2}\right)}{a \eta^{2}} & \frac{\xi \left(a^{2} + 1\right) \left(L \xi - \eta^{2}\right)}{a \eta^{2}}\\\frac{L \left(a^{2} + 1\right)}{a \eta^{2}} & \frac{L a^{2} \xi + L \xi - \eta^{2}}{a \eta^{2}}\end{matrix}\right).
\end{eqnarray}
As an example, considering the case when the beamline is designed such that $L(a^2+1)/(a\eta^2)\ll 1$, the above matrix is approximately 
\begin{eqnarray}\label{sec6:eq:NAMmatrix}
NAM \simeq \left(\begin{matrix}
-a & \xi \left(a+\frac{1}{a}\right) \\
0 & -\frac{1}{a} \end{matrix}\right);
\end{eqnarray}
therefore, the final rms bunch length is 
\begin{multline} \label{eq:bl_deex}
    \sigma_z =[ \sigma_{z,0}^2a^2+\xi^2(a+1/a)^2 \sigma_{\delta,0}^2 \\ -2a\xi(1+a^2)\mean{z_0\delta_0}]^{1/2}.
\end{multline}
 \begin{figure}[htb]
\includegraphics[width=0.45\textwidth]{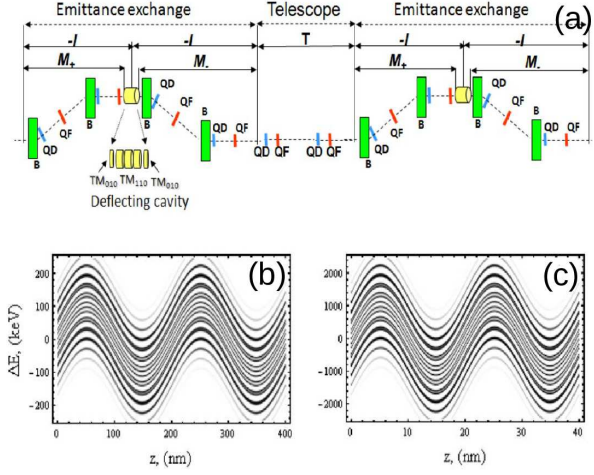}
\caption{Schematic  of  a double-EEX-based  bunch  compressor (a) with a simulated example of an application to provide a 10-fold increase in modulation frequency of an incoming laser-modulated electron beam (b) and (c). From \cite{zholents-2011-a}. \label{sec6:fig:EEX_BC}}
\end{figure}
An important consequence of Eq.~\eqref{eq:bl_deex} is that bunch compression can be accomplished without any longitudinal-phase-space chirp (i.e. $\mean{z_0\delta_0}=0$) and designing the beamline such that $|a^2+\xi^2(a+1/a)^2|<1$. A numerical simulation of a double-EEX-based bunch compressor is shown in Figure~\ref{sec6:fig:EEX_BC}.

\begin{figure}[htb]
\includegraphics[width=0.45\textwidth]{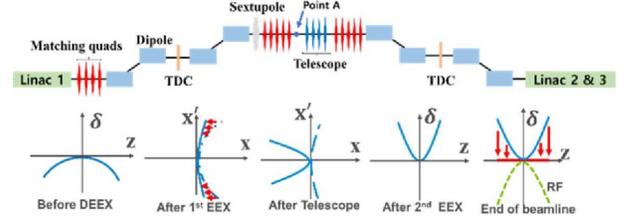}
\caption{Overview of a DEEX-based energy-spread reduction via correction on quadratic nonlinearity with a sextupole magnet. From \cite{seok-2019-b}. \label{fig:sec6:nonlinearDEEX}}
\end{figure}

Additionally, the configuration enables control of nonlinear correlation in the longitudinal-phase-space by using nonlinear magnets between the two EEXs. For instance, ~\cite{seok-2019-b} discusses various configurations to reduce the beam final energy spread via control of the longitudinal phase-space nonlinearities with a DEEX configuration. The method closely follows the technique described in Section~\ref{sec4:sec:nonlinearbeammapping}. Here, the first EEX is applied to the phase space $(\mathbf X_0,\mathbf Z_0)\mapsto (\mathbf X_1\simeq \mathbf Z_0, \mathbf Z_1\simeq \mathbf X_0)$, a then a nonlinear transformation is applied to the horizontal phase space 
\begin{eqnarray}\label{sec6:eq:nonlinearDEEX}
x'_1=f(x_1,y_1),
\end{eqnarray}
and then the second EEX is applied to change back the longitudinal and transverse phase space $(\mathbf X_1, \mathbf Z_1)\mapsto (\mathbf X_2\simeq \mathbf Z_1,\mathbf Z_2\simeq \mathbf X_1)\simeq (\mathbf X_0,\mathbf Z_0)$.
Figure~\ref{fig:sec6:nonlinearDEEX} showcases a configuration capable of introducing a nonlinear phase-space correlation with a DEEX beamline. In this particular case, a sextupole magnet, judiciously located downstream of the first EEX, imposes a quadratic correlation. 
Extension of this technique to control higher-order distortion is straightforward: correction of third order nonlinearties to, e.g., suppress current spike ``horns" sometimes encountered in bright-beam injectors, was demonstrated via numerical simulations~\cite{seok-2019-a} as an alternative to the technique described in Section~\ref{sec4:subsec:modulatorsWlongDisp}; see also Ref.~\cite{charles-2017-a}. 
%This method of controlling the longitudinal phase space is appealing as it can be accomplished with magnets instead of RF cavities. However, it ultimately introduces coupling between the different phase spaces (nonlinear magnets introduce coupling between the two transverse DOFs; see Eq.~\eqref{sec6:eq:nonlinearDEEX}, and improving the technique to ensure preservation of the beam brightness will be critical to its deployment. 

\begin{figure}
\centerline{\includegraphics[width=0.5\textwidth, keepaspectratio=true]{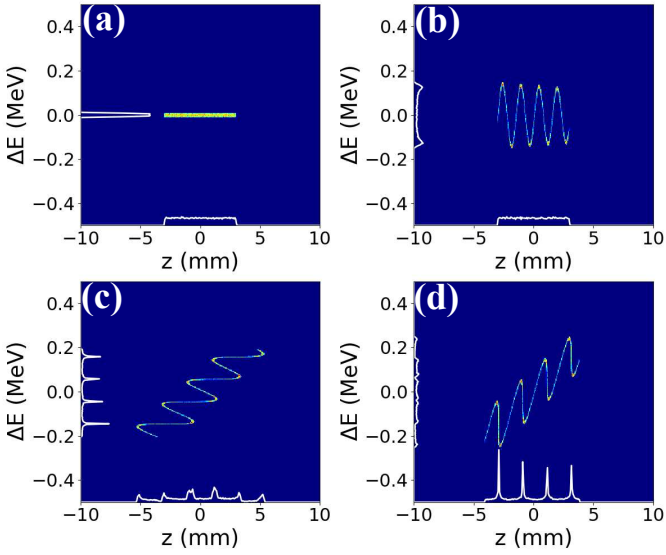}}
\caption{Longitudinal phase spaces from numerical particle tracking. (a) shows the beam's initial phase space and (b) shows the phase space after a modulator. (c) and (d) show the final phase space after a double EEX beamline, and each of them corresponds to spectral and temporal bunching respectively. From \cite{seok-2021-a}. }\label{sec6_fig_multie}
\end{figure}

%While the dechirper directly used a wakefield to control the energy distribution, \cite{seok-2021-a} introduced 
 Another recent development introduced in \cite{seok-2021-a} is using a DEEX beamline with a phase space modulator, such as wakefield structure or transverse wiggler discussed in Section \ref{sec4} and \ref{sec5}. Similar to generating a bunch train from the energy modulation, this method can generate density spikes on both the time and energy axes. If we only consider a small fraction of the sinusoidal modulation that builds up the density spike, we can write this microbunch's longitudinal correlation as $\delta=hz$ (i.e., locally linear correlation with chirp $h$); see Fig.~\ref{sec6_fig_multie}(b). In general, we can define the compression factor for this microbunch as $1/(R_{55}+hR_{56})$. Appropriate beamline parameters for a given local chirp can maximize the compression factor and build up density spikes. We can imagine the same situation for the energy distribution. The energy compression factor will be simply $1/(R_{65}+hR_{66})$. In the case of the chicane, $R_{55}$ is 1 and $R_{56}$ is non-zero, so there is a corresponding $h$. However, $R_{65}$ is zero and $R_{66}$ is 1, so no controls are allowed on the energy distribution. On the other hand, a double EEX beamline that provides non-zero $R_{65}$ and $R_{66}$ would be able to generate energy spikes. This beamline provides control over $R_{65}$ and $R_{66}$ using the quadrupole magnets between two EEX beamlines; see \cite{zholents-2011-a} and \cite{ha-2017-a}. 

Figure~\ref{sec6_fig_multie} showcases numerical tracking results. Similar to the density modulation case, a dielectric structure in front of a double EEX beamline imprints a sinusoidal modulation on the longitudinal phase space. Then, this modulation is converted into energy spikes, controlled with quadrupole magnets located in the middle of the beamline.  These energy spikes can be considered as several beams, each with small energy spread. Because the EEX beamline provides control of $R_{55}$, $R_{56}$, $R_{65}$, and $R_{66}$, it can be used to generate a multi-energy beam with controllable time separation from a single bunch. This may be a useful technique to generate multi-color radiation for various spectroscopy methods or pump-probe experiments.

%%%
%{\color{red}
%\subsubsection{Other EEX applications: chirp control}
%%
%}
%%%%%
\subsection{Generalized phase-space repartitioning between the three degrees of freedom}
\subsubsection{Flat beam transformation combined with emittance exchange}
Future electron-positron linear colliders (LCs) require unprecedented ultra-low transverse emittances. Additionally, the transverse emittance ratio should be high to mitigate beamsstrahlung effects~\cite{yokoya-1992-a}. The present requirements from high-energy physics call for 80\% spin-polarized electron beams. The electron bunch charge range from fC to nC depends on the LC technology choice. The vertical emittance required to mitigate beamsstrahlung effects is attained by injecting the beam into a damping ring to decrease its transverse emittance via radiative cooling, yielding a ``flat” beam with a transverse emittance ratio ranging from 300 to 500, depending on the LC concept. Taking the example from the International Linear Collider (ILC)~\cite{adolphsen-2013-a}, the final emittance partition of $(\varepsilon_x^n,\varepsilon_y^n,\varepsilon_z^n)= (5.5, 20\times 10^{-3}, 6.5\times 10^4)$~$\upmu$m corresponding to a six-dimensional emittance $\varepsilon_{6D}^n=\varepsilon_x^n \varepsilon_y^n \varepsilon_z^n \simeq 7150~\upmu$m$^3$ is generally larger than values typically achieved by state-of-the-art photoinjectors (e.g., for 3.2 nC bunch charges). Consequently, there has been some effort to develop a damping-ring-free electron injector for a future linear collider by combining the round-to-flat-beam transformer (RFTB) and EEX beamlines. Specifically,~\cite{xu-2021-a} considers a 3.2-nC bunch produced in an L-band photoinjector and demonstrates via numerical simulation that a transverse intrinsic emittance partition of  $(\varepsilon_+^n,\varepsilon_-^n,\varepsilon_z^n)\sim (500, 10\times 10^{-3}, 10)$~$\upmu$m assumes an ideal mapping of the intrinsic emittances  $(\varepsilon_x^n,\varepsilon_y^n)= (\varepsilon_1^n,\varepsilon_2^n)$ using an RFBT beamline. The author suggests that the downstream EEX beamline would produce an emittance partition $(\varepsilon_x^n,\varepsilon_y^n,\varepsilon_z^n)\sim (10, 10\times 10^{-3}, 500)$~$\upmu$m. 

\subsubsection{Coupling between the longitudinal and transverse phase spaces}
So far we have discussed beamlines capable of phase-space swapping between the two transverse phase space planes [$(x,x') \leftrightarrow (y,y')$] or between one of the transverse phase space planes and the longitudinal plane [$(x,x') \leftrightarrow (z,\delta)$]. We also presented a method to repartition the phase space between $(x,x')$ and $(y,y')$ using a magnetized beam and an RFBT beamline. We now examine whether a similar transformation is possible between, e.g., $(x,x')$ and $(z,\delta)$ following~\cite{carlsten-2011-a}. In this work, the authors noticed that the matrix of a deflecting cavity in $(x,x',z,\delta)$ under the thin lens approximation (i.e., described by Eq.~\eqref{TDC}  with $L=0$) is similar to the matrix of a skew-quadrupole magnet in $(x,x',y,y')$. Consequently, introducing an initial correlation similar to the one described for the case of a magnetized beam (i.e., by Eq.~\eqref{sec7:cathodeBeamMatrix}) in the $(x,x',z,\delta)$ plane combined with several deflecting cavities separated by ``drift" (with matrix described by Eq.~\eqref{sec2:eq:Dogleg} with $\eta=0$) could repartition the emittances between the $(x,x')$ and $(z,\delta)$ planes. Specifically, \cite{carlsten-2011-a} consider the practical case of a tilted-front laser pulse impinging on a photocathode so that the initial correlation is imparted in the $(x,x',z,\delta)$ phase space of the form 
\begin{eqnarray}
z_c=z +\tau x,
\end{eqnarray}
where $\tau$ is a parameter representing the $x-z$ correlation introduced by the tilt. The corresponding beam matrix at the cathode location is written as
\begin{eqnarray}
\Sigma_c= 
\left(\begin{matrix} 
\sigma^2_x & 0 & \tau \sigma^2_x & 0\\
0 & {\sigma'}^2_x & 0 & 0\\
\tau \sigma^2_x & 0 & \sigma^2_z + \tau^{2} \sigma^2_x & 0\\
0 & 0 & 0 & \sigma^2_{\delta}\end{matrix}\right).
\end{eqnarray} 
The corresponding intrinsic emittances are 
\begin{eqnarray}
\begin{matrix} 
\varepsilon_x &\simeq {\sigma}_x \sigma_{\delta} \tau, \\ 
\varepsilon_z &\simeq \frac{{\sigma'}_x \sigma_{z}}{\tau},
\end{matrix} 
\end{eqnarray} 
where it is assumed that $\tau^2 \gg \frac{{\sigma'}_x \sigma_z }{\sigma_x \sigma_{\delta}}$.
Then ~\cite{carlsten-2011-a} shows that this initial coupling can be removed by locating four deflecting cavities downstream of the electron source to map the intrinsic emittance to conventional emittances. The cavities are separated with drifts with longitudinal dispersion (e.g. magnetic chicane described by Eq.~\eqref{sec2:eq:Dogleg} with $\eta=0$). 

This approach can be further expanded by introducing both a transverse and longitudinal tilt to introduce arbitrary emittance repartitioning within the three degrees of freedom~\cite{yampolsky-2010-a}.

%PP: tilt on laser with three TDC for x-z re-partitioning; also wedge based; Reference B. Carlsten]

\section{Future directions\label{sec7}}
The techniques described in this paper along with that discussed in~\cite {hemsing-2014-a} have open the path to finer control over the beam's phase-space distribution beyond the traditional ensemble-averaged techniques. Further development of phase-space tailoring methods will ultimately aim to provide full six-dimensional control of the phase-space distributions, possibly enabling the design of a tailored beam at the single-particle level. 

Furthering the shaping methods discussed in this review will increase the shaping resolution. In addition, examples of possible expansions based on recent developments are presented in this section for each of the beam-shaping classes considered. The list below is by no means exhaustive given the vigorous on-going research on the topic. 

\subsection{Ab-initio shaping}
Controlling the final beam distribution by properly programming the initial conditions is widespread owing to its versatility and easy implementation, as discussed in Section~\ref{sec3}. Further expanding the resolution of this class of methods will benefit from new electron-source emission concepts, integrated photonic progress, and advances in ultrafast-laser systems. 

Over the last decade, new ultra-cold electron sources based on trapped atoms have emerged~\cite{claessens-2005-a,zolotorev-2007-a}. Although the primary motives of this work were related to the generation of electron beam close to quantum degeneracy, an experiment demonstrated a high degree of control over the beam distribution~\cite{mcculloch-2011-a} via shaping of the ionization laser using a spatial light modulator similar to the one discussed in Section~\ref{sec3}. These sources typically produce electron beams with keV energies, and their integration into relativistic electron sources remains a challenge~\cite{geer-2014-a}. Yet they will provide another toolkit for ab-initio shaping of the beam distribution. 

Likewise, the development of a novel laser architecture, such as discussed in ~\cite{lemons-2021-a}, that splits an incoming laser pulse over multiple channels, each with its own set of controls (phase shifters and polarization control), and then coherently recombines it. The formed laser pulse could provide finer spatiotemporal control over the photoemission process although it is ultimately limited by cathode response time and intrinsic emittance. 

Finally, recent progress in integrated nanophotonic fabrication~\cite{komljenovic-2016-a} has opened the path to the fabrication of pixelized cathodes~\cite{blankemeier-2019-a}. This type of cathode could ultimately support initial control over the emitted beam spatiotemporal distribution with unprecedented spatial and temporal scales. 

Likely, a combination of these developments will eventually be required to form electron beams with the required tailored distributions. 

%%%%%%%%%%%%%%%%%%%%%%%%%%%%%%%%%%%%%%%%%%%%%%%%%%%%%%%%%%%%%%%%%
\subsection{Controlling the beam via external fields} 
Over the years the possibility of shaping the electromagnetic-field distribution using  metamaterials has flowed in many applications, most notably optics, and has led to the development of ``transformation electromagnetics"~\cite{werner-2014-a}, a branch of electromagnetics devoted to the development of transformations capable of spatially tailoring electromagnetic fields. These techniques have recently been applied to the design of electromagnets with unusual properties (e.g., producing a negative magnetic permeability) that could expand the methods described in Section~\ref{sec4}. For instance, \cite{mach-batlle-2020-a} experimentally demonstrated that an active magnetic metamaterial can emulate the field of a straight current wire at a distance. Such a demonstration opens the way to manipulating magnetic fields in inaccessible regions. It is expected that these emerging technologies will be critical to further the development of beam-control techniques based on external electromagnetic fields. Finally, the discovery of knotted solutions to Maxwell's equations~\cite{ranada-1990-a,kedia-2013-a} could also have applications to beam shaping as they provide faster (e.g., localized) spatiotemporal variations of the field compared to conventional ``plane-wave" synthesis solutions. Experimentally producing these knotted solutions is challenging as the solutions can be mathematically formulated as a summation of spherical-harmonic functions. A possible approximate experimental implementation discussed in~\cite{irvine-2008-a} uses tightly focused circularly-polarized laser pulses. 

\subsection{Shaping the beam using collective effects}
Similar to the discussion in the previous section, we expect engineered new materials to play a critical role in fostering more precise control over the beam using the beam's self field.  The main advantages of beam shaping techniques based on wakefield-driven structures reside in their ability to $(i)$ support electromagnetic fields with wavelengths comparable to the bunch length (which is challenging in conventional RF cavities) and $(ii)$ ensure these fields are carrier-envelope-phase (CEP) locked with the bunch, thereby alleviating the need for precise external synchronization and consistently mitigating possible issues associated with shot-to-shot jitter. Over the last two decades, wakefield structures based on meta-material have been developed~\cite{antipov-2007-a} and  tested~\cite{antipov-2008-a,duan-2017-a,lu-2019-a}. Likewise, photonic band-gap (PBG) structures have been introduced in accelerators~\cite{smirnova-2004-a}. An appealing feature of PBG structures is their ability to control the distribution of excited modes, which so far has been employed to suppress noxious modes in a wakefield accelerator, as demonstrated in~\cite{simakov-2016-a}. It is conceivable that PBG structures designed to introduce a well-defined superposition of modes could find applications in bunch shaping (e.g., to synthesize the desired longitudinal or transverse force), expanding on techniques presented in Section~\ref{sec5}.  

\subsection{Redistributing phase space between planes}
Exchanging phase space between two or three degrees of freedom has facilitated the generation of shaped beams for, e.g., advanced acceleration concepts as showcased in Section~\ref{sec6}. We note that so far the implementation of phase-space exchanging beamlines has often been based on a simple configuration and further optimization, e.g., to mitigate sources of 6D-emittance dilution would be critical to the deployment of this class of methods. Likewise, and similar to a laser-based method described in~\cite{xiang-2010-a}, one could consider simpler (and cheaper) versions of phase-space exchangers where the required time-dependent deflecting fields are introduced using transverse wakefields. Likewise, extending the concept of phase-space exchange to control the beam at smaller time scales along the ideas discussed in~\cite{graves-2019-a} could have ground-breaking consequences for the development of room-sized coherent X-ray sources.  
\section*{List of symbols}
\begin{longtable}{ll}
    $s$, $\bar s$, $s_2$, $\tau$ & distance along the reference trajectory\\
        & of the reference particle\\
    $x$, $y$, $z$ & coordinates relative to the reference particle \\
    $x'$, $y'$ & transverse angles\\
    $\delta$ & relative momentum deviation\\
    $\pmb{\mathcal{Z}}$ & 6D canonical phase-space variables\\
    $\zeta_i$  &  ith component of $\pmb{\mathcal {Z}}$ \\
    $p_x$, $p_y$, $p_s$ & momenta\\
    $M$ & map corresponds to position $s$ to $\bar s$\\
    {$R$} & Jacobian  or linear transformation matrix\\
    $e$ & electron charge\\
   {$\mathbf{E}$} & electric field\\
    {$\mathbf{B}$} & magnetic field\\
    $E_0$ & peak electric field strength \\
   {$K_q$} & focusing strength of quadrupole magnet\\
%   {$\rho$} & bending radius \\
%    $C_{s\tau}$, $S_{s\tau}$ & cosine-like and sine-like solution\\
%                             & of betatron motion from $\tau$ to $s$\\
    $\eta$ & dispersion\\
%    $\theta$ & bending angle\\
%    \textcolor{red}{$L$} & $R_{12}$ of dogleg matrix\\
    $\xi$ & $R_{56}$ of dogleg matrix\\
    {$r$}, {$\phi$}, $s$ & cylindrical coordinates \\
    $\omega$ & angular frequency\\
   {$k$} & wave number\\
    $\lambda$ & wavelength\\
    $\Delta_p$ & fractional momentum increase  \\
%   {$\kappa_T$} & kick strength of a transverse deflecting cavity  \\
    $\mathbf A$ & vector potential\\
   {$\kappa_L$} & (spatial) Larmor frequency\\
    ${\mathcal L}$ & canonical angular momentum \\
%    $\Omega$ & phase-space volume\\
    $f({\pmb{\mathcal{Z}};s)}$ & normalized phase-space distribution\\
                       & of particles at $s$\\
   $\ell$ & bunch length of uniform bunch\\
    $\sigma_i$ & root-mean square (rms) value of $i\in \mathcal{Z}$\\
   $\lambda(z)$ & normalized line charge density\\
  $\lambda_\perp(x)$ & normalized transverse charge density\\
 $C$ & compression factor\\
    $\Sigma$ & second-order-moment beam matrix\\
    $\varepsilon_{i}^{proj}$ & RMS emittance along $i\in [x,y,z]$\\
    $\varepsilon_{i}$ & intrinsic emittances where $i\in [1,2,3]$\\
    $\varepsilon_i^n$ & normalized emittance along $i\in [x,y,z]$\\
    $\varepsilon_{Ri}$ & intrinsic emittance of the beam\\
                       & created in the magnetic field where $i \in[1,2]$\\
%    $\mean{x^n} $ & statistical $n$-th order moment of $x$ \\
%    \textcolor{red}{$w(z)$} & longitudinal wakefield\\
%    \textcolor{red}{$\Delta E$} & energy loss\\
    $Q$ & bunch charge\\
    $Z(k)$ & impedance\\
    $Z_0$ & free-space impedance\\
    $b(k;s)$ & bunching factor\\
    $r_e$ & classical electron radius\\
%    \textcolor{red}{$L$} & dechirper length\\
%    $\rho_<$, $\rho_>$ & min($\rho$,$\rho'$), max($\rho$,$\rho'$)\\
    $\lambda_D$ & Debye length\\
    $\omega_p$ & plasma frequency\\
%    \textcolor{red}{$J$} & current density\\
%    $A_{FN}$, $A_{RLD}$, $A_{PC}$ & material constant\\
%    $F$ & electric field strength at local tip \\
   $\phi$ & local work-function\\
%    \textcolor{red}{T} & temperature\\
%    \textcolor{red}{k} & Boltzmann constant\\
 %   \textcolor{red}{$h$} & Planck constant\\
%    \textcolor{red}{$\varepsilon_{6D}^0$, $\varepsilon_i^0$, $\sigma_i^0$} & intrinsic values\\
 %   \textcolor{red}{$\varepsilon_{x,n}$} & normalized emittance\\
%    \textcolor{red}{$E_k$} & mean kinetic energy\\
 %   \textcolor{red}{$\sigma_x^0$} & rms spot radius on the cathode\\
 %   \textcolor{red}{$\sigma_{x,i}$} & rms beam size at the beginning of beamline\\
 %   \textcolor{red}{$\sigma_{x,0}$} & rms beam size at the beginning of beamline\\
%    \textcolor{red}{$\sigma_{x,f}$} & rms beam size at the end of beamline\\
%    \textcolor{red}{$L_{crystal}$} & length of crystal\\
%    $v_{g,e}$, $v_{g,o}$ & group velocities along extraordinary and ordinary axes\\
$h$ & longitudinal-phase-space chirp strength\\
%    \textcolor{red}{$F_{i+1}$} & some force\\
%    \textcolor{red}{\cal M} & map\\
%    $G_i(\zeta_{i+1,m})$ & momentum-dependent position change\\
%    \textcolor{red}{$\Phi_0$} & phase-space density distribution\\
%   \textcolor{red}{$P_i(\zeta_i)$} & profile of position direction\\
%   $H(\zeta)$ & correlation function by external field\\
%   \textcolor{red}{T} & transmission function\\
%   \textcolor{red}{$\xi$} & longitudinal disperson \\
%   \textcolor{red}{'} & d/dt? or other variable\\
%    \textcolor{red}{\st{$K(\kappa_D)$}} & \st{matrix of a TDC with strength $\kappa_d$} \\
%    $m_{ij}$ & matrix element\\
%    S & transfer matrix\\
%    V, $\hat{V}$ & accelerating voltage\\
%    $\varphi$ & RF phase\\
%    $\Theta()$ & Heaviside function\\
%    $R_{z|\delta}$ & longitudinal matrix\\
%    $D$ & distance between element\\
%   \textcolor{red}{$\alpha$,$\beta$,$\gamma$} & Courant-Snyder parameters\\
   $K_{2n}$ & magnetic multipole strength\\
 $K_w$ & wiggler parameter\\
    $\epsilon_0$ & permittivity in vacuum\\
    $N_b$ & total number of particles in bunch\\
%  {$\rho$ & longitudinal charge density\\
  $I$ & current\\
    $I_A$ & Alfven current\\
%    $A$ & modulation amplitude in current profile\\
%    $G$ & modulation gain\\
    $\mathcal{K}$ & loss factor\\
   $w$ & wake function\\
   $W$ & wakefield of bunch\\
    $P_{\theta}$ & canonical angular momentum\\
%    {K} & generalized perveance\\
 %   \textcolor{red}{I} & identity matrix\\
%   \textcolor{red}{J} & 2x2 unit symplectic matrix\\
    
%    $\Leftrightarrow$ & Bar Foo
\end{longtable}

\section{Acknowledgments}
This work was supported by the U.S. Department of Energy (DOE), Office of Science, under award No. DE-AC02-06CH11357 with Argonne National Laboratory. PP also acknowledges support from DOE award No. DE-SC0018656 with Northern Illinois University and from the Center for Bright Beams via NSF award PHY-1549132. 
%
%%%%%%%%%%%%%%%%%%%%%%%%%%%%%%%%%%%%%%%%%%%%%%%%%%%%%%%%%%%%%%%%%%%%%%%%%%
\onecolumngrid
%\pagebreak
\bibliographystyle{plain}
\bibliography{RPMbiblio}
%\appendix
%\input{response_to_comment_1}
\end{document}